\magnification=1150
%
%
%
%
%

\catcode`\Ä=13 \def Ä{\"A}
\catcode`\Å=13 \def Å{\AA }
\catcode`\Ç=13 \def Ç{\c C}
\catcode`\É=13 \def É{\'E}
\catcode`\Ñ=13 \def Ñ{\~N}
\catcode`\Ö=13 \def Ö{\"O}
\catcode`\Ü=13 \def Ü{\"U}
\catcode`\á=13 \def á{\'a}
\catcode`\à=13 \def à{\`a}
\catcode`\â=13 \def â{\^a}
\catcode`\ä=13 \def ä{\"a}
\catcode`\ã=13 \def ã{\~a}
\catcode`\å=13 \def å{\aa }
\catcode`\ç=13 \def ç{\c c}
\catcode`\é=13 \def é{\'e}
\catcode`\è=13 \def è{\`e}
\catcode`\ê=13 \def ê{\^e}
\catcode`\ë=13 \def ë{\"e}
\catcode`\í=13 \def í{\'\i }
\catcode`\ì=13 \def ì{\`\i }
\catcode`\î=13 \def î{\^\i }
\catcode`\ï=13 \def ï{\"\i }
\catcode`\ñ=13 \def ñ{\~n}
\catcode`\ó=13 \def ó{\'o}
\catcode`\ò=13 \def ò{\`o}
\catcode`\ô=13 \def ô{\^o}
\catcode`\ö=13 \def ö{\"o}
\catcode`\õ=13 \def õ{\~o}
\catcode`\ú=13 \def ú{\'u}
\catcode`\ù=13 \def ù{\`u}
\catcode`\û=13 \def û{\^u}
\catcode`\ü=13 \def ü{\"u}
\catcode`\§=13 \def §{\S }
\catcode`\€=13 \def €{\hbox{$\bullet$}}
\catcode`\¶=13 \def ¶{\P }
\catcode`\ß=13 \def ß{\ss }
\catcode`\Æ=13 \def Æ{\AE }
\catcode`\Ø=13 \def Ø{\O }
\catcode`\æ=13 \def æ{\ae }
\catcode`\ø=13 \def ø{\o }
\catcode`\¿=13 \def ¿{{?`}}
\catcode`\«=13 \def «{\ll }
\catcode`\»=13 \def »{\gg }
\catcode`\Š=13 \def Š{\dots }
\catcode`\À=13 \def À{\`A}
\catcode`\Ã=13 \def Ã{\~A}
\catcode`\Õ=13 \def Õ{\~O}
\catcode`\'=13 \def '{\OE }
\catcode`\¦=13 \def ¦{\oe }
\catcode`\³=13 \def ³{{``}}
\catcode`\²=13 \def ²{{''}}


\catcode`\@=11
\def\landscape{\relax} 

\font\tenmib=cmmib10
\font\sevenmib=cmmib7
\font\fivemib=cmmib5
\newfam\mibfam 
\def\mib{\fam\mibfam\tenmib}
\textfont\mibfam=\tenmib
\scriptfont\mibfam=\sevenmib
\scriptscriptfont\mibfam=\fivemib
\mathchardef\alphaB="080B
\mathchardef\betaB="080C
\mathchardef\gammaB="080D
\mathchardef\deltaB="080E
\mathchardef\epsilonB="080F
\mathchardef\zetaB="0810
\mathchardef\etaB="0811
\mathchardef\thetaB="0812
\mathchardef\iotaB="0813
\mathchardef\kappaB="0814
\mathchardef\lambdaB="0815
\mathchardef\muB="0816
\mathchardef\nuB="0817
\mathchardef\xiB="0818
\mathchardef\piB="0819
\mathchardef\rhoB="081A
\mathchardef\sigmaB="081B
\mathchardef\tauB="081C
\mathchardef\upsilonB="081D
\mathchardef\phiB="081E
\mathchardef\chiB="081F
\mathchardef\psiB="0820
\mathchardef\omegaB="0821
\mathchardef\varepsilonB="0822
\mathchardef\varthetaB="0823
\mathchardef\varpiB="0824
\mathchardef\varrhoB="0825
\mathchardef\varsigmaB="0826
\mathchardef\varphiB="0827
%
\font\tenss=cmss10
\font\sevenss=cmss10 at 7pt
\font\fivess=cmss10 at 5pt
\newfam\ssfam 
\textfont\ssfam=\tenss
\scriptfont\ssfam=\sevenss
\scriptscriptfont\ssfam=\fivess
\def\SS{\fam\ssfam\tenss}

\font\teneuf=eufm10
\font\seveneuf=eufm7
\font\fiveeuf=eufm5
\newfam\euffam 
\textfont\euffam=\teneuf
\scriptfont\euffam=\seveneuf
\scriptscriptfont\euffam=\fiveeuf

\font\tenmsy=msbm10
\font\sevenmsy=msbm7
\font\fivemsy=msbm5
\newfam\msyfam 
\textfont\msyfam=\tenmsy
\scriptfont\msyfam=\sevenmsy
\scriptscriptfont\msyfam=\fivemsy
\def\blackB{\fam\msyfam\tenmsy}

\font\tencalb=cmbsy10
\font\sevencalb=cmbsy7
\font\fivecalb=cmbsy5
\newfam\calbfam 
\def\calb{\fam\calbfam\tencalb}
\textfont\calbfam=\tencalb
\scriptfont\calbfam=\sevencalb
\scriptscriptfont\calbfam=\fivecalb
%
\font\tensc=cmcsc10
\font\sevensc=cmcsc10 at 7pt
\font\fivesc=cmcsc10 at 5pt
\newfam\scfam 

\textfont\scfam=\tensc
\scriptfont\scfam=\sevensc
\scriptscriptfont\scfam=\fivesc
%
\font\tencali=eusm10
\font\sevencali=eusm7
\font\fivecali=eusm5
\newfam\califam 
\def\cali{\fam\califam\tencali}
\textfont\califam=\tencali
\scriptfont\califam=\sevencali
\scriptscriptfont\califam=\fivecali
%

\font\twelverm=cmr12
\font\twelvei=cmmi12
\font\twelveex=cmex10 at 12pt
\font\twelvebf=cmbx12
\font\twelveit=cmti12
\font\twelvesy=cmsy10 at 12pt
\font\twelvesl=cmsl12
\font\twelvett=cmtt10 at 12pt
\font\twelvemib=cmmib10 at 12pt
\font\twelvess=cmss10 at 12pt
\font\twelveeuf=eufm10 at 12pt
\font\twelvemsy=msbm10 at 12pt
\font\twelvecali=eusm10 at 12pt

\font\ninerm=cmr9
\font\ninei=cmmi9

\font\ninebf=cmbx9
\font\ninesy=cmsy10 at 9pt

\font\ninemib=cmmib10 at 9pt
\font\niness=cmss10 at 9pt
\font\nineeuf=eufm10 at 9pt
\font\ninemsy=msbm10 at 9pt
\font\ninecali=eusm10 at 9pt


\def\twelvepoint{\def\rm{\fam0\twelverm}%
\textfont0=\twelverm\scriptfont0=\ninerm\scriptscriptfont0=\sevenrm%
\textfont1=\twelvei\scriptfont1=\ninei\scriptscriptfont1=\seveni%
\textfont2=\twelvesy\scriptfont2=\ninesy\scriptscriptfont2=\sevensy%
\textfont3=\twelveex\scriptfont3=\tenex\scriptscriptfont3=\tenex%
\textfont\itfam=\twelveit\def\it{\fam\itfam\twelveit}%
\textfont\slfam=\twelvesl\def\sl{\fam\slfam\twelvesl}%
\textfont\ttfam=\twelvett\def\tt{\fam\ttfam\twelvett}%
\textfont\bffam=\twelvebf\def\bf{\fam\bffam\twelvebf}%
\scriptfont\bffam=\ninebf\scriptscriptfont\bffam=\sevenbf%
\textfont\mibfam=\twelvemib\def\mib{\fam\mibfam\twelvemib}%
\scriptfont\mibfam=\ninemib\scriptscriptfont\mibfam=\sevenmib%
\textfont\ssfam=\twelvess\def\SS{\fam\ssfam\twelvess}%
\scriptfont\ssfam=\niness\scriptscriptfont\ssfam=\sevenss%
\textfont\euffam=\twelveeuf\def\euf{\fam\euffam\twelveeuf}%
\scriptfont\euffam=\nineeuf\scriptscriptfont\euffam=\seveneuf%
\textfont\msyfam=\twelvemsy\def\msy{\fam\msyfam\twelvemsy}%
\scriptfont\msyfam=\ninemsy\scriptscriptfont\msyfam=\sevenmsy%
\textfont\califam=\twelvecali\def\cali{\fam\califam\twelvecali}%
\scriptfont\califam=\ninecali\scriptscriptfont\califam=\sevencali%
\rm%
}

\def\tenpoint{\def\rm{\fam0\tenrm}%
\textfont0=\tenrm\scriptfont0=\sevenrm\scriptscriptfont0=\fiverm%
\textfont1=\teni\scriptfont1=\seveni\scriptscriptfont1=\fivei%
\textfont2=\tensy\scriptfont2=\sevensy\scriptscriptfont2=\fivesy%
\textfont3=\tenex\scriptfont3=\tenex\scriptscriptfont3=\tenex%
\textfont\itfam=\tenit\def\it{\fam\itfam\tenit}%
\textfont\slfam=\tensl\def\sl{\fam\slfam\tensl}%
\textfont\ttfam=\tentt\def\tt{\fam\ttfam\tentt}%
\textfont\bffam=\tenbf\def\bf{\fam\bffam\tenbf}%
\scriptfont\bffam=\sevenbf\scriptscriptfont\bffam=\fivebf%
\textfont\mibfam=\tenmib\def\mib{\fam\mibfam\tenmib}%
\scriptfont\mibfam=\sevenmib\scriptscriptfont\mibfam=\fivemib%
\textfont\ssfam=\tenss\def\SS{\fam\ssfam\tenss}%
\scriptfont\ssfam=\sevenss\scriptscriptfont\ssfam=\fivess%
\textfont\euffam=\teneuf\def\euf{\fam\euffam\teneuf}%
\scriptfont\euffam=\seveneuf\scriptscriptfont\euffam=\fiveeuf%
\textfont\msyfam=\tenmsy\def\msy{\fam\msyfam\tenmsy}%
\scriptfont\msyfam=\sevenmsy\scriptscriptfont\msyfam=\fivemsy%
\textfont\califam=\tencali\def\cali{\fam\califam\tencali}%
\scriptfont\califam=\sevencali\scriptscriptfont\califam=\fivecali%
\rm%
}



\input epsf      

\tenpoint
\newbox\leftpage \newdimen\fullhsize
\tolerance=1000\hfuzz=2pt
\def\almostshipout#1{\if L\lr \count1=1 \message{[\the\count0.\the\count1]}
	\global\setbox\leftpage=#1 \global\let\lr=R
	\else \count1=2
	\shipout\vbox{\landscape
	\hbox to\fullhsize{\box\leftpage\hfil#1}}  \global\let\lr=L\fi
}
\def\LANDSCAPE{
\voffset=-.27truein\vsize=7truein
\hoffset=-.0truein 
\hsize=4.5truein
\fullhsize=9.5truein
\parskip=10pt
\font\chapfont=cmbx12
\font\secfont=cmbx12
\font\subsecfont=cmb10
\font\subsubsecfont=cmr10
\font\headingfont=cmr12
\font\titlefont=cmbx12
\font\headfont=cmsl10 at 9pt
\normalwidth=4.5in
\double=13pt
\single=11pt 
\captionindent=5mm
\let\lr=L
\output={\ifnum\pageno=0
	\shipout\vbox{\landscape\makeheadline
	\hbox to \fullhsize{\hfill\pagebody\hfill}}\advancepageno
	\else
	\almostshipout{\leftline{\vbox{\makeheadline\pagebody\makefootline}}}
	\advancepageno\fi}
}

\newdimen\normalwidth
\newdimen\double
\newdimen\single
\newdimen\indentation \indentation=1cm
\newskip\stdskip \stdskip=.26truein
\newdimen\captionindent \captionindent=1truein

\newif\ifdraft \draftfalse

\font\chapfont=cmbx12 at 14pt
\font\secfont=cmbx12
\font\subsecfont=cmb10
\font\subsubsecfont=cmr10
\font\headingfont=cmr12
\font\headfont=cmsl10 at 9pt
\font\titlefont=cmbx12 at 14pt
\normalwidth=6truein
\double=.26truein
\single=.19truein 
\hsize=\normalwidth
\hfuzz=2pt
\vsize=8.4truein
\hoffset=0.48truein
\voffset=0.1truein
\baselineskip=\double plus 1pt minus 1pt
\xspaceskip=.33em plus 1em minus .1em
\spaceskip=.5em plus 2em minus .1em
\parskip=\stdskip
\parindent=\indentation
\clubpenalty=10000
\widowpenalty=10000
\displaywidowpenalty=500
\overfullrule=0pt
\tolerance=1000

\def\raggedcenter{
	\leftskip=1cm plus 4em \rightskip=\leftskip
	\parfillskip=0pt \spaceskip=.5em \xspaceskip=.5em \pretolerance=9999
	\tolerance=9999 \hyphenpenalty=9999 \exhyphenpenalty=9999\parindent=0pt
}

\def\aujourdhui{\number\day.\number\month.\number\year}
\def\hanging#1{\setbox99\hbox{#1}\par\hangindent\wd99\noindent\box99}
\def\newpage{\vfill\supereject}
\def\newline{\hfill\break}

\def\raggedright{
	\rightskip=0cm plus 3cm
	\spaceskip=.333em \xspaceskip=.5em \pretolerance=9999
	\tolerance=9999 \hyphenpenalty=9999 \exhyphenpenalty=9999\parindent=0pt
}


\headline={\ifdraft\hfill\ninerm\aujourdhui\else\hfill\fi}
\def\pagination{\pageno=1\footline={{\rm\hfil\folio\hfil}}}

\def\normalhead{
\headline={%
\headfont\ifblankhead\hss\global\blankheadfalse\else%
\hfill\botmark\fi}
}
\def\Roman#1{%
\ifcase#1\or A\or B\or C\or D\or E\or F\or G\or H\or I\or J\or K\or L\or M%
\else *\fi
}
\def\roman#1{%
\ifcase#1\or a\or b\or c\or d\or e\or f\or g\or h\or i\or j\or k\or l\or m%
\else *\fi
}

\newtoks\tdmtoks
\def\margefolio#1{\rlap{\hbox to 7mm{#1}\ignorespaces}}
\def\dfil{\leaders\hbox to 1em{\hss.\hss}\hfill}
\def\pageNO#1{\dfil\margefolio{~\hfil #1}\par}

\def\itemitemitem{\par\indent\indent\hangindent3\parindent\textindent}

\newif\iftdm \tdmfalse
\newwrite\tdmP
\newwrite\tdmA
\newwrite\tdmL
\newwrite\tdmT
\newwrite\tdmF

\newif\iftdmA \tdmAfalse
\newif\iftdmT \tdmTfalse
\newif\iftdmF \tdmFfalse

\def\TdM{\tdmtrue
\immediate\openout\tdmP=tdm.aux
\immediate\openout\tdmL=tdmL.aux\write\tdmL{\chapno=0\parskip=6pt}
\let\tdm=\tdmP
}

\def\TdMitem#1{\iftdm%
\write\tdm{\par\string\noindent{#1}\string\pageNO{\folio}}%
\else\relax\fi}%

\def\tdmChap#1#2{%
\global\advance\chapno by1\secno=0\subsecno=0%
\medskip\par\noindent%
\iftdmA \gdef\mainID{\Roman{\chapno}}%
\hanging{\bf Annexe \mainID :\ }%
\else%
\gdef\mainID{\the\chapno}%
\hanging{\bf Chapitre \mainID :\ }\fi%
\uppercase{#1}\pageNO{#2}\medskip}

\def\tdmSec#1#2{\global\advance\secno by1\subsecno=0%
\gdef\currID{\mainID.\the\secno}%
\par\item{\currID} {#1}\pageNO{#2}}

\def\tdmSubsec#1#2{\global\advance\subsecno by1\subsubsecno=0%
\itemitem{\currID.\the\subsecno} {#1}\pageNO{#2}}

\def\tdmSubsubsec#1#2{\global\advance\subsubsecno by1%
\itemitemitem{\currID.\the\subsecno.\the\subsubsecno} {#1}\pageNO{#2}}

\def\tdmFig#1#2{\global\advance\figno by1%
\item{\the\figno.} {#1}\pageNO{#2}}

\def\tdmTable#1#2{\global\advance\tableno by1%
\item{\the\tableno.} {#1}\pageNO{#2}}




\everydisplay={\global\envir=8}
\newcount\eqnum	\eqnum=0
\newwrite\eqfile
\newif\ifEqStored\EqStoredfalse
\def\storeEqs{
	\EqStoredtrue
	\immediate\openout\eqfile=eqs.aux
}
\def\eq{\global\advance\eqnum by1\eqno(\mainID.\the\eqnum)}

\newcount\eqsnum	\eqsnum=0
\def\mathLetters{\global\envir=11\global\advance\eqnum by1\global\eqsnum=0}

\def\eqN{\global\advance\eqnum by1(\mainID.\the\eqnum)}
\def\eqL{%
\global\advance\eqsnum by1%
(\mainID.\roman{\eqnum})}



\def\boxit#1{\leavevmode\kern5pt\hbox{
	\vrule width.5pt\vtop{\vbox{\hrule height.5pt\kern5pt
        \hbox{\kern5pt{#1}\kern5pt}}
      \kern5pt\hrule height.5pt}\vrule width.5pt}\kern5pt}
\def\boxEq#1{\boxit{$\displaystyle #1$}}


\def\makeatletter{\catcode`\@=11\relax}
\def\makeatother{\catcode`\@=12\relax}

\makeatletter
\def\mainID{}

\def\namedef#1{\expandafter\xdef\csname #1\endcsname}
\def\nameuse#1{\csname #1\endcsname}
\newcount\envir \global\envir=0 
\def\thelabel{%
	\ifcase\envir\relax%
	\or\mainID
	\or\mainID.\the\secno
	\or\mainID.\the\secno.\the\subsecno
	\or\mainID.\the\secno.\the\subsecno.\the\subsubsecno
	\or\the\tableno
	\or\the\figno
	\or\mainID
	\or\mainID.\the\eqnum
	\or\the\bibno
	\or\mainID.\the\eqnum
	\or\mainID.\the\eqnum
	\else{**}%
	\fi%
}
\def\nom#1{
	\ifdraft\ifcase\envir\relax%
	\or\leftline{\tt [\string#1]}
	\or\leftline{\tt [\string#1]}
	\or\leftline{\tt [\string#1]}
	\or\leftline{\tt [\string#1]}
	\or\leftline{\tt [\string#1]}
	\or\leftline{\tt [\string#1]}
	\or\leftline{\tt [\string#1]}
	\or\global\advance\eqnum by1%
	\eqno{\matrix{(\mainID.\the\eqnum)\cr{\scriptstyle[\string#1]}\cr}}
	\or\relax
	\or\global\advance\eqnum by1%
	{\scriptstyle[\string#1]}(\mainID.\the\eqnum)
	\or\leftline{\tt [\string#1]}
	\else{**}\fi
	\else\ifcase\envir\relax%
	\or\relax
	\or\relax
	\or\relax
	\or\relax
	\or\relax
	\or\relax
	\or\relax
	\or\eq
	\or\relax
	\or\eqN
	\or\relax
	\else\relax\fi
	\fi%
	\namedef{#1}{\thelabel}
 \ifEqStored\immediate\write\eqfile{
	\string\namedef{#1}{\thelabel}}\fi
}
\def\ifundefined#1{\expandafter\ifx\csname#1\endcsname\relax}
\def\ref#1{\ifundefined{#1}{[#1]}%
\immediate\write16{WARNING!!! "#1" is undefined}%
\else\nameuse{#1}\fi}
\makeatother


\newcount\bibno	\bibno=0
\newwrite\bibfile
\newtoks\bibtoks
\def\initBib{
	\immediate\openout\bibfile=bib.aux\write\bibfile{\parskip=6pt}
}
\def\inputBib{
 \immediate\closeout\bibfile\newpage
 \mark{Bibliographie}
	\centerline{\bf BIBLIOGRAPHIE}\bigskip
	\let\tdm=\tdmP
	\TdMitem{BIBLIOGRAPHIE}
	\input bib.aux
}

\def\bib#1#2{%
{\global\envir=9 \global\advance\bibno by1\nom{#1}}\bibtoks={#2}%
\immediate\write\bibfile{\string\item{[\the\bibno]} \the\bibtoks}}%



\def\chapitreSN#1{
\vfill\supereject%
\global\envir=1%
\eqnum=0
{\par\raggedcenter\baselineskip=18pt\chapfont\uppercase{#1}\par}
\def\mainID{}
\TdMitem{\uppercase{#1}}
\blankheadtrue%
\normalhead%
\mark{#1}
}


\newcount\chapno	\chapno=0
\newif\ifblankhead \blankheadtrue

\def\chapitre#1{
\vfill\supereject%
\global\envir=1%
\global\advance\chapno by1 \secno=0 \subsecno=0 \eqnum=0
\blankheadtrue%
\normalhead%
\mark{#1}
\centerline{\chapfont
CHAPITRE~~\uppercase\expandafter{\romannumeral\the\chapno}}
{\par\raggedcenter\baselineskip=18pt\chapfont\uppercase{#1}\par}
\def\mainID{\the\chapno}
\iftdm\write\tdm{\string\tdmChap{#1}{\folio}}\else\relax\fi
}


\newif\ifhtext
\everypar{\htexttrue}
\newcount\secno	\secno=0
\newskip\secskip \secskip=4mm plus 4mm
\newskip\aftersecskip\aftersecskip=-2mm plus 1mm

\def\section#1{
\ignorespaces
\global\envir=2
\global\advance\secno by1 \subsecno=0 
\vskip\secskip
\penalty-1000
{\twelvepoint\secfont\par\raggedright\baselineskip=15pt
\hanging{\mainID.\the\secno\ } #1}
\ignorespaces
\iftdm\write\tdm{\string\tdmSec{#1}{\folio}}\else\relax\fi
\penalty10000\vskip\aftersecskip\penalty10000
\global\htextfalse\par
}


\newcount\subsecno	\subsecno=0
\newskip\subsecskip \subsecskip=2mm plus 1mm minus 1mm
\newskip\aftersubsecskip \aftersubsecskip=-2mm plus 1mm
\def\soussection#1{
\ignorespaces
\global\envir=3
\global\advance\subsecno by1 \subsubsecno=0 
\ifhtext\vskip\subsecskip\penalty-1000\fi
\iftdm\write\tdm{\string\tdmSubsec{#1}{\folio}}\else\relax\fi
{\subsecfont\par\raggedright\baselineskip=14pt
\hanging{\mainID.\the\secno.\the\subsecno\ }#1}
\ignorespaces
\hyphenpenalty=50
\vskip\aftersubsecskip\par
}


\newcount\subsubsecno	\subsubsecno=0

\def\soussoussection#1{
\global\envir=4
\global\advance\subsubsecno by1 
\goodbreak
\noindent
{\subsubsecfont\par\raggedright\baselineskip=14pt
\hanging{\mainID.\the\secno.\the\subsecno\the\subsubsecno\ }#1\newline}
\iftdm\write\tdm{\string\tdmSubsubsec{#1}{\folio}}\else\relax\fi
}



\newif\ifapp \appfalse
\newcount\appno	\appno=0

\def\annexe#1{
\global\apptrue
\iftdmA\relax\else\tdmAtrue\iftdm
\immediate\openout\tdmA=annexes.aux\let\tdm=\tdmA\fi\fi
\newpage
\global\envir=7
\blankheadtrue
\global\advance\appno by1 \secno=0\subsecno=0\eqnum=0
\def\mainID{\Roman{\appno}}
\mark{Annexe~\mainID}
\centerline{\chapfont ANNEXE \mainID}
{\par\raggedcenter\baselineskip=18pt\chapfont #1\par}
\iftdm\write\tdm{\string\tdmChap{#1}{\folio}}\else\relax\fi
}


\newcount\tableno \tableno=0

\def\begintable#1#2{
\iffloat\midinsert\fi
\global\envir=5
\global\advance\tableno by 1
\iftdmT\relax\else\global\tdmTtrue
\iftdm\immediate\openout\tdmT=tables.aux
\write\tdmT{\tableno=0}\fi\fi
\iftdm
\write\tdmT{\string\tdmTable{#2}{\folio}}%
\fi
\par\begingroup\parindent=0pt\leftskip=\captionindent\rightskip=\captionindent
\baselineskip=\single
{\bf Tableau \the\tableno:} #1
\par\endgroup
\offinterlineskip
\openup1\jot
}

\def\endtable{\iffloat\endinsert\fi\par\bigskip\global\envir=5}

\def\tablerule{\noalign{\vskip3pt\hrule height .5pt \vskip3pt}}
\def\tableruleD{\noalign{
	\vskip3pt\hrule height .5pt\vskip1pt\hrule height .5pt\vskip3pt}}


\def\figbox#1#2{\global\epsfxsize=#2\vcenter{\hbox to #2{\epsfbox{#1}}}}

\newif\iffloat \floattrue
\newcount\figno \figno=0

\def\figPS#1#2#3#4{
\iffloat\midinsert\fi
\par\begingroup\parindent=0pt\leftskip=\captionindent\rightskip=\captionindent
\global\envir=6
\baselineskip=\single
\global\advance\figno by 1
\vglue 1 cm
{\bf Figure \the\figno:} #1\par
\iftdmF\relax\else\global\tdmFtrue\iftdm
\immediate\openout\tdmF=figs.aux\write\tdmF{\figno=0}\fi\fi
\iftdm%
\write\tdmF{\string\tdmFig{#4}{\folio}}%
\fi
\endgroup\par
\iffloat\endinsert\fi
}

\def\pagefigPS#1#2#3#4{
\pageinsert\vfill
\par\begingroup\parindent=0pt\leftskip=\captionindent\rightskip=\captionindent
\global\envir=6
\baselineskip=\single
\global\advance\figno by 1
\epsfxsize=#3
\centerline{\epsfbox{#2}}
{\bf Figure \the\figno:} #1\par
\iftdmF\relax\else\global\tdmFtrue\iftdm
\immediate\openout\tdmF=figs.aux\write\tdmF{\figno=0}\fi\fi
\iftdm%
\write\tdmF{\string\tdmFig{#4}{\folio}}%
\fi
\endgroup\par\vfill
\endinsert
}


\def\figBLANC#1#2#3{
\iffloat\midinsert\fi
\par\begingroup\parindent=0pt\leftskip=\captionindent\rightskip=\captionindent
\baselineskip=\single
\global\envir=6
\global\advance\figno by 1
\vglue#2
{\bf Figure \the\figno:} #1\par
\iftdmF\relax\else\global\tdmFtrue\iftdm
\immediate\openout\tdmF=figs.aux\write\tdmF{\figno=0}\fi\fi
\iftdm=%
\write\tdmF{\string\tdmFig{#3}{\folio}}%
\fi
\endgroup\par
\iffloat\endinsert\fi
}

\catcode`\@=12


\newif\ifPhD \PhDfalse

\PhDtrue 

\def\title{Approximation auto-coh\'erente \`a deux particules, pseudogap et
supraconductivit\'e dans le mod\`ele de Hubbard attractif}
\def\auteur{Steve Allen} 
\def\departement{Physique} 
\def\mois{Ao\^ut 2000} 

%
\footline={}
\vglue 1cm
{\par\raggedcenter\titlefont\title\par}
\vglue 3cm
\centerline{par}
\vglue 2cm
\centerline{\auteur}
\vglue 2cm
\centerline{\ifPhD Th\`ese pr\'esent\'ee\else M\'emoire pr\'esent\'e\fi\
au d\'epartement de \departement\ en vue de}
\centerline{l'obtention du grade de \ifPhD Docteur \`es sciences (Ph.D.)
\else Ma\^\i tre \`es Sciences (M.Sc.)\fi}

\vfill
\vglue 2cm
\centerline{FACULT\'E DES SCIENCES}
\centerline{UNIVERSIT\'E DE SHERBROOKE}
\vglue 2cm
\centerline{Sherbrooke, Qu\'ebec, \mois}
\newpage

%
%
%

\fontdimen16\tensy=2.5pt
\fontdimen17\tensy=2.5pt









\def\thetah{{\bf\hat{\kern -0.1em\thetaB}}}





%

\def\IT{\item{€}}

\def\frac#1#2{{#1\over #2}}
\def\ffrac#1#2{{\textstyle{#1\over #2}}}

\def\text#1{\quad\hbox{#1}\quad}

\newcount\itemno \itemno=0
\newcount\itemitemno \itemitemno=0
\def\IT{
\advance\itemno by 1
\itemitemno=0
\item{\the\itemno.}
}
\def\ITIT{
\advance\itemitemno by 1
\itemitem{\ifcase\itemitemno\or a\or b\or c\or d\or e\or f\or g\or h\else *\fi)}
}

\def\index#1{\relax}


\pagination         
\TdM               	

\pageno=1

 \namedef{sechubbard}{1.3}
 \namedef{hubbard}{1.1}
 \namedef{symhubbard}{1.2}
 \namedef{transhub}{1.3}
 \namedef{eqop}{2.1}
 \namedef{evolution}{2.2}
 \namedef{eqgreen}{2.3}
 \namedef{moyennechamp}{2.4}
 \namedef{eqpartition}{2.5}
 \namedef{eqchamp}{2.6}
 \namedef{eqderiv1}{2.7}
 \namedef{eqmatdrv}{2.8}
 \namedef{eqmouvement}{2.9}
 \namedef{eqmouhub}{2.10}
 \namedef{self}{2.11}
 \namedef{eqdyson}{2.12}
 \namedef{eqdyson2}{2.13}
 \namedef{dembs}{2.2}
 \namedef{correlation}{2.14}
 \namedef{selfchi}{2.15}
 \namedef{inverseg}{2.16}
 \namedef{glinear}{2.17}
 \namedef{gilinear}{2.18}
 \namedef{bsprel}{2.19}
 \namedef{legendre}{2.20}
 \namedef{legendrecond}{2.21}
 \namedef{fonctiong}{2.22}
 \namedef{vertex}{2.23}
 \namedef{bs}{2.24}
 \namedef{svertex}{2.3}
 \namedef{approxim}{2.25}
 \namedef{localplus}{2.26}
 \namedef{localmoins}{2.27}
 \namedef{facteurlocal}{2.28}
 \namedef{selfun}{2.29}
 \namedef{vertexun}{2.30}
 \namedef{corrpaires}{2.3.2}
 \namedef{greenun}{2.31}
 \namedef{mu1}{2.32}
 \namedef{chiun}{2.33}
 \namedef{chi0un}{2.34}
 \namedef{doubleoccup}{2.3.3}
 \namedef{corrlocal}{2.35}
 \namedef{selfbs}{2.36}
 \namedef{selfdeux}{2.37}
 \namedef{gdeux}{2.38}
 \namedef{reglesdesomme}{2.4}
 \namedef{corrlocalm}{2.39}
 \namedef{susceppaires}{2.40}
 \namedef{repsspectral}{2.41}
 \namedef{devhfm}{2.42}
 \namedef{moment0}{2.43}
 \namedef{sommef}{2.44}
 \namedef{sumchi0}{2.45}
 \namedef{devhf}{2.46}
 \namedef{conditionfsumrule}{2.47}
 \namedef{npoles}{2.4.2}
 \namedef{cmcq}{3}
 \namedef{devtrotter}{3.1}
 \namedef{rpavstpsc}{1}
 \namedef{docc}{2}
 \namedef{pqptc}{3.2}
 \namedef{spectralg}{3.3}
 \namedef{rcr}{3.4}
 \namedef{conditionrcr}{3.4}
 \namedef{structure}{3.5}
 \namedef{pschi}{3.6}
 \namedef{fcaract}{3}
 \namedef{selfret}{3.7}
 \namedef{selfim}{3.8}
 \namedef{fliquide}{3.9}
 \namedef{nfliquide}{3.10}
 \namedef{rpasp}{3.11}
 \namedef{corrp}{3.12}
 \namedef{longueportee}{3.13}
 \namedef{lcorrinf}{3.14}
 \namedef{correntemp}{3.15}
 \namedef{corrscaling}{3.16}
 \namedef{effettaille}{3.17}
 \namedef{effetspectre}{3.6.2}
 \namedef{selfint}{3.18}
 \namedef{psint}{3.19}
 \namedef{conditionpg}{3.20}
 \namedef{conditioncorr}{3.21}
 \namedef{longueur}{3.6.3}
 \namedef{lotdb}{3.22}
 \namedef{mumcq}{4}
 \namedef{tc0}{5}
 \namedef{acalcul}{A}
 \namedef{derivch}{A.1}
 \namedef{inverse}{A.2}
 \namedef{goung}{A.3}
 \namedef{gmoung}{A.4}
 \namedef{phioung}{A.5}
 \namedef{gounp}{A.6}
 \namedef{gmounp}{A.7}
 \namedef{gdeuxg}{A.8}
 \namedef{gmdeuxg}{A.9}
 \namedef{gaoung}{A.10}
 \namedef{coungdev}{A.11}
 \namedef{s3fctg3}{A.12}
 \namedef{anex2}{B}
 \namedef{autocoherence}{C}
 \namedef{soung}{C.1}
 \namedef{coung}{C.2}
 \namedef{soungfin}{C.3}
 \namedef{bsgamma}{C.4}
 \namedef{s3approx}{C.5}
 \namedef{g3approx}{C.6}
 \namedef{vren}{C.7}
 \namedef{approxcoh}{C.8}
 \namedef{mem}{D}
 \namedef{eqprolong}{D.1}
 \namedef{proldis}{D.2}
 \namedef{noyau}{D.3}
 \namedef{probA}{D.4}
 \namedef{gorkov}{1}
 \namedef{allen1}{2}
 \namedef{allen2}{3}
 \namedef{anderson}{4}
 \namedef{gw}{5}
 \namedef{assaraf}{6}
 \namedef{attouch}{7}
 \namedef{bcs}{8}
 \namedef{baym3}{9}
 \namedef{baym1}{10}
 \namedef{baym2}{11}
 \namedef{bednorz}{12}
 \namedef{berezinskii}{13}
 \namedef{bss}{14}
 \namedef{bogoliubov}{15}
 \namedef{rg}{16}
 \namedef{bryan}{17}
 \namedef{bulut}{18}
 \namedef{mass}{19}
 \namedef{mass2}{20}
 \namedef{chakra}{21}
 \namedef{chen}{22}
 \namedef{chitra}{23}
 \namedef{coleman}{24}
 \namedef{corson}{25}
 \namedef{xxzmc}{26}
 \namedef{curro}{27}
 \namedef{neutron}{28}
 \namedef{dare}{29}
 \namedef{dare2}{30}
 \namedef{dare3}{31}
 \namedef{flexpg}{32}
 \namedef{tros}{33}
 \namedef{arpes}{34}
 \namedef{parquet}{35}
 \namedef{boson}{36}
 \namedef{emery}{37}
 \namedef{encyclo}{38}
 \namedef{conforme}{39}
 \namedef{fronsdal}{40}
 \namedef{fye}{41}
 \namedef{mcmc}{42}
 \namedef{geman}{43}
 \namedef{dmft}{44}
 \namedef{flex}{45}
 \namedef{hammersley}{46}
 \namedef{haussmann}{47}
 \namedef{hirsch}{48}
 \namedef{hohenberg}{49}
 \namedef{dca}{50}
 \namedef{ihara}{51}
 \namedef{iwamatsu}{52}
 \namedef{janis}{53}
 \namedef{jarrell}{54}
 \namedef{kadanoff}{55}
 \namedef{kagan}{56}
 \namedef{moment}{57}
 \namedef{kalos}{58}
 \namedef{lnp475}{59}
 \namedef{kohn}{60}
 \namedef{kt}{61}
 \namedef{bumsoo}{62}
 \namedef{frank}{63}
 \namedef{letz}{64}
 \namedef{levin}{65}
 \namedef{linden}{66}
 \namedef{loh}{67}
 \namedef{loktev}{68}
 \namedef{cv}{69}
 \namedef{luttinger1}{70}
 \namedef{luttinger2}{71}
 \namedef{luttinger3}{72}
 \namedef{mahan}{73}
 \namedef{mancini}{74}
 \namedef{jumpmu}{75}
 \namedef{martin}{76}
 \namedef{mermin}{77}
 \namedef{meshkov}{78}
 \namedef{bcsbec}{79}
 \namedef{samuel}{80}
 \namedef{samuel2}{81}
 \namedef{negele}{82}
 \namedef{nozieres}{83}
 \namedef{strongcoupling}{84}
 \namedef{nozieres2}{85}
 \namedef{stephane}{86}
 \namedef{pedersen}{87}
 \namedef{pines}{88}
 \namedef{pines2}{89}
 \namedef{hanke}{90}
 \namedef{randeria}{91}
 \namedef{tunnel}{92}
 \namedef{rickayzen}{93}
 \namedef{rick}{94}
 \namedef{robert}{95}
 \namedef{robert2}{96}
 \namedef{raman}{97}
 \namedef{superfluid}{98}
 \namedef{illposed}{99}
 \namedef{momenthub}{100}
 \namedef{schrieffer}{101}
 \namedef{setlur}{102}
 \namedef{singer}{103}
 \namedef{singer2}{104}
 \namedef{singwi}{105}
 \namedef{rmn}{106}
 \namedef{timusk}{107}
 \namedef{david}{108}
 \namedef{mcq}{109}
 \namedef{ncorps}{110}
 \namedef{trotter}{111}
 \namedef{tsallis}{112}
 \namedef{sfvstc}{113}
 \namedef{uemura}{114}
 \namedef{white}{115}
 \namedef{vilk1}{116}
 \namedef{vilk2}{117}
 \namedef{white}{118}
 \namedef{stripes}{119}
 \namedef{bethe}{120}
 \namedef{symetrie}{121}
 \namedef{so5}{122}
 \namedef{diaghtc}{1}
 \namedef{diagtx}{2}
 \namedef{diagtu}{3}
 \namedef{psth}{4}
 \namedef{rpaq}{5}
 \namedef{rpaw}{6}
 \namedef{pqptcfig}{7}
 \namedef{aw2}{8}
 \namedef{aw4}{9}
 \namedef{dstat}{10}
 \namedef{bruit}{11}
 \namedef{gscaling}{12}
 \namedef{ascaling}{13}
 \namedef{ratioxvss}{14}
 \namedef{chiw}{15}
 \namedef{sfluid}{16}
 \namedef{spectre8}{17}
 \namedef{corr}{18}
 \namedef{fspfsc}{19}
 \namedef{aw95}{20}


\storeEqs
\initBib

\chapitreSN{Introduction}

Dans les ann\'ees 50, lorsque Bardeen, Cooper et Schrieffer ont
d\'evelopp\'e une th\'eorie pour expliquer la supraconductivit\'e (th\'eorie
BCS), ils
ont montr\'e que celle-ci permettait d'expliquer l'effet Meissner observ\'e
lorsque l'on soumet un supraconducteur \`a un champ \'electromagn\'etique
[\ref{bcs}]. Leur d\'erivation fut effectu\'ee en utilisant la jauge de
Landau. Cependant, lorsque leur calcul \'etait reproduit dans une autre jauge,
on obtenait une distribution de courant diff\'erente ne respectant pas
la conservation du nombre de charge [\ref{rick}].
En d'autres mots, leur th\'eorie brisait l'invariance de
jauge. Ce probl\`eme fut r\'esolu simultan\'ement par Anderson [\ref{anderson}]
et par Bogoliubov et al. [\ref{bogoliubov}], puis de fa\c con plus compl\`ete
par Rickayzen [\ref{rickayzen}]. L'explication repose sur le fait qu'il faut
inclure dans le calcul des courants la contribution provenant des modes
collectifs
longitudinaux afin de restaurer l'invariance de jauge. Sans cette contribution,
les parties longitudinales des courants paramagn\'etiques et diamagn\'etiques
ne se compensent pas lors d'un changement de jauge.

Suite \`a la solution de ce probl\`eme, Baym et Kadanoff
[\ref{baym1},\ref{baym2}] ont propos\'e un formalisme permettant d'assurer
qu'une solution m\^eme approximative d'un mod\`ele \`a N-corps puisse
satisfaire les lois de conservation globales du nombre de particule (invariance
de jauge), de l'\'energie et de la quantit\'e de mouvement. Ce formalisme
consiste \`a assurer une certaine coh\'erence entre la description des
excitations \'el\'ementaires du syst\`eme et ses propri\'et\'es de transport.

Encore de nos jours, beaucoup de recherches ont pour objet le d\'eveloppement
de nouvelles m\'ethodes analytiques visant \`a pr\'edire les propri\'et\'es de
mod\`eles relativement simples comme le mod\`ele de Hubbard.
L'int\'er\^et pour ce domaine s'est accentu\'e depuis la d\'ecouverte de
mat\'eriaux \`a propri\'et\'es exotiques. La forte anisotropie de la plupart de
ces syst\`emes a motiv\'e beaucoup d'\'etudes sur les syst\`emes
unidimensionnels et bidimensionnels.
En particulier, pour les syst\`emes quasi-unidimensionnels, plusieurs techniques
telles la bosonisation [\ref{boson}], le groupe de renormalisation [\ref{rg}],
la th\'eorie des champs conformes [\ref{conforme}],
l'ansatz de Bethe [\ref{bethe}] ont permis de donner une description
d\'etaill\'ee de l'\'etat fondamental, des \'etats excit\'es et des fonctions de
corr\'elation de diff\'erents mod\`eles et ce pour des cas \`a forte comme
\` a faible interaction. Plus r\'ecemment, la th\'eorie du champ moyen dynamique
[\ref{dmft}] a permis de r\'esoudre exactement des mod\`eles \`a dimension
infinie.

Pour l'\'etude du cas bidimensionnel, \`a faible couplage la phase ordonn\'ee
est assez bien d\'ecrite par une th\'eorie de type BCS.
De m\^eme, la limite \`a fort couplage (limite atomique) du mod\`ele de
Hubbard est assez bien comprise et son comportement critique correspond
\`a la condensation de bosons\footnote{${}^1$}{Dans la limite fort couplage, les
\'electrons forment des paires locales \`a haute temp\'erature. \`A la temp\'erature
critique, on observe une condensation des bosons dans l'\'etat fondamental
appel\'ee condensation de Bose-Einstein.}[\ref{strongcoupling}]. Cependant, en
l'absence de th\'eories couvrant les deux limites beaucoup de recherches
ont \'et\'e men\'ees dans le domaine [\ref{bcsbec}].
En plus des tentatives pour g\'en\'eraliser les m\'ethodes fructueuses \`a une
dimension, plusieurs techniques nouvelles ont
\'et\'e d\'evelopp\'ees: approche par les moments (approximation de type
n-p\^oles) [\ref{momenthub}], approximation dynamique par amas [\ref{dca}],
approche fonctionnelle [\ref{haussmann}], approche GW (Approximation
auto-coh\'erente pour la fonction de Green, G, faisant appel au potentiel
\'ecrant\'e, W) [\ref{gw}],
approximation de parquet [\ref{parquet}], approximation FLEX
(``Fluctuation exchange'')[\ref{flex}],
approximation TPSC (``Two-particle self-consistent'') [\ref{vilk1},\ref{vilk2}].
On a vu aussi le d\'eveloppement
d'approches diagrammatiques \`a partir de la limite fort couplage
[\ref{stephane}].

Dans cette th\`ese, nous d\'eveloppons une g\'en\'eralisation de
l'approximation TPSC [\ref{vilk1},\ref{vilk2}] qui permet de
traiter le mod\`ele de Hubbard attractif.
Parmi les m\'ethodes mentionn\'ees pr\'ec\'edemment, celle-ci semble \^etre
l'une des plus prometteuses. En effet des comparaisons avec les donn\'ees Monte
Carlo pour de petits syst\`emes ont donn\'e des r\'esultats tr\`es
satisfaisants [\ref{samuel}]. En particulier, l'application \`a des
r\'egions du diagramme de phase peu sujettes aux effets de taille a mis en
\'evidence la pr\'esence d'un pseudogap dans le mod\`ele de Hubbard
r\'epulsif [\ref{vilk1},\ref{samuel}].

L'approche est bas\'ee entre autres sur
des r\`egles de somme qui permettent d'assurer une coh\'erence
entre un vertex irr\'eductible approximatif et les fonctions de
corr\'elation \`a deux particules, d'o\`u le nom de l'approche: approximation
auto-coh\'erente \`a deux particules (``Two-particle
self-consistent''). Par la suite, on peut \'evaluer
la contribution de ces fluctuations \`a la self-\'energie.
Ainsi, par l'application de cette m\'ethode nous pourrons obtenir les
propri\'et\'es \`a une particule, dont le poids spectral. Cette m\'ethode
sera appliqu\'ee entre autres \`a une \'etude d\'etaill\'ee du probl\`eme
du pseudogap dans les supraconducteurs \`a haute temp\'erature critique
(SC hTc). Il y a d\'ej\`a un grand nombre de m\'ethodes qui ont \'et\'e
appliqu\'ees \`a ce probl\`eme, la plupart (matrice-T auto-coh\'erente,
FLEX) ne pr\'edisant pas de pseudogap pour le mod\`ele de Hubbard.

Cette th\`ese est divis\'ee de la fa\c con suivante.
Dans le premier chapitre, nous expliquons comment notre \'etude du mod\`ele
de Hubbard attractif en deux dimensions
pourrait aider \`a comprendre la pr\'esence du pseudogap observ\'e dans les
SC hTc. Nous abordons l'effet des fluctuations supraconductrices
sur le poids spectral des excitations fermioniques.
Afin d'expliquer la pr\'esence d'un large
domaine de temp\'erature o\`u un pseudogap pourrait \^etre pr\'esent,
nous mettons l'accent sur l'effet de
l'augmentation de la sym\'etrie du param\`etre d'ordre. L'existence d'un
tel m\'ecanisme dans les SC hTc a \'et\'e pr\'edite par Zhang avec sa
th\'eorie ${\rm SO}(5)$ [\ref{so5}]. Cette th\'eorie suppose que
l'\'elargissement de la sym\'etrie du param\`etre d'ordre serait pr\'esent
entre la r\'egion avec ordre antiferromagn\'etique et la r\'egion avec
ordre supraconducteur l\`a o\`u le pseudogap est observ\'e sur un vaste
domaine de temp\'erature.

Le chapitre suivant constitue le coeur de la th\`ese. Nous y d\'ecrivons
l'approche fonctionnelle d\'evelopp\'ee. Notre point de d\'epart est un
ensemble de relations exactes: l'\'equation du mouvement de la fonction
de Green, l'\'equation de Dyson et l'\'equation de Bethe-Salpeter.
\'Etant donn\'e le formalisme matriciel propos\'e ici, nous red\'erivons
ces formules qui prennent des aspects quelque peu diff\'erents de la
forme habituelle. Par la suite, nous suivons le formalisme de Baym et
Kadanoff qui nous permet d'obtenir le vertex irr\'eductible \`a partir
d'une forme approximative pour la fonctionnelle de self-\'energie.
Le syst\`eme d'\'equations, ainsi \'etabli, peut \^etre ferm\'e par l'imposition
d'une r\`egle de somme d\'eduite du th\'eor\`eme de fluctuation-dissipation.
Ainsi, notre th\'eorie donne une solution approximative du mod\`ele de Hubbard
sans l'ajout d'aucun param\`etre arbitraire. De plus, nous d\'emontrons
que notre approximation reproduit exactement le premier moment de la
susceptibilit\'e de paires et, de fa\c con auto-coh\'erente, le second moment.

Finalement, dans le troisi\`eme chapitre, nous v\'erifions la validit\'e
de notre approximation et comparons nos r\'esultats aux donn\'ees Monte Carlo.
Nous montrons que notre th\'eorie pr\'evoit la pr\'esence d'un pseudogap
dans le poids spectral pr\`es du demi-remplissage et ce pour des temp\'eratures
bien sup\'erieures \`a la temp\'erature critique. Nous montrons que
l'apparition de ce pseudogap est associ\'ee \`a la pr\'esence d'un r\'egime
classique renormalis\'e. Ces r\'esultats sont obtenus dans un r\'egime o\`u
les effets de la taille finie du syst\`eme sont contr\^ol\'es
et ce m\^eme pour les tailles accessibles par simulation Monte Carlo.

Afin d'aider \`a la compr\'ehension de certains passages de la th\`ese,
nous avons ajout\'e quelques annexes \`a la fin de celle-ci.
Par exemple, dans la premi\`ere
annexe, le lecteur trouvera une table de formules utiles pour l'application
du formalisme pr\'esent\'e au deuxi\`eme chapitre. \'Egalement en annexe,
nous avons donn\'e une description de la technique de prolongement analytique
par maximisation d'entropie. Cette technique est la m\'ethode la plus efficace
permettant d'extraire le poids spectral de la fonction de Green en temps
imaginaire mesur\'ee par les simulations Monte Carlo.


\chapitre{Motivation et mise en contexte}

\section{La supraconductivit\'e \`a haute temp\'erature}

Depuis la d\'ecouverte par Bednorz et M\"uller [\ref{bednorz}] de la
pr\'esence d'une phase supraconductrice \`a haute temp\'erature dans le
m\'elange ${\rm Ba}_x{\rm La}_{5-x}{\rm Cu}_5{\rm O}_{5(3-y)}$, on a
d\'ecouvert toute une famille d'oxydes de cuivre pr\'esentant cette
m\^eme phase. Ces compos\'es sont caract\'eris\'es par une forte conductivit\'e
dans les plans cuivre-oxyg\`ene (${\rm CuO}_2$) et une faible conductivit\'e
entre ces plans. Cette anisotropie explique pourquoi on consid\`ere ces
compos\'es comme des syst\`emes quasi-bidimensionnels\footnote{${}^1$}{La
pr\'esence d'une interaction entre les plans cuivre-oxyg\`ene est toutefois
importante car elle d\'etermine la valeur de la temp\'erature critique}.

Pour l'\'etude du
diagramme de phase, la variation de la densit\'e
de porteurs dans ces plans se fait par subtitution d'atomes hors des plans
ou par modification du nombre d'atomes d'oxyg\`ene apical [\ref{lnp475}].
Typiquement, le diagramme de phase est caract\'eris\'e par une phase
antiferromagn\'etique (AFM) dans la r\'egion g\'en\'eralement consid\'er\'ee
comme faiblement dop\'ee (${\rm Nd}_{2-x}{\rm Ce}_x{\rm CuO}_{4}$ avec $x<0.13$
${\rm La}_{2-x}{\rm Sr}_x{\rm CuO}_4$ avec $x<0.02$). \`A plus fort dopage,
assez pr\`es de la disparition de la phase AFM, se trouve
une phase supraconductrice (voir figure \ref{diaghtc}). La r\'egion
entre ces deux phases est diff\'erente d'un compos\'e \`a un autre.
Certains pr\'esentent une phase de verres de spin, d'autres des rayures
(``stripes'') o\`u se pr\'esente une alternance de r\'egions avec ordre
local AFM et de r\'egions riches en porteurs de charges [\ref{stripes}].

Une caract\'eristique commune au diff\'erents compos\'es est la pr\'esence
d'un dopage o\`u la temp\'erature critique entre la phase ``normale'' et
la phase supraconductrice est maximale. Ce dopage est appel\'e dopage
optimale. La section comprenant la phase AFM jusqu'au dopage optimale
est appel\'ee le r\'egime sous-dop\'ee et l'autre section le r\'egime
surdop\'ee.

Une autre caract\'eristique du diagramme de phase est la pr\'esence
de propri\'et\'es anormales pour un vaste domaine de temp\'eratures
sup\'erieures \`a la temp\'erature critique. Ces propri\'et\'es sont
dites anormales car elles ne peuvent \^etre d\'ecrites par la th\'eorie
des liquides de Fermi d\'evelopp\'ee par Landau.
Une des caract\'eristiques ayant attir\'e beaucoup d'attention est
la pr\'esence
d'un abaissement de la densit\'e d'\'etats au niveau de Fermi.
Sous la temp\'erature $T^*$
identifi\'ee sur la figure \ref{diaghtc},
l'abaissement est tel qu'on observe la pr\'esence d'un minimum local \`a
ce point.
C'est ce que l'on appelle le pseudogap (PG).
Ce ph\'enom\`ene a \'et\'e observ\'e dans un grand nombre
d'exp\'eriences: mesures de chaleur sp\'ecifique [\ref{cv}], de conductivit\'e
optique [\ref{corson}], r\'esonance magn\'etique nucl\'eaire [\ref{rmn}],
diffusion Raman [\ref{raman}], diffusion in\'elastique de neutrons
[\ref{neutron}], spectroscopie r\'esolue en temps [\ref{tros}],
spectroscopie par effet tunnel [\ref{tunnel}], photo\'emission [\ref{arpes}]
(voir Timusk et Statt pour une revue des diff\'erents r\'esultats
exp\'erimentaux [\ref{timusk}]).


Le diagramme de phase pr\'esent\'e \`a la figure \ref{diaghtc} est sch\'ematique
et change l\'eg\`erement d'un compos\'e \`a un autre. La pr\'esence du pseudogap
a \'et\'e observ\'e dans plusieurs de ces compos\'es que ce soit ceux de la
famille des YBaCuO, de la famille des lanthanes ou des bismuths. Le pseudogap
a \'et\'e observ\'e au-dessus de la phase AFM aussi bien qu'au-dessus de la
phase supraconductrice. Typiquement, la temp\'erature o\`u celui-ci appara\^\i t
diminue au fur et \`a mesure que l'on s'\'eloigne de la phase AFM, pour tendre
vers la temp\'erature critique dans la phase surdop\'ee.

Notre \'etude du mod\`ele attractif visera \`a expliquer l'existence de
ce r\'egime avec pseudogap. Mais avant de pr\'esenter notre approche, nous
exposons quelques th\'eories propos\'ees pour expliquer ce pseudogap. La
plupart de ces th\'eories sont
pr\'esent\'ees dans l'article de revue de Loktev et al. [\ref{loktev}].

\section{Pseudogap}

Pour la suite de ce travail, nous d\'efinirons le r\'egime pseudogap par
la r\'egion du diagramme de phase o\`u on observe dans le poids spectral
\`a une particule un minimum local au niveau de Fermi (sans toutefois
qu'il n'y ait un vrai gap, c'est-\`a-dire au-dessus de la temp\'erature
critique). D'un point de vue quantitatif le pseudogap sera donn\'e par
la mesure de la s\'eparation en \'energie des maxima locaux situ\'es
de part et d'autre du niveau de Fermi.

Partant de l'observation que le pseudogap \'evolue de fa\c con continue,
devenant le gap dans l'\'etat ordonn\'e (pour un dopage au-dessus de la
phase supraconductrice),
plusieurs th\'eories sont bas\'ees sur l'existence d'un m\'ecanisme
permettant d'abaisser la temp\'erature critique sans toutefois modifier la
temp\'erature \`a laquelle apparaissent les fluctuations supraconductrices.
Le gap dans les excitations fermioniques ne se forme que dans la phase
ordonn\'ee, donc
\`a basse temp\'erature. L'\'ecart important entre ces deux temp\'eratures
laisse donc place \`a un r\'egime o\`u des propri\'et\'es anormales peuvent
se pr\'esenter. Pour mieux identifier ces propri\'et\'es, il faut comprendre
quel serait le m\'ecanisme responsable de l'abaissement de la temp\'erature
critique. Une deuxi\`eme question importante \`a laquelle on doit r\'epondre
concerne le d\'eveloppement du pseudogap dans ce r\'egime. Jusqu'\`a
maintenant, plusieurs th\'eories ont propos\'e diff\'erentes r\'eponses \`a
ces questions.

Kivelson et Emery [\ref{emery}] furent parmi les pionniers de l'\'etude
th\'eorique du pseudogap avec leur sc\'enario bas\'e sur l'effet des
fluctuations supraconductrices.
Ils ont propos\'e que la faible densit\'e de
porteurs dans les supraconducteurs sous-dop\'es donne un
r\^ole important aux fluctuations de phase et provoque l'apparition du
pseudogap. Supposant que le comportement critique appartient \`a la classe du
mod\`ele XY, leur th\'eorie
a obtenu un accord quantitatif int\'eressant avec un certain nombre
d'observations.

Tout r\'ecemment, Nozi\`eres et Pistolesi ont propos\'e une th\'eorie faisant
aussi appel \`a une faible densit\'e de porteurs [\ref{nozieres2}].
Ils ont \'etudi\'e la possibilit\'e qu'un semiconducteur puisse avoir
une transition vers un \'etat supraconducteur.
Le pseudogap r\'esulte alors d'une comp\'etition entre le gap de la phase
supraconductrice et celui associ\'e \`a la phase semiconductrice.

Par ailleurs, Randeria a, tout comme Emery et Kivelson, fait appel \`a
l'observation d'une faible rigidit\'e du superfluide dans l'\'etat ordonn\'e
pour proposer une nouvelle th\'eorie [\ref{randeria}].
Cependant, il a not\'e
le fait que pour le mod\`ele XY, la temp\'erature critique champ moyen
et la transition Berezinsk\u\i i-Kosterlitz-Thouless
(BKT) sont tr\`es rapproch\'ees et qu'un tel mod\`ele ne
peut donc pas pr\'esenter une vaste r\'egion avec un pseudogap. Dans sa
th\'eorie, le pseudogap est un vestige de la phase supraconductrice et provient
de l'existence de paires \`a courte port\'ee au-dessus de la temp\'erature
critique.
Cette th\'eorie requiert donc l'existence d'une interaction assez forte entre
les particules. Comme les autres th\'eories faisant appel \`a la force de
l'interaction, la
dimension du syst\`eme ne joue pas, ici, de r\^ole primordial pour l'existence
du pseudogap.

La grande diff\'erence entre la temp\'erature critique et la temp\'erature
o\`u appara\^\i t le comportement anormal, $T_{\rm cr}\simeq 6T_c$
[\ref{curro}], a aussi
\'et\'e utilis\'ee par Pines pour critiquer la th\'eorie de Emery et Kivelson
[\ref{pines}]. Il a propos\'e avec ses collaborateurs, un sc\'enario o\`u
c'est la formation de pr\'ecurseurs d'ondes de densit\'e de spin qui induit
un pseudogap dans les excitations fermioniques pr\`es des singularit\'es de
Van Hove (``hot spots''). Cette th\'eorie est une extension de la th\'eorie des
liquides de Fermi quasi-antiferromagn\'etiques d\'evelopp\'ee par le groupe de
Pines.

On ne peut passer sous silence la th\'eorie du groupe de Levin [\ref{levin}].
Leur th\'eorie consiste \`a faire une distinction entre le gap dans les
excitations fermioniques et le param\`etre d'ordre. Ils associent la
diff\'erence
entre ces param\`etres \`a la pr\'esence d'excitations de paires de faible
\'energie. Au-dessus de la temp\'erature critique, le param\`etre d'ordre
est nul, mais l'existence de paires pr\'eform\'ees pourrait faire appa\^\i tre
le pseudogap.
Un pr\'ealable important de leur th\'eorie est l'existence
d'une r\'esonance dans la fonction de corr\'elation de paires.
Cependant, la cr\'eation de cette r\'esonance par un processus
microscopique n'a pas \'et\'e d\'emontr\'ee. Le groupe de Levin l'obtient
par une approximation de type matrice-T o\`u les particules coupl\'ees par
l'interaction ont des niveaux de Fermi diff\'erents. Cette diff\'erence
\'etant li\'ee \`a la force du couplage, $\delta\mu\propto\Sigma$,
ils obtiennent qu'au-dessus d'une
valeur critique un pic de r\'esonance appara\^\i t dans la susceptibilit\'e
de paires.
C'est cette r\'esonance qui permet l'apparition d'un pseudogap
au-dessus de la temp\'erature critique. La taille de ce dernier est
proportionnelle \`a l'\'energie de la r\'esonance et donc \`a la diff\'erence
des niveaux de Fermi: $\Sigma\propto\Delta_{\rm pg}^2$. Il ne faut cependant
pas trop pousser la critique de cette th\'eorie et garder un esprit ouvert.
Certains effets, tel la pr\'esence du niveau de Fermi hors de la bande
d'\'energie permise, pourraient amener l'apparition d'une r\'esonance.

En comparaison, la th\'eorie que nous proposons est bas\'ee sur la pr\'esence
de fluctuations de phase supraconductrices. Cependant, nous proposons
que, dans la r\'egion sous-dop\'ee,
ces fluctuations n'appartiennent pas \`a la classe BKT tel que
propos\'e par Loktev et al. [\ref{loktev}] mais \`a une classe \`a
sym\'etrie plus \'elev\'ee.

La th\'eorie ${\rm SO}(5)$ de Zhang [\ref{so5}] fournit un exemple de
th\'eorie o\`u la sym\'etrie joue un r\^ole important dans les SC hTc.
Partant de la
proximit\'e entre la phase supraconductrice et la phase antiferromagn\'etique,
Zhang a sugg\'er\'e que dans la partie du diagramme entre ces
deux phases, le param\`etre d'ordre poss\'ederait une sym\'etrie approximative
de type ${\rm SO}(5)$. Par ailleurs, il est bien connu qu'une augmentation
de la sym\'etrie du param\`etre d'ordre a pour effet de faciliter la
pr\'esence de fluctuations de phase\footnote{$^2$}{\`A basse sym\'etrie, deux
configurations du param\`etre d'ordre minimisant l'\'energie peuvent \^etre
topologiquement distinctes.
Il faut alors surmonter un co\^ut \'energ\'etique important pour passer
de l'une \`a l'autre. Une telle transformation est donc improbable.
L'augmentation de la sym\'etrie du param\`etre d'ordre am\`ene de nouvelles
avenues par lesquelles nous pouvons transformer les configurations.
Il s'ensuit que deux configurations qui \'etaient topologiquement distinctes
\`a basse sym\'etrie peuvent devenir topologiquement \'equivalentes lorsque
la sym\'etrie est augment\'ee.}
et donc d'abaisser la temp\'erature
critique (particuli\`erement pour un syst\`eme bidimensionnel). Ainsi, il
devient possible d'observer une vaste r\'egion de fluctuations de phase
au-dessus de la temp\'erature critique dans la r\'egion \`a sym\'etrie
\'elev\'ee. Ce sont ces fluctuations de phase qui seraient
responsables de l'apparition du pseudogap.

Il est \`a noter que m\^eme si la pr\'esence ou l'absence d'une sym\'etrie
suppl\'ementaire devrait \^etre d\'etectable sans ambiguit\'e,
l'effet que l'on veut mettre en lumi\`ere pourrait se
pr\'esenter dans un cas o\`u la sym\'etrie n'est pas exacte. Ainsi,
l'apparition de cet effet ne se ferait pas de fa\c con brusque mais serait
plut\^ot
quelque chose de continu. Illustrons ce que l'on veut dire en consid\'erant
l'exemple du mod\`ele XXZ avec un terme d'anisotropie, $\lambda$,
faible [\ref{xxzmc}]. \`A une certaine temp\'erature ce mod\`ele pr\'esente
le m\^eme comportement critique que le mod\`ele de Heisenberg, comme si le terme
d'anisotropie n'\'etait pas pr\'esent. Ce n'est qu'\`a plus basse
temp\'erature que le syst\`eme suit un crossover vers le comportement critique
du mod\`ele XY. Plus le terme d'anisotropie est faible plus le crossover
se fera pr\`es de la temp\'erature critique.

Il pourrait en \^etre de m\^eme pour les SC hTc. \`A temp\'erature \'elev\'ee
des fluctuations \`a sym\'etrie \'elev\'ee apparaissent. Cette forte sym\'etrie
implique que la phase du param\`etre d'ordre ne peut s'ordonner qu'\`a basse
temp\'erature. Il peut alors subsister un large domaine de temp\'erature o\`u,
sous l'effet des fluctuations, le syst\`eme pr\'esenterait
des propri\'et\'es anormales. Cet effet serait accentu\'e par la
quasi-bidimensionalit\'e du syst\`eme.
Ainsi, le pseudogap qui serait associ\'e \`a des fluctuations \`a sym\'etrie
\'elev\'ee, pourrait se d\'evelopper
jusqu'\`a une temp\'erature o\`u l'on observerait un crossover vers
un r\'egime critique \`a plus basse sym\'etrie. Ce crossover permettrait
d'expliquer l'\'evolution continue du pseudogap vers le gap qui est associ\'e
\`a un param\`etre d'ordre \`a sym\'etrie r\'eduite, celui de la phase
supraconductrice.

Afin de v\'erifier cette th\'eorie,
nous avons choisi d'\'etudier le mod\`ele de Hubbard
attractif \`a deux dimensions. Il est clair que ce mod\`ele n'est pas un
mod\`ele r\'ealiste pour d\'ecrire la physique des supraconducteurs \`a
haute temp\'erature critique. Cependant, la pr\'esence de r\'egions avec
diff\'erentes
sym\'etries pour le param\`etre d'ordre de ce mod\`ele nous permettra de
v\'erifier si l'\'elargissement de la sym\'etrie du param\`etre d'ordre
pourrait \^etre un m\'ecanisme responsable de l'agrandissement du domaine
o\`u un pseudogap appara\^\i t.

Apr\`es avoir discut\'e du mod\`ele de Hubbard dans la section suivante, nous
reviendrons sur la question de l'existence du pseudogap dans ce mod\`ele.

\section{Mod\`ele de Hubbard attractif}\nom{sechubbard}

Le mod\`ele de Hubbard sur r\'eseau d\'ecrit par la relation (\ref{hubbard}),
pr\'esente deux
param\`etres: le terme de saut, $t$, qui est l'amplitude de probabilit\'e
qu'un \'electrons passe d'un site du r\'eseau \`a un autre site et l'interation
entre des
\'electrons se trouvant sur un m\^eme site, $U$. \`A ces param\`etres
s'ajoutent ceux de l'ensemble grand canonique que nous utiliserons: la
temp\'erature, $T$, et le potentiel chimique, $\mu$, qui sont les param\`etres
de Lagrange associ\'es  respectivement \`a l'\'energie moyenne et au nombre
moyen de particules du syst\`eme.
En plus de ces param\`etres, s'ajoute une libert\'e dans le choix du r\'eseau.
Afin de d\'emontrer l'argumentation avanc\'ee, nous choississons un r\'eseau
qui permet d'assurer la variation de la sym\'etrie du param\`etre d'ordre.
Nous consid\'ererons donc le cas du r\'eseau carr\'e bidimensionnel.
Nous \'etudierons le cas o\`u l'int\'egrale de saut est
non nulle que pour les sites voisins,
$t_{\imath\jmath}=t(\delta_{\imath,\jmath\pm\hat{x}}+\delta_{\imath,\jmath\pm
\hat{y}})$.
Ce choix nous permettra d'utiliser la propri\'et\'e
bipartite du r\'eseau. Il est \`a noter que les indices $\imath$ et $\jmath$
indiquent
un site du r\'eseau, quant aux vecteurs $\hat{x}$ et $\hat{y}$, ils indiquent
les vecteurs de la base du r\'eseau cristallin. De plus, la condition que
$t_{\imath\jmath}=t_{\jmath\imath}$ est une condition essentielle pour
l'hermiticit\'e du Hamiltonien. Comme unit\'e
de mesure pour l'\'energie nous utiliserons le terme de saut, $t$.

$$\eqalign{
\boxEq{{\cal H}=-\sum_{\imath,\jmath,\sigma}t_{\imath\jmath}
\psi_{\imath,\sigma}^\dagger\psi_{\jmath,\sigma}+U\sum_\imath
n_{\imath,\uparrow}n_{\imath,\downarrow}}\mskip 5mu&\cr
\hbox{\rm\bf Hamiltonien du mod\`ele de Hubbard}&\cr
}\nom{hubbard}$$

Le diagramme de phase du mod\`ele \'etudi\'e d\'ependra donc de trois
param\`etres. De fa\c con
sch\'ematique, il sera donn\'e par les figures \ref{diagtx} et \ref{diagtu}.
Pour mettre en \'evidence la ressemblance avec celui des supraconducteurs
de type oxyde de cuivre, nous l'avons repr\'esent\'e en fonction du dopage
$x=1-n$ qui repr\'esente l'\'ecart entre un remplissage donn\'e, $n$, et le
cas du demi-remplissage, $n=1$. Nous avons repr\'esent\'e sur la figure
\ref{diagtx}, la temp\'erature critique, $T_{\rm BKT}$, la temp\'erature
critique pr\'edite par un calcul champ moyen, $T^o$, la temp\'erature
\`a laquelle appara\^\i trait le pseudogap, $T^*$ ainsi que la temp\'erature
o\`u l'on observerait un crossover entre le comportement critique \`a
haute sym\'etrie et le comportement critique $O(2)$. Alors que la temp\'erature
$T^o$ peut \^etre \'evalu\'e num\'eriquement, la temp\'erature critique
$T_{\rm BKT}$ n'est pas connue. Ici, nous avons repr\'esent\'e ces deux courbes
de fa\c con sch\'ematique. Le crossover quant \`a lui n'est pas une ligne bien
d\'efinie. Sa repr\'esentation est \'egalement approximative. \'Etant donn\'e
que le crossover est associ\'e au remplissage, la temp\'erature o\`u celui-ci
se produit augmente en s'\'eloignant du cas demi-rempli ($x=0$). Pour un
remplissage plus grand que le remplissage optimal, cette temp\'erature
d\'ecro\^\i t car alors le comportement critique ne se manifeste qu'\`a basse
temp\'erature.
Enfin, l'existence
d'une temp\'erature o\`u appara\^\i t un pseudogap n'est que sp\'eculative et
nous allons tenter de montrer son existence.


Nous avons donn\'e
\'egalement le diagramme de phase en fonction de l'interaction,
figure \ref{diagtu}.
On y a repr\'esent\'e les deux temp\'eratures dont on conna\^\i t
l'\'evolution:
la temp\'erature critique, $T_{\rm BKT}$ qui augmente de fa\c con exponentielle
\`a faible interaction puis passe par un maximum pour d\'ecro\^\i tre en
suivant une loi de puissance \`a forte interaction; et la temp\'erature
critique champ moyen, $T^o$, qui augmente de fa\c con exponentielle \`a
faible interaction puis tend vers une constante \`a forte
interaction. Les deux courbes sont repr\'esent\'ees de fa\c con sch\'ematique.
Physiquement, la temp\'erature champ moyen correspond \`a la formation de
paires de Cooper.
Lorsque le couplage est faible celle-ci est \`a peu pr\`es identique \`a la
temp\'erature critique $T_{\rm BKT}$. Cependant \`a fort couplage, cette
temp\'erature est bien diff\'erente de la temp\'erature critique. Cette
derni\`ere correspond \`a la temp\'erature o\`u se fait la condensation
des paires locales (condensation de Bose-Einstein). \'Etant
donn\'e que l'\'energie de
liaison des paires est proportionnelle \`a la force du couplage, il en
sera de m\^eme pour la masse des paires. Du fait que des bosons de grande
masse ne peuvent condenser qu'\`a basse temp\'erature ceci explique pourquoi
la temp\'erature critique d\'ecroit ($T_{\rm BKT}\propto 1/\vert U\vert$) \`a
fort couplage [\ref{strongcoupling}]. Ainsi dans cette r\'egion on observe
une grande diff\'erence entre $T_{\rm BKT}$ et $T^o$.
Cette diff\'erence explique pourquoi diff\'erentes th\'eories du pseudogap
font appel \`a la pr\'esence d'une forte interaction. Nous avons identifi\'e
sur le graphique la valeur de couplage o\`u se trouve la valeur maximale
de la temp\'erature critique telle qu'estim\'ee par Singer et al. [\ref{singer2}:
figure 3]
(Estimation par simulation Monte Carlo). Dans notre travail, nous
utiliserons une valeur de $U=-4$. Celle-ci est situ\'ee sur dans la r\'egion
o\`u la temp\'erature critique cro\^\i t en fonction du couplage. Ainsi,
le syst\`eme sera dans la limite couplage faible \`a interm\'ediaire.


Il est possible de montrer qu'\`a
demi-remplissage notre mod\`ele pr\'esente une sym\'etrie plus \'elev\'ee qu'en
dehors
du demi-remplissage. La particularit\'e du cas demi-rempli vient du potentiel
chimique qui est alors connu de fa\c con exacte, $\mu=U/2$. Cette valeur
particuli\`ere de $\mu$ permet d'avoir une sym\'etrie suppl\'ementaire.
Pour le d\'emontrer consid\'erons la transformation canonique suivante:
$$\psi_{\imath\downarrow}\rightarrow(-1)^\imath\psi_{\imath\downarrow}^\dagger
\mskip 50mu\psi_{\imath\downarrow}^\dagger\rightarrow(-1)^\imath\psi_{\imath
\downarrow}\nom{symhubbard}$$
Consid\'erant la transformation de chacun des termes du hamiltonien, on
peut montrer comme r\'esultat interm\'ediaire que:
$$\eqalign{\psi_{\imath\downarrow}^\dagger\psi_{\imath\pm\hat{x}\downarrow}
\rightarrow&\psi_{\imath\pm\hat{x}\downarrow}^\dagger\psi_{\imath\downarrow}\cr
\psi_{\imath\downarrow}^\dagger\psi_{\imath\downarrow}\rightarrow&1-
\psi_{\imath\downarrow}^\dagger\psi_{\imath\downarrow}\cr}$$
De ces r\'esultats on obtient que le terme de saut aux premiers voisins
demeure inchang\'e sous la transformation tandis que le terme d'interaction
change de signe et fait appara\^\i tre un terme suppl\'ementaire:
$${\cal H}_0+U\sum_\imath n_{\imath\uparrow}n_{\imath\downarrow}-\mu\hat{N}
\rightarrow{\cal H}_0-U\sum_\imath n_{\imath\uparrow}n_{\imath\downarrow}+
{U\over 2}\hat{N}-(\mu-U/2)\sum_\imath(n_{\imath\uparrow}-n_{\imath\downarrow}
)\nom{transhub}$$

Ainsi, le mod\`ele de Hubbard attractif ($U<0$) \`a remplissage quelconque
se transforme en mod\`ele de Hubbard r\'epulsif \`a demi-remplissage, $n=1$,
sous
champ magn\'etique, $h=\mu-U/2$. \'Etant donn\'e que pour le cas demi-rempli
le potentiel chimique est donn\'e par
$\mu=U/2$, on voit que dans ce cas le champ magn\'etique obtenu par la
transformation est nul.

Le changement du hamiltonien sous la transformation (\ref{symhubbard}) pour le
cas demi-rempli est associ\'e \`a une sym\'etrie ${\rm SU}(2)$ du mod\`ele.
\`A cette sym\'etrie s'ajoute
la sym\'etrie de spin (voir [\ref{symetrie}]). Ce qui donne une sym\'etrie
${\rm SU}(2)\times{\rm SU}(2)/{\rm Z}_2={\rm SO}(4)$\footnote{$^3$}{La
premi\`ere sym\'etrie ${\rm SU}(2)$ couvre en partie l'invariance
sous rotation du spin. Ainsi, pour le calcul de la sym\'etrie totale du
mod\`ele, il faut enlever le facteur $Z_2$.}.
Hors demi-remplissage, la
sym\'etrie sous rotation de spin ${\rm SU}(2)$ est toujours pr\'esente,
cependant, l'ajout d'un champ magn\'etique sous la transformation
(\ref{symhubbard}) brise la sym\'etrie suppl\'ementaire qui est r\'eduite au
groupe ${\rm U}(1)$. Ainsi la sym\'etrie totale est r\'eduite \`a
${\rm SU}(2)\times{\rm U}(1)$.

La sym\'etrie qui nous int\'eresse est celle associ\'ee au param\`etre d'ordre
et non pas toute celle pr\'esente dans le hamiltonien. Pour le cas
attractif,
on sait qu'\`a basse temp\'erature il y aura apparition de fluctuations
supraconductrices. Ainsi le param\`etre d'ordre sera associ\'e aux op\'erateurs
de cr\'eation et de destruction de paires. Sous la transformation
(\ref{symhubbard}), ces op\'erateurs se transforment de la fa\c con suivante:
$$\eqalign{\Delta_\imath\equiv&\psi_{\imath\uparrow}\psi_{\imath\downarrow}
\rightarrow(-1)^\imath\ffrac{1}{2}(S_\imath^x-iS_\imath^y)\cr
\Delta_\imath^\dagger\equiv&\psi_{\imath\downarrow}^\dagger
\psi_{\imath\uparrow}^\dagger\rightarrow(-1)^{\imath+1}\ffrac{1}{2}(S_\imath^x
+iS_\imath^y)\cr\rho_\imath\equiv&\psi_{\imath\uparrow}^\dagger\psi_{\imath
\uparrow}+\psi_{\imath\downarrow}^\dagger\psi_{\imath\downarrow}\rightarrow
S_\imath^z+1\cr}$$
Par ailleurs, on sait que pour le mod\`ele de Hubbard r\'epulsif en l'absence
de champ magn\'etique, les trois composantes de spin ont des fluctuations
critiques. Donc, en juxtaposant le r\'esultat pr\'ec\'edent avec la
transformation du hamiltonien (\ref{transhub}), on observe que pour le cas
attractif \`a demi-remplissage, des fluctuations
de densit\'e de charge s'ajoutent aux fluctuations supraconductrices.
Sous ces conditions, le param\`etre d'ordre pr\'esente une
sym\'etrie ${\rm SU}(2)$ qui est suffisamment importante qu'il soit
impossible d'avoir un point critique
\`a temp\'erature finie pour ce syst\`eme bidimensionnel\footnote{${}^4$}{Avec
une sym\'etrie ${\rm SU}(2)$, il n'y a plus de configurations topologiquement
distinctes en deux dimensions (plus de vortex).}.
En comparaison, pour le cas hors demi-remplissage,
la transformation du hamiltonien fait appara\^\i tre un champ magn\'etique
qui brise la sym\'etrie de spin pour $U>0$ et donc la sym\'etrie du
param\`etre d'ordre du cas $U<0$ est r\'eduite \`a une sym\'etrie ${\rm U}(1)$.
Dans ce dernier cas il existe un point critique de type (BKT)
\`a temp\'erature finie pour un
syst\`eme en deux dimensions. Ceci nous emm\`ene \`a la conclusion que la
temp\'erature critique BKT qui augmente
avec le remplissage de la bande doit atteindre un maximum puis redescendre
vers la temp\'erature nulle au fur et \`a mesure que l'on s'approche du
demi-remplissage tel que pr\'esent\'e \`a la figure \ref{diagtx}.
L'abaissement de la temp\'erature critique pr\`es du demi-remplissage alors
que la temp\'erature o\`u
apparaissent les fluctuations est \'elev\'ee, permet d'avoir un vaste domaine
de temp\'erature o\`u existerait un pseudogap.

Voici comment nous pr\'evoyons que pourrait se d\'evelopper le gap dans le
poids spectral. Nous nous int\'eresserons \`a quatre points pr\`es du
demi-remplissage tels qu'identifi\'es sur la figure \ref{diagtx}.
\`A la figure \ref{psth} nous avons repr\'esent\'e de fa\c con sch\'ematique
quelle serait, selon notre conjecture, la forme du poids spectral pour chacun
de ces points.
\`A temp\'erature suffisamment \'elev\'ee (mais petite par rapport \`a
l'\'energie de Fermi), point a, la
th\'eorie des liquides de Fermi peut s'appliquer et le poids spectral au
niveau de Fermi a la forme d'une lorentzienne dont la largeur diminue avec
la temp\'erature. En abaissant la temp\'erature sous $T^*$, les
fluctuations supraconductrices deviennent importantes, mais la sym\'etrie
\'elev\'ee du param\`etre et la bidimensionalit\'e du syst\`eme emp\^echent
l'av\`enement d'une transition de phase. C'est alors, qu'appara\^\i trait
un pseudogap dans le poids spectral, point b. Il est possible qu'avant que
ne se d\'eveloppe le pseudogap,
le pic pr\'esent \`a haute temp\'erature s'\'elargisse avec
l'abaissement de la
temp\'erature contrairement aux caract\'eristiques des liquides de
Fermi. Un tel ph\'enom\`ene peut amener la pr\'esence d'un pseudogap
\`a faible fr\'equence dans la densit\'e total d'\'etats sans qu'il n'y
en ait dans le poids spectral. Il peut aussi y avoir des singularit\'es
associ\'ees \`a la forme embo\^\i t\'ee de la surface de Fermi [\ref{frank}].

Une fois que l'on passe sous la temp\'erature critique BKT, le poids
spectral s'annule au niveau de Fermi, point c. On sait qu'\`a partir de ce
point et jusqu'\`a temp\'erature nulle le param\`etre d'ordre prend 
g\'en\'eralement une valeur non-nulle mais avec une phase diff\'erente d'une
r\'egion de l'espace \`a une autre. Il peut aussi exister des r\'egions o\`u il demeure
nul (coeur des paires vortex-antivortex). Globalement, il ne prendra une
valeur non nulle
qu'\`a temp\'erature nulle. Qu'en est-il pour le gap (c'est-\`a-dire, le
domaine de fr\'equences o\`u le poids spectral s'annule)?
Il pourrait
prendre une valeur finie pour toutes temp\'eratures sous $T_{\rm BKT}$,
mais ce n'est pas s\^ur.
Ce que l'on peut affirmer, c'est qu'\`a temp\'erature nulle, point d,
le gap est fini.


Dans la section suivante, nous verrons que les deux premiers spectres
pr\'esent\'es \`a la figure \ref{psth} ont \'et\'e obtenus par certaines
\'etudes du mod\`ele de Hubbard \`a demi-remplissage.

\section{Quelques \'etudes ant\'erieures du mod\`ele de Hubbard}

En terminant ce chapitre, nous pr\'esentons les r\'esultats obtenus par
diff\'erentes m\'ethodes qui ont \'et\'e appliqu\'ees au probl\`eme du
pseudogap dans le mod\`ele de Hubbard attractif.

\'Etant donn\'e la difficult\'e \`a construire une approximation analytique
satisfaisante pour d\'ecrire le mod\`ele de Hubbard, certaines \'etudes
ont utilis\'e des techniques Monte Carlo afin d'\'etudier l'importance
relative des diff\'erents diagrammes obtenus d'un d\'eveloppement perturbatif
[\ref{bulut},\ref{chen}]. Ces \'etudes
\'etant appliqu\'ees au mod\`ele r\'epulsif, on peut \`a demi-remplissage les
extrapoler au mod\`ele attractif. Parmi les r\'esultats obtenus, Chen et al.
[\ref{chen}] ont montr\'e qu'une approximation de type RPA avec un vertex
renormalis\'e permet de bien expliquer le facteur de stucture magn\'etique
observ\'e par simulations Monte Carlo pour une interaction interm\'ediaire
$U\simeq 4$.
On verra au chapitre 3 que cet accord est \'egalement assez bon pour la
susceptibilit\'e de paires du mod\`ele attractif. L'approximation TPSC
a l'avantage de bien pr\'edire la valeur de ce vertex pour le mod\`ele
r\'epulsif.

L'accord entre la susceptibilit\'e magn\'etique d'une approximation RPA
et les r\'esultats Monte Carlo a aussi \'et\'e observ\'e par Bulut et al.
[\ref{bulut}]. Ils ont aussi montr\'e que ce genre d'approximation permet
de bien d\'eterminer la self-\'energie en fr\'equences imaginaires mettant
ainsi en importance les corrections de vertex. Ils ont compar\'e la valeur
effective de l'interaction particule-particule avec un calcul du second
ordre o\`u la correction de ce vertex provient de l'\'echange de
fluctuations de spin. Leur r\'esultat montre que l'accord n'est pas
tr\`es bon et n'est pas am\'elior\'e de fa\c con significative par l'ajout du
troisi\`eme ordre.

Malgr\'e ces r\'esultats peu encourageants, plusieurs auteurs ont utilis\'e
l'approximation FLEX pour l'\'etude du mod\`ele de Hubbard. Il faut
mentionner que cette approximation va au-del\`a de l'unique contribution
de l'\'echange de fluctuations de spin, elle tient compte \'egalement des
fluctuations de charge et des fluctuations de paires de Cooper. R\'ecemment,
Deisz, Hess et Serene [\ref{flexpg}] ont appliqu\'e cette approximation
au probl\`eme du pseudogap dans le mod\`ele de Hubbard attractif. Ils ont
observ\'e que cette approximation pr\'edit des propri\'et\'es anormales
dans le phase normale, tel la suppression du poids spectral \`a faibles
\'energies avec l'abaissement de la temp\'erature. Cependant, aucune
\'evidence pour l'apparition d'un pseudogap n'a \'et\'e observ\'ee et ce
m\^eme pour des syst\`emes de taille $128\times 128$.

La technique Monte Carlo permet d'aller plus loin qu'une simple comparaison
avec les approches analytiques en fr\'equences imaginaires. Utilisant des
m\'ethodes pour faire un prolongement analytique des donn\'ees Monte Carlo,
on peut ainsi v\'erifier si ces donn\'ees pr\'edisent ou non la pr\'esence
d'un pseudogap dans le poids spectral des excitations fermioniques.
Un r\'esultat
reproduit par toutes les simulations Monte Carlo est la pr\'esence d'un
pseudogap pour
les syst\`emes de petites tailles. Le comportement en taille observ\'e
par Veki\'c et White [\ref{white}] tend \`a montrer que ce pseudogap
dispara\^\i t
lorsque le syst\`eme est de taille suffisamment grande. Plus r\'ecemment,
nous avons publi\'e une \'etude du pseudogap dans le mod\`ele de Hubbard
r\'epulsif \`a demi-remplissage [\ref{samuel}]. La sym\'etrie
entre ces mod\`eles montre que ce r\'esultat serait retrouv\'e pour le
mod\`ele attractif. Nous avons observ\'e la pr\'esence d'un pseudogap
pour des syst\`emes de diff\'erentes tailles et notre \'etude en taille a
montr\'e que celui-ci n'est d\^u \`a des effets de taille finie. Nous avons
montr\'e que l'approche
TPSC d\'evelopp\'ee par Vilk et Tremblay [\ref{vilk1},\ref{vilk2}] montre
un accord parfait avec les r\'esultats Monte Carlo pour les syst\`emes
de petites tailles. Cette observation sugg\`ere d'utiliser cette m\'ethode
afin de d\'eterminer s'il existe ou non un pseudogap dans la limite
thermodynamique. Notre conclusion est qu'il en existe un. Dans cet article, nous
expliquons pourquoi le r\'esultat de Veki\'c et White diff\`ere du
notre. En effet, nous avons montr\'e que plus le bruit est important
dans les donn\'ees Monte Carlo, moins on observe de d\'etails dans le poids
spectral obtenu par l'application de la technique de maximisation d'entropie
aux donn\'ees Monte Carlo (cette technique est expliqu\'ee en annexe
\ref{mem}). Ainsi, l'observation que les r\'esultats de
Veki\'c et White sont plus pr\'ecis pour les syst\`emes de petites tailles
que pour les syst\`emes de grande taille renforce notre argumentation selon
lequel il y aurait bien un r\'egime pseudogap dans la limite thermodynamique.

D'autres approches analytiques pr\'evoient \'egalement l'existence d'un
pseudogap. Parmi celles-ci,
il y a une approche par transformation de Hubbard-Stratonovich
[\ref{mass},\ref{mass2}] o\`u l'on observe l'apparition d'un pseudogap
\`a une temp\'erature l\'eg\`erement sup\'erieure \`a la temp\'erature
critique. Il y a aussi l'approximation de type matrice-T
[\ref{kagan},\ref{letz}]. On remarque pour cette derni\`ere approximation
que lorsque le calcul est enti\`erement auto-coh\'erent le pseudogap n'est
plus observ\'e. Tout r\'ecemment, on a vu le d\'eveloppement de
l'approximation d'amas dynamique (DCA). Celle-ci a \'egalement \'et\'e
appliqu\'ee au probl\`eme du pseudogap [\ref{dca}]. Il semble qu'en
am\'eliorant la pr\'ecision de cette approximation, la pr\'esence d'un
pseudogap dans le spectre devient de plus en plus marqu\'ee. Ainsi, cette
m\'ethode renforce notre r\'esultat.


\chapitre{Formalisme}

Dans ce chapitre, nous explicitons les diff\'erentes \'etapes
menant \`a l'approximation auto-coh\'erente \`a deux particules. Pour
notre d\'erivation nous adoptons un formalisme d\'evelopp\'e dans les ann\'ees
50 et 60. Nous r\'ef\'erons le lecteur aux articles de Martin et Schwinger
[\ref{martin}], de Kadanoff et Martin [\ref{kadanoff}], de Baym et Kadanoff
[{\ref{baym1},\ref{baym2}] et de Luttinger [\ref{luttinger1},\ref{luttinger2},%
\ref{luttinger3},\ref{kohn}]. Pour des ouvrages
p\'edagogiques nous sugg\'erons le livre de Mahan [\ref{mahan}] ainsi que
les notes de cours de Tremblay [\ref{ncorps}].

Pour r\'esumer les principales particularit\'es de notre formalisme,
mentionnons que nous utilisons l'approche en temps imaginaire avec
int\'egration sur le contour de Matsubara. De plus, nous employons
la notation compacte [\ref{martin}] o\`u les chiffres arabes sont utilis\'es
pour repr\'esenter un point dans l'espace \`a trois dimensions form\'e par le
r\'eseau cristallin et le temps imaginaire. \'Etant donn\'e que l'on
s'int\'eresse au mod\`ele de Hubbard attractif, nous faisons usage de
la notation matricielle de Nambu.

Le chapitre est s\'epar\'e en quatre sections. Dans la premi\`ere section
nous obtenons des r\'esultats exacts et d\'efinissons les diff\'erentes
quantit\'es utilis\'ees par la suite. Nous consid\'erons la fonction de
Green qui permet de d\'ecrire l'\'etat thermodynamique d'un syst\`eme.
En outre, en pr\'esence d'un champ externe, cette fonction
permet de d\'eduire les propri\'et\'es de transport du syst\`eme.
On peut alors affirmer qu'elle contient toute l'information n\'ecessaire
pour d\'ecrire le syst\`eme. La premi\`ere relation exacte que nous
\'etudierons est l'\'equation du mouvement de la fonction de Green. Pour le
mod\`ele
de Hubbard, cette \'equation nous emm\`ene \`a consid\'erer le hamitonien
en deux parties. La premi\`ere partie est diagonalisable de fa\c con
analytique et assez facilement. La deuxi\`eme introduit la self-\'energie
qui est li\'ee \`a la fonction de Green par la relation de Dyson. En fait,
cette derni\`ere \'equation peut \^etre consid\'er\'ee comme une d\'efinition
de la self-\'energie et est donc applicable \`a tout hamiltonien.

L'\'equation du mouvement permet de faire un lien entre les
propri\'et\'es dites \`a une particule (fonction de Green, self-\'energie)
et la fonction de corr\'elation \`a deux particules. C'est pourquoi
nous pr\'esentons dans la deuxi\`eme section une relation exacte permettant
d'\'evaluer cette fonction de corr\'elation en l'absence de champ externe
et dans la phase d\'esordonn\'ee.
Cette relation est l'\'equation de Bethe-Salpeter. Celle-ci fait appel \`a
un nouvel objet qui est le vertex irr\'eductible. Il sera donc essentiel
d'approximer ce vertex pour \'evaluer la fonction de corr\'elation. Le
lecteur pourra noter que pour la d\'erivation de cette formule nous ne
faisons appel qu'\`a la relation de Dyson et pas \`a l'\'equation du
mouvement. Ainsi l'\'equation de Bethe-Salpeter pourrait s'appliquer \`a
d'autres hamiltoniens.

La troisi\`eme section contient le d\'eveloppement de l'approximation. Pour
ce faire nous suivons un formalisme qui s'apparente \`a celui de Baym et
Kadanoff
[\ref{baym1},\ref{baym2}] en ce sens que nous d\'eveloppons une approximation
pour la self-\'energie en pr\'esence du champ externe. De cette m\^eme
approximation nous estimons la fonction de Green et le vertex irr\'eductible.
\'Etant donn\'e que l'on s'int\'eresse aux propri\'et\'es du mod\`ele
en l'absence de champ externe, notre approximation n'est d\'evelopp\'ee
qu'au premier ordre afin de d\'eterminer le vertex.
Notre approximation pour la self-\'energie fait appel \`a une quantit\'e
inconnue: la valeur moyenne de double occupation (deux \'electrons sur un
m\^eme site du r\'eseau). Suivant la proposition de Vilk et Tremblay
[\ref{vilk1},\ref{vilk2}], nous d\'eterminons cette valeur de fa\c con
auto-coh\'erente en imposant une r\`egle de somme. Par la suite,
utilisant la relation de Bethe-Salpeter et l'\'equation du mouvement, nous
pouvons raffiner l'estimation de la self-\'energie.

Enfin, \`a la derni\`ere section, nous montrons que notre
approximation permet de satisfaire une nouvelle r\`egle de somme, qui est
l'analogue de la r\`egle de somme f, et le th\'eor\`eme
de Mermin-Wagner, selon lequel il est impossible d'avoir une brisure
spontan\'ee de sym\'etrie
continue \`a temp\'erature finie en deux dimensions. Le fait que notre approximation
respecte les moments de la susceptibilit\'e de paires nous permet
de faire un parall\`ele entre l'approche des moments et la r\`egle
de somme choisie pour d\'eterminer la valeur moyenne de double occupation.

\section{Relations exactes}

\soussection{D\'efinitions}

Nous cherchons une approximation pour d\'ecrire le mod\`ele
de Hubbard attractif. Nous savons qu'\`a basse temp\'erature ce mod\`ele
pr\'esente des fluctuations supraconductrices importantes. Il est alors
appropri\'e d'adopter la notation de Nambu o\`u on d\'efinit des
op\'erateurs vectoriels:
$${\mib\Psi}(1)=\pmatrix{\psi_\uparrow(1)\cr\psi_\downarrow^\dagger(1)\cr}
\mskip 50mu{\mib\Psi}^\dagger(1)=\pmatrix{\psi_\uparrow^\dagger(1)\mskip 3mu, &
\psi_\downarrow(1)\cr}\nom{eqop}$$
Les fl\`eches $\uparrow$ et $\downarrow$ servent \`a indiquer les deux
orientations possibles du spin \'electronique. La d\'ependance temporelle de
ces op\'erateurs est d\'efinie de fa\c con conventionnelle par l'op\'erateur
d'\'evolution en temps imaginaire:
$$\eqalign{{\mib\Psi}(1)&\equiv{\mib\Psi}({\bf r}_1,\tau_1)\equiv
e^{\tau_1({\cal H}-\mu\hat{N})}{\mib\Psi}({\bf r}_1)e^{-\tau_1({\cal H}-
\mu\hat{N})}\cr{\mib\Psi}^\dagger(1)&\equiv{\mib\Psi}^\dagger({\bf r}_1,\tau_1)
\equiv e^{\tau_1({\cal H}-\mu\hat{N})}{\mib\Psi}^\dagger({\bf r}_1)
e^{-\tau_1({\cal H}-\mu\hat{N})}\cr}\nom{evolution}$$
o\`u $\hat{N}$ est l'op\'erateur du nombre de particules, $\mu$, le
potentiel chimique et les variables de temps imaginaire sont d\'efinies
dans l'intervalle $(0,\beta)$, $\beta$ \'etant l'inverse de la temp\'erature.
Suivant les d\'efinitions (\ref{evolution}), l'op\'erateur de cr\'eation
${\mib\Psi}^\dagger(\tau)$ ne sera pas le conjugu\'e hermitique de l'op\'erateur
d'annihilation ${\mib\Psi}(\tau)$. Pris au temps $\tau=0$, il le sera et ces
deux op\'erateurs ob\'eiront \`a la r\`egle d'anticommutation:
$$\{{\mib\Psi}_i({\bf r}_1),{\mib\Psi}_j^\dagger({\bf r}_2)\}=\delta_{i,j}
\delta({\bf r}_1-{\bf r}_2)$$

Dans la notation de Nambu la fonction de Green prend une forme matricielle.
$$\eqalign{{\bf G}(1,2;\phi)&=-\langle{\mib\Psi}(1){\mib\Psi}^\dagger(2)
\rangle_\phi\cr
&=-\pmatrix{\langle\psi_\uparrow(1)\psi_\uparrow^\dagger(2)\rangle_\phi &
\langle\psi_\uparrow(1)\psi_\downarrow(2)\rangle_\phi\cr \langle
\psi_\downarrow^\dagger(1)\psi_\uparrow^\dagger(2)\rangle_\phi & \langle
\psi_\downarrow^\dagger(1)\psi_\downarrow(2)\rangle_\phi\cr} \cr}
\nom{eqgreen}$$
On reconna\^\i t le long de la diagonale la fonction usuelle
pour les deux composantes de spin. Hors diagonale, on retrouve les
fonctions de Green anormales pr\'esent\'ees par Abrikosov et al. [\ref{gorkov}].
La moyenne thermodynamique a \'et\'e d\'efinie en pr\'esence d'un champ
source externe:
$$\langle\circ\rangle_\phi=Z^{-1}[\phi]{\rm Tr}\Bigl\lbrace e^{-\beta({\cal H}-
\mu\hat{N})}T_\tau{\rm exp}\bigl(-{\mib\Psi}^\dagger(\overline{1})\phiB(
\overline{1},\overline{2}){\mib\Psi}(\overline{2})\bigr)\circ\Bigr\rbrace
\nom{moyennechamp}$$
o\`u
$$
Z[\phi]={\rm Tr}\left(e^{-\beta({\cal H}-\mu\hat{N})}
T_\tau{\rm exp}\left(-{\mib\Psi}^\dagger(\overline{1})\phiB(\overline{1},
\overline{2}){\mib\Psi}(\overline{2})\right)\right)
\nom{eqpartition}
$$
Pour indiquer que l'on int\`egre sur un point, on place une ligne au-dessus
du chiffre associ\'e \`a ce point (ex.: $\overline{1}$).
L'op\'erateur d'ordonnance temporelle $T_\tau$ est utilis\'e pour
indiquer que les op\'erateurs doivent \^etre ordonn\'es selon leur
indice de temps imaginaire. Il faut d'abord faire le produit direct
puis ordonner, dans chaque \'el\'ement de matrice, les op\'erateurs en suivant
le contour de Matsubara, c'est-\`a-dire en les pla\c cant de la droite vers la
gauche en ordre croissant de temps imaginaire.

Le champ externe est choisi afin de pouvoir \'evaluer les corr\'elations
supraconductrices, c'est-\`a-dire:

$${\phiB}(1,2)=\pmatrix{0 &\phi_{12}(1,2)\cr\phi_{21}(1,2)&0\cr}
\nom{eqchamp}$$
Les indices inf\'erieurs $12$ et $21$ servent pour la notation des deux
\'el\'ements $(1,2)$ et $(2,1)$ de la matrice $\phiB$. Ainsi, nous avons
introduit deux champs distincts qui seront coupl\'es aux deux termes hors
diagonaux de la fonction de Green.
On peut v\'erifier que la d\'eriv\'ee de la fonction de partition,
(\ref{eqpartition}), permet de g\'en\'erer ces termes hors diagonaux.
$${\delta{\rm ln}Z[\phi]\over\delta\phiB(2,1)}=-\pmatrix{0&G_{12}(1,2;\phi)
\cr G_{21}(1,2;\phi)&0\cr}\nom{eqderiv1}$$
o\`u on a d\'efini:
$${\delta\hfill\over\delta\phiB(2,1)}=\pmatrix{0&{\delta\hfill\over\delta
\phi_{21}(2,1)}\cr{\delta\hfill\over\delta\phi_{12}(2,1)}&0\cr}\nom{eqmatdrv}$$
Les indices $21$ et $12$ servent \`a indiquer clairement par rapport \`a
quelle composante du champ $\phi$ la diff\'erentiation est faite. Cette d\'efinition
de la matrice de d\'erivation nous permet d'\'ecrire:
$${\delta\hfill\over\delta\phiB(4,3)}\phiB(1,2)=\delta(1-4)\delta(3-2)
\pmatrix{1&0\cr0&1}$$
o\`u pour appliquer la diff\'erentiation nous devons faire le produit matriciel
et d\'eriver chaque composante \`a tour de r\^ole.

Avant de terminer cette section, il faut remarquer qu'il reste un point \`a
d\'eterminer pour compl\'eter la d\'efinition de la fonction de
Green. Le lecteur aura not\'e, par la
pr\'esence du terme $\mu\hat{N}$, que nous utilisons l'ensemble grand
canonique o\`u le potentiel chimique est d\'etermin\'e afin d'obtenir
le remplissage d\'esir\'e. Cependant, que devient cette condition en
pr\'esence d'un champ externe? Nous choisissons de traiter le potentiel
chimique de la m\^eme fa\c con que la temp\'erature (les deux \'etant des
param\`etres de Lagrange permettant d'ajuster le nombre moyen de particules
et l'\'energie moyenne respectivement). Ainsi, il n'est pas modifi\'e par
la pr\'esence du champ externe et il est ajust\'e pour que le nombre
d'occupation moyen en l'absence de ce champ ait la valeur d\'esir\'ee.

\soussection{\'Equation du mouvement}

Afin d'\'evaluer la fonction de Green d'un mod\`ele donn\'e, il est
courant de consid\'erer l'\'equation du mouvement de cette fonction.
On proc\`ede donc en diff\'erentiant ${\bf G}(\phi)$
par rapport au temps. Trois termes sont g\'en\'er\'es par
cette d\'erivation. Les deux premiers proviennent de l'ordonnance temporelle:
il y a l'ordonnance d'un vecteur de Nambu par rapport \`a l'autre (voir
\'equation (\ref{eqgreen})) et il y
a l'ordonnance temporelle par rapport au champ externe. Le
troisi\`eme terme vient
de la d\'ependance temporelle de l'op\'erateur (\ref{evolution}).
On peut \'ecrire la d\'erivation ainsi:
$$\eqalign{{\partial{\bf G}(1,2;\phi)\over\partial\tau_1}=&-\langle\{{\mib\Psi}
({\bf r}_1,\tau_1),{\mib\Psi^\dagger}({\bf r}_2,\tau_1)\}\rangle_\phi\delta
(\tau_1-\tau_2)\cr&{}+\bigl\langle[{\mib\Psi^\dagger}(\overline{3})\phiB
(\overline{3},\overline{4}){\mib\Psi}(\overline{4}),{\mib\Psi}(1)]
{\mib\Psi^\dagger}(2)\bigr\rangle_\phi\delta(\tau_1-\tau_{\overline{3}})\cr&{}
-\Bigl\langle{\partial{\mib\Psi}(1)\over\partial\tau_1}{\mib\Psi^\dagger}(2)
\Bigr\rangle_\phi\cr}\nom{eqmouvement}$$
La notation utilis\'ee
dans l'\'equation (\ref{eqmouvement}) est un peu ambigu\"e. Il faut bien
comprendre ce que
l'on veut repr\'esenter par l'anticommutation entre bra et ket. Il faut dans un
premier temps faire le produit direct des deux \'el\'ements dans l'ordre o\`u
ils apparaissent dans l'anticommutateur, puis appliquer
l'anticommutation \'el\'ement par \'el\'ement. Cette
ambigu\"\i t\'e n'appara\^\i t que dans le premier terme du c\^ot\'e droit
de l'\'equation (\ref{eqmouvement}).

Le troisi\`eme terme peut \^etre \'evalu\'e en utilisant l'\'equation
de Heisenberg:
$${\partial{\mib\Psi}(1)\over\partial\tau_1}=[{\cal H}-\mu\hat{N},{\mib\Psi}(1)]
$$

En appliquant (\ref{eqmouvement}) au mod\`ele de Hubbard (\ref{hubbard}), on
obtient:
$$\eqalign{-{\partial{\bf G}(1,2;\phi)\over\partial\tau_1}=&\delta(1-2)\cr&{}
+\biggl[\delta(
\tau_1-\tau_{\overline{3}})\pmatrix{-t({\bf r}_1,{\bf r}_{\overline{3}})-\mu&
0\cr0&t({\bf r}_{\overline{3}},{\bf r}_1)+\mu}+\phiB(1,\overline{3})\biggr]
{\bf G}(\overline{3},2)\cr&{}+U\pmatrix{\langle\psi_\uparrow(1)
\psi_\downarrow(1)\psi_\downarrow^\dagger(1^+)\psi_\uparrow^\dagger(2)
\rangle_\phi&\langle\psi_\uparrow(1)\psi_\downarrow(1)\psi_\downarrow^\dagger
(1^+)\psi_\downarrow(2)\rangle_\phi\cr\langle\psi_\downarrow^\dagger(1)
\psi_\uparrow^\dagger(1)\psi_\uparrow(1^-)\psi_\uparrow^\dagger(2)
\rangle_\phi&\langle\psi_\downarrow^\dagger(1)\psi_\uparrow^\dagger(1)
\psi_\uparrow(1^-)\psi_\downarrow(2)\rangle_\phi}\cr}\nom{eqmouhub}$$
\'Etant donn\'e que diff\'erents op\'erateurs apparaissant dans une moyenne
sont \'evalu\'es \`a une m\^eme valeur de temps imaginaire, il devient
n\'ecessaire d'indiquer explicitement l'ordre selon lequel il faut ordonner
ces op\'erateurs. Pour ce faire nous pla\c cons un indice sup\'erieur
${}^+$ ou ${}^-$ qui indique que l'op\'erateur doit \^etre plac\'e plus
\`a gauche ou plus \`a droite lors de l'arrangement en ordre.

Le dernier terme de la relation (\ref{eqmouhub}) provient du terme d'interaction
dans le mod\`ele de Hubbard. En l'absence de ce terme, il serait facile de
d\'eterminer la fonction de Green. C'est pourquoi nous allons traiter
s\'epar\'ement ce dernier terme en introduisant la self-\'energie: 
$${\mib\Sigma}(1,2;\phi)=U\pmatrix{\langle\psi_\uparrow(1)\psi_\downarrow(1)
\psi_\downarrow^\dagger(1^+)\psi_\uparrow^\dagger(\overline{3})\rangle_\phi&
\langle\psi_\uparrow(1)\psi_\downarrow(1)\psi_\downarrow^\dagger(1^+)
\psi_\downarrow(\overline{3})\rangle_\phi\cr\langle\psi_\downarrow^\dagger(1)
\psi_\uparrow^\dagger(1)\psi_\uparrow(1^-)\psi_\uparrow^\dagger(\overline{3})
\rangle_\phi&\langle\psi_\downarrow^\dagger(1)\psi_\uparrow(1)\psi_\uparrow
(1^-)\psi_\downarrow(\overline{3})\rangle_\phi}{\bf G}^{-1}
(\overline{3},2;\phi)\nom{self}$$
On voit par cette relation que le calcul de la self-\'energie fera appel
aux fonctions de corr\'elation.
Cette d\'efinition, avec (\ref{eqmouhub}), nous donne la relation de Dyson:
$$\boxEq{\bigl({\bf G}_0^{-1}(1,\overline{3})-\phiB(1,\overline{3})-
{\mib\Sigma}(1,\overline{3};\phi)\bigr){\bf G}(\overline{3},2;\phi)=
\deltaB(1-2)}\nom{eqdyson}$$
o\`u on a d\'efini ${\bf G}_0^{-1}$ par les autres termes pr\'esents dans
l'\'equation du mouvement:
$${\bf G}_0^{-1}(1,2)=-\delta(1-2){\partial\hfill\over\partial\tau_1}-
\delta(\tau_1-\tau_2)\pmatrix{-t({\bf r}_1,{\bf r}_2)-\mu&0\cr0&t({\bf r}_2,
{\bf r}_1)+\mu}$$

Ce r\'esultat justifie notre choix selon lequel le potentiel chimique est
fix\'e par le remplissage en l'absence du champ externe. Ainsi, $G_0$ est
ind\'ependant de ce champ externe.

Jusqu'\`a maintenant, nous nous sommes employ\'e \`a regarder ce que l'on
obtenait en \'etudiant la diff\'erentiation de ${\bf G}$ par rapport \`a un de
ses arguments. Si l'on avait consid\'er\'e la diff\'erentiation par rapport
\`a l'autre argument, la relation de Dyson aurait pris la forme suivante:
$${\bf G}(1,\overline{3};\phi)\bigl({\bf G}_0^{-1}(\overline{3},2)-\phiB(
\overline{3},2)-{\mib\Sigma}(\overline{3},2;\phi)\bigr)=\deltaB(1-2)
\nom{eqdyson2}$$

\'Etant donn\'e que la relation de Dyson nous servira \`a estimer la
fonction de Green \`a partir d'une approximation de la self-\'energie, on
peut se demander si le r\'esultat serait le m\^eme si au lieu de consid\'erer
la relation (\ref{eqdyson}), nous utilisions (\ref{eqdyson2}). La r\'eponse
\`a cette question nous vient simplement de l'alg\`ebre lin\'eaire: tant
que la fonction de Green ne sera pas singuli\`ere son inverse sera unique.

\section{\'Equation de Bethe-Salpeter}\nom{dembs}

\soussection{D\'efinition}

On a vu de l'\'equation du mouvement que l'\'evaluation de la self-\'energie
passe par le calcul d'une fonction de corr\'elation \`a quatre points. Or,
il est possible de montrer que cette fonction de corr\'elation est en fait
la r\'eponse de la fonction de Green au champ externe. Ainsi, on d\'efinit
$$\eqalign{\chiB(1,2,3,4;\phi)\equiv&-{\delta\hfill\over\delta\phiB(2,1)}
{\bf G}(3,4;\phi)\cr=&\pmatrix{\langle\psi_\uparrow(1)\psi_\downarrow(2)
\psi_\downarrow^\dagger(3)\psi_\uparrow^\dagger(4)\rangle_\phi&
\langle\psi_\uparrow(1)\psi_\downarrow(2)\psi_\downarrow^\dagger(3)
\psi_\downarrow(4)\rangle_\phi\cr\langle\psi_\downarrow^\dagger(1)
\psi_\uparrow^\dagger(2)\psi_\uparrow(3)\psi_\uparrow^\dagger(4)
\rangle_\phi&\langle\psi_\downarrow^\dagger(1)\psi_\uparrow^\dagger(2)
\psi_\uparrow(3)\psi_\downarrow(4)\rangle_\phi}\cr&{}-{\bf G}_{\rm HD}
(1,2;\phi){\bf G}(3,4;\phi)\cr}
\nom{correlation}$$
o\`u l'indice ${}_{\rm HD}$ indique que seule la partie hors diagonale de
la matrice est conserv\'ee, les \'el\'ements de la partie diagonale \'etant
remplac\'es par z\'ero. Nous utiliserons aussi l'indice
${}_{\rm D}$ qui ne conserve que la partie diagonale de la matrice.
Le dernier terme de l'\'equation (\ref{correlation})
vient de la pr\'esence de $Z^{-1}[\phi]$ dans la d\'efinition de la
moyenne thermodynamique (\ref{moyennechamp}). La matrice de d\'erivation
utilis\'ee est la m\^eme que celle introduite pr\'ec\'edemment (\ref{eqmatdrv}).
La fa\c con de l'appliquer consiste \`a faire la multiplication matricielle
dans un premier temps puis d'appliquer la d\'erivation appropri\'ee \`a chaque
\'el\'ement. Nous illustrons par l'exemple de la d\'erivation d'une matrice
$A(\phi)$ quelconque:
$${\delta\hfill\over\delta\phiB(1,2)}A(\phi)\equiv\pmatrix{{\delta A_{21}(\phi)
\over\delta\phi_{21}(1,2)}&{\delta A_{22}(\phi)\over\delta\phi_{21}(1,2)}\cr
{\delta A_{11}(\phi)\over\delta\phi_{12}(1,2)}&{\delta A_{12}(\phi)\over
\delta\phi_{12}(1,2)}}$$

La comparaison avec (\ref{self}), nous montre que la self-\'energie sera
alors donn\'ee par:
$$\boxEq{{\mib\Sigma}(1,2;\phi)=U\chiB(1,1,1^\pm,\overline{3};\phi)
{\bf G}^{-1}(\overline{3},2;\phi)+U\delta(1^\pm-2){\bf G}_{\rm HD}(1,1;\phi)}
\nom{selfchi}$$
Nous avons utilis\'e comme indice sup\'erieur ${}^\pm$ pour indiquer que
dans la ligne sup\'erieure de $\chiB$ le premier op\'erateur sera consid\'er\'e
\`a un temps infinit\'esimalement sup\'erieur \`a $\tau_1$ tandis que dans
la deuxi\`eme ligne de la matrice, il sera consid\'er\'e comme \'etant \`a
un temps inf\'erieur.

La d\'efinition de la fonction de corr\'elation (\ref{correlation}), comme
\'etant la r\'eponse de la fonction de Green au champ externe, nous indique
comment nous pourrions l'estimer. De la d\'efinition:
$${\bf G}(1,\overline{3};\phi){\bf G}^{-1}(\overline{3},2;\phi)=\delta(1-2)
\nom{inverseg}$$
nous voyons que la d\'ependance de ${\bf G}$ sur $\phi$ devrait compenser
exactement celle de ${\bf G}^{-1}$. Au premier ordre en $\phi$, la fonction
de Green est donn\'ee par:
$${\bf G}(1,2;\phi)={\bf G}(1,2)-\phiB(\overline{4},\overline{3})\chiB(
\overline{3},\overline{4},1,2)+O(\phiB^2)\nom{glinear}$$
Il est \`a noter que le terme correspondant \`a la transpos\'ee de $\phiB$
n'appara\^\i t pas dans cette expression car nous sommes int\'eress\'es
par la phase normale o\`u
$$\biggl({\delta\hfill\over\delta\phiB(4,3)}\biggr)^T{\bf G}(1,2)\Biggr
\vert_{\phi=0}=0$$
Ainsi, la relation de Bethe-Salpeter que nous d\'erivons ne
sera valide que dans l'\'etat d\'esordonn\'e o\`u les termes hors diagonaux
des matrices ${\bf G}$ et $\chiB$ sont nuls.

Pour ${\bf G}^{-1}$, on peut utiliser l'\'equation de Dyson (\ref{eqdyson}),
pour montrer qu'au premier ordre on a:
$${\bf G}^{-1}(1,2;\phi)={\bf G}^{-1}(1,2)-\phiB(1,2)-\phiB(\overline{4},
\overline{3})\biggl[{\delta\hfill\over\delta\phiB(\overline{4},\overline{3})}
\Sigma(1,2;\phi)\biggr]_{\phi=0}+O(\phi^2)\nom{gilinear}$$
Encore une fois, il devrait appara\^\i tre un terme suppl\'ementaire.
Cependant, on peut montrer que ce terme s'annule au premier ordre en $\phi$:
$$\biggl({\delta\hfill\over\delta\phiB(4,3)}\biggr)^T\Sigma(1,2)
\biggr\vert_{\phi=0}=0$$
o\`u ${}^T$ indique la transpos\'ee de la matrice.

Introduisant (\ref{glinear}) et (\ref{gilinear}) dans (\ref{inverseg}), on
trouve une relation pour $\chi$ en annulant le coefficient du terme
proportionnel \`a $\phi$:
$$\eqalign{
\chiB(1,2,3,4)=&\sigma_x{\bf G}(3,2)\sigma_x{\bf G}(1,4)\cr&{}-\sigma_x
{\bf G}(3,\overline{5})\sigma_x\biggl[{\partial\hfill\over\partial
\phiB(2,1)}{\mib\Sigma}(\overline{5},\overline{6})\biggr]_{\phi=0}
{\bf G}(\overline{6},4)\cr}\nom{bsprel}$$
o\`u $\sigma_x$ est la matrice de Pauli correspondant \`a la composante $x$
du spin. Cette
matrice provient de la commutation de la matrice diagonale ${\bf G}$
et de la matrice hors diagonale $\phiB$:
$${\bf G}(1,2)\phiB(4,3)=\phiB(4,3)\sigma_x{\bf G}(1,2)\sigma_x$$
La matrice ${\bf G}$ est diagonale par le fait qu'on
l'\'evalue en champ externe nul et \`a temp\'erature finie (pas
d'ordre \`a longue port\'ee).

\soussection{Transformation de Legendre}

On voit de la relation (\ref{bsprel}) que la d\'etermination de la
fonction de corr\'elation passe par l'estimation de la variation de
la self-\'energie par rapport au champ externe. C'est ce terme qui
permet d'introduire le vertex irr\'eductible.
Mais pour y arriver, nous allons consid\'erer une transformation de
Legendre. L'utilit\'e de cette transformation s'explique par le fait que la
d\'ependance de la self-\'energie sur $\phi$ se fait au travers de la
fonction de Green ${\mib\Sigma}(\phi)={\mib\Sigma}({\cal G})\vert_{{\cal G}=
G_{\rm HD}(\phi)}$. Ainsi, de fa\c con g\'en\'erale, on peut d\'efinir
la transformation de Legendre de la fonction de partition:
$$W[{\cal G}]=\Bigl(-{\rm ln}Z[\phi]-{\rm Tr}\bigl({\bf G}_{\rm HD}
(\overline{1},\overline{2};\phi)\phiB(\overline{2},\overline{1})\bigr)
\Bigr)_{G_{\rm HD}[\phi]={\cal G}}\nom{legendre}$$
avec la condition
$${\delta{\rm ln}Z[\phi]\over\delta\phiB(2,1)}\biggr\vert_{G_{\rm HD}(\phi)=
{\cal G}}=-{\calb G}(1,2)\nom{legendrecond}$$
L'indice apr\`es la derni\`ere parenth\`ese du c\^ot\'e droit de l'\'egalit\'e
signifie que la fonction \`a l'int\'erieur des parenth\`eses est \'evalu\'ee
\`a une valeur de $\phiB$ telle que les composantes hors diagonales de ${\bf G}
(\phi)$ soient \'egales au nouveau champ introduit ${\calb G}$. Cette
condition ne s'applique que sur les composantes hors diagonales car comme
le champ $\phiB$ ne poss\'edait que deux composantes non nulles, il doit en
\^etre de m\^eme du champ ${\calb G}$, celui-ci \'etant introduit par une
transformation de Legendre (Les d\'etails de cette transformation sont donn\'es
en annexe \ref{anex2}).
Pour montrer que toute la d\'ependance
de ${\mib\Sigma}$ sur $\phiB$ peut s'exprimer \`a travers ${\calb G}$ (voir
annexe \ref{anex2}), il
suffit d'examiner la d\'erivation de $W$ par rapport \`a ${\calb G}$:
$$\eqalign{{\delta W[{\cal G}]\over\delta{\calb G}(2,1)}=&-{\mib\Phi}(1,2;
{\cal G})-\biggl({\delta\hfill\over\delta{\calb G}(2,1)}{\mib\Phi}(
\overline{3},\overline{4};{\cal G})\biggr)\biggl({\delta\hfill\over\delta\phiB
(\overline{3},\overline{4})}{\rm ln}Z[\phi]+{\calb G}(\overline{4},
\overline{3})\biggr)_{\phi=\Phi({\cal G})}\cr{}=&-{\mib\Phi}(1,2;{\cal G})\cr}
\nom{fonctiong}$$
o\`u la matrice de d\'erivation a \'et\'e d\'efinie par:
$${\delta\hfill\over\delta{\calb G}(2,1)}=\pmatrix{0&{\delta\hfill\over
\delta{\cal G}_{21}(2,1)}\cr{\delta\hfill\over\delta{\cal G}_{12}(2,1)}&0}$$
et ${\mib\Phi}({\cal G})$ est la valeur du champ $\phi$ obtenue en posant la
condition
(\ref{legendrecond}). Cette derni\`ere pr\'esuppose que pour une valeur de
$\phiB$ donn\'ee une seule valeur possible de ${\bf G}_{\rm HD}(\phi)$ soit
donn\'ee par la
d\'erivation de la fonction de partition (\ref{eqderiv1}) et que pour une valeur
de ${\bf G}_{\rm HD}(\phi)={\calb G}$ donn\'ee une seule et unique valeur de
$\phiB$ permettre d'obtenir cette fonction de Green.

Utilisant la transformation de Legendre introduite, on est \`a m\^eme
d'\'evaluer la variation de la self-\'energie. Nous appliquons la r\`egle de
d\'erivation en cha\^\i ne qui est d\'emontr\'ee en annexe \ref{acalcul}
$$\biggl[{\delta\hfill\over\delta\phiB(4,3)}{\mib\Sigma}(1,2;\phi)\biggr]_{
\phi=0}=\biggl[{\delta\hfill\over\delta\phiB(4,3)}{\bf G}_{\rm HD}(\overline{5},
\overline{6};\phi)\biggr]_{\phi=0}\biggl[{\delta\hfill\over\delta{\calb G}
(\overline{5},\overline{6})}{\mib\Sigma}(1,2;{\cal G})\biggr]_{{\cal G}=0}$$
Il semble alors opportun d'introduire le vertex irr\'eductible qui est
d\'efini par:
$${\mib\Gamma}(1,2,3,4;{\cal G})={\delta\hfill\over\delta{\calb G}(2,1)}{
\mib\Sigma}(3,4;{\cal G})\nom{vertex}$$
Ceci nous emm\`ene \`a la relation de Bethe-Salpeter:
$$\boxEq{\eqalign{\chiB(1,2,3,4)=&-\sigma_x{\bf G}(3,2)\sigma_x{\bf G}(1,4)
\cr&{}+\sigma_x{\bf G}(3,\overline{5})\sigma_x\chiB(1,2,\overline{7},
\overline{8}){\mib\Gamma}(\overline{8},\overline{7},\overline{5},
\overline{6}){\bf G}(\overline{6},4)\cr}}\nom{bs}$$

\section{Estimation des vertex irr\'eductibles}\nom{svertex}

\soussection{Approximation de la self-\'energie}

Afin d'estimer les vertex irr\'eductibles nous allons utiliser une approche
qui s'apparente au formalisme
de Baym [\ref{baym2}] et faire dans un premier temps une approximation
de la self-\'energie en pr\'esence d'un champ externe.
Puis le vertex sera estim\'e en utilisant la d\'efinition (\ref{vertex}).
Il faudra donc exprimer ${\mib\Sigma}$ en fonction de ${\calb G}$.
\'Etant donn\'e que l'on s'int\'eresse uniquement
\`a la phase normale o\`u les termes hors diagonaux des matrices sont nuls,
on peut se contenter d'\'evaluer ${\mib\Sigma}$ au premier ordre en
${\calb G}$.

Comme approximation,
Vilk et Tremblay [\ref{vilk2}] ont propos\'e dans l'esprit de l'approche de
Singwi et al. [\ref{singwi}], de consid\'erer la self-\'energie au premier
ordre en d\'eveloppement perturbatif et de corriger cette approximation
par un facteur multiplicatif qui soit tel que l'approximation puisse, en
principe, \^etre exacte lorsque l'on consid\`ere les corr\'elations locales.
Pour bien indiquer que ce sera l\`a notre premier
niveau d'approximation nous placerons un indice sup\'erieur ${}^{(1)}$ \`a
toutes les quantit\'es approxim\'ees. L'approximation est la suivante:
$${\mib\Sigma}^{(1)}(1,\overline{3};{\cal G}){\bf G}(\overline{3},2;{\cal G})
\approx U{\mib\Lambda}(1,{\cal G}){\bf G}(1,2;{\cal G})\nom{approxim}$$
La matrice ${\mib\Lambda}$ sera d\'etermin\'ee par la valeur
des fonctions de corr\'elations locales. Nous pouvons par exemple poser:
$${\mib\Sigma}^{(1)}(1,\overline{3};{\cal G}){\bf G}^{(1)}(\overline{3},1^+;
{\cal G})=U\chiB^{(1)}(1,1,1^\pm,1^{++};{\cal G})+U{\calb G}(1,1)
{\bf G}^{(1)}(1,1^+)\nom{localplus}$$
L'indice sup\'erieur ${}^{++}$ sert \`a indiquer l'ordonnance de l'op\'erateur
correspondant par rapport aux op\'erateurs \'evalu\'es au m\^eme temps $\tau_1$
et \'egalement \`a $\tau_1^+$.
Nous avons clairement indiqu\'e qu'il faut exprimer $\chiB$ et ${\bf G}$
comme des fonctions
de ${\calb G}$ afin d'obtenir $\mib\Lambda$ comme une fonction de
${\calb G}$. Encore une fois on peut se contenter du premier ordre.
Pour la fonction de Green, c'est assez simple:
$${\bf G}^{(1)}(1,2;{\cal G})={\bf G}^{(1)}(1,2)+{\calb G}(1,2)$$
Pour la fonction de corr\'elation,
on utilise le principe de Pauli consid\'er\'e localement qui, par le fait que
l'on ne s'int\'eresse qu'aux termes locaux, permet de
simplifier l'expression:
$$\eqalign{\chiB^{(1)}(1,1,1^\pm,1^{++};{\cal G})=&\pmatrix{\langle n_\uparrow
n_\downarrow\rangle_{\cal G}&-\langle\psi_\uparrow(1)\psi_\downarrow(1)
\rangle_{\cal G}\cr0&-\langle n_\uparrow(1-n_\downarrow)\rangle_{\cal G}}-
{\calb G}(1,1){\bf G}^{(1)}(1,1^+)\cr{}=&\pmatrix{\langle n_\uparrow
n_\downarrow\rangle&0\cr0&\langle n_\uparrow n_\downarrow\rangle-\ffrac{n}{2}}
+{\calb G}_{12}(1,1)\sigma_{+}-{\calb G}(1,1){\bf G}^{(1)}(1,1^+)}$$

Ainsi, la relation (\ref{localplus}) nous donne une relation permettant
d'approximer ${\mib\Lambda}$ apparaissant en (\ref{approxim}) et donc la
self-\'energie ${\mib\Sigma}$.
Cependant, on peut se demander si l'approximation serait la m\^eme si l'on
imposait la relation suivante:
$${\mib\Sigma}^{(1)}(1,\overline{3};{\cal G}){\bf G}^{(1)}(\overline{3},1^-;
{\cal G})=U\chiB^{(1)}(1,1,1^\pm,1^{--};{\cal G})+U{\calb G}(1,1)
{\bf G}^{(1)}(1,1^-)\nom{localmoins}$$
On peut montrer que ces deux approximations donnent le m\^eme r\'esultat
pour ce qui est de la partie diagonale du vertex irr\'eductible obtenu par
d\'eriv\'ee fonctionnelle. Cependant, m\^eme apr\`es
avoir \'eteint le champ externe, la partie diagonale de la self-\'energie
demeure diff\'erente selon l'approximation choisie. Afin de d\'eterminer
l'approximation qui serait la plus appropri\'ee,
consid\'erons ces relations exprim\'ees comme des sommes des composantes
de Fourier. On obtient pour la partie gauche de l'\'egalit\'e de
(\ref{localplus}):
$$\lim_{\eta\rightarrow 0^+}{T\over N}\sum_{\tilde{k}}{\mib\Sigma}^{(1)}
(\tilde{k};{\cal G}){\bf G}^{(1)}(\tilde{k};{\cal G})e^{-ik_n\eta}$$
Le facteur de convergence est essentiel car aux hautes fr\'equences la
self-\'energie est
constante, $\lim_{k_n\rightarrow\infty}\Sigma({\bf k},ik_n)\sim U\ffrac{n}{2}$
tandis que la fonction de Green d\'ecro\^\i t en loi de puissance
$\lim_{k_n\rightarrow\infty} G({\bf k},ik_n)\sim{1\over ik_n}$. Cette lente
convergence de la somme nous indique que les composantes \`a hautes
fr\'equences contribuent de fa\c con importante \`a la somme.
De m\^eme, lorsqu'on consid\`ere
la transform\'ee de Fourier de (\ref{localmoins}), on obtient un r\'esultat
identique \`a l'exception du signe dans l'exponentiel du facteur de
convergence. Cette diff\'erence de signe nous indique que si l'on
consid\'erait la somme
de (\ref{localplus}) et de (\ref{localmoins}), le facteur de convergence
ne serait plus n\'ecessaire. On le v\'erifie par le fait que:
$$\lim_{k_n\rightarrow\infty}\Bigl({\mib\Sigma}({\bf k},ik_n;{\cal G})
{\bf G}({\bf k},ik_n;{\cal G})+{\mib\Sigma}({\bf k},-ik_n;{\cal G}){\bf G}
({\bf k},-ik_n;{\cal G})\Bigr)\propto{1\over(ik_n)^2}$$
Ainsi, pour la somme sur les fr\'equences de Matsubara, une contribution
plus importante vient des basses fr\'equences lorsqu'on consid\`ere
la somme de (\ref{localplus}) et de (\ref{localmoins}).
C'est pourquoi nous choisissons ${\mib\Lambda}$ tel que:
$$\eqalign{{\mib\Lambda}(1,{\cal G})=&{\calb G}(1,1)+\ffrac{1}{2}\Bigl(
\chiB^{(1)}(1,1,1^\pm,1^{++};{\cal G})\bigl({\bf G}^{(1)}(1,1^+;{\cal G})
\bigr)^{-1}\hfill\cr&\hfill{}+\chiB^{(1)}(1,1,1^\pm,1^{--};{\cal G})\bigl(
{\bf G}^{(1)}(1,1^-;{\cal G})\bigr)^{-1}\Bigr)\cr}\nom{facteurlocal}$$

De ce r\'esultat, on est \`a m\^eme de d\'eduire la self-\'energie et sa
d\'ependance au premier ordre en ${\calb G}$. Ainsi la d\'efinition
du vertex irr\'eductible (\ref{vertex}) nous permet de cr\'eer un lien
entre ce vertex et les corr\'elations locales. Introduisant l'\'equation
(\ref{facteurlocal}) dans l'approximation pour la self-\'energie
(\ref{approxim}), on obtient, apr\`es avoir pos\'e $\phiB=0$
$$\boxEq{\eqalign{{\mib\Sigma}^{(1)}(1,2)=&U\delta(1-2){(\ffrac{1}{2}-
\langle n_\uparrow\rangle)\langle n_\uparrow n_\downarrow\rangle+
\ffrac{1}{2}\langle n_\uparrow\rangle\langle n_\downarrow\rangle\over
\langle n_\uparrow\rangle(1-\langle n_\uparrow\rangle)}\sigma_z\cr{}=&
\ffrac{1}{2}\Bigl(U-(1-n)U_{pp}\Bigr)\delta(1-2)\sigma_z}}
\nom{selfun}$$
et en utilisant la d\'efinition du vertex irr\'eductible (\ref{vertex}):
$$\boxEq{\eqalign{{\mib\Gamma}^{(1)}(1,2,3,4)=&U\delta(1-2)\delta(1-3)
\delta(1-4){\langle n_\uparrow(1-n_\downarrow)\rangle\over\langle
n_\downarrow\rangle\langle 1-n_\uparrow\rangle}{\bf 1}\cr{}=&U_{pp}
\delta(1-2)\delta(1-3)\delta(1-4){\bf 1}}}
\nom{vertexun}$$
o\`u $\sigma_z$ est la matrice de Pauli correspondant \`a la composante $z$
du spin. On a introduit la variable $U_{pp}$ qui correspond \`a la valeur
renormalis\'ee du vertex.

Malgr\'e que nous ayons d\'emontr\'e une fa\c con de d\'eriver une approximation
pour la self-\'energie et le vertex, ces relations devraient \^etre
consid\'er\'ees comme des ansatz qui seront utilis\'es dans le calcul des
fonctions de Green \`a une et \`a deux particules.

\soussection{D\'etermination de la fonction de corr\'elation}\nom{corrpaires}

Comme on s'int\'eresse
\`a la phase normale sans champ externe, toutes les matrices sont diagonales
et nous pouvons \'evaluer les composantes $(1,1)$ ind\'ependamment des
composantes $(2,2)$. C'est pourquoi \`a partir de cette section et pour le
reste du chapitre nous nous concentrerons sur les composantes $(1,1)$ de chaque
matrice. Nous ne l'indiquons pas de fa\c con explicite afin de ne pas
alourdir la notation. Toutefois, le lecteur attentif, remarquera que nous
n'utiliserons plus les caract\`eres gras.

Le calcul de $\Sigma$ et de $\Gamma$
nous permet d'estimer la fonction de corr\'elation.
Mais avant de pouvoir le faire nous devons d\'eterminer la fonction de Green
en utilisant la relation de Dyson, (\ref{eqdyson}). Pour la d\'eterminer,
nous proc\'ederons de la m\^eme fa\c con que dans le formalisme de Baym et
Kadanoff [\ref{baym1},\ref{baym2}] en utilisant la m\^eme self-\'energie
que celle que nous avons utilis\'ee pour d\'eriver le vertex irr\'eductible.
On peut ainsi dire que l'on consid\`ere la fonction de Green et le vertex
irr\'eductible \`a un m\^eme niveau d'approximation.

On peut remarquer que
suivant notre premi\`ere approximation, la self-\'energie n'est en fait
qu'une correction du potentiel chimique.
En appliquant une
transformation de Fourier on peut diagonaliser les termes sans
interactions. Ce faisant, on  obtient la fonction de Green:
$$\boxEq{G^{(1)}(\tilde{k})={1\over ik_n-\epsilon({\bf k})+\mu^{(1)}-
\Sigma^{(1)}}}\nom{greenun}$$
o\`u $\tilde{k}=({\bf k},ik_n)$ et $k_n=(2n+1)\pi T$ est une fr\'equence
de Matsubara fermionique avec $n\in(-\infty,+\infty)$. Nous avons plac\'e
un indice sup\'erieur ${}^{(1)}$ au potentiel chimique pour indiquer que
celui-ci sera ajust\'e afin que la fonction de Green $G^{(1)}$ ait le bon
remplissage, c'est-\`a-dire tel que\footnote{${}^1$}{Cette condition accompagn\'ee
du fait que la self-\'energie, $\Sigma^{(1)}$ n'est qu'un d\'eplacement
du potentiel
chimique nous montre que la valeur exacte de cette derni\`ere quantit\'e n'aura
pas d'importance dans l'estimation de $G^{(1)}$.}:
$$\lim_{\eta\rightarrow 0^-}{T\over N}\sum_{\tilde{k}}G^{(1)}(\tilde{k})
e^{-ik_n\eta}=n\nom{mu1}$$

Il est int\'eressant de v\'erifier le cas demi-rempli
o\`u la valeur du potentiel chimique est connue exactement, $\mu=U/2$.
On sait que pour que le nombre moyen de particules correspondant \`a $G^{(1)}$
soit $n=1$, il faut avoir $\mu^{(1)}-\Sigma^{(1)}=0$. Or, \`a demi-remplissage
la self-\'energie obtenue, \'eq.(\ref{selfun}), est donn\'ee par
$\Sigma^{(1)}=U/2$. Donc, le potentiel
chimique prend sa valeur exacte $\mu^{(1)}=U/2$.

Maintenant nous pouvons d\'eterminer la fonction de corr\'elation. \'Etant
donn\'e que dans notre approximation le vertex irr\'eductible est
local et statique, \'eq.(\ref{vertexun}),
nous pouvons nous restreindre \`a l'\'evaluation des fonctions de corr\'elation
\`a deux points:
$\chiB_{11}(1,1,2,2)=\chi({\bf r}_1-{\bf r}_2,\tau_1-\tau_2)$. Cette fonction
correspond \`a la fonction de corr\'elation pour les paires locales.
En utilisant la relation de Bethe-Salpeter, (\ref{bs}), on peut montrer
qu'apr\`es avoir appliqu\'e la transformation de Fourier, on obtient:
$$\boxEq{\chi^{(1)}(\tilde{q})={\chi_0^{(1)}(\tilde{q})\over 1+\Gamma^{(1)}
\chi_0^{(1)}(\tilde{q})}}\nom{chiun}$$
o\`u
$$\chi^{(1)}(\tilde{q})=\int d^2{\bf r}\int_0^\beta d\tau\chi({\bf r},\tau)
e^{iq_n\tau-i{\bf q}\cdot{\bf r}}$$
avec $q_n=2n\pi T$ une fr\'equence de Matsubara bosonique. La fonction
de corr\'elation obtenue est de la m\^eme forme que celle obtenue d'une
approximation de type matrice-T \`a l'exception du fait que le vertex est
renormalis\'e. De plus, le num\'erateur est donn\'e par
$$\chi_0^{(1)}(\tilde{q})={T\over N}\sum_{\tilde{k}}G^{(1)}(-\tilde{k}+
\tilde{q})G^{(1)}(\tilde{k})\nom{chi0un}$$
Comme on peut le remarquer, $\chi_0^{(1)}$ a la m\^eme forme que la fonction de
corr\'elation en l'absence d'interaction. Cependant, il est \`a noter que
la fonction de Green qui y appara\^\i t est la fonction de Green corrig\'ee
par la premi\`ere estimation de la self-\'energie.

\soussection{D\'etermination de la double occupation}\nom{doubleoccup}

Le d\'eveloppement pr\'esent\'e pr\'ec\'edemment nous a permis d'\'ecrire
le vertex irr\'eductible comme une fonction de la valeur moyenne d'occupation
double (\ref{vertexun}). Cependant, cette quantit\'e est inconnue.
Diff\'erentes approches permettent de r\'egler ce probl\`eme: on peut, par
exemple, consid\'erer la valeur obtenue par simulation Monte Carlo, ou,
comme nous le verrons dans le chapitre suivant, celle
obtenue d'un calcul champ moyen \`a temp\'erature
nulle. La solution propos\'ee ici consiste \`a consid\'erer une r\`egle de
somme afin de fermer le syst\`eme d'\'equations sans appel \`a d'autre
information ext\'erieure. Ce choix nous assure ainsi
que notre approximation respectera cette r\`egle de somme.

La r\`egle de somme que nous allons consid\'erer donne une autre relation
liant le vertex irr\'eductible \`a la valeur moyenne de double occupation.
Cette r\`egle de somme d\'ecoule du th\'eor\`eme de fluctuation-dissipation.
$$\lim_{\eta\rightarrow 0^-}{T\over N}\sum_{\tilde{q}}
\chi(\tilde{q})e^{-iq_n\eta}=\langle n_\uparrow n_\downarrow\rangle$$
\`A ce niveau, on substitue des expressions approximatives qui permettent
d'obtenir $\langle n_\uparrow n_\downarrow\rangle$ de fa\c con auto-coh\'erente.
$$\boxEq{
\lim_{\eta\rightarrow 0^-}{T\over N}\sum_{\tilde{q}}{\chi_0^{(1)}
(\tilde{q})\over 1+\Gamma^{(1)}\chi_0^{(1)}(\tilde{q})}e^{-iq_n\eta}=
\langle n_\uparrow n_\downarrow\rangle}\nom{corrlocal}$$
\'Etant donn\'e le comportement de la susceptibilit\'e \`a haute fr\'equence,
${\rm lim}_{q_n\rightarrow\infty}\chi_0(iq_n)\propto 1/(iq_n)$, le facteur
$e^{-iq_n\eta}$ est essentiel pour assurer la convergence de la somme.
Cette r\`egle de somme est en fait la
s\'erie de Fourier-Matsubara et ce facteur de convergence sert \`a indiquer
l'ordonnance temporelle des op\'erateurs.

Ainsi, les relations (\ref{vertexun}), (\ref{greenun}), (\ref{mu1}),
(\ref{chi0un}) et (\ref{corrlocal})
forment un syst\`eme d'\'equations qui nous permet de d\'eterminer, de fa\c con
approximative, les propri\'et\'es \`a deux particules du mod\`ele de Hubbard.
Parfois, on a caract\'eris\'e
cette approche par ``imposition d'une r\`egle de somme exacte''. Cependant,
il est \`a noter qu'il serait plus pr\'ecis de parler de r\`egle de somme
auto-coh\'erente car comme nous l'avons mentionn\'e pr\'ec\'edemment la double
occupation n'est pas connue de fa\c con exacte.
Les valeurs obtenues ainsi pour la double occupation sont pr\'esent\'ees au
chapitre suivant o\`u l'on compare avec les r\'esultats obtenus par simulation
Monte Carlo.

\soussection{Raffinement de l'approximation}

Connaissant les propri\'et\'es \`a deux particules il est possible de raffiner
notre approximation pour les propri\'et\'es \`a une particule. Utilisant
l'\'equation du mouvement, (\ref{self}), et l'\'equation de Bethe-Salpeter,
(\ref{bs}), on peut d\'emontrer la relation exacte:
$$\boxEq{{\mib\Sigma}(1,2)=-U\delta(1-2)\sigma_x{\bf G}(1^\mp,1)\sigma_x+U
\sigma_x{\bf G}(1,\overline{5})\sigma_x\chiB(1,1,\overline{7},\overline{8})
{\mib\Gamma}(\overline{8},\overline{7},\overline{5},2)}\nom{selfbs}$$
On voit qu'\'etant donn\'e qu'\`a notre premier niveau d'approximation le
vertex irr\'eductible est local, la seule fonction de corr\'elation
contribuant \`a la self-\'energie sera la fonction de corr\'elation de
paires \`a deux points que nous avons \'evalu\'ee \`a la section
(\ref{corrpaires}). Utilisant la transformation de Fourier, on obtient
ce que l'on appellera le deuxi\`eme niveau d'approximation pour la
self-\'energie:
$$\boxEq{\Sigma^{(2)}(\tilde{k})=U\ffrac{n}{2}-U{T\over N}\sum_{\tilde{q}}
\Gamma^{(1)}\chi^{(1)}(\tilde{q})G^{(1)}(-\tilde{k}+\tilde{q})}\nom{selfdeux}$$
La relation (\ref{selfdeux}) est utile car elle permet de s\'eparer la
self-\'energie en deux termes. Le premier terme est constant et est donc
pr\'esent aux grandes fr\'equences comme aux basses fr\'equences. Le
deuxi\`eme terme contient les contributions basses fr\'equences que nous
\'evaluons au premier niveau d'approximation.

On peut remarquer que ce r\'esultat diff\`ere du r\'esultat obtenu par
Vilk et Tremblay [\ref{vilk1}] par le fait qu'il n'appara\^\i t dans
l'expression de la self-\'energie qu'une seule fonction de corr\'elation
et un seul vertex renormalis\'e. On devrait s'attendre qu'\`a demi-remplissage
on retrouve leur r\'esultat (\`a cause de la sym\'etrie particule-trou pour
ce remplissage). Cette diff\'erence vient du fait que notre calcul
est fait dans le canal transverse et leur calcul a \'et\'e fait dans le
canal longitudinal. Il serait possible de faire le calcul du cas attractif
dans le canal longitudinal, on retrouverait alors la formule de Vilk et
Tremblay pour la self-\'energie. Pour le cas attractif, c'est la
susceptibilit\'e de charge qui serait critique \`a demi-remplissage
(voir section \ref{sechubbard}). Ainsi la sym\'etrie entre les
corr\'elations de charge et celles de paires qui semble \^etre bris\'ee
dans notre approximation pourrait \^etre retrouv\'ee en traitant le
canal longitudinal et le canal transverse de la m\^eme fa\c con [\ref{samuel}].

La fonction de Green se d\'etermine simplement par l'utilisation de
l'\'equation de Dyson
$$G^{(2)}(\tilde{k})={1\over ik_n-\epsilon({\bf k})+\mu^{(2)}-\Sigma^{(2)}
(\tilde{k})}\nom{gdeux}$$
En accord avec l'argument de Luttinger [\ref{luttinger2}], le potentiel
chimique est \'evalu\'e au m\^eme niveau d'approximation que la self-\'energie.
Nous devons donc \'egalement raffiner l'estimation de ce param\`etre.
$$\eqalign{\lim_{\eta\rightarrow 0^-}{T\over N}\sum_{\tilde{k}}
G(\tilde{k})e^{-ik_n\eta}=n&\cr\boxEq{\lim_{\eta\rightarrow 0^-}
{T\over N}\sum_{\tilde{k}}{e^{-ik_n\eta}\over ik_n-\epsilon({\bf k})+\mu^{(2)}
-\Sigma^{(2)}(\tilde{k})}=n}\mskip -18mu &\cr}$$

Avec le formalisme pr\'esent\'e jusqu'\`a maintenant, il semble difficile
de poursuivre les it\'erations et d'obtenir une meilleure estimation du
vertex irr\'eductible. Pour y arriver, il faudrait am\'eliorer notre
estimation des termes hors-diagonaux de la self-\'energie. De plus,
l'\'equation de Bethe-Salpeter d\'eriv\'ee n'est valide que pour les
termes diagonaux (Nous expliquons en annexe \ref{autocoherence} comment il
serait possible de le faire).
Ainsi, l'approximation de la self-\'energie
par la relation (\ref{selfdeux}) sera notre approximation finale.
Nous pr\'esentons au chapitre suivant la fonction de Green et la
self-\'energie obtenues par les relations (\ref{gdeux}) et (\ref{selfdeux}).

\section{R\`egles de somme}\nom{reglesdesomme}

Maintenant que nous avons termin\'e le d\'eveloppement de notre approximation
nous allons v\'erifier si celle-ci permet de satisfaire certaines r\`egles
de somme ainsi que le th\'eor\`eme de Mermin-Wagner.

\soussection{Moments de la susceptibilit\'e}

On peut se demander si notre r\'esultat serait le m\^eme si au lieu de prendre
la r\`egle de somme (\ref{corrlocal}) on avait d\'etermin\'e la valeur moyenne
de double occupation avec la r\`egle de somme exacte suivante\footnote{$^2$}%
{La diff\'erence par rapport \`a la r\`egle de somme (\ref{corrlocal}) vient
du changement de la limite. Ici, on pose $\eta\rightarrow 0^+$.}:
$$\lim_{\eta\rightarrow 0^+}{T\over N}\sum_{\tilde{q}}\chi(\tilde{q})
e^{-iq_n\eta}=\langle(1-n_\uparrow)(1-n_\downarrow)\rangle=1-n+\langle
n_\uparrow n_\downarrow\rangle\nom{corrlocalm}$$
On peut montrer en fait que la r\'eponse est affirmative si notre approximation
satisfait le moment d'ordre z\'ero de la susceptibilit\'e de paires.
En fait, nous irons plus loin, nous allons d\'emontrer que notre approximation
satisfait les deux premiers moments de la susceptibilit\'e
de paires.

Pour la d\'emonstration, nous allons, dans un premier temps, d\'eterminer les
r\'esultats exacts. D\'efinissons la partie imaginaire de la susceptibilit\'e
de paires:
$$\chi''({\bf q},t)={1\over 2}\langle[\Delta({\bf q},t),\Delta^\dagger({\bf q}
,0)]\rangle\nom{susceppaires}$$
Utilisant la repr\'esentation de Lehmann [\ref{ncorps}] pour la fonction de
corr\'elation et pour la susceptibilit\'e, on peut d\'emontrer la
repr\'esentation spectrale suivante
$$\chi({\bf q},iq_n)=\int_{-\infty}^\infty{d\omega\over\pi}{\chi''({\bf q},
\omega)\over\omega-iq_n}\nom{repsspectral}$$

Le d\'eveloppement haute fr\'equence de $\chi(\tilde{q})$ permet de faire
le lien entre cette fonction et les moments de la fonction spectrale
$\chi''(q,\omega)$.
$$\lim_{q_n\rightarrow\infty}\chi({\bf q},iq_n)={-1\over iq_n}
\int_{-\infty}^\infty{d\omega\over
\pi}\chi''({\bf q},\omega)-{1\over(iq_n)^2}\int_{-\infty}^\infty{d\omega\over
\pi}\omega\chi''({\bf q},\omega)+O(1/q_n^3)\nom{devhfm}$$
On voit que les deux premiers termes du d\'eveloppement seront donn\'es par
les deux premiers moments de la partie imaginaire de la susceptibilit\'e.
On peut v\'erifier que le moment d'ordre z\'ero correspond
exactement \`a la diff\'erence entre (\ref{corrlocal}) et (\ref{corrlocalm}):
$$\eqalign{\int_{-\infty}^\infty{d\omega\over\pi}\chi''({\bf q},\omega)=&
\langle[\Delta({\bf q}),\Delta^\dagger({\bf q})]\rangle=1-n\cr=&
\lim_{\eta\rightarrow 0^+}T\sum_{q_n}\Bigl(e^{-iq_n\eta}-e^{iq_n\eta}\Bigr)
\chi({\bf q},iq_n)\cr}\nom{moment0}$$
Ainsi, \'etant donn\'e que nous avons utilis\'e (\ref{corrlocal}) comme
condition pour d\'eterminer la double occupation, on voit que si notre
approximation satisfait le moment
d'ordre z\'ero, la double occupation que nous aurions obtenue aurait \'et\'e
la m\^eme si nous avions impos\'e la r\`egle de somme (\ref{corrlocalm}).

Pour ce qui est du premier moment, qui est une g\'en\'eralisation de la
r\`egle de somme f au cas du mod\`ele de Hubbard attractif,
il sera donn\'e par:
$$\eqalign{\int_{-\infty}^\infty{d\omega\over\pi}\chi''({\bf q},\omega)\omega
=&-\langle[[H,\Delta({\bf q})],\Delta^\dagger({\bf q})]\rangle\cr=&{1\over N}
\sum_{\bf k}(\epsilon({\bf k})+\epsilon(-{\bf k}+{\bf q})-2\mu+U)(1-
2n({\bf k}))\cr}\nom{sommef}$$
o\`u $n({\bf k})=\sum_\sigma\langle\psi_\sigma^\dagger({\bf k})\psi_\sigma(
{\bf k})\rangle$ est la densit\'e d'\'etats renormalis\'ee.

On voit de par les expressions (\ref{moment0}) et (\ref{sommef})
que la principale contribution \`a ces sommes
vient du comportement haute fr\'equence de la fonction de corr\'elation.
C'est pourquoi, afin de v\'erifier si notre approximation satisfait les
moments, nous proc\'edons au d\'eveloppement en $1/iq_n$ de l'expression
(\ref{chiun}). Regardons dans un premier temps le
comportement de $\chi_0^{(1)}$ qui est donn\'ee par l'\'equation (\ref{chi0un}).
$$\eqalign{\lim_{q_n\rightarrow\infty}\chi_0^{(1)}({\bf q},iq_n)&=
\lim_{q_n\rightarrow\infty}{1\over N}\sum_{\bf k}{n_F(\varepsilon({\bf k}))+
n_F(\varepsilon(-{\bf k}+{\bf q}))-1\over iq_n-\varepsilon({\bf k})-
\varepsilon(-{\bf k}+{\bf q})}
\cr&\mskip -40mu={n-1\over iq_n}+{1\over(iq_n)^2}{1\over N}\sum_{\bf k}(
2n_F(\varepsilon({\bf k}))-1)(\varepsilon({\bf k})+\varepsilon(-{\bf k}+
{\bf q}))+O(1/q_n^3)\cr}$$
o\`u $\varepsilon({\bf k})=\epsilon({\bf k})-\mu^{(1)}+\Sigma^{(1)}$ et
$n_F$ est la
fonction de distribution de Fermi-Dirac. Ce r\'esultat vient du fait que
le potentiel chimique $\mu^{(1)}$ est ajust\'e de fa\c con \`a ce que
$G^{(1)}$ ait le bon remplissage $n$. On peut en effet v\'erifier que si
le potentiel chimique est ainsi ajust\'e, $\chi_0^{(1)}$ satisfait la
condition:
$$\lim_{\eta\rightarrow 0^+}{T\over N}\sum_{\tilde{q}}\chi_0^{(1)}(
\tilde{q})\Bigl(e^{-iq_n\eta}-e^{iq_n\eta}\Bigr)=1-n\nom{sumchi0}$$

On peut maintenant montrer en d\'eveloppant le d\'enominateur apparaissant dans
$\chiB^{(1)}$ que la limite pour la fonction de corr\'elation
renormalis\'ee sera:
$$\eqalign{\lim_{q_n\rightarrow\infty}\chi^{(1)}({\bf q},iq_n)=&
{n-1\over iq_n}\cr&\mskip -130mu{}
+{1\over(iq_n)^2}\Bigl({1\over N}\sum_{\bf k}(2n_F(\varepsilon({\bf k}))-1)
(\varepsilon({\bf k})+\varepsilon(-{\bf k}+{\bf q}))-(1-n)^2\Gamma^{(1)}
\Bigr)+O(1/q_n^3)\cr}\nom{devhf}$$
Comparant la relation (\ref{devhfm}) au r\'esultat (\ref{devhf}), on peut
\'evaluer les deux premiers moments de la susceptibilit\'e. Pout le moment
d'orde z\'ero on a
$$\int_{-\infty}^\infty{d\omega\over\pi}\chi''^{(1)}({\bf k},\omega)=1-n$$
On satisfait donc de fa\c con exacte le moment d'ordre z\'ero de la
susceptibilit\'e de paires. On voit qu'une condition importante vient
du comportement \`a haute fr\'equence du num\'erateur dans la forme de type
RPA que prend la susceptibilit\'e dans cette approximation.

Pour ce qui est du premier moment le r\'esultat est le suivant:
$$\eqalign{\int_{-\infty}^\infty{d\omega\over\pi}\chi''^{(1)}({\bf q},\omega)
\omega=&{1\over N}\sum_{\bf k}(1-2n_F(\varepsilon({\bf k})))(\epsilon({\bf k})+
\epsilon({\bf k})-2(\mu^{(1)}-\Sigma^{(1)}))\cr&{}+(1-n)^2\Gamma^{(1)}\cr}$$
En comparant avec le r\'esultat exact (\ref{sommef}), on voit qu'on
satisfait de fa\c con autocoh\'erente ce moment car dans notre approximation
la fonction de distribution $n({\bf k})$ est r\'eduite \`a la
distribution de Fermi-Dirac. De plus, on utilise le fait que dans cette
approximation le potentiel chimique est approxim\'e par $\mu\simeq\mu^{(1)}$
et que la correction du potentiel chimique provenant de notre approximation
de la self-\'energie (\ref{selfun}) permet de satisfaire la condition:
$$U=2\Sigma^{(1)}+(1-n)\Gamma^{(1)}\nom{conditionfsumrule}$$
Donc, la
d\'emonstration est faite que le comportement hautes fr\'equences de la
fonction de corr\'elation sera tel qu'il permet de satisfaire
notre g\'en\'eralisation de la r\`egle de somme f.

\soussection{Comparaison avec l'approche des moments}\nom{npoles}

Il est int\'eressant de noter la ressemblance
entre la m\'ethode utilis\'ee pour d\'eterminer le vertex et les approches
de type approximation par les moments [\ref{moment},\ref{mancini}]. Dans ces
approches, on consid\`ere une approximation de type n-p\^oles pour la fonction
de Green. Ce qui correspond \`a une approximation o\`u apparaissent $2n$
param\`etres associ\'es \`a la position et au poids de chaque p\^ole.
Ces param\`etres sont d\'etermin\'es par la connaissance de $2n$ moments
spectraux. Une telle r\'ealisation est possible par le fait que les moments
du poids spectral contiennent toute l'information n\'ecessaire pour construire
la fonction de Green. Lonke [\ref{moment}] a montr\'e que la connaissance de
$n$ de ces moments permet de d\'efinir un domaine ferm\'e du plan complexe
auquel appartiendra la fonction de Green.

Le probl\`eme auquel nous nous sommes confront\'es pour la d\'etermination
de la fonction de corr\'elation est diff\'erent pour trois raisons.
Primo, dans notre cas, on ne tronque pas le nombre de p\^oles. Nous
utilisons les p\^oles que nous pouvons d\'eterminer de fa\c con exacte
pour la fonction $\chi_0(\omega)$ (exacte pour une valeur de $\mu$ donn\'ee).
Les p\^oles de $\chi$ seront diff\'erents de ceux de $\chi_0$ \`a cause du
param\`etre $\Gamma$ que nous avons introduit. Mais nous gardons un nombre
infini de p\^oles. Ce qui nous donne un certain avantage sur la
m\'ethode habituelle de l'approche des moments.

Secundo, les moments de la susceptibilit\'e de paires ne sont pas
connus de fa\c con exacte, contrairement \`a ceux du poids spectral \`a une
particule.
Il faut donc ajouter d'autres relations pour fermer le syst\`eme d'\'equations.
Ici, on a utilis\'e la relation (\ref{vertexun}) que l'on a d\'eriv\'ee
d'une approximation du type Baym-Kadanoff.

Tertio, la susceptibilit\'e de paires n'est pas un poids spectral car elle
n'est pas positive dans tout son domaine de d\'efinition. En effet, elle est
positive pour les fr\'equences positives et n\'egative pour les fr\'equences
n\'egatives, c'est-\`a-dire qu'elle m\`ene \`a une dissipation positive.
Si on voulait utiliser la m\'ethode des moments, il serait pr\'ef\'erable
de d\'efinir une fonction de distribution qui soit positive.

Afin de s'en d\'efinir une, consid\'erons
la repr\'esentation spectrale de la fonction de corr\'elation en temps
imaginaire.
$$\eqalign{\chi({\bf q},\tau)=&\int_{-\infty}^\infty{d\omega\over\pi}\chi''
({\bf q},\omega){e^{-\tau\omega}\over 1-e^{-\beta\omega}}\cr=&
\int_{-\infty}^\infty{d\omega\over 2\pi}{\chi''({\bf q},\omega)\over{\rm tanh}
(\beta\omega/2)}{e^{(\ffrac{\beta}{2}-\tau)\omega}\over{\rm cosh}(\beta\omega/2
)}\cr\equiv&\int_{-\infty}^\infty{d\omega\over 2\pi}A_\chi({\bf q},\omega)
{e^{(\ffrac{\beta}{2}-\tau)\omega}\over{\rm cosh}(\beta\omega/2)}\cr}$$
De par la positivit\'e de la dissipation, $\chi''({\bf q},\omega)=
-\chi''({\bf q},-\omega)$, il est
facile de montrer que le poids spectral, $A_\chi$ est positif pour toutes
les fr\'equences. De plus, on sait
que la susceptibilit\'e s'annule \`a fr\'equence nulle. On en d\'eduit que
$A_\chi$ sera int\'egrable \`a ce point. On peut montrer que le
premier moment de ce poids correspond \`a la condition que nous avions
pos\'ee pour d\'eterminer le vertex irr\'eductible. Plus pr\'ecis\'ement,
nous obtenons la somme de (\ref{corrlocal}) et de (\ref{corrlocalm}), mais
comme mentionn\'e pr\'ec\'edemment si l'on satisfait l'une de ces conditions
l'autre est satisfaite aussi.
$$\eqalign{\int{d\omega\over\pi}A_\chi({\bf q},\omega)=&T\sum_{q_n}\chi(
{\bf q},iq_n)\Bigl(e^{-iq_n0^+}+e^{-iq_n0^-}\Bigr)\cr=&\langle n_\uparrow
n_\downarrow\rangle+\langle(1-n_\uparrow)(1-n_\downarrow)\rangle\cr}$$
Ainsi, on peut pr\'esenter la r\`egle de somme choisie pour d\'eterminer la
valeur moyenne de double occupation (\ref{corrlocal}), comme une fa\c con de
d\'eterminer le d\'eplacement des p\^oles de $\chi^{(1)}$ par rapport \`a ceux
de $\chi_0^{(1)}$ en imposant le respect du premier moment du poids spectral
$A_\chi$.

\soussection{Th\'eor\`eme de Mermin-Wagner}

Il est int\'eressant de v\'erifier si notre syst\`eme d'\'equations
permet de satisfaire le th\'eor\`eme de Hohenberg-Mermin-Wagner-Coleman
[\ref{hohenberg},\ref{mermin},\ref{coleman}] selon lequel
aucun param\`etre d'ordre invariant sous sym\'etrie continue peut acqu\'erir une
valeur moyenne non-nulle \`a temp\'erature finie
dans un syst\`eme bidimensionnel. Il est \`a noter que ceci n'est pas en
contradiction avec la pr\'esence d'une transition de type
Berezinskii-Kosterlitz-Thouless [\ref{berezinskii},\ref{kt}] \`a temp\'erature
finie car comme nous l'avons mentionn\'e pr\'ec\'edemment cette transition
se fait sans apparition d'ordre \`a longue port\'ee.

Consid\'erons l'expression pour le vertex irr\'eductible (\ref{vertexun}),
pour la fonction de corr\'elation (\ref{chiun}) et la r\`egle de somme
(\ref{corrlocal}). Ces trois relations sont suffisantes pour d\'emontrer
le respect du th\'eor\`eme de Mermin-Wagner.

Consid\'erons d'abord la limite haute temp\'erature o\`u l'on
conna\^\i t approximativement la valeur moyenne de double occupation.
On peut montrer que dans cette limite
$\langle n_\uparrow n_\downarrow\rangle$ est pr\`es de la valeur Hartree,
$n/4$. Alors, le vertex est faiblement renormalis\'e $\Gamma\approx U$
($U$ \'etant n\'egatif le vertex le sera \'egalement). Donc,
si la temp\'erature est suffisamment \'elev\'ee par rapport \`a $U$, le
d\'enominateur dans (\ref{chiun}) est toujours positif. En abaissant, la
temp\'erature $\chi_0(\tilde{q}=0)$ augmente, il en sera donc de m\^eme
de $\chi^{(1)}(\tilde{q}=0)$
et de $\langle n_\uparrow n_\downarrow\rangle$. L'expression du vertex
(\ref{vertexun}) nous indique que celui-ci devrait alors diminuer (en valeur
absolue).

Pour la d\'emonstration de l'absence de point critique, on proc\'edera par
l'absurde. Supposons donc, qu'\`a une temp\'erature donn\'ee l'augmentation de
$\chi_0(\tilde{q}=0)$ est suffisante pour qu'il y ait une singularit\'e:
$\Gamma\chi_0(\tilde{q}=0)=-1$. Alors,
\'etudions le comportement autour de ce point singulier:
$1+\Gamma\chi({\bf q},iq_n=0)\propto{\bf q}^2$.
Un tel comportement n'est pas int\'egrable pour un syst\`eme bidimensionnel
(\'etant donn\'e qu'il y a au num\'erateur un facteur $q$ provenant de la
mesure). On
obtient que pour toute temp\'erature finie, la somme (\ref{corrlocal}) ne
convergera pas. Ce qui indiquerait un taux de double occupation divergeant.
De par la relation (\ref{vertexun}), on aurait un changement de signe
du vertex qui serait infiniment r\'epulsif ce qui est en contradiction
avec la supposition de d\'epart. On en conclut que notre formalisme
emp\^eche l'existence de point critique et qu'au fur et \`a mesure que
la temp\'erature est abaiss\'ee, la renormalisation du vertex devient de
plus en plus importante de telle fa\c con que sa valeur absolue soit
inf\'erieure \`a la valeur critique. La seule supposition que nous avons
faite est qu'il existe bien une valeur de $\langle n_\uparrow
n_\downarrow\rangle$ qui permette de satisfaire simultan\'ement la r\`egle de
somme (\ref{corrlocal}) et l'expression pour le vertex irr\'eductible
(\ref{vertexun}).

Il faut porter attention au cas d'un r\'eseau fini car alors la
fonction $\chi_0({\bf q},iq_n=0)$ prend un nombre fini de valeurs distinctes.
Il est alors possible de trouver une valeur de $\Gamma$ telle que
pour certaines valeurs de ${\bf q}$, le d\'enominateur de (\ref{chiun})
puisse \^etre n\'egatif alors que la somme (\ref{corrlocal}) demeure
finie et positive (le d\'enominateur ne s'annulant pour aucune des valeurs
possibles de ${\bf q}$). Le syst\`eme d'\'equations pourrait alors \^etre
enti\`erement satisfait.\footnote{${}^3$}{Il est \`a noter qu'une \'etude en
temp\'erature devrait alors soit montrer que le vertex admet une
discontinuit\'e ou qu'\`a basse temp\'erature il existerait deux valeurs
distinctes de $\Gamma$ qui permettent de satisfaire simultan\'ement la r\`egle
de somme (\ref{corrlocal}) et l'expression du vertex irr\'eductible
(\ref{vertexun}) dont une o\`u $\Gamma\chi_0(\tilde{q}=0)
<1$ et l'autre o\`u $\Gamma\chi_0(\tilde{q}=0)>1$. Des r\'esultats
pr\'eliminaires obtenus par Kyung semble montrer que cette deuxi\`eme solution
serait observ\'ee pour des interactions interm\'ediaires [\ref{bumsoo}].}

Notre preuve du respect du th\'eor\`eme de Mermin-Wagner est bas\'e sur le
fait que dans notre approximation la susceptibilit\'e de paires ne peut pas
\^etre singuli\`ere. Ainsi, on en d\'eduit que notre approximation ne peut
pas pr\'edire la pr\'esence d'un point critique BKT et ce m\^eme hors du
demi-remplissage. Ceci sera donc une limite de notre approximation qui ne
pourra pas s'appliquer dans la limite tr\`es basse temp\'erature hors
demi-remplissage.

\soussection{Est-ce que notre approche est conservative?}

La ressemblance entre notre approche et une approche de type Baym-Kadanoff
va plus loin que ce que nous avons mentionn\'e \`a la section \ref{svertex}.
En effet, tout comme la fonction de partition en pr\'esence d'un
champ externe permettait de g\'en\'erer les termes hors diagonaux de la
fonction de Green, la fonctionnelle $W[{\cal G}]$ permet de g\'en\'erer
les termes hors diagonaux de la self-\'energie (voir annexe \ref{anex2}).
Ainsi, le r\'esultat d\'emontr\'e par Baym et Kadanoff
[\ref{baym1},\ref{baym2}],
se g\'en\'eralise dans notre cas \`a la d\'efinition d'un vertex
irr\'eductible dont les termes diagonaux peuvent s'exprimer comme une
double d\'erivation d'une fonctionnelle $\Theta$ par rapport \`a la fonction
de Green. Pour d\'eterminer cette fonctionnelle,
on doit consid\'erer la partie hors diagonale de la self-\'energie
approxim\'ee par (\ref{approxim}) et (\ref{facteurlocal}). Au premier
ordre en ${\calb G}$, on obtient:
$$\eqalign{\Sigma_{12}(1,2;{\cal G})=&{\delta\Theta[{\cal G}]\over\delta
{\cal G}_{21}(2,1)}={\cal G}_{12}(1,1)\Gamma_{22}^{(1)}\delta(1-2)\cr
\Sigma_{21}(1,2;{\cal G})=&{\delta\Theta[{\cal G}]\over\delta{\cal G}_{12}
(2,1)}={\cal G}_{21}(1,1)\Gamma_{11}^{(1)}\delta(1-2)\cr}$$
Or, nous avons obtenu $\Gamma_{11}=\Gamma_{22}$. Ainsi, il est facile de
d\'evelopper une fonctionnelle qui permet de g\'en\'erer les parties
hors diagonales de la self-\'energie au premier ordre en ${\calb G}$
et de g\'en\'erer la partie diagonale du vertex irr\'eductible. Celle-ci
est donn\'ee par:
$$\Theta[{\cal G}]=\ffrac{1}{2}{\rm Tr}\biggl\{{\calb G}(\overline{1},
\overline{2}){\mib\Gamma}^{(1)}(\overline{2},\overline{1},\overline{4},
\overline{3}){\calb G}(\overline{3},\overline{4})\biggr\}$$
Il est \`a noter que la fonctionnelle $\Theta$ ne permet de g\'en\'erer
que les termes non-diagonaux de la self-\'energie.

Il est pertinent de se demander si notre approche conserve
de fa\c con globale
l'\'energie, la quantit\'e de mouvement et le nombre de particules.
Par conservation globale, on veut dire que les fonctions de corr\'elation
seront telles qu'elles pourront bien \'evaluer la variation de la
fonction de Green par rapport \`a l'introduction d'un
champ infinit\'esimal appropri\'e. Par exemple, si l'on fait une petite
variation du champ de jauge, on peut v\'erifier si l'approximation
satisfait la loi de conservation globale du nombre de particules [\ref{baym2}].
Si l'on se r\'ef\`ere \`a la d\'erivation de Baym [\ref{baym2}] pour
le d\'eveloppement d'une approximation conservative, il faut montrer
que la self-\'energie (partie diagonale dans notre cas) peut s'\'ecrire comme
la variation d'une fonctionnelle
par rapport \`a une variation de la fonction de Green. Cependant, dans notre
cas, le champ que nous avons introduit est coupl\'e \`a des fonction
de Green anormales. Ainsi la fonction de r\'eponse que nous \'evaluons,
$\chi(\tilde{q})$ ne peut pas d\'ecrire la r\'eponse du syst\`eme \`a une
variation globale du nombre de particules ou de l'\'energie.

Pour pouvoir affirmer si notre approximation est conservative, il faut
d'abord d\'eterminer comment nous d\'ecidons d'\'evaluer les fonctions
de r\'eponse pertinentes. Nous pourrions consid\'erer la relation de
Bethe-Salpeter, (\ref{bs}), que nous avons utilis\'ee pour \'evaluer
la fonction de corr\'elation de paires, et, pr\'esumant que les
sym\'etries de croisement sont satisfaites, \'evaluer les fonctions
de r\'eponses d'int\'er\^et. Malgr\'e la ressemblance entre notre
approche et le formalisme prescrit par Baym, il n'y a aucune raison de
penser qu'une telle approximation pourrait \^etre conservative.

La meilleure fa\c con de proc\'eder serait de consid\'erer notre formalisme
en pr\'esence d'un autre champ externe qui permettrait de d\'ecrire la
r\'eponse du syst\`eme \`a une variation globale du nombre de particule
(changement de jauge), de l'\'energie ou de la quantit\'e de mouvement.
Ainsi, on pourrait s'assurer d'avoir une estimation des fonctions de r\'eponse
qui soit conservative. Cette \'etude a \'et\'e d\'evelopp\'ee par Kyung
[\ref{bumsoo}] pour le calcul de la susceptibilit\'e magn\'etique et la
susceptibilit\'e de charge du mod\`ele de Hubbard attractif.


\chapitre{Comparaison aux donn\'ees Monte Carlo}\nom{cmcq}

Maintenant que nous avons d\'evelopp\'e une approximation pour l'\'etude
du mod\`ele de Hubbard, nous pouvons v\'erifier la validit\'e de cette
approximation et \'etudier les propri\'et\'es pr\'edites par celle-ci.
Nous pouvons entre autre v\'erifier ce que nous avions pr\'esum\'e
au chapitre 1, c'est-\`a-dire la pr\'esence d'un large r\'egime avec
pseudogap pr\`es du demi-remplissage.

Pour notre \'etude nous utiliserons une valeur d'interaction situ\'e
du c\^ot\'e des faibles couplages par rapport au couplage maximisant
l'interaction (voir figure \ref{diagtu}), soit $U=-4$. Ainsi, notre
\'etude ne devrait pas pr\'esenter des effets de couplage fort.
Malgr\'e que le cas \`a demi-remplissage a d\'ej\`a \'et\'e
\'etudi\'e pour le mod\`ele r\'epulsif dans le cadre de l'approximation
TPSC [\ref{vilk1}] nous l'\'etudierons \`a nouveau car cela nous permettra
de comparer les r\'esultats du canal longitudinal avec ceux du  canal
transverse.
La raison du choix de ce cas est qu'il est alors beaucoup plus facile
de faire une \'etude en taille par simulation Monte Carlo quantique
sans avoir des effets parasites tel que l'absence de points sur la
surface de Fermi. De plus, dans notre \'etude nous pr\'esenterons
de nouvelles quantit\'es telle que la densit\'e de suprafluide. Malgr\'e
le choix du cas demi-rempli, le lecteur pourra noter que les
\'equations d\'eriv\'ees dans le cadre de la m\'ethode TPSC ne
seront pas restreintes \`a ce cas.

Ce dernier chapitre est divis\'e en cinq parties. Dans la premi\`ere
partie, nous pr\'esentons la technique Monte Carlo quantique (MCQ)
utilis\'ee pour le calcul des propri\'et\'es du mod\`ele de Hubbard.
Nous expliquons les sources d'erreurs qui sont reli\'ees \`a cette
technique.

La technique MCQ est utile car elle permet de v\'erifier la validit\'e
de notre approximation. C'est ce qui est pr\'esent\'e dans la deuxi\`eme
partie du chapitre, sections 2 et 3. Nous y d\'emontrons que la
forme de type RPA utilis\'ee pour la susceptibilit\'e de paires est
bien observ\'ee dans les r\'esultats MCQ et que le vertex renormalis\'e
est bien approxim\'e par notre approche analytique. Dans la section 3,
nous montrons que le spectre \`a une particule est bien estim\'e par
notre approche analytique. Nous montrons que ce spectre pr\'esente
un pseudogap au niveau de Fermi et ce \`a des temp\'eratures bien
sup\'erieures \`a la temp\'erature critique.

Dans la troisi\`eme partie, section 4 et 5, nous \'etudions d'autres
propri\'et\'es mesur\'ees par MCQ afin d'\'etudier le m\'ecanisme
qui serait responsable de l'apparition du pseudogap. Comme nous le
verrons, les donn\'ees MCQ d\'emontre clairement qu'aux temp\'eratures
o\`u le pseudogap est observ\'e, les fluctuations de paires dominantes
sont des fluctuations classiques. De plus, nous montrons que la
densit\'e de suprafluide y demeure tr\`es faible et qu'ainsi le
m\'ecanisme \`a la base du r\'egime pseudogap n'est pas li\'e \`a la
pr\'esence d'ordre \`a longue port\'ee.
Nous argumentons que le pseudogap
correspond \`a un pr\'esurseur de l'\'etat supraconducteur pr\'esent
\`a plus basse temp\'erature. Ce r\'esultat montre bien que les
fluctuations supraconductrices \`a longue port\'ee jouent un r\^ole
important dans la disparition du liquide de Fermi.

La partie suivante, section 6, pr\'esente l'\'etude du m\'ecanisme
pour la formation du pseudogap tel que pr\'edit par notre approximation.
Nous voyons qu'il correspond aux propri\'et\'es que l'on a observ\'e dans
l'\'etude MCQ. La longueur de corr\'elation de paires devient tr\`es
importante \`a basse temp\'erature amenant l'apparition d'un r\'egime
classique renormalis\'e. Nous montrons que lorsque cette longueur devient
plus importante que la longueur d'onde de de Broglie, il peut appara\^\i tre
un pseudogap dans le spectre \`a une particule.

Dans la derni\`ere section, nous montrons que ce que nous observ\'e par
l'application des m\'ethodes de MCQ au mod\`ele de Hubbard demi-rempli
peut s'extrapoler en dehors du demi-remplissage. On y observe \'egalement
un pseudogap ainsi que l'apparition d'un r\'egime classique renormalis\'e.

\section{Monte Carlo quantique}

Les m\'ethodes Monte Carlo visent \`a r\'esoudre des probl\`emes complexes
par l'utilisation des nombres al\'eatoires et des statistiques. De fa\c con
g\'en\'erale, elles consistent \`a r\'esoudre des probl\`emes d'int\'egration
ou de somme de fa\c con partielle en s\'electionnant par un processus
al\'eatoire un certains nombre d'\'el\'ements de la somme. C'est pourquoi
ces m\'ethodes font parties d'une branche des math\'ematiques dite
exp\'erimentale. \'Etant donn\'e que l'on ne fait qu'un \'echantillonage
plus ou moins repr\'esentatif de la totalit\'e des observations possibles,
la r\'eponse obtenue par l'application des m\'ethodes Monte Carlo est
statistique dans sa nature et est
donc entach\'ee d'une certaine incertitude. Cette derni\`ere donne \`a
l'exp\'erimentateur une mesure de la vraisemblance de la r\'eponse observ\'ee.
Comme ouvrage g\'en\'eral sur les techniques Monte Carlo
nous recommandons la lecture des ouvrages de Hammersley et Handscomb
[\ref{hammersley}], de Kalos et Whitlock [\ref{kalos}] et de Robert
et Casella [\ref{robert}].

En m\'ecanique statistique, il est courant d'utiliser les m\'ethodes Monte Carlo
pour l'\'evaluation des moyennes thermodynamiques.
Il faut dans un premier temps choisir des variables al\'eatoires permettant
d'explorer l'espace des phases. Il faut s'assurer que le le domaines de
valeurs admises par les variables permettent d'explorer l'ensemble de l'espace
des phases de fa\c con \`a permettre le calcul des moyennes thermodynamiques
d'int\'er\^et. Pour un syst\`eme quantique, il n'est pas simple d'y parvenir.
Nous utilisons le d\'eveloppement de Trotter [\ref{trotter}] qui permet
de transformer notre syst\`eme quantique bidimensionnel en un syst\`eme
classique tridimensionnel:
$$\eqalign{e^{-\beta({\cal H}_0+{\cal H}_I)}=&{\rm exp}^{-\int_0^\beta d\tau(
{\cal H}_0+{\cal H}_I)}\cr=&\lim_{N_\tau\rightarrow\infty}
\prod_{l=1}^{N_\tau}e^{-\Delta\tau{\cal H}_0}e^{-\Delta
\tau{\cal H}_I}\cr\approx&\prod_{l=1}^{N_\tau}\biggl(e^{-\Delta\tau{\cal H}_0}
e^{-\Delta\tau{\cal H}_I}+O(\Delta\tau^2)\biggr)\cr}\nom{devtrotter}$$
o\`u $N_\tau\Delta\tau=\beta$.
La troisi\`eme dimension correspond au temps imaginaire. La discr\'etisation
de ce temps introduit une erreur de l'ordre de $\Delta\tau^2$ dans le
calcul des moyennes thermodynamiques [\ref{fye}]. Typiquement, nous choisirons
$\Delta\tau\simeq 1/10$, ce qui implique que cette approximation induira
une erreur syst\'ematique de l'ordre de $1\%$ qui sera environ, pour notre
\'echantillonage, du m\^eme ordre de grandeur ou inf\'erieure \`a l'erreur
statistique.
On voit de notre r\'esultat (\ref{devtrotter}), que la dimension
suppl\'ementaire sera discr\`ete avec $N_\tau$ points.

Ayant ajout\'e cette dimension, on peut introduire la variable
al\'eatoire qui nous permettra d'explorer l'espace des phases. On peut y
parvenir par l'utilisation d'une transformation de Hubbard-Stratonovich o\`u
la variable auxiliaire [\ref{negele}] correspond \`a la variable al\'eatoire.
Appliquant cette transformation de fa\c con \`a d\'ecoupler le terme
d'interaction, ${\cal H}_I$, on peut alors faire la trace sur les fermions
de fa\c con analytique. La somme sur les configurations du champ al\'eatoire
sera quant \`a elle faite par la m\'ethode Monte Carlo.
Pour l'application de la transformation de Hubbard-Stratonovich au mod\`ele
de Hubbard, nous
r\'ef\'erons le lecteur aux papiers originaux [\ref{bss},\ref{hirsch},%
\ref{white}] ainsi qu'\`a la pr\'esentation de Tremblay [\ref{mcq}].
Le lecteur y trouvera aussi l'expression
permettant d'\'evaluer les observables en fonction des configurations
du champ al\'eatoire $x_i$ o\`u $i$ correspond aux points du r\'eseau
tridimensionnel espace et temps imaginaire. Il pourra noter que la variable
al\'eatoire choisie permet
d'\'evaluer directement un nombre important d'observables.
Les d\'etails de l'application des m\'ethodes Monte Carlo
au mod\`ele de Hubbard peuvent aussi \^etre retrouv\'es dans le rapport
technique de Touchette et Poulin [\ref{david}].

\`A l'erreur syst\'ematique s'ajoute
l'erreur statistique mentionn\'ee pr\'ec\'edemment.
Celle-ci est li\'ee \`a l'efficience de la simulation \`a
explorer l'espace des phases. Afin de conna\^\i tre la vraisemblance
de nos simulations, il sera important d'\'evaluer l'erreur statistique.
L'importance de cette erreur d\'epend, entre autres, de la
technique d'\'echantillonage choisie. Par exemple, si l'on fait $N$
observations d'une observable $O$, la moyenne de notre \'echantillonage
sera donn\'ee par
$$\mu={\sum_{i=1}^NO_ip_i\over\sum_{i=1}^Np_i}$$
o\`u $p_i$ est la densit\'e de probabilit\'e associ\'ee \`a l'\'ev\`enement
$O_i$. Si les observations sont ind\'ependantes les unes des autres,
il est alors justifi\'e d'utiliser comme estimateur de l'incertitude sur
la moyenne la variance de cette observable r\'eduit par un facteur li\'e
au nombre d'observations:
$$\delta\mu^2={1\over N-1}\Bigl({\sum_{i=1}^NO_i^2p_i\over\sum_{i=1}^Np_i}
-\mu^2\Bigr)$$

Dans notre cas, on utilise un
processus du genre marche al\'eatoire (cha\^\i ne de Markov) o\`u le
point de l'espace des phases consid\'er\'e \`a une \'etape donn\'ee est
enti\`erement d\'etermin\'e par le point pr\'ec\'edent. Dans un tel
processus, il y a une forte corr\'elation entre des observations rapproch\'ees.
Cette corr\'elation entre les diff\'erentes mesures complique
l'estimation de la variance d'une observable. Afin de se ramener \`a des
observables ind\'ependantes, on regroupe les $N$ observations en sous-groupes
de $M$ observations ($N$ \'etant un multiple de $M$). On peut alors d\'efinir
de nouvelles variables:
$$\tilde{O}_l={\sum_{j=1+(l-1)M}^{lM}O_jp_j\over\sum_{j=1+
(l-1)M}^{lM}p_j}\mskip 2mu ,\mskip 30mu l\in[1,N/M]$$
Si $M$ est suffisamment
grand par rapport \`a la longueur des corr\'elations entre les mesures, on
peut alors supposer que les $N/M$ observations de la variable $\tilde{O}$
sont ind\'ependantes les unes des autres. Dans ces conditions, nous
pouvons utiliser les estimateurs suivants pour la moyenne de l'observable $O$
et pour son incertitude:
$$\eqalign{\mu=&{\sum_{l=1}^{N/M}\tilde{O}_l\tilde{p}_l\over
\sum_{l=1}^{N/M}\tilde{p}_l}={\sum_{i=1}^NO_ip_i\over\sum_{i=1}^Np_i}\cr
\delta\mu^2=&{1\over N/M-1}\biggl({\sum_{l=1}^{N/M}\tilde{O}_l^2
\tilde{p}_l\over\sum_{l=1}^{N/M}\tilde{p}_l}-\mu^2\biggr)\cr}$$
o\`u, on a introduit une nouvelle densit\'e de probabilit\'e
$$\tilde{p}_l=\sum_{j=1+(l-1)M}^{lM}p_j\mskip 2mu ,\mskip 30mu l\in
[1,N/M]$$
Typiquement, les corr\'elations entre les mesures couvrent 50 mesures
pour une temp\'erature de $T=1/5$ (moins \'etendues \`a plus haute
temp\'erature).
C'est pourquoi nous avons regroup\'e les mesures en blocs de 125 \`a 350
mesures. Il est \`a noter que l'\'evaluation de la probabilit\'e $p_i$
(ou de $\tilde{p}_{i,j}$)
n'est pas faite de fa\c con explicite. Celle-ci est partiellement
compens\'ee par l'utilisation d'un \'echantillonage biais\'ee (voir
[\ref{david}]).

En terminant, mentionnons qu'il existe un certain nombre
de conditions o\`u il est difficile de faire des \'echantillonages
efficaces et donc o\`u la convergence vers la moyenne statistique est
lente. Une premi\`ere de ces conditions est li\'ee \`a ce que l'on appelle
le probl\`eme de signe [\ref{mcq}]. Ce probl\`eme est observ\'e
parfois lorsque la convergence de l'\'evaluation de la fonction de
partition est lente et que celle-ci prend une valeur presque nulle.
Cependant, avec la variable al\'eatoire que nous utilisons,
ce probl\`eme ne sera pas pr\'esent dans le mod\`ele de Hubbard attractif.
Un autre probl\`eme important est le probl\`eme du collement. Dans le
mod\`ele de Hubbard on l'observe \`a basse temp\'erature et pour de forts
couplages. Il r\'esulte de l'incapacit\'e \`a bien explorer l'espace des
phases, la variable al\'eatoire prenant des valeurs toutes situ\'ees
dans une m\^eme r\'egion. Pour surmonter ce probl\`eme, nous faisons
plusieurs simulations avec des conditions initiales diff\'erentes
ce qui devrait permettre, pour un nombre de simulations suffisamment
\'elev\'e, d'explorer une bonne partie de l'espace des phases.

\section{Validation des approximations de la m\'ethode analytique}

Maintenant que nous avons d\'ecrit bri\`evement la technique Monte Carlo
et ses sources d'erreurs ou d'incertitude, nous pouvons comparer les
r\'esultats obtenus lorsque appliqu\'ee au mod\`ele de Hubbard avec les
r\'esultats de l'approche \`a N-corps d\'evelopp\'ee au chapitre 2.
Dans le reste de ce chapitre nous r\'ef\'ererons aux r\'esultats Monte
Carlo quantique par l'abbr\'eviation MCQ et \`a l'approche analytique du
chapitre 2 par l'abbr\'eviation TPSC (``Two-particle Self-consistent'').

Dans un premier temps, nous allons v\'erifier la validit\'e des approximations
faites.
La premi\`ere approximation \`a v\'erifier est l'approximation du
vertex irr\'eductible. Est-il appropri\'e de le supposer constant?
Pour le v\'erifier regardons la fonction de corr\'elation obtenue
par simulation Monte Carlo et comparons-la \`a une approximation
de type RPA avec une interaction renormalis\'ee. Nous pr\'esentons
\`a la figure \ref{rpaq} la susceptibilit\'e \`a fr\'equence nulle en
fonction du vecteur d'onde. Les donn\'ees proviennent d'un cas avec
interaction interm\'ediaire $U=-4$, \`a demi-remplissage $n=1$ et
l'inverse de la temp\'erature $\beta=4$. Le vertex renormalis\'e, $\Gamma$,
a \'et\'e ajust\'e de fa\c con \`a ce que la valeur \`a ${\bf q}=(0,0)$
obtenue de la forme RPA, repr\'esent\'ee par une croix sur la figure,
concorde avec celle mesur\'ee par Monte Carlo, repr\'esent\'ee par un carr\'e.
On voit que cette valeur du
vertex permet de bien retrouver la susceptibilit\'e pr\`es de
${\bf q}=(\pi,\pi)=(4,4)$. Cependant, pour les points au voisinage de $(0,0)$,
les donn\'ees Monte Carlo seraient plus approch\'ees avec un vertex plus
\'elev\'e en valeur absolue. On en d\'eduit qu'un vertex avec une d\'ependance
vectorielle telle que sa valeur est plus petite pr\`es de ${\bf q}=(0,0)$
qu'ailleurs pourrait donner une meilleure estimation de la susceptibilit\'e
de paires.


\`A la figure \ref{rpaw}, on pr\'esente la d\'ependance en fr\'equences de
Matsubara de la susceptibilit\'e de paires uniforme, ${\bf q}=(0,0)$ dans
les m\^emes conditions. Les donn\'ees Monte Carlo ont \'et\'e obtenues par une
transformation de Fourier des r\'esultats obtenus en temps imaginaire.
Il faut noter que les barres d'incertitude sur ces
donn\'ees sont approximatives car nous n'avons pas mesur\'e la fonction
de covariance de la susceptibilit\'e en temps imaginaire.
On remarque que la valeur du vertex
ajust\'ee par le r\'esultat MCQ \`a fr\'equence nulle est trop petite pour
les fr\'equences voisines mais a peu d'effet \`a haute fr\'equence.
Il est \`a noter que l'\'ecart observ\'e est relativement insignifiant par
rapport \`a la valeur \`a fr\'equence nulle qui n'a pas \'et\'e plac\'ee
sur la figure car elle est environ 20 fois plus importante que la fr\'equence
voisine. Le fait que la susceptibilit\'e \`a fr\'equence nulle soit beaucoup
plus importante qu'aux autres fr\'equences de Matsubara est d\^u au fait
que le syst\`eme est dans un r\'egime classique renormalis\'e comme nous le
verrons \`a la section \ref{rcr}.


On d\'eduit de nos r\'esultats qu'une approximation de type RPA avec un
vertex constant n'est pas si mauvaise. Il est alors int\'eressant de
comparer la valeur du vertex ajust\'ee par valeur de la susceptibilit\'e
uniforme et
statique mesur\'ee par MCQ et celle obtenue par notre approche. Dans le
tableau \ref{rpavstpsc}, on pr\'esente les r\'esultats pour un r\'eseau
$8\times 8$ \`a demi-remplissage avec interaction interm\'ediaire pour
trois temp\'eratures diff\'erentes.

\begintable{Comparaison de la valeur du vertex choisie afin que la
susceptibilit\'e de paires d'une approximation du type RPA concorde
avec les r\'esultats Monte Carlo (MCQ) avec celle obtenue par l'approximation
propos\'ee (TPSC) pour un r\'eseau de taille $8\times 8$ \`a demi-remplissage,
$n=1$ et $U=-4$.}{Vertex irr\'eductible en approximation locale}
$$\vbox{\halign{
#&&\hskip 2em#\cr
\tableruleD
$\beta$&MCQ&TPSC\cr
\tablerule
3&-2.084&-2.1183\cr
4&-1.940&-1.9977\cr
5&-1.784&-1.8189\cr
\tableruleD}}$$
\endtable
\nom{rpavstpsc}

L'accord est assez bon. Cependant, on doit faire remarquer que dans la
forme RPA, lorsque la valeur du vertex devient telle que le d\'enominateur
est petit \`a faible fr\'equence, une petite variation sur la valeur du
vertex peut avoir un effet important sur la susceptibilit\'e de paires.
Il est donc int\'eressant de comparer d'autres donn\'ees telle la valeur
moyenne de double occupation. \'Etant donn\'e que l'\'evaluation de celle-ci
est beaucoup plus rapide que celle de la susceptibilit\'e de paires, nous
pr\'esentons au tableau \ref{docc} la valeur mesur\'ee pour une plus grande
diversit\'e de conditions.

\begintable{Comparaison des valeurs moyennes de double occupation pr\'edites
par l'approximation propos\'ee (TPSC) et celles mesur\'ees par Monte Carlo
(MCQ) pour des r\'eseaux $8\times 8$. Les chiffres entre parenth\`eses servent \`a
indiquer l'incertitude sur le dernier chiffre significatif.}{Valeurs moyennes de
double occupation}
$$\vbox{\halign{
#&&\hskip 2em#\cr
\tableruleD
U&n&$\beta$&MCQ&TPSC\cr
\tablerule
-4&1.&2&$0.3675(2)$&0.364\cr
-4&1.&3&$0.3671(2)$&0.370\cr
-4&1.&4&$0.3689(2)$&0.377\cr
-4&1.&5&$0.380(5)$&0.386\cr
-4&0.8&4&$0.2671(1)$&0.273\cr
-4&0.8&6&$0.2703(1)$&0.288\cr
-4&0.55&4&$0.1561(1)$&0.157\cr
-4&0.55&6&$0.1590(1)$&0.169\cr
-6&1.&3&$0.4227(5)$&0.4038\cr
-6&1.&4&$0.4258(8)$&0.4140\cr
-6&1.&5&$0.4267(4)$&0.4233\cr
\tableruleD}}$$
\endtable\nom{docc}

\`A premi\`ere vue, il semble y avoir un assez bon accord entre ces valeurs.
Pour les cas \`a plus faible remplissage, on peut d\'enoter que l'accord n'est
pas aussi bon \`a basse temp\'erature. En terminant, on peut remarquer que
les donn\'ees MCQ pr\'esent\'ees au tableau \ref{docc} sont tr\`es pr\'ecises,
en fait, l'erreur statistique est inf\'erieure \`a l'erreur syst\'ematique
qui doit \^etre de l'ordre de 1\% (car $\Delta\tau=0.1$).

\section{Propri\'et\'es \`a une particule}

\soussection{Poids de quasiparticules}

Pour l'\'etude des propri\'et\'es \`a une particule, une observation directe
des donn\'ees MCQ nous montre que notre syst\`eme n'a pas le comportement
habituel des liquides de Fermi. Cette observation d\'ecoule de la mesure
du poids de
quasiparticule tel qu'introduit par Vilk et Tremblay [\ref{vilk1}].
G\'en\'eralement, lorsqu'on parle du poids de quasiparticule on parle du
saut observ\'e \`a temp\'erature nulle dans la fonction de distribution au
niveau de Fermi. Luttinger a propos\'e une premi\`ere g\'en\'eralisation
\`a temp\'erature finie qui consiste en le r\'esidu du p\^ole de la fonction
de Green au niveau de Fermi. Cette quantit\'e \'etant difficile \`a extraire
des simulations MCQ, Vilk et Tremblay ont propos\'e de prendre une mesure
du poids spectral \`a basse fr\'equence donn\'ee par:
$$\tilde{z}({\bf k})\equiv\int_{-\infty}^{\infty}{d\omega\over 2\pi}{A({\bf k},
\omega)\over{\rm cosh}(\beta\omega/2)}=-2G({\bf k},\tau=\beta/2)\nom{pqptc}$$

La derni\`ere \'egalit\'e se d\'emontre par la repr\'esentation spectrale de
la fonction de Green en temps imaginaire (\ref{spectralg}). On peut montrer
que pour un liquide de Fermi, on retrouve \`a temp\'erature nulle la
d\'efinition usuelle du poids de quasiparticules.

On a repr\'esent\'e \`a la figure (\ref{pqptcfig}) le poids spectral \`a basse
\'energie,
$\tilde{z}$, mesur\'e par MCQ en fonction de la temp\'erature. On a utilis\'e
les conditions $U=-4$, $n=0.95$ et $\Delta\tau=1/10$. Il est possible de
v\'erifier que le r\'esultat est sensiblement identique \`a demi-remplissage.
Afin de bien
montrer l'effet de taille nous avons pr\'esent\'e \`a basse temp\'erature les
r\'esultats obtenus pour des syst\`emes de diff\'erentes tailles (voir les
indications dans la l\'egende). Dans la partie du haut de la m\^eme figure nous
avons pr\'esent\'e le facteur de structure de paires au vecteur d'onde
${\bf q}=(0,0)$. Celui-ci a \'egalement \'et\'e obtenue par simulation
MCQ. Pour chaque point, nous avons fait entre 5000 et 20000 mesures (plus de
mesures \`a plus basse temp\'erature et pour les plus petits syst\`emes).

Pour un liquide de Fermi, on devrait observer une augmentation du poids
spectral au niveau de Fermi, $z({\bf k}_F)$, avec l'abaissement de
la temp\'erature indiquant que le poids spectral est piqu\'e
pr\`es des fr\'equences nulles \`a basse temp\'erature. Cependant,
les donn\'ees pr\'esent\'ees \`a la figure \ref{pqptcfig} montrent
que le poids spectral diminue lorsque l'on abaisse la temp\'erature
d\'emontrant que notre syst\`eme n'est pas du type liquide de Fermi.
On observe que l'abaissement du poids spectral
s'accompagne d'une augmentation des corr\'elations de paires. Il est donc
normal de penser que ce sont les fluctuations de paires qui sont responsables
de la disparition du liquide de Fermi et que s'il y a formation d'un
pseudogap dans le syst\`eme celui-ci corresponde \`a un pr\'ecurseur
de l'\'etat supraconducteur pr\'esent \`a temp\'erature nulle.


\soussection{Spectre \`a une particule}

L'abaissement de poids spectral \`a basse fr\'equence accompagnant
une diminution de
temp\'erature \'etant r\'ev\'elateur d'un comportement diff\'erent
des liquides de Fermi, n'est cependant pas suffisant pour d\'emontrer
la pr\'esence d'un pseudogap. Pour le d\'emontrer il faut obtenir le
poids spectral. Celui-ci s'obtient par prolongement analytique de
la fonction de Green en temps imaginaire. Une des techniques les
plus efficaces pour le prolongement analytique des donn\'ees MCQ est la
m\'ethode de maximisation d'entropie (MME) d\'ecrite en annexe \ref{mem}.
Celle-ci se fait par l'utilisation de la repr\'esentation spectrale de
la fonction de Green en temps imaginaire:
$$G({\bf k},\tau)=-\int_{-\infty}^\infty{d\omega\over 2\pi}A({\bf k},\omega)
{e^{-\tau\omega}\over 1+e^{-\beta\omega}}\nom{spectralg}$$

Nous pr\'esentons aux figures \ref{aw2} et \ref{aw4} le poids spectral
obtenu par le prolongement des donn\'ees MCQ (partie gauche de chaque
figure) et par prolongement des donn\'ees obtenues par notre approche
analytique (partie droite de chaque figure). Les r\'esultats analytiques
ont \'et\'e obtenu par l'utilisation des \'equation de $G^{(2)}$ et
de $\Sigma^{(2)}$, \'equation (\ref{gdeux}) et (\ref{selfdeux}).
Pour ces r\'esultats nous avons
appliqu\'e la m\^eme technique de prolongement analytique que pour les
donn\'ees MCQ. De plus, \'etant
donn\'e que cette technique est sensible \`a la pr\'ecision des donn\'ees
nous avons ajout\'e un bruit gaussien dans nos donn\'ees analytiques permettant
ainsi d'ajuster leur pr\'ecision \`a celle des
simulations MCQ. Il est \`a noter que l'ajout de ce bruit diminue le
nombre de d\'etails que l'on peut observer dans le spectre et qu'un bruit
trop important pourrait \'eventuellement camoufler l'existence d'un pseudogap.
C'est pourquoi nous avons fait des simulations MCQ tr\`es pr\'ecises avec
de 125 000 \`a 160 000 mesures pour chaque vecteur d'onde et chaque
temp\'erature.
Pour chaque poids spectral pr\'esent\'e nous avons indiqu\'e le vecteur d'onde
correspondant du c\^ot\'e gauche du graphique.

On observe qu'\`a la temp\'erature la plus \'elev\'ee aucun pseudogap n'est
observ\'e et que la relation de dispersion n'est que l\'eg\`erement modifi\'ee
par rapport \`a celle des \'electrons libres. Les diff\'erences deviennent plus
importantes au fur et \`a mesure que l'on s'\'eloigne du niveau de Fermi.
Lorsqu'on abaisse la temp\'erature un pseudogap apparait au niveau de Fermi.
Ce pseudogap correspond \`a un minimum local \`a basse \'energie tel
qu'on observe au centre de la figure \ref{aw4}. Le pseudogap est plus important
pr\`es du vecteur d'onde $(0,\pi)$ que dans la r\'egion du vecteur
$(\pi/2,\pi/2)$. Cette caract\'eristique n'est pas reli\'ee \`a la sym\'etrie
du gap qui est bien de type ``s-wave'' dans le mod\`ele de Hubbard attractif
mais est due \`a la variation de la vitesse de Fermi. Comme on l'expliquera
\`a la section \ref{effetspectre} il est plus facile de cr\'eer un pseudogap
dans une r\'egion de la surface de Fermi
caract\'eris\'ee par une plus petite vitesse de Fermi.

Il faut remarquer que la discr\'etisation du temps imaginaire limite notre
capacit\'e de prolonger le poids spectral \`a haute fr\'equence. Pour s'en
convaincre, il suffit d'observer la forme du noyau dans la relation
(\ref{spectralg}). De par sa forme, on conclut que tr\`es peu de points en
temps imaginaire nous permettent de d\'eduire le poids spectral \`a haute
\'energie, d'o\`u la difficult\'e.

On peut remarquer que l'abaissement du poids de quasiparticules pr\'ec\`ede
l'ouverture du pseudogap. Ce r\'esultat nous indique qu'avant l'apparition
du pseudogap, le pic dans le poids spectral au niveau de Fermi (voir par
exemple $\beta=2$ et $k=(0,\pi)$) s'\'elargit avec l'abaissement de
temp\'erature. Ce ph\'enom\`ene semble \^etre plus prononc\'e pr\`es du
niveau de Fermi. Ainsi, on pourrait voir l'apparition d'un pseudogap dans
la densit\'e totale d'\'etats avant d'en voir dans le poids spectral.
\`A la figure \ref{dstat}, on pr\'esente la densit\'e d'\'etats obtenus
par maximisation d'entropie. La partie de gauche de la figure nous montre
le r\'esultat obtenu des donn\'ees MCQ et la partie de droite celui de notre
approche analytique. La densit\'e d'\'etats, $D(\omega)$ \'etant simplement la
somme sur les vecteurs d'onde du poids spectral, on peut l'obtenir par
l'application de la m\'ethode de maximisation d'entropie \`a la formule:
$${1\over N}\sum_{\bf k}G({\bf k},\tau)=\int_{-\infty}^\infty{d\omega\over 2\pi}
D(\omega){e^{-\tau\omega}\over 1+e^{-\beta\omega}}$$
o\`u $N$ est le nombre de vecteur d'onde du r\'eseau. Pour l'estimation de
l'incertitude sur les points en temps imaginaire, nous avons suppos\'e que
la matrice de covariance \'etait diagonale dans l'espace de Brillouin.

On voit qu'au fur et \`a mesure que l'on abaisse la
temp\'erature (courbe du bas \`a celle du haut) un pseudogap appara\^\i t
dans la densit\'e d'\'etats \`a basse fr\'equence. Celui-ci appara\^\i t \`a
des temp\'eratures o\`u le poids spectral ne montre pas de signe de la
pr\'esence d'un pseudogap. Il est \`a noter que la d\'ependance en taille de
cet effet n'est pas compl\`etement v\'erifi\'ee est qu'une \'etude plus
d\'etaill\'ee devrait \^etre publi\'ee sous peu [\ref{bumsoo}]. Diff\'erentes
raisons peuvent expliquer le fait que le pseudogap est l\'eg\`erement
surestim\'e par la m\'ethode analytique. Premi\`erement, il n'est pas toujours
facile d'ajuster le bruit des donn\'ees analytiques \`a celui des donn\'ees
MCQ particuli\`erement pour la densit\'e d'\'etats o\`u on doit sommer
la fonction de Green sur les diff\'erents vecteurs d'onde avant le prolongement
analytique. De plus, une comparaison des fonctions de Green obtenues des deux
approches en temps imaginaire
semble montrer que notre approche analytique surestime l'importance des
corr\'elations (voir figure \ref{gscaling}).


\soussection{\'Etude en taille}

Avant de terminer cette section, il nous faut parler de l'effet de diff\'erents
facteurs sur la pr\'esence ou l'absence d'un pseudogap dans le poids spectral.
Il y a un facteur que nous avons d\'ej\`a mentionn\'e soit l'effet de la
pr\'ecision de nos mesures. On montre \`a la figure \ref{bruit} l'effet de 
cette pr\'ecision.
Pour bien d\'emontrer le fait qu'un manque de pr\'ecision att\'enue les
d\'etails dans le poids spectral, nous avons consid\'er\'e la fonction de
Green au vecteur d'onde $(0,\pi)$ tel qu'obtenue par MCQ \`a temp\'erature
$T=1/4$. Nous avons ajout\'e
un bruit gaussien plus ou moins important \`a nos donn\'ees et avons fait
le prolongement analytique par MME. Dans la partie de gauche de la figure
nous montrons la fonction de Green utilis\'ee pour chacun des prolongement
analytique, le bruit \'etant de plus en plus important du bas au haut de
la figure. La premi\`ere courbe correspond aux donn\'ees obtenus par nos
simulations sans l'ajout de bruit. La partie droite montre les poids
spectraux correspondant. On voit que plus le bruit est important, moins le
pseudogap est prononc\'e. Ce r\'esultat explique pourquoi lors de comparaison
des r\'esultats MCQ et ceux de la m\'ethode analytique, il est important
d'ajouter du bruit dans les donn\'ees de la m\'ethode analytique. De plus,
ce r\'esultat renforce notre conclusion selon laquelle il y a bien un
pseudogap dans le poids spectral du mod\`ele de Hubbard.


Cependant, une \'etude en taille montre que le pseudogap est moins prononc\'e
lorsque l'on consid\`ere un r\'eseau de taille plus importante. Ainsi, il
est important de faire une \'etude en taille afin de pr\'edire s'il existe
ou non un pseudogap dans la limite thermodynamique. White et al.
[\ref{white}] ont conclu que l'effet de taille finie serait responsable de la
pr\'esence du pseudogap et que pour un r\'eseau de taille suffisamment
importante le pseudogap disparaitra\^\i t. Cependant, ils ont n\'eglig\'e
l'effet de la pr\'ecision des mesures et les donn\'ees pour les r\'eseaux de
grandes tailles \'etaient peu pr\'ecises. Nous pr\'esentons ici une \'etude
plus pr\'ecise de l'\'evolution du pseudogap avec la taille du r\'eseau.

\`A la figure \ref{gscaling}, nous pr\'esentons la fonction de Green
en temps imaginaire \`a temp\'erature $T=1/4$ au vecteur d'onde
${\bf k}=(0,\pi)$ pour diff\'erentes tailles de r\'eseau.
La partie de gauche nous montre les r\'esultats MCQ
et la partie de droite le calcul analytique (TPSC). La convergence \'etant
plus rapide pour les plus grands syst\`emes\footnote{$^1$}{La convergence
est plus rapide pour les syst\`emes de plus grandes tailles car pour ces syst\`emes
il y a plus de configurations distinctes dans une plage donn\'e d'\'energie.
De plus, pour de plus grands syst\`emes le probl\`eme
de collement semble moins important.},
le nombre de mesures n\'ecessaires
en MCQ d\'epend de cette taille: 300 000 mesures pour le $6\times 6$, 150 000
pour le $8\times 8$ et 90 000 pour le $10\times 10$.
Nous voyons que l'accord entre les r\'esultats MCQ et notre approche
n'est pas parfait, la fonction de Green du calcul analytique \'etant
g\'en\'eralement plus petite en valeur absolue \`a $\tau=\beta/2=2$ que celle
obtenue par MCQ%
\footnote{$^2$}{Nous avons montr\'e que l'accord pouvait \^etre am\'elior\'e
lorsque
le calcul est fait dans le canal longitudinal. Ainsi, on s\'epare le calcul de
la self-\'energie entre les contributions provenant d'une susceptibilit\'e
critique et d'une susceptibilit\'e non-critique. Pour la s\'eparation entre
les diff\'erentes contributions, la forme sym\'etris\'ee donne les meilleurs
r\'esultats \`a demi-remplissage [\ref{samuel}]}.
Cependant, l'\'evolution en taille semble \^etre sensiblement identique
(par exemple, on peut comparer les r\'eseaux de taille $8\times 8$ et
$10\times 10$ de part et d'autre).


Afin de s'assurer que l'\'evolution en taille soit bien \'evalu\'ee nous
avons ajust\'e le bruit dans les donn\'ees MCQ et
les r\'esultats de la m\'ethode analytique \`a un ordre de grandeur moyen
de l'ordre de $0.002\simeq 0.5\%-1\%$ identique pour toutes les tailles de
r\'eseau consid\'er\'ees. L'ajout de bruit se fait par
l'utilisation d'un g\'en\'erateur de nombre al\'eatoire pour g\'en\'erer
un bruit gaussien. L'ordre de grandeur du bruit est environ trois fois
plus important que l'estimation de l'incertitude sur les simulations les
moins pr\'ecises. Nous avons fait ce choix car l'autocorr\'elation entre les
mesures peut induire une sous-estimation des erreurs sur les donn\'ees
MCQ. Pour s'assurer de la pr\'ecision respective de nos donn\'ees, nous
avons \'evalu\'e pour chacune le $\chi^2$ entre la fonction de Green
correspondant au poids spectral obtenu et la fonction de Green de d\'epart.
Le r\'esultat \'etant du m\^eme ordre de grandeur pour chaque taille du
r\'eseau nous assure ainsi que le bruit \'etait de la m\^eme importance.

Regardons maintenant ce que l'on obtient suite au prolongement analytique.
La figure \ref{ascaling} nous montre l'\'evolution en taille du poids
spectral au vecteur d'onde ${\bf k}=(0,\pi)$ et \`a une temp\'erature $T=1/4$.
Comme \`a l'habitude, la partie de gauche montre les r\'esultats MCQ et la
partie de droite le traitement analytique. Une premi\`ere observation est
que l'approche analytique surestime l'importance du pseudogap. Cependant,
l'\'evolution avec la taille du r\'eseau semble se faire de fa\c con
sensiblement identique dans les deux cas. Ce r\'esultat sugg\`ere que le
pseudogap n'est pas d\^u \`a un effet de taille finie mais serait bien
pr\'esent dans le poids spectral et ce \`a des temp\'eratures bien
sup\'erieures \`a la temp\'erature critique. De plus, il est possible
d'\'evaluer le poids spectral dans la limite thermodynamique pour notre
approche analytique. Tel que publi\'e [\ref{samuel}], ce r\'esultat montre
qu'il y a bien un pseudogap dans cette limite \`a une temp\'erature $T=1/5$.


\section{R\'egime classique renormalis\'e}\nom{rcr}

\soussection{Comportement critique de la susceptibilit\'e de paires}

Comme nous avons vu, l'abaissement du poids de quasiparticule est
accompagn\'e d'une hausse du facteur de structure de paires. C'est
pourquoi nous allons d\'etailler l'\'etude du comportement de la
susceptibilit\'e de paires dans le r\'egime de temp\'erature o\`u le
pseudogap est observ\'e. Une observation directe des r\'esultats Monte Carlo
nous indique que la susceptibilit\'e de paires devient tr\`es piqu\'ee
autour de la fr\'equence
de Matsubara $q_n=0$. Pour mettre cet effet en \'evidence nous pouvons
observer le rapport
$$T\chi({\bf q},iq_n=0)/S_\Delta({\bf q})\nom{conditionrcr}$$
entre la
susceptibilit\'e de paires \`a $q_n=0$ et
le facteur de structure de paires donn\'e par la relation suivante:
$$S_\Delta({\bf q})=T\lim_{\eta\rightarrow0^+}\sum_{q_n}\chi({\bf q},iq_n)
e^{iq_n\eta}\nom{structure}$$

\`A la figure \ref{ratioxvss}, nous pr\'esentons le r\'esultat observ\'e pour
le cas uniforme ${\bf q}=0$. Pour chaque point, nous avons \'evalu\'e entre
5000 et 30000 mesures par la technique MCQ.
Nous pr\'esentons le comportement haute temp\'erature en m\'edaillon.
On observe que les effets quantiques deviennent
de plus en plus importants au fur et \`a mesure que l'on abaisse la
temp\'erature. La contribution au facteur de structure provenant de la 
composante \`a fr\'equence nulle diminue ainsi de fa\c con lin\'eaire avec
l'abaissement de la temp\'erature jusqu'\`a une temp\'erature o\`u on
observe un changement brusque dans ce comportement. Le minimum est observ\'e
\`a $T=1/2$ soit \`a une temp\'erature l\'eg\`erement sup\'erieure \`a la
temp\'erature o\`u appara\^\i t le pseudogap dans la densit\'e d'\'etats.
\`A partir de ce minimum, l'importance de la contribution provenant de
la susceptibilit\'e \`a fr\'equence nulle augmente tr\`es rapidement avec
l'abaissement de la temp\'erature. Ce comportement repr\'esente un retour
vers un r\'egime classique \`a basse temp\'erature. \`A tr\`es basse
temp\'erature, on voit l'apparition d'un maximum puis un retour vers un
r\'egime quantique. On peut remarquer que la temp\'erature \`a laquelle
appara\^\i t ce maximum d\'epend de la taille du syst\`eme. On argumentera
\`a la section \ref{longueur} que cet effet est probablement d\^u \`a un effet
de taille finie et qu'\`a tr\`es basse temp\'erature, lorsque la longueur
de corr\'elation est du m\^eme ordre de grandeur que la taille du syst\`eme,
les effets quantiques se manifestent \`a nouveau.


Le fait que la principale contribution au facteur de structure de paires
viennent de la susceptibilit\'e \`a fr\'equence de Matusbara nulle pour $T<1/2$
est une caract\'eristique qui montre que, pour ce qui est des propri\'et\'es
\`a deux particules, le syst\`eme se comporte comme s'il \'etait
classique\footnote{${}^3$}{On peut le comprendre par le fait que si le syst\`eme
est classique on peut supposer que l'op\'erateur $\Delta$ commute avec
l'hamiltonien. Dans ces conditions, cet op\'erateur est ind\'ependant du
temps imaginaire $\Delta(\tau)=\Delta$ et la susceptibilit\'e n'est non nulle
que pour la fr\'equence de Matsubara nulle, $\chi({\bf q},iq_n)\propto\delta_{n,0}$}. En fait, nous qualifierons de r\'egime classique renormalis\'e
le r\'egime basse temp\'erature o\`u le rapport
$T\chi(\tilde{q}=0)/S_\Delta({\bf q}=0)$ augmente en diminuant la
temp\'erature.

On observe un comportement critique classique lorsque l'\'ecart
entre les niveaux d'\'energie est petit par rapport \`a la temp\'erature.
Ainsi, le syst\`eme se comporte comme s'il y avait un continuum d'\'etats
disponibles, comme s'il \'etait classique [\ref{chakra}]. On pr\'ecise
que le r\'egime
est renormalis\'e pour indiquer que les param\`etres d'\'ecrivant
la physique du syst\`eme sont diff\'erents de ceux apparaissant dans
le mod\`ele et donc de ceux que l'on observe \`a haute temp\'erature.

\soussection{Fr\'equences caract\'eristiques}

Il y a une deuxi\`eme quantit\'e importante qui peut nous r\'ev\'eler que le
r\'egime o\`u on observe le pseudogap est un r\'egime classique renormalis\'e,
c'est la fr\'equence caract\'eristique des fluctuations de paires. Celle-ci
\'etant mesur\'ee en fr\'equence r\'eelle, on doit encore une fois faire appel
au prolongement analytique pour mesurer cette quantit\'e. On proc\`ede de la
m\^eme fa\c con que pour le calcul du poids spectral \`a une particule:
on utilise une m\'ethode de maximisation d'entropie pour d\'eduire la partie
imaginaire
de la susceptibilit\'e en fr\'equence r\'eelle de la susceptibilit\'e en
temps imaginaire mesur\'ee par MCQ. Ces deux quantit\'es sont reli\'ees
par une relation spectrale\footnote{$^4$}{En utilisant cette relation, on
peut montrer que si la fr\'equence caract\'eristique est petite par rapport
\`a la temp\'erature alors $\chi(\tau)$ est \`a peu pr\`es constante et
le rapport (\ref{conditionrcr}) vaut pr\`es de $1$.}:
$$\chi({\bf q},\tau)=\int_{-\infty}^\infty{d\omega\over\pi}\chi''({\bf q},
\omega){e^{-\tau\omega}\over 1-e^{-\beta\omega}}$$
Contrairement au cas de la fonction de Green \`a une particule, le noyau
apparaissant ici a un p\^ole sur le domaine d'int\'egration. L'int\'egrale
n'est pas pour autant ind\'efinie, car on peut montrer que la partie imaginaire
de la susceptibilit\'e s'annule \`a ce p\^ole qui est du premier ordre.
Cependant, la pr\'esence de ce p\^ole rend le calcul num\'erique difficile.
On surmonte facilement ce probl\`eme en utilisant le poids spectral introduit
\`a la section \ref{npoles}:
$$A_\chi({\bf q},\omega)={\chi''({\bf q},\omega)\over{\rm tanh}(\beta\omega/2)}
\nom{pschi}$$
Ce poids spectral peut \^etre d\'eduit de la susceptibilit\'e en temps
imaginaire par l'utilisation de la maximisation d'entropie et du noyau
appropri\'e. On peut par la suite ais\'ement inverser la relation
(\ref{pschi}) pour d\'eduire la partie imaginaire de la susceptibilit\'e.
Tout comme pour le poids spectral \`a une particule, la discr\'etisation
du temps imaginaire limite la capacit\'e de d\'eduire le poids spectral
qu'au domaine de fr\'equence autour de la fr\'equence nulle. Mais comme
on pourra le voir, tout le poids spectral associ\'e \`a la susceptibilit\'e
de paires est concentr\'e dans ces fr\'equences.

On a repr\'esent\'e \`a la figure \ref{chiw} la partie imaginaire de la
susceptibilit\'e divis\'ee par la fr\'equence tel qu'obtenue par l'application
de la MME aux donn\'ees MCQ. Nous avons utilis\'e les conditions habituelles:
$U=-4$, $n=1$, $\Delta\tau=1/10$ avec un r\'eseau de taille $8\times 8$.
Pour chaque temp\'erature, nos simulations contiennent entre 125 000 et
160 000 mesures afin d'assurer une pr\'ecision suffisamment \'elev\'ee
pour avoir le plus de d\'etails possibles dans le poids spectral.


On voit qu'au fur et \`a mesure que l'on abaisse la temp\'erature la
susceptibilit\'e de paires devient de plus en plus piqu\'ee autour de
la fr\'equence nulle. Voyons quel effet cela a sur la fr\'equence
caract\'eristique des fluctuations de paires. La fonction de corr\'elation
de paires est li\'ee \`a la partie imaginaire de la susceptibilit\'e
par la relation:
$$\chi({\bf q},iq_n=0)=\int_{-\infty}^\infty{d\omega\over\pi}{\chi''({\bf q},
\omega)\over\omega}$$
On voit que pour le cas uniforme, ${\bf q}=0$, $\chi''(0,\omega)/\omega$
a la forme d'une lorentzienne centr\'ee \`a fr\'equence nulle, il est donc
justifi\'e de prendre comme mesure de la fr\'equence caract\'eristique la
demi-largeur \`a mi-hauteur de ce pic. Ce choix nous donne les fr\'equences
pr\'esent\'ees au tableau \ref{fcaract}.

\begintable{Fr\'equences caract\'eristiques des fluctuations de paires du
mod\`ele de Hubbard \`a $U=-4$, $n=1$ tel que mesur\'e par MCQ pour un
r\'eseau $8\times 8$ avec une discr\'etisation du temps imaginaire
$\Delta\tau=1/10$.}{Fr\'equences caract\'eristiques des fluctuations de paires.}
$$\vbox{\halign{
#&\hskip 2em#\cr
\tableruleD
T&$\omega_c$\cr
\tablerule
1/3&0.37\cr
1/4&0.24\cr
1/5&0.18\cr
\tableruleD}}$$
\endtable\nom{fcaract}

On voit qu'\`a haute temp\'erature la fr\'equence caract\'eristique est
sup\'erieure \`a la temp\'erature. Avec l'abaissement de la temp\'erature,
la fr\'equence caract\'eristique diminue rapidement et devient \'egale
\`a la temp\'erature lorsque le pseudogap s'ouvre dans le poids spectral
\`a une particule. \`A temp\'erature plus basse, $T=1/5$, la fr\'equence
devient inf\'erieure \`a la temp\'erature. Cependant, on doit noter que nos
r\'esultats sont entach\'es d'une certaine incertitude. \'Etant donn\'e que la
fr\'equence caract\'eristique s'obtient par un prolongement analytique, il est
difficile d'estimer cette incertitude. On devine que si nos r\'esultats
MCQ \'etaient plus pr\'ecis, on obtiendrait des pics un peu plus \'etroit sur
la figure \ref{chiw}. On en d\'eduit que nos valeurs de fr\'equences
caract\'eristiques sont une limite sup\'erieure. Par ailleurs, le fait que nos
donn\'ees sont un peu plus pr\'ecise \`a plus haute temp\'erature, nous permet
de d\'eduire que le fait que $\omega_c$ varie plus rapidement que la temp\'erature,
$T$, est bien observ\'e dans nos simulations.

Vilk et Tremblay [\ref{vilk1}]
ont argument\'e que le fait que la fr\'equence caract\'eristique devienne
inf\'erieure \`a la temp\'erature est directement li\'e \`a l'apparition
du pseudogap. Leur explication s'applique aussi bien pour le mod\`ele de
Hubbard attractif. Pour le montrer, nous devons consid\'erer l'expression
de la partie imaginaire de la self-\'energie au niveau de Fermi en fonction
de la susceptibilit\'e de paires.
On peut d\'eduire de l'\'equation (\ref{selfdeux}) la self-\'energie
retard\'ee en faisant le prolongement analytique $ik_n\rightarrow\omega+i\eta$.
On obtient la relation suivante:
$$\eqalign{\Sigma({\bf k}_F,\omega+i\eta)=&U{n\over 2}\cr&\mskip -40mu{}-UU_{pp}
\int{d^dq\over(2\pi)^d}\int{d\omega'\over\pi}[n(\omega')+f(\varepsilon(-{\bf k}_F+
{\bf q}))]{\chi''({\bf q},\omega')\over\omega-\omega'+\varepsilon(-{\bf k}_F
+{\bf q})+i\eta}}\nom{selfret}$$
o\`u $\varepsilon({\bf k})$ est la relation de dispersion (dans notre approche
analytique $\varepsilon({\bf k})=\epsilon({\bf k})-\mu^{(1)}+\Sigma^{(1)}$,
voir \'equation (\ref{greenun})),
$f(\varepsilon)$ est la fonction de distribution de Fermi-Dirac et $n(\omega)$
est celle de Bose-Einstein. On peut en extraire la partie imaginaire en
proc\'edant \`a l'int\'egration sur une composante du vecteur d'onde.
La limite $\eta\rightarrow 0^+$ fait appara\^\i tre une fonction delta \`a
l'int\'erieure des int\'egrales pour la partie imaginaire de la self-\'energie.
On peut alors int\'egrer une des composantes du vecteur d'onde. S\'eparons
donc le vecteur d'onde en ses deux composantes: $q_\parallel$ et $q_\perp$.
Afin de simplifier le calcul lorsque $\chi''$ est piqu\'e pr\`es de
${\bf q}=0$, on approxime
$\varepsilon(-{\bf k}_F+{\bf q})\approx v_Fq_\parallel$. On obtient alors pour la
partie imaginaire de la self-\'energie retard\'ee:
$$\Sigma_R''(k_F,\omega)\approx-{UU_{pp}\over 2v_F}\int{dq_\perp\over 2\pi}
\int{d\omega'\over\pi}[n(\omega')+f(\omega'-\omega)]\chi''(q_\perp,q_\parallel,
\omega')\nom{selfim}$$
o\`u $q_\parallel\approx(\omega'-\omega)/v_F$. Comme not\'e par Vilk et
Tremblay [\ref{vilk1}], le comportement de type liquide de Fermi ou
non-liquide de Fermi est d\'etermin\'e par la largeur relative en fr\'equence
de la fonction $g(\omega')=\chi''({\bf q},\omega')/\omega'$ par rapport \`a
celle de $h(\omega')=n(\omega')+f(\omega'-\omega)$. Si $g(\omega')$ est
\`a peu pr\`es constante sur le domaine de fr\'equence o\`u $h(\omega')$
est non-nulle, alors on obtient un comportement caract\'eristique des
liquides de Fermi:
$$\Sigma_R''(k_F,\omega)\propto\omega^2+(\pi T)^2\nom{fliquide}$$

Cependant, lorsque la fr\'equence caract\'eristique des fluctuations de paires
est petite par rapport \`a la temp\'erature, on ne peut plus approximer
$g(\omega')$ par une constante dans le domaine o\`u $h(\omega')$ est non-nulle.
Si la fonction $g(\omega')$ est tellement piqu\'ee pour $\omega'<T$ que
l'on puisse approximer $h(\omega')$ par son comportement basse fr\'equence
$T/\omega'$, on obtient alors pour le comportement \`a
basse fr\'equence de la self-\'energie:
$$\eqalign{\Sigma_R''(k_F,\omega=0)\approx&-{UU_{pp}\over 2v_F}
\int{dq_\perp\over 2\pi}\int{d\omega'\over\pi}{T\over\omega'}
\chi''(q_\perp,q_\parallel,\omega')\cr\approx&-{UU_{pp}T\over 2v_F}\int
{dq_\perp\over 2\pi}\chi(q_\perp,q_\parallel=0,iq_n=0)\cr}
\nom{nfliquide}$$
Une contribution importante \`a l'int\'egrale sur les vecteurs d'onde vient
du voisinage de $q_\perp\simeq 0$. Par nos r\'esultats MCQ, figure \ref{chiw},
il semble \`a premi\`ere vue que la partie imaginaire de la self \`a
fr\'equence nulle augmente avec un abaissement de la temp\'erature. Ce
qui d\'emontre que celle-ci ne correspond pas \`a un comportement du type
liquide de Fermi. Cependant, on peut noter qu'il y a un facteur $T$ devant
l'int\'egrale dans l'\'equation (\ref{nfliquide}) et que notre d\'eduction
d\'ependra de la dimension du syst\`eme (\`a plus haute dimension les
vecteurs d'onde au voisinage de $q_\perp=0$ ont moins d'importance dans
le r\'esultat de l'int\'egral et le comportement de type liquide de Fermi
peut \^etre retrouv\'e).

\section{M\'ecanisme pour la formation du pseudogap}

\soussection{Densit\'e de suprafluide}

Jusqu'\`a maintenant, nous avons montr\'e que le spectre \`a une particule
est caract\'eris\'e par un pseudogap \`a une temp\'erature sup\'erieure
\`a la temp\'erature critique, et que ce r\'egime o\`u
appara\^\i t le pseudogap est de type classique renormalis\'e. Cependant,
on peut toujours se questionner sur le m\'ecanisme qui serait responsable
de la formation du pseudogap: Est-ce d\^u aux fluctuations de paires?
A-t-on affaire \`a des paires locales pr\'eform\'ees?

La densit\'e de suprafluide est une quantit\'e qui est nulle dans l'\'etat
normal et ne prend une valeur finie que pour des temp\'eratures inf\'erieures
ou \'egales \`a la temp\'erature critique (c'est \'egalement le cas pour
une transition de type BKT). Pour ce qui est de la limite fort couplage,
la densit\'e de suprafluide prend une valeur finie que sous la temp\'erature
critique m\^eme si
les \'electrons forment des paires \`a haute temp\'erature (nous les
appellerons paires locales pr\'eform\'ees).
\`A la temp\'erature critique BKT, ces paires condensent et on observe alors
une valeur finie de la densit\'e de suprafluide.

Pour un r\'eseau de taille finie, on ne s'attend pas \`a avoir un saut tr\`es
prononc\'e dans la densit\'e de suprafluide \`a la temp\'erature critique.
Ce saut devrait plut\^ot \^etre \'etal\'e sur une certaine plage de
temp\'erature. Ainsi, selon la d\'efinition de la temp\'erature critique,
la densit\'e de suprafluide pourrait appara\^\i tre \`a une temp\'erature
l\'eg\`erement sup\'erieure \`a cette temp\'erature critique.
C'est pourquoi nous \'etudierons la densit\'e de suprafluide afin de bien
montrer que la temp\'erature o\`u appara\^\i t le pseudogap est bien sup\'erieure
\`a la temp\'erature critique renfor\c cant ainsi notre argumentation pr\'esent\'ee
par l'\'etude des effets de taille.

Scalapino et al. [\ref{superfluid}]
ont sugg\'er\'e une m\'ethode permettant d'\'evaluer la densit\'e de
suprafluide
par simulation MCQ. Cette m\'ethode fait appel \`a la limite grande longueur
d'onde et basse fr\'equence de la fonction de corr\'elation du courant.
Utilisant la th\'eorie de r\'eponse lin\'eaire,
ils ont montr\'e que la densit\'e de suprafluide sera donn\'ee par:
$${n_s\over m^*}=\Lambda_{xx}(q_x\rightarrow 0,q_y=0,iq_n=0)-\Lambda_{xx}
(q_x=0,q_y\rightarrow 0,iq_n=0)$$
o\`u $m^*$ est la masse effective et $\Lambda_{xx}$ est la fonction de
corr\'elation courant-courant dans la direction $x$:
$$\Lambda_{xx}({\bf q},iq_n)={1\over N}\sum_{{\bf l},{\bf n}}\int_0^\beta
d\tau e^{iq_n\tau-i{\bf q}({\bf l}-{\bf n})}\langle j_x(l,\tau)j_x(n,0)\rangle
$$
avec le courant d\'efini sans le facteur de charge habituel:
$$j_x({\bf l},\tau)=\sum_\sigma\Bigl(\psi_\sigma^\dagger({\bf l}+\hat{x},\tau)
\psi_\sigma({\bf l},\tau)-\psi_\sigma^\dagger({\bf l},\tau)\psi_\sigma(
{\bf l}+\hat{x},\tau)\Bigr)$$

\`A la figure \ref{sfluid}, on pr\'esente la fonction de corr\'elation
courant-courant telle que mesur\'ee par MCQ pour diff\'erents vecteurs
d'onde \`a fr\'equence nulle. Cette quantit\'e \'etant dynamique, nous
avons fait 112 000 mesures pour atteindre une pr\'ecision suffisante:
\'Etant donn\'e le peu de points disponibles,
il est difficile d'estimer les valeurs limites d'int\'er\^et.
Une simple extrapolation lin\'eaire semble montrer que la densit\'e de
suprafluide est tr\`es petite. \'Etant donn\'e que l'on ne conna\^\i t pas
la masse effective des porteurs, il est difficile d'avoir une
estimation quantitative de cette densit\'e. Ce que l'on peut faire est de
comparer l'importance du poids de suprafluide $D_s=n_s\pi e^2/m^*$ au poids
de Drude $D=n_p\pi e^2/m^*$ o\`u $n_p$ est la densit\'e de porteur libre.
Le poids de Drude s'obtient en consid\'erant d'autres limites de la
fonction de corr\'elation de paires:
$${n_p\over m^*}=\Lambda_{xx}(q_x\rightarrow 0,q_y=0,iq_n=0)-\Lambda_{xx}
(q_x=0,q_y=0,iq_n\rightarrow 0)$$


Dans la figure \ref{sfluid}, on a repr\'esent\'e la limite basse fr\'equence
de la fonction de corr\'elation courant-courant uniforme par des cercles.
Pour ces donn\'ees nous avons fait une extrapolation \`a fr\'equence nulle en
utilisant un lissage quadratique.
Ainsi, nous pouvons conclure que le poids de Drude est beaucoup
plus important que le poids de suprafluide et d'apr\`es la th\'eorie
d\'evelopp\'ee par Scalapino et al. [\ref{superfluid}], on aurait donc affaire
\`a un m\'etal et non \`a un supraconducteur. Ce r\'esultat sugg\`ere
que la pr\'esence d'un pseudogap dans le poids spectral ne serait pas li\'ee
\`a la pr\'esence d'un \'etat ordonn\'e. D'autre part, si l'on \'etait
en pr\'esence de paires locales pr\'eform\'ees (comme dans les th\'eories
de couplage fort), on devrait observer une faible densit\'e de porteurs libres
et donc un poids de Drude assez faible. \'Etant donn\'e que l'on a choisi
une interaction interm\'ediaire $U=-4$, nous ne sommes pas tr\`es \'etonn\'es
d'observer un poids de Drude relativement important et, donc, qu'il n'y a pas
de pr\'esence de paires locales dans le r\'egime avec pseudogap.

\soussection{Pr\'ecurseurs de l'\'etat ordonn\'e}

On peut argumenter que le pseudogap est en fait un pr\'ecurseur de l'\'etat
ordonn\'e. Nous avons montr\'e au chapitre 1 qu'\`a demi-remplissage le
point critique est \`a temp\'erature nulle. Cependant, pour un syst\`eme
de taille finie, lorsque la longueur de corr\'elation atteint la taille du
syst\`eme, celui-ci devrait pr\'esenter certaines caract\'eristiques de
la phase ordonn\'ee. Par exemple, on peut s'attendre \`a ce que le poids
spectral pr\'esente un gap tel que pr\'evu dans la phase supraconductrice.
C'est pourquoi nous avons consid\'er\'e un syst\`eme de taille $4\times 4$ \`a
$\beta=8$ afin de v\'erifier si le pseudogap est un pr\'ecurseur de l'\'etat
ordonn\'e.

\`A la figure \ref{spectre8}, on pr\'esente le spectre \`a une particule
obtenue par prolongement analytique des donn\'ees MCQ \`a basse temp\'erature.
On observe qu'il y a alors un
gap s\'eparant les deux pics caract\'eristiques de l'\'etat ordonn\'e.
On voit que ces deux pics sont situ\'es aux m\^emes fr\'equences que
les maxima observ\'es dans le r\'egime caract\'eris\'e par un pseudogap.
Ce r\'esultat sugg\`ere que le pseudogap soit un pr\'ecurseur de l'\'etat
supraconducteur et que celui-ci serait d\^u \`a la pr\'esence de fluctuations
supraconductrices. Sur la figure \ref{spectre8}, on voit la subsistance d'un
petit pic \`a fr\'equence nulle. L'existence de ce pic est \`a v\'erifier.
Celui-ci pourrait \^etre d\^u \`a des effets de taille finie.


\section{Longueurs caract\'eristiques}

\soussection{Longueur de corr\'elation de paires}

Nous avons observ\'e que le pseudogap est peu sujet \`a l'effet de la taille
du syst\`eme aux temp\'eratures plus \'elev\'ees. Malgr\'e tout, il serait
int\'eressant de voir comment se comporte cette caract\'eristique avec un
changement de la taille du r\'eseau. Nous avons trop peu de donn\'ees MCQ
pour pouvoir d\'eterminer l'\'evolution en taille.
Cependant, l'accord entre notre approche et le MCQ sugg\`ere d'utiliser le
comportement en taille de notre approche pour pr\'edire les r\'esultats
dans la limite thermodynamique.

Pour bien d\'etailler notre \'etude en taille, il faut trouver les longueurs
caract\'eristiques du syst\`eme. Le comportement critique de la
susceptibilit\'e de paires nous indique qu'une premi\`ere longueur
\`a prendre en compte est la longueur de corr\'elation de paires.
Pour l'introduire, consid\'erons notre approximation pour la fonction de
corr\'elation:
$$\chi^{(1)}(\tilde{q})={\chi_0^{(1)}(\tilde{q})\over 1+U_{pp}
\chi_0^{(1)}(\tilde{q})}\nom{rpasp}$$

On aurait un point critique si l'on pouvait obtenir
$U_{pp}=U_c\equiv-1/\chi_0^{(1)}(\tilde{q}=0)$. La valeur $U_c$ est
appel\'ee valeur critique du vertex. On sait qu'\`a un point critique
on observe une divergence d'une longueur de corr\'elation, ainsi il semble
appropri\'e de d\'efinir la longueur de corr\'elation de paires par:
$$\xi_L^2=\xi_0^2{U_{pp}\over U_c-U_{pp}}\nom{corrp}$$
o\`u
$$\xi_0^2=-{1\over 2\chi_0(\tilde{q}=0)}{\partial^2\chi_0(\tilde{q}=0)\over
\partial{\bf q}^2}$$
On a plac\'e un indice $L$ \`a la longueur de corr\'elation pour indiquer de
fa\c con explicite sa d\'ependance sur la taille du syst\`eme.
Comme on a vu, pr\'ec\'edemment, \`a basse temp\'erature, $\beta\simeq 4$,
la susceptibilit\'e de paires est piqu\'ee en fr\'equences de Matsubara et
en vecteurs d'onde autour du point $\tilde{q}=(0,0,0)$ qui est caract\'eristique
de l'approche d'un point critique supraconducteur. On peut alors d\'evelopper
la fonction de corr\'elation autour de ce point. Utilisant la longueur de
corr\'elation introduite pr\'ec\'edemment, eq. (\ref{corrp}), on peut
\'ecrire:
$$\chi^{(1)}({\bf q},iq_n=0)\approx{-U_{pp}^{-1}\xi_0^{-2}\over\xi_L^{-2}+
{\bf q}^2}\nom{longueportee}$$

Il est facile d'\'evaluer le comportement en temp\'erature de la longueur
de corr\'elation dans le r\'egime classique renormalis\'e. \'Etant donn\'e
que celle-ci est reli\'ee \`a la valeur du vertex, on doit consid\'erer
la relation utilis\'ee pour \'evaluer ce vertex, soit l'\'equation
(\ref{corrlocal}). Incorporant
l'approximation (\ref{longueportee}) dans cette condition on obtient:
$${T\over U_{pp}\xi_0^2}\int{d^2q\over(2\pi)^2}{1\over\xi_\infty^{-2}+{\bf q}^2}
+C=\langle n_\uparrow n_\downarrow\rangle\nom{lcorrinf}$$
o\`u $C$ contient toutes les contributions des fr\'equences de Matsubara
non-nulles de (\ref{corrlocal}). Ces contributions sont non-singuli\`eres
et d\'ependent faiblement de la temp\'erature.
On peut maintenant d\'emontrer que le comportement critique correspond \`a
celui du mod\`ele sph\'erique [\ref{dare2}]:
$$\xi_\infty=\Lambda^{-1}{\rm exp}\bigl(c/T\bigr)\nom{correntemp}$$
On a introduit un param\`etre, $\Lambda^{-1}$, pour assurer la convergence de
l'int\'egrale. Celui-ci correspond \`a peu pr\`es \`a la plus petite dimension
du syst\`eme: la longueur de coh\'erence $\xi_0$. La variable $c$  est \`a peu
pr\`es constante en temp\'erature et est donn\'ee par:
$$c=\pi\vert U_{pp}\vert\xi_0^2(\langle n_\uparrow n_\downarrow\rangle-C)$$
La faible d\'ependance en temp\'erature de $c$ \`a $n=1$ a \'et\'e analys\'ee
dans [\ref{dare2}]. Il est possible de faire un lien entre la fr\'equence
caract\'eristique et la longueur de corr\'elation. Dar\'e et al. [\ref{dare3}]
ont trouv\'e que dans le cas r\'epulsif, l'approximation TPSC pr\'evoit
que la fr\'equence caract\'eristique des fluctuations magn\'etiques suit le
comportement suivant $\omega_c\propto\xi^{-z}$ avec $z=2$ ($z$ est appel\'e
exposant dynamique). La sym\'etrie entre notre approximation et celle
du cas r\'epulsif montre que dans notre cas la fr\'equence caract\'eristique
de la susceptibilit\'e de paires a un exposant critique avec la m\^eme 
valeur, $z=2$.
Ainsi, la fr\'equence carat\'eristique d\'ecro\^\i t plus
rapidement que la temp\'erature.

Pour un r\'eseau de taille finie le r\'esultat est diff\'erent.
Supposons que le r\'eseau est un carr\'e de taille $L\times L$. Ici en traitant
s\'epar\'ement le vecteur ${\bf q}=0$ et en rempla\c cant le reste de la
somme discr\`ete sur les vecteurs d'onde par une int\'egrale, la
relation (\ref{lcorrinf}) donne:
$$\langle n_\uparrow n_\downarrow\rangle\approx{T\over\pi U_{pp}\xi_0^2}
\biggl[{\xi_L^2\over L^2}+{\rm ln}\Bigl({1+\xi_L^2\Lambda^2\over 1+\xi_L^2/L^2}
\Bigr)\biggr]$$
Supposant que la valeur moyenne de double occupation est la m\^eme pour
un r\'eseau de taille finie que pour celui de taille infinie (ce qui est
une bonne approximation car cette quantit\'e est peu sujette aux effets de
taille), on obtient la relation:
$${\rm ln}\bigl(1+\xi_\infty^2\Lambda^2\bigr)={\xi_L^2\over L^2}+{\rm ln}
\biggl({1+\xi_L^2\Lambda^2\over 1+\xi_L^2/L^2}\biggr)\nom{corrscaling}$$
Ce r\'esultat nous montre que si la longueur de corr\'elation est petite par
rapport \`a la taille du syst\`eme, $\xi_L<<L$, elle est \`a peu pr\`es
identique \`a celle du r\'eseau de taille infinie $\xi_L\approx\xi_\infty$.
Sinon, on aura $\xi_L<\xi_\infty$ car $x>{\rm ln}(1+x)$. Dans ce dernier
cas, la longueur de corr\'elation perd son comportement exponentiel
et suit plut\^ot un comportement en loi de puissance:
$$\xi_L\propto{L\over T^{1/2}}\nom{effettaille}$$
Ce comportement pourrait persister tant que le syst\`eme demeure dans un
r\'egime classique et qu'il est alors justifi\'e de n\'egliger la somme
sur les fr\'equences de Matsubara dans la r\`egle de somme (\ref{corrlocal}).
Cependant, on sait par la diminution de l'importance relative de la susceptibilit\'e
statique observ\'ee \`a basse temp\'erature (figure \ref{ratioxvss}) que le syst\`eme
redevient quantique. On montrera que
le comportement en temp\'erature obtenu (\ref{effettaille}) d\'emontre
aussi un retour vers un r\'egime quantique lorsque les effets de taille
sont importants (voir section \ref{effetspectre}).

\soussection{Effet des fluctuations sur le spectre \`a une particule}%
\nom{effetspectre}

Voyons comment l'augmentation de la longueur de corr\'elation pourrait
induire un pseudogap dans le poids spectral. Consid\'erons donc l'expression
de la self-\'energie retard\'ee en fonction de la susceptibilit\'e de paires,
eq. (\ref{selfdeux}). Comme le r\'egime est classique, il est justifi\'e,
\`a basse fr\'equence,
de ne conserver que la contribution \`a la self-\'energie provenant de la
susceptibilit\'e statique:
$$\Sigma({\bf k}_F,ik_n)=U{n\over 2}-UU_{pp}{T\over N}\sum_{\bf q}\chi
({\bf q},0){1\over -ik_n-\epsilon(-{\bf k}_F+{\bf q})+\mu}$$

Utilisant le d\'eveloppement longue port\'ee pr\'esent\'ee \`a l'\'equation
(\ref{longueportee}), nous pouvons faire le prolongement analytique de
la self-\'energie.
$$\Sigma_R({\bf k}_F,\omega+i\eta)\approx U{n\over 2}-{UT\over\xi_0^2}
\int{d^2q\over(2\pi)^2}{1\over{\bf q}^2+\xi^{-2}}{1\over\omega+i\eta+
\epsilon(-{\bf k}_F+{\bf q})-\mu}$$

Supposant que l'on puisse lin\'eariser la relation de dispersion, on \'ecrit
$\epsilon(-{\bf k}_F+{\bf q})\simeq\epsilon(-{\bf k}_F)-{\bf q}\cdot{\bf v}_F$.
Ainsi, on peut calculer de fa\c con analytique la partie r\'eelle et la
partie imaginaire de la self-\'energie retard\'ee. On obtient une forme
semblable \`a ce qui fut obtenu par Vilk et Tremblay dans le cas r\'epulsif
[\ref{vilk2}]:
$$\eqalign{\Sigma_R'({\bf k}_F,\omega)\simeq&U{n\over 2}-{UT\over
4\pi\xi_0^2}{1\over\sqrt{v_F^2\xi^{-2}+\omega^2}}{\rm ln}\biggl\vert{\sqrt{
\omega^2+v_F^2\xi^{-2}}+\omega\over\sqrt{\omega^2+v_F^2\xi^{-2}}-\omega}
\biggr\vert\cr\Sigma_R''({\bf k}_F,\omega)\simeq&{UT\over 4\xi_0^2}{1\over
\sqrt{\omega^2+\xi^{-2}v_F^2}}\cr}\nom{selfint}$$
On est alors \`a m\^eme d'\'evaluer le poids spectral. Celui-ci est reli\'e
\`a la self-\'energie retard\'ee par la relation habituelle:
$$A({\bf k},\omega)=-2{\rm Im}G_R({\bf k},\omega)={-2\Sigma_R''({\bf k},\omega)
\over\bigl(\omega-\epsilon({\bf k})-\Sigma_R'({\bf k},\omega)\bigr)^2+
\bigl(\Sigma_R''({\bf k},\omega)\bigr)^2}\nom{psint}$$

V\'erifions quelle est la condition n\'ecessaire pour l'apparition du
pseudogap. La pr\'esence d'un pseudogap \`a basse fr\'equence signifie
que le poids spectral \`a fr\'equence nulle doit \^etre convexe, c'est-\`a-dire
que:
$${\partial^2A({\bf k}_F,\omega)\over\partial\omega^2}\biggr\vert_{\omega=0}>0$$

Pour v\'erifier comment on peut satisfaire une telle condition, \'etudions le
comportement de la self-\'energie autour de la fr\'equence nulle. Par la
d\'efinition du niveau de Fermi [\ref{luttinger2}], on a:
$$\epsilon({\bf k}_F)-\mu+\Sigma_R'({\bf k}_F,\omega=0)=0$$
On peut \'egalement montrer que la partie imaginaire de la self-\'energie
est optimale \`a fr\'equence nulle (pour un liquide de Fermi on y observe
un maximum alors que d'apr\`es la relation (\ref{selfint}), on y observe
un minimum) et que la partie r\'eelle y poss\`ede
un point d'inflexion. Ainsi, on obtient:
$$\eqalign{{\partial^2A({\bf k},\omega)\over\partial\omega^2}\biggr\vert_{
\omega=0}=&{1\over\Sigma_R''^2({\bf k},0)}\biggl[{\partial^2\Sigma_R''
({\bf k}_F,\omega)\over\partial\omega^2}\Bigr\vert_{\omega=0}+{2\over\Sigma_R''
({\bf k}_F,0)}\Bigl(1-{\partial\Sigma_R'({\bf k}_F,\omega)\over\partial
\omega}\Bigr\vert_{\omega=0}\Bigr)^2\biggr]\cr=&{1\over\Sigma_R''^2({\bf k}_F
,0)}\biggl[{(\pi^2-8)\over 4\pi^2}{\vert U_{pp}\vert T\xi^3\over\xi_0^2v_F^3
}+{8\xi\over\pi v_F}-{8\xi_0^2v_F\over\vert U_{pp}\vert T\xi}\biggr]\cr}
\nom{conditionpg}$$
o\`u la deuxi\`eme ligne est obtenue en utilisant les r\'esultats analytiques
pour la self-\'energie (\ref{selfint}). Regardons dans un premier temps,
l'information
que l'on peut d\'eduire de la premi\`ere ligne. Le premier terme \`a
l'int\'erieur des crochets du c\^ot\'e droit est positif tandis que le
deuxi\`eme est n\'egatif (la partie imaginaire de la self-\'energie retard\'ee
\'etant n\'egative). On reconna\^\i t que le terme entre parenth\`ese est
l'inverse du poids de quasiparticule (d\'efinition usuelle). Ainsi, la
disparition du poids de quasiparticule ne repr\'esente pas l'apparition
d'un pseudogap. Au contraire, il indique une augmentation de la concavit\'e
du poids spectral \`a basse fr\'equence. Afin qu'il puisse y avoir un
pseudogap la partie imaginaire de la self-\'energie doit \^etre \'elev\'ee
\`a fr\'equence nulle et y avoir une forte convexit\'e.

L'\'etude de la deuxi\`eme ligne de l'\'equation (\ref{conditionpg}), nous
r\'ev\`ele que lorsque le troisi\`eme terme est dominant, c'est-\`a-dire
\`a basse temp\'erature et en pr\'esence d'une longueur de corr\'elation
petite, le poids spectral est concave. Cette concavit\'e augmente avec
l'abaissement de la temp\'erature. Cependant, lorsque la longueur de
corr\'elation est suffisamment importante, le terme dominant est le premier
terme. On a alors affaire \`a un poids spectral convexe. Il y aura donc un
pseudogap \`a basse fr\'equence. La condition de convexit\'e repose
principalement sur la comparaison entre le premier et le troisi\`eme terme
(mentionnons que le deuxi\`eme terme contribue \'egalement \`a la pr\'esence
d'un pseudogap):
$$\biggl({\xi\over v_F}\biggr)^4>{32\pi^2\over\pi^2-8}{\xi_0^4\over U^2T^2}
\nom{conditioncorr}$$

La condition (\ref{conditioncorr}), nous r\'ev\`ele que plus la vitesse de
Fermi sera petite plus il sera facile d'ouvrir un pseudogap. Ce r\'esultat
concorde avec notre observation que le pseudogap semble appara\^\i tre
d'abord pr\`es des singularit\'es de van Hove, $(0,\pi)$, o\`u la vitesse
de Fermi est plus petite. Lorsque le comportement de la fonction de
corr\'elation est exponentiel en temp\'erature, tel que pr\'esent\'e \`a la
formule (\ref{correntemp}), il est clair qu'un pseudogap appara\^\i t \`a
basse temp\'erature. Cependant, qu'arrive-t'il lorsque les effets de taille
sont importants? On a montr\'e \`a la relation (\ref{effettaille}), que dans
ces conditions la longueur de corr\'elation cro\^\i t en suivant une loi
de puissance en $1/T^{1/2}$. Alors, les trois termes apparaissant dans la
relation (\ref{conditionpg}), ont un comportement identique en temp\'erature,
c'est-\`a-dire en $1/T^{1/2}$. Ainsi, si un pseudogap est pr\'esent dans le
syst\`eme, il demeurera et deviendra m\^eme plus prononc\'e avec l'abaissement
de la temp\'erature jusqu'\`a donner un gap \`a temp\'erature nulle.
Il faut cependant, noter que notre approximation classique n'est s\^urement
pas valide jusqu'\`a cette temp\'erature.

\soussection{Longueur d'onde thermique de de Broglie}\nom{longueur}

Comme on a montr\'e pr\'ec\'edemment, il est plus facile de cr\'eer un
pseudogap lorsque la vitesse de Fermi est plus faible. Il est donc
int\'eressant d'introduire une nouvelle quantit\'e, la longueur d'onde
thermique de de Broglie [\ref{vilk1}]:
$$\xi_{\rm th}={v_F\over T}\nom{lotdb}$$
En fonction de cette nouvelle quantit\'e, la condition pour l'existence d'un
pseudogap (\ref{conditioncorr}) prend la forme:
$$\biggl({\xi\over\xi_{\rm th}}\biggr)^4>{32\pi^2\over\pi^2-8}{\xi_0^4T^2
\over U^2}$$

Il est possible d'estimer la longueur de corr\'elation de paires et la
longueur d'onde thermique de de Broglie \`a partir des mesures MCQ.
Pour la longueur de corr\'elation, c'est assez simple: consid\'erant
la susceptibilit\'e de paires statique, $\chi({\bf q},iq_n=0)$ nous pouvons
en extraire la longueur de corr\'elation en utilisant le d\'eveloppement
autour du vecteur d'onde ${\bf q}=0$, \'equation (\ref{longueportee}).
Il est \`a noter qu'\'etant donn\'e que nous avons des r\'eseaux de
petite taille, le r\'esultat est moins pr\'ecis \`a basse temp\'erature
o\`u la susceptibilit\'e de paires est fortement piqu\'ee autour de
${\bf q}=0$.

L'\'evaluation de la longueur d'onde thermique de de Broglie est plus
difficile. Nous devons d'abord faire le prolongement analytique de la
fonction de Green pour diff\'erents vecteurs d'onde afin d'en extraire
la vitesse de Fermi. L'\'evaluation de cette derni\`ere n'est pas tr\`es
pr\'ecise. Elle est extraite de la position du pic dans le poids spectral pour
chacun des vecteurs d'onde choisis (voir par exemple la figure \ref{aw4}).
La dispersion de ces pics nous permet
d'\'evaluer les diff\'erentes composantes de la vitesse de Fermi. Par la
suite, nous pouvons \'evaluer la norme de la vitesse de Fermi et la
longueur d'onde thermique par la relation (\ref{lotdb}).

Nous pr\'esentons \`a la figure \ref{corr} la longueur de corr\'elation et
la longueur thermique obtenue pour un r\'eseau de taille $6\times 6$ et de
taille $8\times 8$ en fonction de la temp\'erature. Nous pr\'esentons les
r\'esultats pour une interaction $U=-4$ avec un remplissage $n=0.95$.
Les deux longueurs sont donn\'ees en unit\'e de pas du r\'eseau.
Afin de voir l'effet de taille finie nous avons extrapol\'e la longueur
de corr\'elation pour un r\'eseau de taille infinie en utilisant la relation
(\ref{corrscaling}). \`A certaines temp\'eratures le r\'esultat extrapol\'e
du r\'eseau $6\times 6$ ne s'accorde pas avec celui extrapol\'e du r\'eseau
$8\times 8$. Nous avons alors consid\'er\'e la moyenne des deux r\'esultats.
Le calcul de la longueur de corr\'elation ne doit \^etre consid\'er\'e
que comme une estimation du premier ordre. Une des raisons en est que
le param\`etre $\Lambda$ n'est pas connu. Pour notre calcul, nous avons
utilis\'e $\Lambda=\pi$. On peut toutefois v\'erifier que le changement de
la valeur de $\Lambda$ change peu l'extrapolation au r\'eseau de taille infinie.


Aux temp\'eratures les plus \'elev\'ees, la longueur d'onde thermique est
plus grande que la longueur de corr\'elation et elle augmente avec
l'abaissement de la temp\'erature en suivant la loi de puissance $1/T$.
La longueur de corr\'elation quant \`a elle augmente avec l'abaissement
de la temp\'erature en suivant un comportement exponentiel tel que
pr\'evu $\xi\propto{\rm exp}(c/T)$. Ainsi \`a une certaine temp\'erature
la longueur de corr\'elation devient plus grande que la longueur d'onde
thermique. On peut l'observer \`a une temp\'erature l\'eg\`erement inf\'erieure
\`a la temp\'erature o\`u appara\^\i t le pseudogap dans le poids spectral.
On peut aussi observer que les effets de taille finie changent peu la longueur
de corr\'elation car celle-ci est relativement petite par rapport
\`a la taille du r\'eseau et ce pour un vaste domaine de temp\'erature
($\xi\simeq 1.32$ \`a $T=1/4$.).

En terminant, remarquons que l'on n'observe pas explicitement dans la figure
\ref{corr} le comportement en $\xi\propto 1/T^{1/2}$ pr\'evu \`a tr\`es
basse temp\'erature. La raison en est simplement que les effets de taille
semblent peu importants pour l'ensemble des points pr\'esent\'es.
S'il existe un r\'egime de temp\'erature o\`u un tel comportement est
pr\'esent alors on s'attend \`a ce que
la longueur thermique puisse rattraper la longueur de corr\'elation
(la premi\`ere de ces longueurs ayant un comportement en $1/T$). Ainsi, cet
effet pourrait
provoquer le retour vers un r\'egime quantique observ\'e \`a tr\`es basse
temp\'erature pour des syst\`emes de petite taille (voir section \ref{rcr}).

\section{Hors demi-remplissage}

Il est int\'eressant de se demander si les r\'esultats pr\'esent\'es
pour le cas demi-rempli peuvent s'extrapoler hors du demi-remplissage.
La principale diff\'erence vient du fait que le comportement critique
est alors de type $O(2)$. Ce comportement critique permet l'existence
d'un point critique de type BKT \`a temp\'erature finie. Cependant, il est
\`a noter que les syst\`emes \'etudi\'es par MCQ sont beaucoup trop petits
pour nous r\'ev\'eler la pr\'esence d'un tel point critique. Ce
probl\`eme est encore plus prononc\'e pr\`es du demi-remplissage o\`u
le comportement critique est de type $O(3)$ pour une large plage de
temp\'erature. Dans ces conditions, on peut s'attendre \`a ce que le
crossover entre les comportements critiques $O(2)$ et $O(3)$ ne puisse
avoir lieu car la longueur de corr\'elation devrait \^etre du m\^eme
ordre de grandeur que la taille du syst\`eme \`a une temp\'erature trop \'elev\'ee.
Ceci peut \^etre v\'erifi\'e par
la comparaison du facteur de structure de paires uniforme au facteur de
structure de charge pris au vecteur d'onde ${\bf Q}\equiv(\pi,\pi)$.
Ce dernier est d\'efini par:
$$S_{\rho}({\bf Q})=\langle\rho({\bf Q})\rho({\bf Q})\rangle$$
o\`u la densit\'e de particule est donn\'ee par:
$$\rho({\bf Q})=\sum_{{\bf k},\sigma}\psi_\sigma^\dagger({\bf k})\psi_\sigma
({\bf k}+{\bf Q})$$

La comparaison entre le facteur de structure de charge et celui de paires
est pr\'esent\'ee \`a la figure (\ref{fspfsc}). Sur la partie de gauche
de la figure, on peut voir qu'au remplissage $n=0.95$, ces deux quantit\'es
sont presque identiques. La diff\'erence ne se fait sentir qu'\`a tr\`es basse
temp\'erature o\`u l'incertitude sur nos r\'esultats est importante. Pour fins
de comparaison, nous pr\'esentons dans la m\^eme figure les facteurs de
structure pour le remplissage $n=0.8$. On voit que dans ce dernier cas
le comportement critique est de type $O(2)$ car le facteur de structure de
charge n'est pas critique.


Une autre quantit\'e nous r\'ev\`ele la brisure de la sym\'etrie $O(3)$, c'est
la valeur du potentiel chimique. Comme on a vu au chapitre 1, c'est cette
quantit\'e qui brise la sym\'etrie $O(3)$ dans le mod\`ele de Hubbard.
Dans le tableau \ref{mumcq}, nous pr\'esentons les valeurs du potentiel
chimique utilis\'ees dans les calculs MCQ pour le r\'eseau $8\times 8$.
Aux valeurs pr\'esent\'ees, nous avons soustrait le terme $U/2$. Rappelons
que pour le cas demi-rempli, on a $\mu-U/2=0$.

\begintable{Potentiel chimique du mod\`ele de Hubbard \`a $U=-4$ tel que
mesur\'e par MCQ pour un r\'eseau $8\times 8$ avec une discr\'etisation du
temps imaginaire $\Delta\tau=1/10$.}{Potentiel chimique}
$$\vbox{\halign{
#&\hskip 2em#&\hskip 2em#\cr
\tableruleD
$n$&$T$&$\mu-U/2$\cr
\tablerule
0.95&1/3&-0.082\cr
0.95&1/4&-0.06\cr
0.95&1/5&-0.06\cr
0.95&1/8&-0.07\cr
0.8&1/3&-0.275\cr
0.8&1/4&-0.26\cr
0.8&1/5&-0.25\cr
\tableruleD}}$$
\endtable\nom{mumcq}

On remarque qu'au remplissage $n=0.95$, aux temp\'eratures d'int\'er\^et, le
potentiel chimique est bien inf\'erieur \`a la temp\'erature. En comparant
les valeurs pr\'esent\'es au tableau \ref{mumcq} \`a la figure \ref{fspfsc},
on peut remarquer que la diff\'erence entre le facteur de structure de paires
et le facteur de structure de charge (voir figure \ref{fspfsc}) se fait
sentir lorsque la temp\'erature est du m\^eme ordre de grandeur que le
potentiel chimique ($T\simeq 2\mu$): $T\simeq 1/8$
au remplissage $n=0.95$ et $T\simeq 1/3$ \`a $n=0.8$.

Nous avons publi\'e une \'etude d\'etaill\'ee du remplissage $n=0.95$
[\ref{allen1}]. Nous
avons montr\'e dans ce cas nous observons \'egalement la pr\'esence d'un
pseudogap et ce \`a une temp\'erature bien sup\'erieure \`a la temp\'erature
critique. Dans la figure \ref{aw95} nous montrons le poids spectral obtenu
par l'application de la MME aux donn\'ees MCQ. Pour le remplissage $n=0.95$,
nous montrons le poids spectral au vecteur d'onde ${\bf k}=(0,\pi)$. Le fait
que ce point n'est pas exactement sur la surface de Fermi est responsable de
l'asym\'etrie observ\'ee entre les fr\'equences n\'egatives et les fr\'equences
positives. Dans la partie de droite de cette m\^eme figure nous pr\'esentons
le poids spectral au remplissage $n=0.8$ pour le vecteur d'onde
${\bf k}=(0,3\pi/4)$. On voit que ce point est tr\`es pr\`es de la surface
de Fermi. Pour les deux remplissages nous pr\'esentons les r\'esultats pour
une temp\'erature $T=1/5$ avec une interaction $U=-4$.


Nous voyons qu'\`a la temp\'erature $T=1/5$, pour chacun des remplissages
\'etudi\'es, le poids spectral \`a une particule pr\'esente un pseudogap.
Ceci semble montrer que pr\`es du demi-remplissage, la temp\'erature
\`a laquelle appara\^\i t le pseudogap, $T^*$, d\'epend
faiblement du remplissage. Si l'on compare la valeur du poids de
quasiparticule pour chacun de ces remplissages, il semble que la
temp\'erature $T^*(n)$ diminue l\'eg\'erement au fur et \`a mesure que l'on
s'\'eloigne du demi-remplissage. En fait, ce comportement pourrait tr\`es bien
\^etre similaire au comportement de la temp\'erature critique champ moyen,
$T_c^0$,
en fonction du remplissage. Pour illustrer ce fait, nous pr\'esentons
au tableau \ref{tc0}, la valeur de la temp\'erature critique pr\'edite par
un calcul champ moyen pour diff\'erents remplissages. Nous donnons \'egalement
la valeur du poids spectral basse fr\'equence mesur\'e par MCQ \`a $T=1/4$ (voir
\'equation (\ref{pqptc})). Il faut dire que le poids spectral ne
d\'emontre pas tout \`a fait un comportement monotone en remplissage. Cependant,
il faut d\'enoter que pour le remplissage $n=0.95$, le vecteur d'onde pour
lequel nous avons \'evalu\'e ce poids, ${\bf k}=(0,\pi)$, n'est pas situ\'e
sur la surface de Fermi et se trouve l\'eg\`erement en dehors de la mer de
Fermi. Cela pourrait expliquer le fait que le poids spectral basse
fr\'equence y est
l\'eg\`erement inf\'erieur \`a celui mesur\'e au demi-remplissage. Afin
d'obtenir une comparaison plus d\'etaill\'ee il serait int\'eressant d'avoir
une mesure de la temp\'erature o\`u appara\^\i t le pseudogap en fonction
du remplissage, $T^*(n)$.

\begintable{Temp\'erature critique pr\'edite par une th\'eorie champ moyen
pour diff\'erents remplissages pour un r\'eseau de taille infinie avec une
interaction $U=-4$. Pour fins de comparaison, nous pr\'esentons \'egalement
la valeur du poids spectral basse fr\'equence mesur\'ee \`a $T=1/4$ pour un
r\'eseau de taille $8\times 8$.}{Temp\'erature critique pr\'edite par une
th\'eorie champ moyen.}
$$\vbox{\halign{
#&\hskip 2em#&\hskip 2em#\cr
\tableruleD
$n$&$T_c^0$&$\tilde{z}$\cr
\tablerule
1.&0.7386&$0.446\pm0.001$\cr
0.95&0.7374&$0.441\pm0.003$\cr
0.8&0.7198&$0.514\pm0.001$\cr
\tableruleD}}$$
\endtable\nom{tc0}

Le fait que le comportement en remplissage de $T^*$ serait semblable \`a
celui de $T_c^0$ renforcerait notre argumentation selon quoi
le pseudogap est d\^u aux fluctuations supraconductrices.


\chapitreSN{Conclusion}

\`A basse temp\'erature, lorsque les effets de corr\'elation sont
importants le mod\`ele de Hubbard pr\'esente des caract\'eristiques
pouvant \^etre assimil\'ees \`a un r\'egime classique renormalis\'e.
Pr\`es du demi-remplissage, ce mod\`ele poss\`ede un param\`etre d'ordre
qui pr\'esente une sym\'etrie plus \'elev\'ee. Nous avons montr\'e que,
dans ces conditions, une approximation de type RPA avec un vertex
renormalis\'e concorde avec ce qui est observ\'e par les simulations
Monte Carlo. Ce r\'esultat avait d\'ej\`a \'et\'e observ\'e et utilis\'e
comme justification heuristique de l'approximation TPSC appliqu\'ee au
mod\`ele de Hubbard r\'epulsif.

Ces \'etudes pr\'ealables nous ont fourni une base pour la
g\'en\'eralisation de l'approximation TPSC. En plus d'enrichir le
formalisme \`a la base de cette approximation, nous l'avons appliqu\'ee
au mod\`ele de Hubbard attractif. Pour le cas \`a interaction interm\'ediaire
\'etudi\'e nous avons montr\'e que notre approximation permet de retrouver,
sans param\`etre ajustable, la valeur du vertex observ\'ee par les simulations
MCQ.

Nous avons \'egalement montr\'e que dans le r\'egime classique renormalis\'e
le spectre \`a une particule est caract\'eris\'e par un abaissement du poids
\`a basse \'energie au niveau de Fermi: un pseudogap. Au d\'ebut de cette
th\`ese, nous avons argument\'e que la pr\'esence du pseudogap dans un
large domaine de temp\'erature dans le mod\`ele de Hubbard attractif permet
de mieux comprendre la pr\'esence d'un pseudogap dans les supraconducteurs
\`a haute temp\'erature critique et ce \`a des temp\'eratures beaucoup
plus importantes que la temp\'erature critique. Notre explication
provient de l'\'elargissement de la sym\'etrie du param\`etre d'ordre qui
serait responsable d'un abaissement de la temp\'erature critique sans
toutefois avoir d'effet marqu\'e sur la temp\'erature o\`u apparaissent
les fluctuations critiques. Notre \'etude a \'et\'e pr\'esent\'e pour une
interaction inf\'erieure au couplage maximisant la temp\'erature critique
(r\'egime d'interaction faible \`a interm\'ediaire), $U=-4$. Ainsi
le pseudogap observ\'e n'est pas d\^u \`a des effets d'interactions fortes.

Certains r\'esultats exp\'erimentaux sur les supraconducteurs \`a haute
temp\'erature critique permettent d'appliquer notre \'etude \`a ces
syst\`emes. Corson et al. [\ref{corson}] ont observ\'e par des mesures
de conductivit\'e \`a haute fr\'equence que le r\'egime o\`u appara\^\i t
le pseudogap dans les SC hTc est un r\'egime classique renormalis\'e.
De plus, leur \'etude a montr\'e que la temp\'erature critique est
associ\'ee \`a une disparition de la coh\'erence de phase.

Malgr\'e les r\'esultats obtenus, il semble \'evident que notre approximation
ne permet pas d'expliquer les r\'esultats observ\'es au del\`a du cas
d'interaction interm\'ediaire. Nous avons d\'ecouvert r\'ecemment qu'il est
possible, dans l'esprit de l'approximation TPSC, d'am\'eliorer le formalisme
en ajoutant un vertex dynamique non uniforme (voir l'annexe C). Une telle
approximation
pourrait permettre d'am\'eliorer l'estimation de la susceptibilit\'e critique.
Cependant, il y aurait un effet plus important sur l'estimation de la
self-\'energie. De plus, il serait int\'eressant de v\'erifier si elle
permettrait d'obtenir des r\'esultats int\'eressants \`a plus forte
interaction.

En terminant, mentionnons \'egalement les am\'eliorations possibles de la
m\'ethode Monte Carlo. Dans un premier temps, il serait possible
d'acc\'el\'erer la vitesse de convergence par un choix appropri\'e
d'estimateurs. Assaraf et Caffarel [\ref{assaraf}] ont pr\'esent\'e une
m\'ethode o\`u, par l'utilisation de simulations pr\'ealables, il est possible
de renormaliser les estimateurs pour de nouveaux estimateurs avec une plus
faible variance. Une autre am\'elioration notable pour le cas \`a forte
interaction et \`a basse temp\'erature, consiste \`a am\'eliorer l'ergodicit\'e
de l'algorithme. Une telle am\'elioration a \'et\'e propos\'e par Iwamatsu et
Okabe [\ref{iwamatsu}]. Leur m\'ethode est
bas\'ee sur les travaux de Tsallis et Stariolo [\ref{tsallis}] et ils l'ont
appliqu\'ee \`a l'\'etude par simulation Monte Carlo d'un double puits
quantique.

La technique de prolongement analytique par maximisation d'entropie pourrait
\'egalement \^etre raffin\'ee. Tel que mentionn\'e dans le rapport de
Touchette et Poulin [\ref{david}], nous avons d\'evelopp\'e un nouvel
algorithme bas\'e sur une technique de recuit simul\'e grandement utilis\'ee
dans les probl\`emes d'optimisation [\ref{geman}]. La force de cette m\'ethode
provient de sa capacit\'e de trouver le mimimum absolu d'une fonction lorsque
celle-ci poss\`ede plusieurs minima locaux. Cependant, il est difficile
d'\'evaluer l'incertitude sur les r\'esultats obtenus par cette technique.
Une m\'ethode qui nous permettrait de faire un prolongement analytique tout en
permettant d'\'evaluer l'incertitude sur le r\'esultat obtenu serait une
technique de Monte Carlo bay\'esien [\ref{mcmc},\ref{robert},\ref{robert2}].


\annexe{Table de formules}\nom{acalcul}

Pour faciliter l'utilisation de la technique d\'evelopp\'ee, cette annexe
constitue une table des principales formules utiles.
Aucune des formules pr\'esent\'ees ici ne d\'epend du choix
de l'hamiltonien. Le lecteur int\'eress\'e \`a un hamiltonien particulier
est donc invit\'e \`a \'evaluer les \'equation du mouvement pour son
hamiltonien et \`a y ajouter les formules qui lui seront utiles.

\section{D\'efinitions}

On consid\`ere une fonction de Green de forme matricielle $2\times2$ dont
les \'el\'ements hors diagonaux sont associ\'es aux param\`etres
d'ordre et les termes diagonaux sont les termes normaux. On consid\`ere
le syst\`eme en pr\'esence d'un champ source externe, $\phi$, coupl\'e aux
param\`etres d'ordre.

L'\'equation de Dyson peut \^etre consid\'er\'ee comme une
d\'efinition de la self-\'energie
$$\bigl({\bf G}_0^{-1}(1,\overline{3})-\phiB(1,\overline{3})-{\mib\Sigma}(1,
\overline{3};\phi)\bigr){\bf G}(\overline{3},2;\phi)=\deltaB(1-2)$$

D\'efinition de la fonction de corr\'elation:
$$\chiB(1,2,3,4;\phi)\equiv-{\delta\over\delta\phiB(2,1)}{\bf G}(3,4;\phi)$$
o\`u
$${\delta\hfill\over\delta\phiB(1,2)}\equiv\pmatrix{0&{\delta\hfill\over
\delta\phi_{21}(1,2)}\cr{\delta\hfill\over\delta\phi_{12}(1,2)}&0\cr}$$

D\'efinition du vertex irr\'eductible:
$${\mib\Gamma}(1,2,3,4;{\cal G})\equiv{\delta\hfill\over\delta{\calb G}(2,1)}
{\mib\Sigma}(3,4;{\cal G})$$
o\`u
$${\delta\hfill\over\delta{\calb G}(1,2)}\equiv\pmatrix{0&{\delta\hfill\over
\delta{\cal G}_{21}(1,2)}\cr{\delta\hfill\over\delta{\cal G}_{12}(1,2)}&0\cr}$$

On peut passer d'une expression en fonction du champs $\phiB$ \`a une en
fonction de ${\calb G}$ par un changement de variable. Par exemple,
ce changement pour le champ $\phiB$ peut s'\'ecrire:
$${\mib\Phi}(1,2;{\cal G})\equiv\phiB(1,2)\Big|_{G_{\rm HD}[\phi]={\cal G}}$$

\section{R\`egles exactes}

Il est \`a noter que d\^u \`a la forme matricielle de l'op\'erateur
d\'eriv\'ee, la d\'erivation d'un produit de matrices ne se fait pas
de fa\c con usuelle. En utilisant le fait que cette matrice n'a que
des composantes hors diagonales, on peut d\'emontrer la relation suivante:
$${\delta\hfill\over\delta\phiB(2,1)}{\bf AB}=\biggl({\delta\hfill\over
\delta\phiB(2,1)}{\bf A}\biggr){\bf B}+\sigma_x{\bf A}_{\rm D}\sigma_x
{\delta\hfill\over\delta\phiB(2,1)}{\bf B}+\sigma_x{\bf A}_{\rm HD}
{\delta\hfill\over\delta\phiB(2,1)}\sigma_x{\bf B}$$
Cette relation se d\'emontre par la r\`egle de commutation entre une matrice
diagonale, ${\bf D}$, et une matrice hors diagonale, ${\bf H}$, (matrices
$2\times 2$):
$${\bf HD}=\sigma_x{\bf D}\sigma_x{\bf H}$$
et par la r\`egle de commutation entre deux matrices hors diagonales,
${\bf H}_1$ et ${\bf H}_2$:
$${\bf H}_1{\bf H}_2=\sigma_x{\bf H}_2{\bf H}_1\sigma_x$$

Pour d\'emontrer la r\`egle de d\'erivation en cha\^\i ne lorsque l'on passe
d'une d\'erivation par rapport au champ $\phiB$ \`a une par rapport au champ
${\calb G}$, il suffit de le faire de fa\c con usuelle puis de ramener
le tout sous forme matricielle.
$$\eqalign{{\delta\hfill\over\delta\phiB(2,1)}=&\biggl({\delta{\bf G}_{12}(
\overline{3},\overline{4})\over\delta\phiB(2,1)}\biggr){\delta\hfill\over
\delta{\cal G}_{12}(\overline{3},\overline{4})}+\biggl({\delta{\bf G}_{21}(
\overline{3},\overline{4})\over\delta\phiB(2,1)}\biggr){\delta\hfill\over
\delta{\cal G}_{21}(\overline{3},\overline{4})}\cr=&\biggl(
{\delta\hfill\over\delta\phiB(2,1)}{\bf G}_{\rm HD}(\overline{3},
\overline{4};\phi)\biggr){\delta\hfill\over\delta{\calb G}(\overline{3},
\overline{4})}+\biggl({\delta\hfill\over\delta\phiB(2,1)}\sigma_x
{\bf G}_{\rm HD}(\overline{3},\overline{4};\phi)\biggr)
{\delta\hfill\over\delta{\calb G}(\overline{3},\overline{4})}
\sigma_x}\nom{derivch}$$

La d\'efinition de l'inverse de la fonction de Green permet de faire un lien
entre la d\'ependance de $G^{-1}$ sur $\phi$ et celle de $G$. Il suffit
de consid\'erer la relation:
$${\bf G}(1,\overline{3};\phi){\bf G}^{-1}(\overline{3},2;\phi)=\delta(1-2)
\nom{inverse}$$
Cette relation nous permet de d\'eriver plusieurs relations int\'eressantes.
Par exemple, on peut l'utiliser pour d\'emontrer la relation de Bethe-Salpeter.

\section{Calcul au premier ordre}

La fonction de Green au premier ordre en ${\calb G}$:
$$G(1,2;{\cal G})=G(1,2)+{\calb G}(1,2)+O({\calb G}^2)\nom{goung}$$
nous permet d'\'evaluer ${\bf G}^{-1}$ au premier ordre:
$$G^{-1}(1,2;{\cal G})=G^{-1}(1,2)-G^{-1}(1,\overline{3}){\calb G}(
\overline{3},\overline{4})G^{-1}(\overline{4},2)+O({\calb G}^2)\nom{gmoung}$$
Combinant (\ref{gmoung}) avec la relation de Dyson (\ref{eqdyson}) on
obtient une expression pour le champ externe au premier ordre
en ${\calb G}$:
$$\eqalign{{\mib\Phi}(1,2;{\cal G})=&-{\calb G}(\overline{4},\overline{3})
\biggl[{\delta\hfill\over\delta{\calb G}(\overline{4},\overline{3})}\bigl(
{\bf G}^{-1}(1,2;{\cal G})+{\mib\Sigma}(1,2;{\cal G})\bigr)\biggr]_{{\cal G}=0}
+O({\calb G}^2)\cr=&{\calb G}(\overline{4},\overline{3})\Bigl(\sigma_x
{\bf G}^{-1}(1,\overline{4})\sigma_x{\bf G}^{-1}(\overline{3},2)-{\mib\Gamma}
(\overline{3},\overline{4},1,2)\Bigr)+O({\calb G}^2)\cr}\nom{phioung}$$

Pour ce qui est du d\'eveloppement au premier ordre en $\phi$, on obtient:
$${\bf G}(1,2;\phi)={\bf G}(1,2)-\phiB(\overline{4},\overline{3})\chiB(
\overline{3},\overline{4},1,2)+O(\phiB^2)\nom{gounp}$$
qui d\'ecoule de la d\'efinition de $\chiB$. Pour ${\bf G}^{-1}$, le
r\'esultat correspondant se d\'emontre en consid\'erant l'\'equation de Dyson
(\ref{eqdyson}) avec la d\'efinition du vertex (\ref{vertex}) et la
r\`egle de d\'erivation en cha\^\i ne (\ref{derivch}):
$${\bf G}^{-1}(1,2;\phi)={\bf G}^{-1}(1,2)-\phiB(1,2)+\phiB(\overline{4},
\overline{3})\chiB(\overline{3},\overline{4},\overline{5},\overline{6})
{\mib\Gamma}(\overline{6},\overline{5},1,2)+O(\phiB^2)\nom{gmounp}$$
C'est la combinaison de (\ref{gounp}) avec (\ref{gmounp}) dans (\ref{inverse})
qui permet de d\'emontrer la relation de Bethe-Salpeter.

\section{Calcul au deuxi\`eme ordre}

Pour assurer une autocoh\'erence dans l'estimation du vertex, il faut aller
au deuxi\`eme ordre. Ceci correspond au premier terme de la d\'ependance en
$\phi$ de
la partie diagonale des matrices. L'observation que la partie hors diagonale
des matrices peut s'exprimer comme une somme de puissances impaires de $\phi$
simplifie grandement le probl\`eme. On peut ainsi obtenir ${\bf G}({\cal G})$
au deuxi\`eme ordre en ${\calb G}$ sans introduire de nouvelles fonctions:
$$\eqalign{{\bf G}(1,2;{\cal G})=&{\bf G}(1,2)+{\calb G}(1,2)\cr&\mskip -20mu
{}-{\calb G}(\overline{6},\overline{5})\Bigl(\sigma_x{\bf G}^{-1}(\overline{4},
\overline{6})\sigma_x{\bf G}^{-1}(\overline{5},\overline{3})-{\mib\Gamma}(
\overline{5},\overline{6},\overline{4},\overline{3})\Bigr)\chiB_{\rm HD}
(\overline{3},\overline{4},1,2;{\cal G})+O({\calb G}^3)\cr}\nom{gdeuxg}$$

Pour ${\bf G}^{-1}$, on utilise l'\'equation de Dyson (\ref{eqdyson}),
les r\`egles de la section (A.2) et le fait que la matrice
${\mib\Phi}({\cal G})$ \'etant hors diagonale n'a dans la phase normale
que des puissances impaires en ${\calb G}$.
$$\eqalign{{\bf G}^{-1}(1,2;{\cal G})=&{\bf G}^{-1}(1,2)-{\bf G}^{-1}(1,
\overline{3}){\calb G}(\overline{3},\overline{4}){\bf G}^{-1}(
\overline{4},2)\cr&{}-{\calb G}(\overline{4},\overline{3}){\mib\Gamma}_{
\rm HD}(\overline{3},\overline{4},1,2;{\cal G})+O({\calb G}^3)\cr}
\nom{gmdeuxg}$$

Afin de bien obtenir les relations (\ref{gdeuxg}) et (\ref{gmdeuxg}) \`a
l'ordre quadratique en ${\calb G}$, il nous faut obtenir
${\mib\Gamma}_{\rm HD}$
et $\chiB_{\rm HD}$ \`a l'ordre lin\'eaire en ${\calb G}$. Ceci peut se
faire en utilisant les d\'efinitions (\ref{vertex}) et (\ref{correlation})
et que seule les parties diagonales de ${\bf G}$ et de ${\mib\Sigma}$ ont
des termes de puissances paires dans leur d\'eveloppement en ${\calb G}$.
On obtient pour ${\mib\Gamma}_{\rm HD}$ une expression o\`u appara\^\i t un
vertex \`a trois particules:
$${\mib\Gamma}_{\rm HD}(1,2,3,4;{\cal G})={\calb G}(\overline{5},\overline{6})
\biggl[{\delta\hfill\over\delta{\calb G}(\overline{5},\overline{6})}{\delta
\hfill\over\delta{\calb G}(2,1)}{\mib\Sigma}(3,4;{\cal G})\biggr]_{{\cal G}
=0}+O({\calb G}^3)\nom{gaoung}$$

De la m\^eme fa\c con l'expression pour $\chiB_{\rm HD}$ fait appara\^\i tre
une fonction de corr\'elation \`a trois particules:
$$\eqalign{\chiB_{\rm HD}(1,2,3,4;{\cal G})&=-{\mib\Phi}(\overline{5},
\overline{6};{\cal G})\biggl[{\delta\hfill\over\delta\phiB(\overline{5},
\overline{6})}{\delta\hfill\over\delta\phiB(2,1)}{\bf G}(3,4;\phi)\biggr]_{
\phi=0}\cr&\mskip -120mu={\calb G}(\overline{8},\overline{7})\bigl(-\sigma_x
{\bf G}^{-1}(\overline{5},\overline{8})\sigma_x{\bf G}^{-1}(\overline{7},
\overline{6})+{\mib\Gamma}(\overline{7},\overline{8},\overline{5},\overline{6})
\bigr)\biggl[{\delta\hfill\over\delta\phiB(\overline{5},\overline{6})}{\delta
\hfill\over\delta\phiB(2,1)}{\bf G}(3,4;\phi)\biggr]_{\phi=0}\cr}\nom{coungdev}
$$
o\`u on a fait appel au d\'eveloppement de ${\mib\Phi}$ au premier ordre
en ${\calb G}$, (\ref{phioung}). Introduisant les d\'eveloppement du
deuxi\`eme ordre (\ref{gdeuxg}) et (\ref{gmdeuxg}) dans (\ref{inverse}),
on peut montrer qu'il existe une relation liant le vertex \`a trois
particules et la fonction de corr\'elation \`a trois particules:
$$\eqalign{\biggl[{\delta\hfill\over\delta{\calb G}(6,5)}{\delta\hfill\over
\delta{\calb G}(4,3)}{\mib\Sigma}(1,2;{\cal G})\biggr]_{{\cal G}=0}=&-\sigma_x
{\bf G}^{-1}(3,6)\sigma_x{\bf G}^{-1}(1,4){\bf G}^{-1}(5,2)+\Biggl({\bf G}^{-1}
(1,\overline{7})\cr&{}\times\sigma_x\Bigl(\sigma_x{\bf G}^{-1}(\overline{9},4)
\sigma_x{\bf G}^{-1}(3,\overline{10})-{\mib\Gamma}(3,4,\overline{9},
\overline{10})\Bigr)\sigma_x\cr&{}\times\Bigl(\sigma_x{\bf G}^{-1}(
\overline{11},6)\sigma_x{\bf G}^{-1}(5,\overline{12})-{\mib\Gamma}(5,6,
\overline{11},\overline{12})\Bigr)\cr&{}\times\biggl[{\delta\hfill\over\delta
\phiB(\overline{11},\overline{12})}{\delta\hfill\over\delta\phiB(\overline{9},
\overline{10})}{\bf G}(\overline{7},\overline{8};\phi)\biggr]_{\phi=0}
{\bf G}^{-1}(\overline{8},2)\Biggr)\cr}\nom{s3fctg3}$$

En annexe \ref{autocoherence}, on utilise ces r\'esultats afin de proposer une
fa\c con d'ajouter la coh\'erence entre les propri\'et\'es \`a une et \`a
deux particules.


\annexe{Transformation de Legendre}\nom{anex2}

\section{D\'efinition}

Avant de voir l'application de la transformation de Legendre, voyons la
d\'efinition de cette transformation. La description pr\'esent\'ee ici
s'inspire de celle retrouv\'ee dans {\it Encyclop\ae dia of Mathematics}
[\ref{encyclo}].

Supposons que l'on connaisse une fonction $f:A\rightarrow{\blackB R}$
r\'eguli\`ere sur l'ensemble ouvert $A$ d'un espace norm\'e $X$. On peut
consid\'erer cette fonction comme une surface dans l'espace.
La surface \'etant alors d\'ecrite par l'ensemble des points $(x,f(x))$
o\`u $x\in A$.

La possibilit\'e d'\'evaluer la transformation de Legendre, $f^*$, de la
fonction $f$ repose sur l'existence d'une seconde description de la surface.
Celle-ci correspond \`a la famille des plans tangents \`a la surface. On doit
alors faire appel \`a une nouvelle variable servant \`a param\'etriser les
plans tangents, $x^*\in B$ et
\`a une fonction tangente affine pour la description de ces plans, $f^*(x^*)$.
L'ensemble ouvert $B$ appartient \`a un espace norm\'e $X^*$.
L'\'equivalence entre les deux descriptions provient de l'existence d'une
application biunivoque de $A$ sur $B$. Cette application permet de d\'efinir
la transformation de Legendre. Supposant que $f$ est une fonction sur
${\blackB R}^n$ et que le d\'eterminant ${\rm det}(\partial^2f/\partial x_i
\partial x_j)$ est finie sur $A$, la transformation de Legendre sera donn\'ee
par les formules suivantes:
$$f'(x)=x^*,\mskip 15mu f^*(x^*)=\langle x,f'(x)\rangle -f(x)$$
o\`u
$$\langle x,y\rangle =\sum_{i=1}^nx_iy_i,\mskip 15mu f'(x)=\biggl({\partial
f\over\partial x_1},...,{\partial f\over\partial x_n}\biggr)$$

Ainsi, lorsqu'on s'int\'eresse \`a la transformation de Legendre d'une
fonction, il faut conna\^\i tre la d\'eriv\'ee de Fr\'echet correspondante
$f'(x)$ et la d\'efinition de la norme dans l'espace d'int\'er\^et. La
transformation de Legendre introduit donc une nouvelle variable prenant
des valeurs dans un ensemble $B$ qui peut \^etre reli\'e de fa\c con
biunivoque \`a l'ensemble $A$ des valeurs admises par la variable $x$.

\section{Transformation de Legendre de la fonction de partition}

Dans notre cas, nous nous int\'eressons \`a la transformation de Legendre
de la fonction de partition $Z[\phi]$. Par rapport \`a la description de
la section pr\'ec\'edente nous pouvons identifier
$f$ au logarithme de la fonction de partition ${\rm ln}Z[\phi]$ et la
variable $x$ au champ $\phi$. L'op\'eration $<x,y>$ est quant \`a elle
remplac\'ee par la trace ${\rm Tr}\{xy\}$ telle qu'introduite en m\'ecanique
statistique quantique.
La d\'eriv\'ee de Fr\'echet de la fonction
de partition correspond \`a la partie hors diagonale de la fonction de Green
${\bf G}_{\rm HD}(\phi)$.

Nous devons donc trouver une transformation qui permet d'appliquer
l'ensemble ouvert des valeurs admissibles par la variable $\phi$ de
fa\c con biunivoque sur un ensemble ouvert des valeurs d'une variable
${\calb G}$. Utilisant la d\'efinition pr\'esent\'ee \`a la section
pr\'ec\'edente nous pouvons introduire la transformation de Legendre que
nous avons pr\'esent\'ee \`a la section \ref{dembs}:
$$W[{\cal G}]=\Bigl(-{\rm ln}Z[\Phi]-{\rm Tr}\bigl({\bf G}_{\rm HD}(
\overline{1},\overline{2};\phi)\phiB(\overline{2},\overline{1})\bigr)\Bigr)_{
G_{\rm HD}[\Phi]={\cal G}}$$
o\`u
$${\bf G}_{\rm HD}(1,2;\phi)=-{\delta{\rm ln}Z[\phi]\over\delta\phi(2,1)}$$

Tout comme il nous a \'et\'e possible de montrer que la partie hors diagonale
de la fonction de Green pouvait \^etre obtenue par la variation de la
fonction de partition par rapport au champ externe $\phiB$, eq.(\ref{eqderiv1}),
la partie hors diagonale de la self-\'energie pourra \^etre g\'en\'er\'ee par
la variation d'une fonctionnelle par rapport au champ introduit par la
transform\'ee de Legendre, ${\calb G}$. Nous avons montr\'e (\ref{fonctiong})
que:
$${\partial W[{\cal G}]\over\partial{\calb G}(2,1)}=-{\mib\Phi}(1,2;
{\cal G})$$
Or, de l'\'equation de Dyson, (\ref{eqdyson}), nous avons:
$${\mib\Phi}(1,2;{\cal G})=-{\bf G}_{\rm HD}^{-1}(1,2;{\cal G})-
{\mib\Sigma}_{\rm HD}(1,2;{\cal G})$$

Si l'on consid\`ere l'approximation du premier ordre en terme anormal, on
obtient la relation suivante:
$$\eqalign{{\bf G}_{\rm HD}^{-1}(1,2;{\cal G})=&-{\bf G}^{-1}(1,
\overline{3}){\calb G}(\overline{3},\overline{4}){\bf G}^{-1}(\overline{4},2)
\cr=&-{1\over 2}{\delta\hfill\over\delta{\calb G}(2,1)}{\rm Tr}\Bigl({
\calb G}(\overline{3},\overline{4}){\bf G}^{-1}(\overline{4},\overline{5})
{\calb G}(\overline{5},\overline{6}){\bf G}^{-1}(\overline{6},\overline{3})
\Bigr)\cr}$$

D\`es lors, on peut montrer qu'au premier ordre en terme anormal les termes
hors diagonaux de la self-\'energie seront donn\'es par:
$$\eqalign{{\mib\Sigma}_{\rm HD}(1,2;{\cal G})=&{\delta\hfill\over\delta
{\calb G}(2,1)}\Bigl(W[{\cal G}]+{1\over 2}{\rm Tr}\bigl({\calb G}(
\overline{3},\overline{4}){\bf G}^{-1}(\overline{4},\overline{5}){\calb G}
(\overline{5},\overline{6}){\bf G}^{-1}(\overline{6},\overline{3})\bigr)\Bigr)
\cr\equiv&{\delta\hfill\over\delta{\calb G}(2,1)}\Theta[{\cal G}]\cr}$$

Ainsi les termes diagonaux du vertex irr\'eductible sont donn\'es par la
d\'erivation seconde de cette m\^eme fonctionnelle par rapport aux termes
anormaux de la fonction de Green.
$${\bf\Gamma}(1,2,3,4)=\biggl[{\delta\hfill\over\delta{\calb G}(2,1)}
{\delta\hfill\over\delta{\calb G}(4,3)}\Theta[{\cal G}]\biggr]_{{\cal G}
=0}$$


\annexe{Autocoh\'erence}\nom{autocoherence}

\section{Comparaison \`a un hamiltonien avec interaction non-locale}

Une critique importante de notre approche vient du fait que celle-ci ne
pr\'esente
pas d'autocoh\'erence entre les propri\'et\'es \`a une particule et les
propri\'et\'es \`a deux particules.
Ainsi, la self-\'energie au deuxi\`eme niveau d'approximation
est diff\'erente de celle au premier niveau. Afin d'ajouter de
l'autocoh\'erence dans notre approximation il faudrait pouvoir utiliser
la self-\'energie obtenue au deuxi\`eme niveau pour r\'e\'evaluer le vertex
et la fonction de corr\'elation qui pourrait alors \^etre utilis\'es pour
passer au niveau suivant et ainsi de suite, it\'erer
jusqu'\`a convergence de l'approximation.

Dans cette annexe, nous proposons une m\'ethode pour
ajouter l'autocoh\'erence. Le fait qu'au deuxi\`eme niveau la self-\'energie
est dynamique et d\'elocalis\'e peut sembler compliquer les choses si
l'on veut garder une certaine coh\'erence. Cependant, avant de d\'ebuter
les calculs, on peut se guider en \'etudiant des relations simples mais
appliqu\'ees \`a un hamiltonien un peu plus complexe que celui de Hubbard.
Nous allons consid\'erer un hamiltonien avec interaction non-locale. Ce
hamiltonien pr\'esente deux termes. Le premier \'etant le terme
d'\'energie cin\'etique pour les \'electrons. Le deuxi\`eme pourrait \^etre
consid\'er\'e comme un terme d'\'energie cin\'etique pour les paires
de Cooper.
$${\cal H}=-\psi^\dagger(\overline{1})t(\overline{1},\overline{2})\psi(
\overline{2})+\Delta^\dagger(\overline{1})V(\overline{1},\overline{2})\Delta
(\overline{2})$$

Pour un hamiltonien de ce type, la self-\'energie au premier ordre en
interaction sera donn\'ee par:
$${\mib\Sigma}(1,2;{\cal G})\approx{\bf V}(1,\overline{3}){\calb G}(
\overline{3},\overline{3})\delta(1-2)+\sigma_x{\bf V}(1,2){\bf G}(1,2)\sigma_x
+O({\calb G}^2)$$
o\`u la matrice d'interaction est d\'efinie par:
$${\bf V}(1,2)=\pmatrix{V(1,2)&0\cr0&-V(2,1)}$$

Il semble donc possible d'approximer la renormalisation du vertex de fa\c con
telle que la partie diagonale de la self-\'energie soit non-locale tandis que
la partie hors diagonale demeure locale. Une telle approximation donnerait
un vertex dont l'\'el\'ement $(1,1)$ pourrait \^etre approxim\'e par:
$$\Gamma(1,2,3,4)\approx\sigma_xV(3,1)\sigma_x\delta(1-2)\delta(3-4)$$
Utilisant une telle approximation avec l'\'equation de Bethe-Salpeter
(\ref{bs}), on obtient une fonction de corr\'elation de paires de forme
RPA mais avec un vertex dynamique et d\'elocalis\'e au d\'enominateur.
$$\chi(\tilde{q})={\chi_0(\tilde{q})\over1+\Gamma(\tilde{q})\chi_0(\tilde{q})}$$

\section{\'Equation autocoh\'erente pour le vertex irr\'eductible}

Voyons maintenant comment il serait possible d'ajouter l'autocoh\'erence.
On consid\`ere que l'on a obtenu notre premi\`ere approximation pour
le vertex et la fonciton de corr\'elation: ${\mib\Gamma}^{(1)}$ et
$\chiB^{(1)}$ donn\'e respectivement par les relations (\ref{vertexun}), et
(\ref{chiun}). Ainsi, utilisant l'\'equation du mouvement et l'\'equation
de Bethe-Salpeter nous avons obtenu ${\mib\Sigma}^{(2)}$ et ${\bf G}^{(2)}$
donn\'e par (\ref{selfdeux}) et (\ref{gdeux}). On voit de par sa d\'efinition
que pour r\'e\'evaluer ${\mib\Gamma}$, il nous faudrait conna\^\i tre
${\mib\Sigma}^{(2)}$ au premier ordre en ${\calb G}$. Pour y arriver nous
pouvons proc\'eder de la m\^eme fa\c con que pour le terme ind\'ependant
de ${\calb G}$: en utilisant l'\'equation du mouvement (\ref{eqmouhub}).
$$\eqalign{{\mib\Sigma}(1,2;{\cal G})=&U{\calb G}(1,1)
\delta(1-2)-U\chiB(1,1,1^{\pm},\overline{3}){\bf G}^{-1}(\overline{3},
\overline{4}){\calb G}(\overline{4},\overline{5}){\bf G}^{-1}(\overline{5},
2)\cr&{}+U\chiB(1,1,1^\pm,\overline{3};{\cal G}){\bf G}^{-1}
(\overline{3},2)+O({\calb G}^2)\cr}\nom{soung}$$

Pour estimer le terme $\chiB({\cal G})$ \`a l'ordre lin\'eaire en ${\calb G}$,
nous utilisons les d\'eveloppements du deuxi\`eme ordre pour ${\bf G}({\cal G})$
et ${\bf G}^{-1}({\cal G})$ que nous avons pr\'esent\'es en annexe
\ref{acalcul}. Utilisant le fait que les termes du deuxi\`eme ordre doivent
s'annuler du c\^ot\'e gauche de l'identit\'e suivante:
$${\bf G}^{-1}(1,\overline{3};{\cal G}){\bf G}(\overline{3},2;{\cal G})=
\delta(1-2)$$
on peut montrer que:
$$\eqalign{\chiB(1,2,3,4;{\cal G})=&\chiB(1,2,3,4)-{\mib\cal G}(1,4){\bf G}(3,2)
\cr&{}+\chiB(1,2,\overline{5},\overline{6})\biggl({\mib\Gamma}(\overline{6},
\overline{5},\overline{7},\overline{8}){\mib\cal G}(\overline{8},4)+
{\mib\Gamma}_{\rm HD}(\overline{6},\overline{5},\overline{7},\overline{8};
{\cal G}){\bf G}(\overline{8},4)\biggr){\bf G}(3,\overline{7})\cr&{}+
O({\calb G}^2)\cr}\nom{coung}$$
Incorporant (\ref{coung}) dans (\ref{soung}), nous obtenons:
$${\mib\Sigma}_{\rm HD}(1,2;{\cal G})=U\delta(1-2){\calb G}(1,1)+U\chiB(
1,1,\overline{5},\overline{6}){\mib\Gamma}_{\rm HD}(\overline{6},\overline{5},
\overline{7},2;{\cal G}){\bf G}(1,\overline{7})+O({\calb G}^3)\nom{soungfin}$$

La d\'efinition du vertex irr\'eductible (\ref{vertex}), nous permet
en combinant (\ref{soungfin}) et (\ref{gaoung}) d'obtenir une relation
d\'eterminant la partie diagonale du vertex en l'absence de champ externe en
fonction d'un vertex \`a trois particules:
$$\eqalign{{\mib\Gamma}(1,2,3,4)=&U\delta(1-2)\delta(1-3)\delta(1-4)\cr&{}+U
{\bf G}(3,\overline{7})\sigma_x\chiB(3,3,\overline{5},\overline{6})\sigma_x
\biggl[{\delta\hfill\over\delta{\calb G}(2,1)}{\delta\hfill\over\delta{\calb
G}(\overline{5},\overline{6})}{\mib\Sigma}(\overline{7},4,{\cal G})\biggr]_{
{\cal G}=0}\cr}\nom{bsgamma}$$

Notre sch\'ema d'approximation pourrait donc \^etre d\'evelopp\'e de la
fa\c con suivante: on approxime la self-\'energie au deuxi\`eme ordre en
${\calb G}$. Partant de cette approximation on \'evalue tel que pr\'esent\'e
au chapitre 2,
$\Gamma^{(1)}$, $\chi^{(1)}$, $\Sigma^{(1)}$ et $G^{(1)}$. On y
ajoute le calcul de la d\'eriv\'ee seconde de la self-\'energie. \'Etant
donn\'e que cette derni\`ere correspond \`a un vertex \`a trois particules,
on pourrait garder cette quantit\'e fix\'ee et assurer l'autocoh\'erence
entre les
propri\'et\'es \`a une et \`a deux particules en utilisant les relations
de Dyson, eq.(\ref{eqdyson}), notre expression exacte pour la self-\'energie,
eq.(\ref{selfdeux}), l'\'equation de Bethe-Salpeter, eq.(\ref{bs}), et 
l'expression du vertex irr\'eductible en fonction de vertex plus complexe,
eq.(\ref{bsgamma}). Cette fa\c con de proc\'eder s'apparente \`a la m\'ethode
hi\'erarchique propos\'ee par Kadanoff et Martin [\ref{kadanoff}] qui consiste
\`a utiliser les \'equations du mouvement pour les fonctions de corr\'elation
\`a une et \`a deux particules qui permettent de faire le lien entre elles
et avec la fonction de corr\'elation \`a trois particules. Par la suite,
utilisant une approximation pour la fonction de corr\'elation \`a trois
particules sous forme d'une combinaison de fonctions \`a une et \`a deux
particules, ils proposent une solution autocoh\'erente pour les fonctions
de Green \`a une et deux particules.

\section{Suggestion pour l'ajout de l'autocoh\'erence}

Dans cette derni\`ere section, nous proposons une approximation pour le
vertex \`a trois particules. Il est possible de d\'evelopper une approximation
qui respecte l'esprit de la technique pr\'esent\'ee au chapitre deux.
Nous utilisons la relation (\ref{s3fctg3}). Encore une fois, on introduit
une matrice, ${\mib\Lambda}$, qui sera d\'etermin\'ee par les corr\'elations
locales:
$$\eqalign{U{\bf G}(1,\overline{7})\chiB(3,4,\overline{9},\overline{10})
&\biggl[{\delta\hfill\over\delta{\calb G}(\overline{9},\overline{10}
)}{\delta\hfill\over\delta{\calb G}(\overline{5},\overline{6})}{\mib\Sigma}
(\overline{7},\overline{8};{\cal G})\biggr]_{{\cal G}=0}\sigma_x\chiB(1,1,
\overline{5},\overline{6})\sigma_x{\bf G}(\overline{8},2)\hfill\cr=-&\sigma_x
{\mib\Sigma}(1,\overline{5})\sigma_x\chiB(3,4,\overline{5},2)+U\biggl[{\delta
\hfill\over\delta\phiB(4,3)}{\delta\hfill\over\delta\phiB(1,1)}{\bf G}(1^\mp,2;
\phi)\biggr]_{\phi=0}\cr\approx-&U{\mib\Lambda}\sigma_x{\bf G}(1,\overline{5})
\sigma_x{\bf G}(\overline{6},1)\chiB(3,4,\overline{8},\overline{7}){\mib\Gamma}
(\overline{7},\overline{8},\overline{5},\overline{6}){\bf G}(1,2)\cr}
\nom{s3approx}$$

Pour justifier cette approximation, on peut consid\'erer l'\'equation du
mouvement pour la fonction de corr\'elation \`a deux particules:
$$\eqalign{\sigma_x{\bf G}_0^{-1}(1,\overline{5})\sigma_x\chiB(3,4,\overline{5},
2)=&-\delta(1-4){\bf G}(3,2)+U{\bf G}(1,2)\chiB(3,4,1,1)\cr&{}+U\biggl[
{\delta\hfill\over\delta\phiB(4,3)}{\delta\hfill\over\delta\phiB(1,1)}{\bf G}
(1^\mp,2;\phi)\biggr]_{\phi=0}}$$
On voit appara\^\i tre dans cette relation une fonction de Green \`a trois
particules via le terme de d\'erivation. On propose donc une approximation
pour cette fonction qui permet de r\'esoudre l'\'equation du mouvement \`a
partir des quantit\'es introduites pr\'ec\'edemment.
$$\eqalign{U\biggl[{\delta\hfill\over\delta\phiB(4,3)}{\delta\hfill\over\delta
\phiB(1,1)}{\bf G}(1^\mp,2;\phi)\biggr]_{\phi=0}\approx&\sigma_x{\mib\Sigma}(1,
\overline{5})\sigma_x\chiB(3,4,\overline{5},2)-U\chiB(3,4,1,1){\bf G}(1,2)
\cr&{}-U\sigma_x{\bf G}(1,4)\sigma_x{\bf G}(3,1){\bf G}(1,2)\cr}\nom{g3approx}$$

On peut v\'erifier qu'au premier ordre en $U$ l'\'egalit\'e entre les deux
c\^ot\'e est v\'erifif\'ee de fa\c con exacte.
Ainsi, en introduisant
l'approximation (\ref{g3approx}) dans la premi\`ere ligne de (\ref{s3approx})
on obtient la deuxi\`eme ligne de cette m\^eme \'equation o\`u on a ajout\'e
le facteur multiplicatif ${\mib\Lambda}$. Combinant les
relations (\ref{bsgamma}) et (\ref{s3approx}), il devient alors possible
d'approximer le vertex irr\'eductible.
$${\mib\Gamma}(1,2,3,4)\approx U\delta(1-2)\delta(1-3)\delta(1-4)-U{\mib\Lambda}
\delta(3-4)\sigma_x{\bf G}(3,\overline{5})\sigma_x{\bf G}(\overline{6},3)
{\mib\Gamma}(1,2,\overline{5},\overline{6})\nom{vren}$$

La relation peut se simplifier si l'on consid\`ere en premi\`ere
approximation le r\'esultat obtenu au chapitre deux et que l'on utilise la
relation (\ref{vren}) afin d'am\'eliorer l'approximation.
$${\mib\Gamma}^{(2)}(1,2,3,4)=U\delta(1-2)\delta(1-3)\delta(1-4)-U{\mib\Lambda}
\delta(1-2)\delta(3-4)\sigma_x{\bf G}^{(1)}(3,1)\sigma_x{\bf G}^{(1)}(1,3)
{\mib\Gamma}^{(1)}$$

Supposant que les it\'erations permettent de converger vers une solution unique,
on obtiendrait alors un vertex \`a deux points, ${\mib\Gamma}(1,2,3,4)=
{\mib\Gamma}(1,2,3,4)\delta(1-2)\delta(3-4)\equiv{\mib\Gamma}^*(1,3)\delta(1-2)
\delta(3-4)$, correspondant \`a la solution de la relation:
$$\boxEq{{\mib\Gamma}^*(1,2)=U\delta(1-2)-U{\mib\Lambda}\sigma_x{\bf G}^*(2,
\overline{3})\sigma_x{\bf G}^*(\overline{3},2){\mib\Gamma}^*(1,\overline{3})}$$

On voit que l'on obtient une forme RPA pour la renormalisation du vertex.
\'Etant donn\'e que le coefficient devant le vertex droit est n\'egatif
dans l'espace de Fourier, on remarque que le vertex sera inf\'erieur, en valeur
absolue, \`a l'interaction nue. De plus, il est facile de montrer que le
comportement aux grandes fr\'equences de ce vertex est identique au r\'esultat
exact: ${\rm lim}_{q_n\rightarrow\infty}\Gamma({\bf q},iq_n)\rightarrow U$.
Il ne reste ainsi qu'\`a d\'eterminer le facteur multiplicatif $\Lambda$.

Consid\'erant la relation o\`u nous avons introduit le facteur $\Lambda$. Nous
pouvons la r\'eecrire sous la forme:
$$\eqalign{{\mib\Lambda}\biggl(\chiB(3,4,1,1)-\chiB_0(3,4,1,1)\biggr){\bf G}
(1,2)\approx&\sigma_x{{\mib\Sigma}(1,\overline{5})\over U}\sigma_x\chiB(
3,4,\overline{5},2)\cr&{}+\biggl[{\delta\hfill\over\delta\phiB(4,3)}{\delta
\hfill\over\delta\phiB(1,1)}{\bf G}(1^\pm,2;\phi)\biggr]_{\phi=0}\cr}
\nom{approxcoh}$$
o\`u
$$\chiB_0(3,4,1,1)=-\sigma_x{\bf G}(1,4)\sigma_x{\bf G}(3,1)$$
Le deuxi\`eme terme du c\^ot\'e droit de l'\'egalit\'e est assez simple \`a
\'evaluer de fa\c con locale. Dans un premier temps, on peut \'ecrire
ce terme sous la forme suivante:
$$\eqalign{\biggl[{\delta\hfill\over\delta\phiB(4,3)}{\delta\hfill\over\delta
\phiB(1,1)}{\bf G}(1^\pm,2;\phi)\biggr]_{\phi=0}=&\chiB(3,4,1,1){\bf G}(1,2)\cr
&\mskip -250mu{}+\pmatrix{\langle\psi_\uparrow(3)\psi_\downarrow(4)
\psi_\downarrow^\dagger(1)\psi_\uparrow^\dagger(1)\psi_\uparrow(1^-)
\psi_\uparrow^\dagger(2)\rangle&0\cr0&\langle\psi_\downarrow^\dagger(3)
\psi_\uparrow^\dagger(4)\psi_\uparrow(1)\psi_\downarrow(1)
\psi_\downarrow^\dagger(1^+)\psi_\downarrow(2)\rangle}\cr}$$

Le premier terme du c\^ot\'e droit de l'\'equation (\ref{approxcoh}) est plus
compliqu\'e. De fa\c con locale, on peut trouver une expression simple en
utilisant le fait que la self-\'energie peut s'\'ecrire:
$${\mib\Sigma}(1,2)=-U{\bf G}^{-1}(1,\overline{3})\sigma_x\chiB(2,2,
\overline{3},2^\pm)\sigma_x$$

Comme pr\'esent\'e au chapitre 2 lors du calcul de la self-\'energie,
\'equation (\ref{approxim}) et (\ref{facteurlocal}), nous regardons la
contribution provenant des
diff\'erents termes locaux pour d\'eterminer ${\mib\Lambda}$ de l'\'equation
(\ref{approxcoh}). Il faut faire attention car plusieurs ordonnances
sont possibles: $3=1^\pm$, $2=1^\pm$, $3=2^\pm$ et $4=1^\pm$. Il y a donc
12 termes diff\'erents: 6 pour l'ordonnance respective des indices $1$, $2$
et $3$ multipli\'e par un facteur deux provenant de l'ordonnance des indices
$1$ et $4$. D\`es lors, on obtient l'approximation:
$$\eqalign{{\mib\Lambda}(\langle n_\uparrow n_\downarrow\rangle-\ffrac{n^2}{4}
)=&{1\over 6}\biggl({n^2-4\langle n_\uparrow n_\downarrow+\ffrac{3}{4}(n^2-4n
\langle n_\uparrow n_\downarrow\rangle)\over n}+{(2+3n)\langle n_\uparrow
n_\downarrow\rangle-5n^2/4\over 2-n}\biggr){\blackB I}\cr&{}+{1\over 4}\sigma_x
{{\mib\Sigma}(1,\overline{5})\over U}{\bf G}(\overline{5},\overline{6})\Bigl(
{{\mib\Sigma}(\overline{6},1^+)\over U}+{{\mib\Sigma}(\overline{6},1^-)\over U}
\Bigr)\biggl({2\over n}-{2\over 2-n}\biggr)\sigma_x\sigma_z\cr}$$
o\`u ${\blackB I}$ est la matrice identit\'e.
En fait, il serait sans doute plus juste de pr\'esenter ce r\'esultat comme
\'etant un ansatz plut\^ot qu'une approximation.


\annexe{Prolongement analytique}\nom{mem}


Comme nous l'avons vu au dernier chapitre, il est possible de d\'eterminer
le poids spectral \`a partir de la connaissance de la fonction de Green en
temps imaginaire. On peut \'etablir un lien entre ces deux quantit\'es en
utilisant
la repr\'esentation de Lehmann [\ref{ncorps}]. La relation ainsi obtenue est
appel\'ee repr\'esentation spectrale de la fonction de Green en temps
imaginaire. Celle-ci est donn\'ee par la relation (\ref{eqprolong}).
$$\boxEq{G({\bf k},\tau)=-\int_{-\infty}^\infty {d\omega\over 2\pi}{e^{-\tau
\omega}\over 1+e^{-\beta\omega}}A({\bf k},\omega)}\nom{eqprolong}$$
Son domaine de validit\'e
est restreint \`a $\tau\in\mskip 1mu(0,\beta)$ (L'utilisation de ``('' au
lieu de ``['' signifie que la borne n'est pas incluse dans le domaine).

Donc, nous voulons d\'eterminer le poids spectral, $A(\omega)$, pour un
ensemble de valeurs
de $\omega$ en utilisant la connaissance de la
fonction de Green, $G(\tau)$ pour un ensemble de valeurs de $\tau$. Il est
\`a noter que pour des donn\'ees Monte Carlo, notre connaissance de la
fonction de Green n'est que statistique et qu'\`a chaque valeur de $G(\tau)$
nous devons associer une incertitude, $\sigma(\tau)$.

Le probl\`eme auquel nous faisons face est que l'\'equation (\ref{eqprolong})
n'est pas
r\'eversible. Plus pr\'ecis\'ement on qualifiera ce probl\`eme comme \'etant
``mal pos\'e''. Dans la section qui suit, nous d\'etaillons les diff\'erentes
raisons pour laquelle ce probl\`eme est dit mal pos\'e. Nous mentionnons comment
chacun de ces probl\`emes pourrait \^etre r\'egl\'e. Dans la seconde section
nous d\'ecrivons la fonctionnelle d'entropie introduite afin de
r\'egulariser le probl\`eme. Enfin, dans la derni\`ere section,
nous proposons diff\'erents algorithmes permettant de d\'eterminer le poids
spectral. Une comparaison de deux de ces algorithmes a \'et\'e pr\'esent\'e par
Touchette et Poulin [\ref{david}].

\section{Probl\`emes math\'ematiquement mal pos\'es}

Baym et Mermin [\ref{baym3}] ont
montr\'e que si l'on conna\^\i t la fonction de Green en temps imaginaire
sur toute la plage $(0,\beta)$ alors il n'existe qu'une seule fonction
$A(\omega)$ permettant de satisfaire la relation (\ref{eqprolong}) et ayant
les propri\'et\'es d'un poids spectral [\ref{ncorps}]. Pour le cas qui nous
int\'eresse, on peut dire que nous n'avons pas suffisamment d'information
pour d\'eterminer sans \'equivoque le poids spectral. Ce manque d'information
provient de diff\'erentes raisons: premi\`erement, la fonction de Green n'a
\'et\'e \'evalu\'ee qu'en un certain nombre de points en temps imaginaire,
$\tau_i (i\in[1,N])$; deuxi\`emement, nos
donn\'ees sont caract\'eris\'ees par une incertitude statistique.

Tel que mentionn\'e par Sch\"afer et Sternin [\ref{illposed}], il existe
plusieurs fa\c cons de r\'egulariser le probl\`eme. D'abord, la
discr\'etisation des fr\'equences permet de r\'eduire le nombre de solutions
possibles. Cependant, \c ca n'aurait pas de sens de r\'eduire le nombre de
fr\'equences \`a un point tel que la solution soit unique car la solution
n'aurait probablement pas une allure physiquement acceptable. De plus, elle
ne respecterait peut etre pas certaines propri\'et\'es des poids spectraux
telle la positivit\'e. La discr\'etisation des fr\'equences ram\`ene le
probl\`eme \`a la solution de la relation:
$$G(\tau_i)=\sum_{j=1}^MK(\tau_i,\omega_j)A(\omega_j)\mskip 1mu,\mskip 30mu
i\in[1,N]\nom{proldis}$$
o\`u on a introduit le noyau
$$K(\tau,\omega)={e^{-\tau\omega}\over 1+e^{-\beta\omega}}\nom{noyau}$$

Sous forme matricielle, la relation (\ref{proldis}) prend la forme simple
$G=KA$. Ainsi la solution du probl\`eme devient $A=K^{-1}G$. Cependant, comme
nous avons mentionn\'e, lors de la discr\'etisation on doit garder un nombre
de fr\'equences plus importants que le nombre de donn\'ees en temps imaginaire,
$M>N$. Ainsi, la matrice $K$ est une matrice rectangulaire et son inverse
n'est pas unique.

Afin de d\'eterminer une solution unique,
on doit faire appel \`a d'autres m\'ethodes de r\'egularisation comme la
maximisation
d'entropie. Cette m\'ethode consiste \`a choisir parmi les solutions celle
qui contient le moins d'information possible. Par information contenue
dans une fonction, on parle du comportement de cette fonction. Plus une
fonction est monotone, moins on peut en tirer d'information, le cas limite
\'etant la fonction constante. Pour quantifier cette propri\'et\'e on utilise
la fonctionnelle d'entropie, $S(A)$. Celle-ci mesure l'absence d'information
contenue dans une distribution.

Le choix de la m\'ethode de maximisation
d'entropie repose sur le fait qu'\'etant donn\'e que le poids spectral
est \'evalu\'e en un nombre de points plus \'elev\'e, celui-ci a la capacit\'e
de contenir plus d'information que ne peut en contenir la fonction de Green.
Ce qui n'a pas de sens car on utilise la fonction de Green afin de d\'eterminer
le poids spectral, ainsi l'information contenue dans le poids spectral devrait
\^etre limit\'ee par l'information contenue dans la fonction de Green et
l'information que l'on a sur les propri\'et\'es des poids spectraux.
Ainsi, on cherchera parmi les poids spectraux qui satisfont la relation
(\ref{proldis}) celui qui maximise la fonctionnelle d'entropie $S(A)$.

Il est \`a noter que l'imposition de certaines conditions physiques pour le
poids spectral peut r\'eduire de fa\c con consid\'erable le nombre de
solutions possibles. Par exemple, on peut montrer que le fait que le poids
spectral soit positif en tout point, impose que la fonction de Green ait
un maximum unique dans l'intervalle $(0,\beta)$ et un comportement
monotone de part et d'autre de ce maximum. La pr\'esence de bruit dans nos
donn\'ees MCQ peut faire en sorte que la fonction de Green ne respecte pas
cette propri\'et\'e (Ce probl\`eme sera plus accentu\'e pour des valeurs
de $\Delta\tau$ plus faibles). Il faut alors ajouter parmi l'ensemble des
solutions admises, celles qui ne permettent pas de satisfaire de fa\c con
exacte la relation (\ref{proldis}). Une fa\c con d'y parvenir consiste \`a
faire une d\'ecomposition en valeurs singuli\`eres du noyau, eq. (\ref{noyau}),
et de tronquer les valeurs propres les plus faibles. On obtient alors
un noyau qui permet une plus grande ind\'ependance du poids spectral par
rapport \`a de petites fluctuations de la fonction de Green.

La d\'ecomposition en valeurs singuli\`eres du noyau n'est pas l'unique
m\'ethode permettant d'\'elargir le nombre de solutions. On peut
\'egalement \'elargir l'ensemble des solutions \`a toutes les fonctions
possibles respectant certaines conditions physiques et \`a chercher parmi
ces solutions, celle qui permet de minimiser
$$\chi^2(A,G)=\sum_i{\Bigl(G(\tau_i)-\sum_jK(\tau_i,\omega_j)A(\omega_j)\Bigr)^2
\over\sigma^2(\tau_i)}$$

On fait alors face au probl\`eme de minimisation des moindres carr\'es tout
en maximisant la fonctionnelle d'entropie. En faisant appel aux statistiques
bayesiennes [\ref{robert}], on peut ramener la d\'etermination du poids
spectral \`a la recherche de la fonction permettant de maximiser la
probabilit\'e:
$$P(A|G,\alpha)=e^{-\ffrac{1}{2}\chi^2(A,G)+\alpha S(A)}\nom{probA}$$
Le facteur $\alpha$ a \'et\'e introduit afin de donner plus ou moins
d'importance au terme d'entropie par rapport au terme de moindres carr\'ees.
Le choix d'une valeur pour ce facteur est assez complexe et nous r\'ef\'erons
le lecteur \`a l'article de Jarrell et Gubernatis [\ref{jarrell}].

\section{Fonctionnelle d'entropie}

Pour le choix de la fonctionnelle d'entropie plusieurs auteurs proposent
la fonctionnelle d'entropie introduite par Jaynes (voir [\ref{ihara}]) qui est
mesur\'ee par rapport \`a un mod\`ele par d\'efaut, $m(\omega)$.
$$S(A|m)=\int d\omega\Bigl(A(\omega)-m(\omega)-A(\omega){\rm ln}\bigl(A(\omega)/
m(\omega)\bigr)\Bigr)$$

Le choix de cette fonctionnelle vient de l'entropie conditionnelle d'une
distribution continue. \'Etant donn\'e que le poids spectral n'a pas une
normalisation assur\'ee, on ajoute \`a cette entropie conditionnelle le
terme $A-m$ qui fait en sorte que la normalisation de $A$ devrait \^etre
aussi rapproch\'ee que possible de celle du mod\`ele par d\'efaut. On peut
v\'erifier que notre entropie a un maximum unique donn\'ee par $A=m$ et
qu'elle respecte la propri\'et\'e de convexit\'e de l'entropie. Pour le
d\'emontrer, on suppose que le poids spectral est une quantit\'e positive en
tout point. Il est \`a noter que cette entropie donne l'entropie
de la distribution $A$ mesur\'ee par rapport \`a celle de $m$ et sera donc
n\'egative pour toute fonction $A$ et nulle pour $A=m$. Le mod\`ele par
d\'efaut doit contenir le moins d'information possible, c'est pourquoi on
choisit $m(\omega)=m$. On peut v\'erifier qu'en g\'en\'eral, le terme en
$\chi^2$ dans (\ref{probA}) est dominant pour la maximisation de la
probabilit\'e (\ref{probA}) et qu'un changement du mod\`ele par d\'efaut ou
de l'entropie a peu d'effet sur le poids spectral obtenu.

\section{Algorithme pour la maximisation d'entropie}

Parmi les algorithmes introduits pour trouver le minimum de l'\'equation
(\ref{probA}), on mentionne g\'en\'erallement les algorithmes de Bryan
[\ref{bryan}] et de Meshkov et Berkov [\ref{meshkov}]. Ces deux algorithmes
sont deux m\'ethodes diff\'erentes pour la recherche du maximum de
l'\'equation (\ref{probA}). Il est \`a noter que ces deux m\'ethodes donnent
des r\'esultats quelques peu diff\'erents [\ref{david}]. Pour l'\'etude du
poids spectral dans cette th\`ese notre choix s'est port\'e sur l'algorithme
de Bryan, car cet algorithme s'est av\'er\'e plus rapide que celui de Meshkov.
On a toutefois v\'erifi\'e que les r\'esultats obtenus par chacun des
algorithmes ne sont pas tr\`es diff\'erents de fa\c con quantitative aussi
bien que de fa\c con qualitative. Nous avons \'egalement d\'evelopp\'e un
nouvel algorithme donnant des r\'esultats sensiblement identiques \`a la
m\'ethode de Bryan. L'avantage de ce nouvel algorithme permet de s\'eparer
les diff\'erents probl\`emes reli\'es \`a la technique de prolongement
analytique et ainsi de faire ressortir l'effet de la pr\'ecision des
donn\'ees. Malgr\'e que celui-ci ne fut pas utilis\'e pour les r\'esultats
pr\'esent\'es au troisi\`eme chapitre, nous tenons \`a pr\'esenter ce nouvel
algorithme car nous pouvons ainsi d\'etailler les diff\'erents
\'el\'ements des techniques de r\'egularisation.

S'inspirant
des travaux de Geman et Geman [\ref{geman}], le nouvel algorithme vise \`a
chercher parmi les solutions de l'\'equation (\ref{proldis}), celle qui
maximise l'entropie. La difficult\'e dans la recherche du maximum de la
probabilit\'e (\ref{probA}), vient de la pr\'esence de minimum locaux
dans la partie provenant des moindres carr\'es, $\chi^2$. Cependant, le terme
d'entropie est parfaitement convexe. C'est pourquoi notre algorithme consiste
\`a consid\'erer comme point de d\'epart une valeur de $\alpha$ tr\`es
\'elev\'ee pour ainsi n'accorder que peu d'importance au $\chi^2$. Ce terme
ne devrait induire qu'un petit d\'eplacement dans la position du maximum.
Trouvant la position de ce maximum, on change notre mod\`ele par d\'efaut,
on abaisse l\'eg\`erement la valeur de $\alpha$ et on cherche la nouvelle
position du maximum de la probabilit\'e. Ainsi, on peut introduire peu \`a
peu dans le poids spectral l'information contenue dans la fonction de Green.
Interpr\'etant le param\`etre $\alpha$ comme une temp\'erature que l'on
abaisse, cette approche correspond \`a une m\'ethode de recuit simul\'e. Une
caract\'eristique importante pour l'utilisation de cette m\'ethode vient
du fait que la fonctionnelle d'entropie a un comportement convexe en
fonction du poids spectral [\ref{attouch}].

Voici plus en d\'etail le fonctionnement de la m\'ethode.
Initialement, on consid\`ere une valeur de $\alpha$ tr\`es \'elev\'ee et on
cherche la fonction $A$ maximisant
$$P_0(A|G,\alpha)={\rm exp}\Bigl(-\ffrac{1}{2}\chi^2(A,G)+\alpha_0S(A,m)\Bigr)$$

Le maximum de l'entropie est \`a $A=m$. Le terme $\alpha$ \'etant important,
la solution au probl\`eme de maximisation devrait \^etre tout pr\`es du
point $A=m$ mais l\'eg\`erement d\'eplac\'ee d\^u \`a la pr\'esence du terme
de moindres carr\'es, $\chi^2$. Appelons cette solution $A_1$.
On peut dire que le $\chi^2$ ajoute dans le poids spectral une partie de
l'information contenue dans la fonction de Green. On abaisse alors la valeur
de $\alpha$ et on cherche alors la fonction $A$ qui maximise
$$P_1(A|G,\alpha)={\rm exp}\Bigl(-\ffrac{1}{2}\chi^2(A,G)+\alpha_1S(A,A_1)\Bigr)
$$

On a ainsi encore une entropie convexe mais avec un minimum d\'eplac\'e
\`a $A=A_1$. Le choix d'abaisser la valeur de $\alpha$ vient du fait que
le nouveau mod\`ele par d\'efaut, $A_1$, contient plus d'information que
n'en contenait $m$. Afin d'aller chercher encore une certaine quantit\'e
d'information, on se doit donc d'abaisser la valeur de $\alpha$.
On it\`ere ainsi jusqu'\`a convergence de la fonction $A$ vers une solution
$A_n\simeq A_{n-1}$.

Cette m\'ethode semble donner des r\'esultats concluants. Cependant, il
reste \`a v\'erifier que la solution est unique et n'a pas de comportement
singulier par rapport au choix du point de d\'epart. Il reste aussi
\`a d\'eterminer l'\'evolution optimale pour les valeurs du param\`etre
$\alpha$. Typiquement, nous utilisons comme point de d\'epart
$\alpha_0=\ffrac{1}{2}\chi^2(m,G)$, puis \`a la premi\`ere it\'eration
$\alpha_1=\ffrac{1}{2}\chi^2(A_1,G)$ et ainsi de suite.

Le probl\`eme de l'incertitude statistique est
tenu en compte par le fait que nous utilisons une d\'ecomposition
en valeurs singuli\`eres du noyau et que nous tronquons les valeurs
propres les plus petites. Pour d\'ecire cette proc\'edure nous pouvons utiliser
la notation matricielle o\`u nous avons $N$ valeurs de $\tau$ et supposer
que notre
discr\'etisation des fr\'equences conserve $M>N$ points en fr\'equence.
Ainsi, $A$ est un vecteur de dimension $M$, $G$ est un vecteur de dimension
$N$ et le noyau, $K$, est une matrice de dimension $N\times M$.
On proc\`ede par la diagonalisation en valeurs singuli\`eres du noyau:
$$K=Q_1\Sigma Q_2$$

La matrice $\Sigma$ est une matrice rectangulaire $n\times M$. Seules les
composantes le long de la diagonale donn\'ees par les points $(1,1)$ \`a
$(N,N)$ sont non-nulles. Les derni\`eres composantes \'etant petites, elles
repr\'esentent une forte d\'ependance du poids spectral sur de petites
variations de la fonction de Green. C'est pourquoi il est appropri\'e de
tronquer la matrice $\Sigma$ en rempla\c cant ces derniers \'el\'ements
par des z\'eros. On note la matrice tronqu\'ee par $\Sigma_s$ et le nouveau
noyau par $K_s=Q_1\Sigma_sQ_2$. Cette modification du noyau augmente le
nombre de poids spectraux permettant de satisfaire la repr\'esentation
spectrale (\ref{proldis}).


{

\parindent=0pt
\parskip=7pt
\initBib
\baselineskip=\single


\bib{gorkov}{A. A. Abrikosov et al., {\it Methods of Quantum Field Theory in
Statistical Physics}, Pergamon, Elmsford, N.Y. (1965).}


\bib{allen1}{S. Allen et al., Phys. Rev. Lett. {\bf 83}, pp. 4129-4132 (1999).}

\bib{allen2}{S. Allen et al., en pr\'eparation.}

\bib{anderson}{P. W. Anderson, Phys. Rev. {\bf 112}, pp. 1900-1916 (1958). Aussi
disponible dans {\it The Many-Body Problem}, D. Pines.}

\bib{gw}{F. Aryasetiawan et O. Gunnarsson, Rep. Prog. Phys. {\bf 61},
pp. 237-312 (1998); L. Steinbeck et al., Comp. Phys. Comm. {\bf 125},
pp. 105-118 (2000).}

\bib{assaraf}{R. Assaraf et M. Caffarel, Phys. Rev. Lett. {\bf 83},
pp. 4682-4685 (1999).}

\bib{attouch}{H. Attouch, SIAM J. Optimization {\bf 6}, pp. 769-806 (1996).}

\bib{bcs}{J. Bardeen, L. Cooper et J. Schrieffer, Phys. Rev. {\bf 108},
pp. 1175-1204 (1957). Aussi disponible dans {\it The Many-Body Problem}, D. Pines.}

\bib{baym3}{G. Baym et N. D. Mermin, J. Math. Phys. {\bf 2}, pp. 232-234
(1961).}

\bib{baym1}{G. Baym, L. P. Kadanoff, Phys. Rev. {\bf 124}, pp. 287-299 (1961).}

\bib{baym2}{G. Baym, Phys. Rev. {\bf 127}, pp. 1391-1401 (1962).}


\bib{bednorz}{J. G. Bednorz et K. A. M\"uller, Z. Phys. B {\bf 64}, pp. 189-193
(1986).}


\bib{berezinskii}{V. L. Berezinski\u\i, Sov. Phys. {\bf JETP 32}, p. 493 (1971);
idem {\bf JETP 34}, p. 610 (1972).}

\bib{bss}{R. Blankenbecler, D. J. Scalapino et R. L. Sugar, Phys. Rev. {\bf D24},
p. 2278 (1981).}

\bib{bogoliubov}{N. N. Bogoliubov et al., {\it A New Method in the Theory of
Superconductivity}, Academic of Sciences of USSR Press, Moscou, 1958, traduit
en anglais par Consultants Bureau Inc, New-York, 1959.}

\bib{rg}{C. Bourbonnais et L. G. Caron, Physica {\bf B143}, p. 451 (1986);
J. S\'olyom, Adv. Phys. {\bf 28}, p. 201 (1979).}

\bib{bryan}{R. K. Bryan, {\it Solving oversampled data problems by maximum entropy},
dans {\it Maximum Entropy and Bayesian Methods}, P. Foug\`eres edt., p. 221 (1989).}

\bib{bulut}{N. Bulut et al., Phys. Rev. {\bf B47}, pp. 2742-2753 (1993).}

\bib{mass}{M. Capezzali, {\it Superconductivity in two dimensions}, th\`ese
de doctorat, Institut de Physique, Universit\'e de Neuch\^atel (1998).}

\bib{mass2}{M. Capezzali et al., cond-mat/9806211 publi\'e dans Physica B;
cond-mat/9809349 publi\'e dans Physica C}

\bib{chakra}{S. Chakravarty et al., Phys. Rev. {\bf B39}, pp. 2344-2371 (1989).}

\bib{chen}{L. Chen et al., Phys. Rev. Lett. {\bf 66}, pp. 369-372 (1991).}

\bib{chitra}{R. Chitra et G. Kotliar, cond-mat/9911223}

\bib{coleman}{S. Coleman, Comm. Math. Phys. {\bf 59}, p. 259 (1973).}

\bib{corson}{J. Corson et al., Nature {\bf 398}, pp. 221-223 (1999).}

\bib{xxzmc}{A. Cuccoli et al., Phys. Rev {\bf B52}, pp. 10221-10231 (1995);
L. Capriotti et al., {\it Proceedings of NATO Adv. Research Workshop}, Trieste,
Aug. 1996, Kluwer Academic Publishers, Dordrecht, 1997, pp. 397-404.}

\bib{curro}{N. J. Curro et al., Phys. Rev. {\bf B56}, pp. 877-885 (1997).}

\bib{neutron}{P. Dai et al., cond-mat/9712311.}

\bib{dare}{A.-M. Dar\'e, {\it Le mod\`ele de Hubbard \`a faible densit\'e et
\`a proximit\'e du demi-remplissage}, th\`ese de doctorat, D\'epartement de
Physique, Universit\'e de Sherbrooke (1994).}

\bib{dare2}{A.-M. Dar\'e et al., Phys. Rev. {\bf B49}, pp. 4106-4118 (1994).}

\bib{dare3}{A.-M. Dar\'e et al., Phys. Rev. {\bf B53}, pp. 14236-14251 (1996).}

\bib{flexpg}{J. J. Deisz et al., Phys. Rev. Lett. {\bf 76}, p. 1312 (1996);
J. J. Deisz et al., Phys. Rev. Lett. {\bf 80}, pp. 373-376 (1998);
D. W. Hess et al., cond-mat/9808267/}

\bib{tros}{J. Demsar et al., cond-mat/9812079; cond-mat/9905026.}

\bib{arpes}{H. Ding et al., Nature {\bf 382}, pp. 51-54 (1996); A. G. Loeser
et al., Science {\bf 273}, pp. 325-329 (1996).}


\bib{parquet}{C. De Dominicis et P. C. Martin, J. Math. Phys. {\bf 5}, 14
(1964); V. Jani\u s, cond-mat/9806118.}


\bib{boson}{V. J. Emery in {\it Highly Conducting One-Dimensional Solids},
J. T. Devreese et al. Eds., Plenum (1979).}

\bib{emery}{V. J. Emery et S. A. Kivelson, Nature {\bf 374}, pp. 434-437
(1995); E. W. Carlson et al., cond-mat/9902077.}

\bib{encyclo}{Encyclo\ae dia of Mathematics, Kluwer Academic Publishers,
Dordrecht (1989).}

\bib{conforme}{H. Frahm et V. E. Korepin, Phys. Rev. {\bf B42}, p. 10553
(1990); P. DiFrancesco et al., {\it Conformal Field Theory}, Springer-Verlag,
New-York (1996).}

\bib{fronsdal}{{\bf Lecture Notes on the Many-body Problem}, C. Fronsdal edt.,
Benjamin, N.Y. (1962).}

\bib{fye}{R. M. Fye, Phys. Rev. {\bf B33}, p. 6271 (1986).}

\bib{mcmc}{D. Gamerman, {\it Markov chain Monte Carlo}, coll. Texts in
Statistical Science, Chapman \& Hall, London (1997).}

\bib{geman}{S. Geman et D. Geman, IEEE transactions on PAMI-{\bf 6}, pp.721-741
(1984).}

\bib{dmft}{A. Georges et al., Rev. Mod. Phys. {\bf 68}, p. 13 (1996).}


\bib{flex}{D. R. Hamann, Phys. Rev. {\bf 186}, p. 549 (1969); N. E. Bickers
et S. R. White, Phys. Rev. {\bf B43}, p. 8044 (1991).}

\bib{hammersley}{J. M. Hammersley et D. C. Handscomb {\it Les M\'ethodes de
Monte Carlo}, Dunod, Paris (1967).}

\bib{haussmann}{R. Haussmann, Z. Phys. B {\bf 91}, pp.291-308 (1993); Phys. Rev.
{\bf B49}, pp. 12975-12983 (1994).}

\bib{hirsch}{J. E. Hirsch, Phys. Rev. {\bf B28}, pp. 4059-4061 (1983);
Phys. Rev. {\bf B31}, pp. 4403-4419 (1985).}

\bib{hohenberg}{P. C. Hohenberg, Phys. Rev. {\bf 158}, p. 383 (1967).}

\bib{dca}{C. Huscroft et al., cond-mat/9910226; Th. Maier et al.,
cond-mat/0002352.}

\bib{ihara}{S. Ihara, {\it Information theory for continuous systems},
World Scientific, Singapore (1993).}

\bib{iwamatsu}{M. Iwamatsu et Y. Okabe, Physica A {\bf 278}, pp. 414-427 (2000).}

\bib{janis}{V. Jani\u s, J. Phys. Cond. Mat. {\bf 10}, pp. 2915-2932 (1998);
J. Phys. Cond. Mat. {\bf 8}, p. L173 (1996); cond-mat/9810175; cond-mat/9904069.}

\bib{jarrell}{M. Jarrell et J. E. Gubernatis, Phys. Rep. {\bf 269}, pp. 133-195
(1996); J. E. Gubernatis et al., Phys. Rev. {\bf B44}, pp. 6011-6029 (1991);
R. N. Silver et al., Phys. Rev. {\bf B41}, pp. 2380-2389 (1990).}

\bib{kadanoff}{L. P. Kadanoff et P. C. Martin, Phys. Rev. {\bf 124}, pp. 670-697 (1961).}

\bib{kagan}{M. Y. Kagan et al., Phys. Rev. {\bf B57}, pp. 5995-6002 (1998);
cond-mat/9911337.}

\bib{moment}{O. K. Kalashnikov, E. S. Fradkin, Sov. Phys. JETP {\bf 28}, pp.
317-325 (1969); A. Lonke, J. Math. Phys. {\bf 12}, pp. 2422-2438 (1971).}

\bib{kalos}{M. H. Kalos et P. A. Whitlock {\it Monte Carlo Methods}, John Wiley
\& Sons, New York (1986).}

\bib{lnp475}{{\it Recent Developments in High Temperature Superconductivity},
J. Klanut et al. Eds, Springer-Verlag, LNP 475, Berlin (1996)}

\bib{kohn}{W. Kohn et J. M. Luttinger, Phys. Rev. {\bf 118}, pp.41-45 (1960).}

\bib{kt}{J. M. Kosterlitz et D. J. Thouless, J. Phys. C{\bf 5}, p. L124 (1972);
idem C{\bf 6}, p. 1181 (1973).}

\bib{bumsoo}{B. Kyung (non-publi\'e).}

\bib{frank}{F. Lemay (non-publi\'e).}

\bib{letz}{ M. Letz et al., J. Low Temp. Phys. {\bf 117}, pp. 149-174 (1999);
J. Phys. Chem. Solids {\bf 59}, pp. 1838-1840 (1998); J. Phys. Cond. Mat. {\bf
10}, pp. 6931-6952 (1998).}

\bib{levin}{K. Levin et al., cond-mat/0003133; voir les r\'ef\'erences
incluses.}

\bib{linden}{W. von der Linden et al., J. Phys. Cond. Mat. {\bf 8}, pp.
3881-3888 (1996).}

\bib{loh}{E. Y. Loh et al., Phys. Rev. {\bf B41}, pp. 9301-9307 (1990).}

\bib{loktev}{V. M. Loktev et al., {\it Phase Fluctuations and Pseudogap
Phenomena}, soumis \`a Elsevier preprint (1999). Voir les r\'ef\'erences
incluses.}
 
\bib{cv}{J. W. Loram et al., Phys. Rev. Lett. {\bf 71}, pp. 1740-1743
(1993).}

\bib{luttinger1}{J. M. Luttinger et J. C. Ward, Phys. Rev. {\bf 118}, pp. 1417-1427
(1960).}

\bib{luttinger2}{J. M. Luttinger, Phys. Rev. {\bf 119}, pp. 1153-1163 (1960).}

\bib{luttinger3}{J. M. Luttinger, Phys. Rev. {\bf 121}, pp. 942-949 (1961).}

\bib{mahan}{G. D. Mahan, {\it Many-Particle Physics}, Plenum Press, 2th edt (1990).}

\bib{mancini}{F. Mancini, cond-mat/9803276.}

\bib{jumpmu}{D. van der Marel et G. Rietveld, Phys. Rev. Lett. {\bf 69}, pp.
2575-2577 (1992).}

\bib{martin}{P. C. Martin et J. Schwinger, Phys. Rev. {\bf 115}, pp. 1342-1373 (1959).}


\bib{mermin}{N. D. Mermin et H. Wagner, Phys. Rev. Lett. {\bf 17}, p. 1113
(1966).}

\bib{meshkov}{S. V. Meshkov et D. V. Berkov, Int. J. Mod. Phys. {\bf C5}, pp.
987-995 (1994).}

\bib{bcsbec}{C. P. Moca et al., cond-mat/9904197; J. P. Wallington et J. F.
Annett, Phys. Rev. {\bf B61}, pp. 1433-1445 (2000); L. S. Borkowski et
C. A. R. S\'a de Melo, cond-mat/9810370; P. Pierri et G. C. Strinati,
cond-mat/9811166; E. Babaev et
H. Kleinert, cond-mat/9804206; P. A. Sreeram et S. G. Mishra, cond-mat/9909181.}

\bib{samuel}{S. Moukouri et al., Phys. Rev. {\bf B61}, pp. 7887-7892 (2000).}

\bib{samuel2}{S. Moukouri, Rapport sur les techniques de prolongement
analytique, D\'epartement de Physique, Universit\'e de Sherbrooke, (1998).}

\bib{negele}{J. W. Negele et H. Orland, {\it Quantum Many-Particle Systems},
Frontiers in Physics, Addison-Wesley, New-York (1988).}

\bib{nozieres}{P. Nozi\`eres et J. M. Luttinger , Phys. Rev. {\bf 127}, pp.
1423-1440 (1962).}

\bib{strongcoupling}{P. Nozi\`eres et S. Schmitt-Rink, J. Low Temp. Phys., {\bf 59},
pp. 195-211 (1985).}

\bib{nozieres2}{P. Nozi\`eres et F. Pistolesi, cond-mat/9902273.}

\bib{stephane}{S. Pairault et al., Phys. Rev. Lett. {\bf 80},
pp. 5389-5392 (1998); cond-mat/9905242.}


\bib{pedersen}{M. H. Pedersen, th\`ese de doctorat, Universit\"at Z\"urich
(1996).}

\bib{pines}{D. Pines, Tr. J. of Physics {\bf 20}, pp. 535-547 (1996);
D. Pines, Z. Phys. B {\bf 103}, pp. 129-135 (1997); J. Schmalian et al.,
Phys. Rev. Lett. {\bf 80}, pp. 3839-3842 (1998).}

\bib{pines2}{D. Pines, {\it The Many-Body Problem}, W. A. Benjamin edt., New-York, 1962.}


\bib{hanke}{R. Preuss et al., Phys. Rev. Lett. {\bf 79}, pp. 1122-1125 (1997);
A. Dorneich cond-mat/9909352; C. Gr\"ober et al., cond-mat/0001366.}

\bib{randeria}{M. Randeria, cond-mat/9710223, Varenna Lectures (1997).}

\bib{tunnel}{Ch. Renner et al., Phys. Rev. Lett. {\bf 80}, pp. 149-152 (1998).}

\bib{rickayzen}{G. Rickayzen, Phys. Rev. {\bf 115}, pp. 195-808 (1959).
Disponible dans {\it The Many-Body Problem}, D. Pines.}

\bib{rick}{G. Rickayzen {\it Theory of Superconductivity}, John Wiley \& Sons,
New York (1965).}

\bib{robert}{C. Robert, {\it L'analyse statistique bayesienne}, Economica, Paris
(1992).}

\bib{robert2}{C. Robert et G. Casella, {\it Monte Carlo Statistical Methods},
Springer Verlag, Berlin (1999).}

\bib{raman}{G. Ruani et P. Ricci, Phys. Rev. {\bf B55}, pp. 93-96 (1997);
M. Opel et al., J. Low Temp. Phys. {\bf 117}, pp. 347-352 (1999).}

\bib{superfluid}{D. J. Scalapino et al., Phys. Rev. {\bf B47}, pp. 7995-8007 (1993).}

\bib{illposed}{H. Sch\"afer et E. Stermin, La physique au Canada, mars-avril 1997,
pp.77-85.}

\bib{momenthub}{T. Schneider et al., Z. Phys. B {\bf 100}, pp. 263-276 (1996);
R. Micnas et al., Phys. Rev. {\bf B52}, pp. 16223-15232 (1995); J. J.
Rodr\'\i guez-N\'u$\tilde{\rm n}$ez et al., Int J. Mod. Phys. C{\bf 8}, pp.
1145-1158 (1998); Physica A {\bf 232}, pp.403-416 (1996); M. H. Pedersen et al.,
Acta Phys. Pol. A {\bf 91}, pp. 419-422 (1997).}

\bib{schrieffer}{J. R. Schrieffer, {\it Theory of Superconductivity}, Benjamin,
New York (1964).}

\bib{setlur}{G. S. Setlur, cond-mat/9908269.}

\bib{singer}{J. M. Singer et al., Phys. Rev. {\bf B54}, pp. 1286-1301 (1996);
Eur. Phys. J. B {\bf 7}, pp. 37-51 (1999).}

\bib{singer2}{J. M. Singer et al. Eur. Phys. J. B {\bf 2}, p. 37 (1998)}

\bib{singwi}{K. S. Singwi et M. P. Tosi, in {\it Solid State Physics}, vol. 36
H. Ehrenreich et al. Edt., Academic Press, N.Y. (1981).}

\bib{rmn}{M. Takigawa et al., Phys. Rev. {\bf B43}, p. 247 (1991); H. Alloul
et al., Phys. Rev. Lett. {\bf 70}, p. 1071 (1993).}

\bib{timusk}{T. Timusk et B. Statt, Rep. Progr. Phys. {\bf 62}, pp. 61-122
(1999).}

\bib{david}{H. Touchette et D. Poulin, {\it Aspects num\'eriques des simulations
du mod\`ele de Hubbard: Monte Carlo quantique et m\'ethode d'entropie maximum},
Rapport technique, CRPS (1999).}

\bib{mcq}{A.-M. S. Tremblay, {\it M\'ethode Monte Carlo pour les \'electrons sur
r\'eseau}, dans {\it Simulations num\'eriques en physique}, L. Lewis, J Lopez
edt., vol. 2, pp. 529-582 (1993).}

\bib{ncorps}{A.-M. S. Tremblay, {\it Probl\`emes \`a N-corps}, Notes de cours,
D\'epartement de Physique, Universit\'e de Sherbrooke, version pr\'eliminaire,
mai 1997.}

\bib{trotter}{H. Trotter, Proc. Am. Math. Soc., vol. 10, p. 545 (1959).}

\bib{tsallis}{C. Tsallis et D. A. Stariolo, Physica A {\bf 233}, p. 395 (1996).}

\bib{sfvstc}{Y. J. Uemura, {\it Proceeedings of the 10th ann. HTS Workshop on
Phys.}, Huston, March 1996, World-Sientific, Singapore, 1996, pp. 68-71.}

\bib{uemura}{Y. J. Uemura, Hyperfine Interact. {\bf 105}, pp. 35-46 (1997).}


\bib{white}{M. Veki\'c et S. R. White, Phys. Rev. {\bf B47}, pp. 1160-1163 (1993).}

\bib{vilk1}{Y. M. Vilk et A.-M. S. Tremblay, J. Phys. I (France) {\bf 7}, pp.
1309-1357 (1997); Y. M. Vilk et A.-M. S. Tremblay, Europhys. Lett. {\bf 33} pp.
159-163 (1996).}

\bib{vilk2}{Y. M. Vilk et al., J. Phys. Chem. Solids {\bf 59}, pp.1873-1875 (1998).}

\bib{white}{S. R. White et al., Phys. Rev. {\bf B40}, p. 506 (1989).}

\bib{stripes}{S. R. White et D. J. Scalapino, cond-mat/0006071.}

\bib{bethe}{C. N. Yang, Phys. Rev. Lett. {\bf 19}, p. 1312 (1967); E. H. Lieb
et F. Y. Wu, Phys. Rev. Lett. {\bf 20}, p. 1445 (1968).}

\bib{symetrie}{C. N. Yang et S. C. Zhang, Mod. Phys. Lett. {\bf B4}, p. 759 (1990).}

\bib{so5}{S. C. Zhang, Science {\bf 275}, pp. 1089-1096 (1997).}


\inputBib
}
\figPS{Diagramme de phase des oxydes de cuivre avec phase supraconductrice (SC)
et phase antiferromagn\'etique (AFM). La temp\'erature $T^*$ indique la
temp\'erature \`a partir de laquelle on mesure la pr\'esence d'un pseudogap
(PG) au niveau de Fermi. La ligne au-dessus de $T^*$ identifie une temp\'erature
\`a partir de laquelle certains oxydes d\'emontrent des comportements
diff\'erents de ce qui est pr\'evu par la th\'eorie des liquides de Fermi.}
{diaghtc.eps}{10 cm}{Diagramme de phase des
supraconducteurs \`a haute temp\'erature critique.}\nom{diaghtc}

\figPS{Diagramme de phase en temp\'erature et en remplissage du mod\`ele de
Hubbard 2D (x=1-n). La ligne \'epaisse repr\'esente la temp\'erature critique
$T_{\rm BKT}$, la ligne mince la temp\'erature critique champ moyen, $T^o$,
la ligne hachur\'ee, la temp\'erature \`a laquelle appara\^\i trait le
pseudogap et la ligne pointill\'ee, la temp\'erature o\`u l'on observerait le
crossover dans le comportement critique. Nous pr\'esentons \`a la figure
\ref{psth} la forme sch\'ematique du poids spectral aux quatre points a, b, c
et d.}
{diagtx.eps}{8 cm}{Diagramme de phase du mod\`ele de Hubbard:
Temp\'erature-remplissage}\nom{diagtx}

\figPS{Diagramme de phase sch\'ematique en temp\'erature et en interaction
du mod\`ele
de Hubbard attractif 2D. La ligne pleine repr\'esente la temp\'erature
critique BKT et la ligne hachur\'ee, la temp\'erature critique champ moyen.}
{diagtu.eps}{8 cm}{Diagramme de phase du mod\`ele de Hubbard:
Temp\'erature-interaction}\nom{diagtu}

\figPS{Poids spectral pour quatre points du diagramme de phase du mod\`ele
de Hubbard attractif \`a deux dimensions. $T^*$ repr\'esente la temp\'erature
\`a laquelle appara\^\i trait le pseudogap et $T_{\rm BKT}$, la temp\'erature
du point critique BKT.}{psth.eps}{8 cm}{Poids spectral pr\'evu pour quatre
points du diagramme de phase}\nom{psth}

\figPS{Susceptibilit\'e de paires en fonction du vecteur d'onde. Les carr\'es
repr\'esentent les r\'esultats MCQ tandis que les croix repr\'esentent
les r\'esultats d'une approximation RPA o\`u la valeur de $\Gamma$ est fix\'ee
par la valeur \`a ${\bf q}=(0,0)$. Les vecteurs d'onde sont num\'erot\'es selon
les points du r\'eseau. Ainsi $(0,4)=(0,\pi)$}{rpa4q.eps}{15 truecm}
{Susceptibilit\'e de paires en fonction du vecteur d'onde: comparaison des
donn\'ees MCQ avec une approximation RPA.}\nom{rpaq}

\figPS{Susceptibilit\'e de paires en fonction de la fr\'equence de Matsubara.
Les carr\'es repr\'esentent les r\'esultats MCQ tandis que les croix
repr\'esentent les r\'esultats d'une approximation RPA o\`u la valeur de
$\Gamma$ est fix\'ee par la valeur \`a $q_n=0$. Les chiffres sur l'abscisse
correspondent \`a la valeur de $n$ dans la fr\'equence de Matsubara
$q_n=2\pi nT$.}{rpa4w.eps}{15 truecm}{Susceptibilit\'e de paires en fonction
de la fr\'equence de Matsubara: Comparaison des donn\'ees MCQ avec une
approximation RPA.}\nom{rpaw}

\figPS{La partie du bas pr\'esente le poids de quasiparticules tel que
donn\'e par la formule (\ref{pqptc}) et la partie du haut le facteur de
structure de paires. Ces deux quantit\'es sont pr\'esent\'ees en fonction
de la temp\'erature et ont \'et\'e mesur\'ees par simulation MCQ. Le poids
de quasiparticule est donn\'e au vecteur d'onde ${\bf k}=(0,\pi)$ et le
facteur de structure au vecteur ${\bf q}=(0,0)$. Nous pr\'esentons les
r\'esultats pour diff\'erentes tailles de r\'eseau avec interaction $U=-4$
et pr\`es du demi-remplissage, $n=0.95$.}{pqptc.eps}{115 truemm}{Poids de
quasiparticule et facteur de structure de paires en fonction de la
temp\'erature.}\nom{pqptcfig}

\figPS{Poids spectral obtenu par MME pour diff\'erents vecteurs d'onde
d'un r\'eseau $8\times 8$ en fonction de la fr\'equence. Conditions
usuelles $U=-4$, $n=1$ \`a une temp\'erature $T=1/2$. La partie de
gauche pr\'esente les donn\'ees MCQ obtenue avec une discr\'etisation
$\Delta\tau=1/10$. La partie de droite provient du r\'esultat de l'approche
analytique (TPSC). On a inclu dans ce dernier un bruit gaussien du m\^eme
ordre de grandeur que celui des donn\'ees MCQ.}{ps2v.eps}{15 truecm}
{Relation de dispersion du mod\`ele de Hubbard \`a une temp\'erature $T=1/2$.}
\nom{aw2}

\figPS{Poids spectral obtenu par MME pour diff\'erents vecteurs d'onde
d'un r\'eseau $8\times 8$ \`a $T=1/4$ en fonction de la fr\'equence.
Les r\'esultats pr\'esent\'es ont \'et\'e mesur\'es \`a interaction $U=-4$
pour un  syst\`eme demi-rempli $n=1$.
Comme la figure pr\'ec\'edente, la partie de gauche montre les r\'esultats
provenant des donn\'ees MCQ tandis que celle de droite provient des
r\'esultats de l'approche analytique.}{ps4v.eps}{15 truecm}{Relation de
dispersion su mod\`ele de Hubbard \`a $T=1/4$.}\nom{aw4}

\figPS{Densit\'e d'\'etats en fonction de la fr\'equence pour trois
temp\'eratures diff\'erentes: $T=1/2$, $1/3$ et $1/4$. Nous avons utilis\'e
les conditions suivantes:$U=-4$, $n=1.$, $\Delta\tau=0.1$ pour un r\'eseau
$8\times 8$. La partie de gauche provient du prolongement analytique des
donn\'ees MCQ et la partie de celles de la m\'ethode analytique}{dstat.eps}
{15 truecm}{Densit\'e d'\'etats en fonction de la fr\'equence.}\nom{dstat}

\figPS{Effet de la pr\'ecision des donn\'ees MCQ sur le poids spectral
obtenue par la m\'ethode de maximisation d'entropie. Nous consid\'erons
la fonction de Green mesur\'ee par MCQ au vecteur $(0,\pi)$ avec les
conditions habituelles: $U=-4$, $n=1$, $\Delta\tau=0.1$ sur un r\'eseau
de taille $8\times 8$ \`a temp\'erature $T=1/4$. La partie de gauche montre
la fonction de Green
consid\'er\'ee pour le prolongement. La courbe du bas montre les donn\'ees
obtenue et pour les deux autres nous avons ajout\'e un bruit gaussien
(d'amplitude plus \'elev\'ee pour celle du haut). La partie de droite de
la figure nous montre le r\'esultat du prolongement analytique.}{bruit.eps}
{15 truecm}{Effet de la pr\'ecision des donn\'ees MCQ sur le poids spectral
obtenue par MME.}\nom{bruit}

\figPS{Fonction de Green en temps imaginaire pour des syst\`emes de
diff\'erentes tailles. On pr\'esente les donn\'ees pour le cas $U=-4$,
$n=1$, ${\bf k}=(0,\pi)$ avec pour le MCQ $\Delta\tau=1/10$. La partie de
gauche montre les donn\'ees MCQ tandis que la partie de droite a \'et\'e
obtenue suite l'utilisation des s\'eries de Fourier sur l'approche analytique
\'evalu\'ee en fr\'equence de Matsubara.}{gt.eps}{15 truecm}{Fonction de
Green en temps imaginaire pour des syst\`emes de diff\'erentes tailles.}
\nom{gscaling}

\figPS{Poids spectral pour des syst\`emes de diff\'erentes tailles et au
vecteur d'onde ${\bf k}=(0,\pi)$. La partie de gauche de la figure montre
les r\'esultats MCQ et la partie de droite les r\'esultats de l'approche
analytique (TPSC). Ces r\'esultats viennent du cas $U=-4$, $n=1$ avec pour
les donn\'ees MCQ $\Delta\tau=1/10$.}{awsc.eps}{15 truecm}{Poids spectral
pour des syst\`emes de diff\'erentes tailles.}\nom{ascaling}

\figPS{Contribution de la susceptibilit\'e de paires uniforme \`a fr\'equence
nulle au facteur de structure de paires en fonction de la temp\'erature.
La valeur $1$ correspond au cas o\`u
le facteur de structure serait totalement d\'etermin\'e par cette fr\'equence.
Les points proviennent des mesures MCQ pour le cas $U=-4$, $n=0.95$ et
sont pr\'esent\'es pour diff\'erentes tailles de r\'eseaux.}{xpvsfsp.eps}
{15 truecm}{Ratio de la susceptibilit\'e de paires uniforme \`a fr\'equence
nulle et du facteur de structure de paires en fonction de la temp\'erature.}
\nom{ratioxvss}

\figPS{Partie imaginaire de la susceptibilit\'e de paires uniforme divis\'ee
par la fr\'equence et en fonction de la fr\'equence. Les r\'esultats proviennent
du mod\`ele de Hubbard avec $U=-4$, $n=1$ sur un r\'eseau carr\'e $8\times 8$
avec une discr\'etisation en temps imaginaire de $\Delta\tau=1/10$. On
pr\'esente les r\'esultats aux temp\'eratures $T=1/3$, $1/4$ et $1/5$.}
{chiw.eps}{12 truecm}{Partie imaginaire de la susceptibilit\'e de paires
uniforme divis\'ee par la fr\'equence en fonction de la fr\'equence.}
\nom{chiw}

\figPS{Fonction de corr\'elation courant-courant dans la direction $\hat{x}$
pour diff\'erents vecteurs d'onde. Les carr\'es repr\'esentent $q_y=0$, $iq_n=0$
avec diff\'erentes valeurs de $q_x$, les losanges repr\'esentent le r\'esultat
pour diff\'erentes valeurs de $q_y$ avec $q_x=0$ et $iq_n$=0 et les cercles
repr\'esente le cas limite $iq_n\rightarrow 0$ avec $q_x=q_y=0$. Les
r\'esultats pr\'esent\'es sont pour les conditions habituelles, $U=-4$, $n=1$,
$\Delta\tau=1/10$ pour un r\'eseau $8\times 8$ \`a temp\'erature $T=1/4$.}
{drude.eps}{12 truecm}{Fonction de corr\'elation courant-courant dans la
direction $\hat{x}$ pour diff\'erents vecteurs d'onde}\nom{sfluid}

\figPS{Spectre \`a une particule pour un r\'eseau de taille $4\times 4$ \`a
basse temp\'erature $T=1/8$ et au vecteur d'onde $(0,\pi)$. En comparaison,
on montre le spectre obtenu \`a $T=1/5$. Les valeur des param\`etres du
mod\`ele sont $U=-4$, $n=1.$ avec une discr\'etisation du temps imaginaire
de $\Delta\tau=1/10$.}{awb8.eps}{12 truecm}{Spectre \`a une particule d'un
r\'eseau $4\times 4$ \`a $T=1/8$.}\nom{spectre8}

\figPS{Longueur de corr\'elation de paires compar\'ee \`a la longueur d'onde
de de Broglie en fonction de la temp\'erature. Les donn\'ees sont extraites
des mesures MCQ pour un cas \`a interaction $U=-4$, et remplissage $n=0.95$.
Les losanges repr\'esentent la longueur de corr\'elation  pour un r\'eseau
de taille $6\times 6$, les triangles, celle d'un r\'eseau $8\times 8$,
les croix celle extrapol\'ee pour la limite thermodynamique
et les cercles pleins, la longueur d'onde de de Broglie pour un r\'eseau
$8\times 8$.}{corr.eps}{12 truecm}{Longueur de corr\'elation compar\'ee \`a la
longueur d'onde de de Broglie en fonction de la temp\'erature.}\nom{corr}

\figPS{Comparaison entre le facteur de structure de paires (symboles gras)
et le facteur de structure de charge (symboles vides) en fonction de la
temp\'erature. Les carr\'es pr\'esentent les r\'esultats pour le r\'eseau
$6\times 6$ et les triangles pour le r\'eseau $8\times 8$. La partie de
gauche montre les r\'esultats au remplissage $n=0.95$ et celle de droite
pour $n=0.8$. Pour les deux remplissages nous avons utilis\'e une interaction
$U=-4$ et une discr\'etisation de $\Delta\tau=1/10$.}{fspfsc.eps}
{15 truecm}{Comparaison entre le facteur de structure de charge et le
facteur de structure de paires en fonction de la temp\'erature.}\nom{fspfsc}

\figPS{Poids spectral au remplissage $n=0.95$ et au vecteur d'onde
${\bf k}=(0,\pi)$ et au remplissage $n=0.8$ pour le vecteur
${\bf k}=(0,3\pi/4)$. Les deux courbes ont \'et\'e obtenues par l'application
de la technique de maximisation d'entropie aux donn\'ees MCQ. Nous pr\'esentons
les r\'esultats \`a temp\'erature $T=1/5$ avec interaction $U=-4$.}{aw95.eps}
{15 truecm}{Poids spectral aux remplissages $n=0.95$ et $n=0.8$.}\nom{aw95}




\newpage
\pageno=-2
\headline={\hss}
\centerline{SOMMAIRE}
\bigskip
\let\tdm=\tdmL
\TdMitem{Sommaire}

Dans cette th\`ese nous pr\'esentons une nouvelle m\'ethode non perturbative
pour le calcul des propri\'et\'es d'un syst\`eme de fermions. Notre m\'ethode
g\'en\'eralise l'approximation auto-coh\'erente \`a deux particules
propos\'ee par Vilk et Tremblay pour le mod\`ele de Hubbard r\'epulsif.
Notre m\'ethode peut
s'appliquer \`a l'\'etude du comportement pr\'e-critique lorsque la sym\'etrie
du param\`etre d'ordre est suffisamment \'elev\'ee. Nous appliquons la
m\'ethode au probl\`eme du pseudogap dans le mod\`ele de Hubbard attractif.
Nos r\'esultats montrent un excellent accord avec les donn\'ees Monte Carlo
pour de petits syst\`emes. Nous observons que le r\'egime o\`u appara\^\i t
le pseudogap dans le poids spectral \`a une particule est un r\'egime
classique renormalis\'e caract\'eris\'e par
une fr\'equence caract\'eristique des fluctuations supraconductrices
inf\'erieure \`a la temp\'erature. Une autre caract\'eristique est la faible
densit\'e de superfluide de cette phase d\'emontrant que nous ne sommes pas
en pr\'esence de paires pr\'eform\'ees.

Les r\'esultats obtenus semblent montrer que la haute sym\'etrie du param\`etre
d'ordre et la bidimensionalit\'e du syst\`eme \'etudi\'e \'elargissent le
domaine de temp\'erature pour lequel le r\'egime pseudogap est observ\'e.
Nous argumentons que ce r\'esultat est transposable aux supraconducteurs
\`a haute temp\'erature critique o\`u le pseudogap appara\^\i t \`a des
temp\'eratures beaucoup plus grandes que la temp\'erature critique.
La forte sym\'etrie dans ces syst\`emes pourraient \^etre reli\'ee \`a la
th\'eorie $SO(5)$ de Zhang.

En annexe, nous d\'emontrons un r\'esultat tout r\'ecent qui permettrait
d'assurer l'auto-coh\'erence entre les propri\'et\'es \`a une et \`a deux
particules par l'ajout d'une dynamique au vertex irr\'eductible. Cet ajout
laisse entrevoir la possibilit\'e d'\'etendre la m\'ethode au cas d'une
forte interaction.

\newpage

\centerline{REMERCIEMENTS}
\bigskip
\TdMitem{Remerciements}

Toute la r\'ealisation de ce travail n'aurait pu se faire sans
l'aide et l'appui pr\'ecieux d'Andr\'e-Marie Tremblay. Je le
remercie pour m'avoir propos\'e ce projet de recherche et avoir eu la patience
de m'expliquer la physique li\'ee \`a ce projet. Son aide pour
la r\'edaction de la th\`ese a \'et\'e tr\`es importante.

Je tiens aussi \`a exprimer un remerciement tout particulier
\`a mon fr\`ere Dave. Ses connaissances et sa compr\'ehension
de la physique m'ont \'et\'e d'une grande aide. Ses
commentaires lors de la r\'edaction de la th\`ese ont \'et\'e
grandement b\'en\'efiques et ont permis de la rendre un peu plus lisible.

Je remercie aussi mes parents, ma famille et mes amis pour leur
appui moral dans les moments plus difficiles, comme dans les
moments plus agr\'eables. Vous avez su me donner le courage de mener
\`a terme mes \'etudes.
Je remercie aussi mes coll\`egues de bureau et tous les membres
du d\'epartement qui ont rendu mon s\'ejour agr\'eable \`a
Sherbrooke.

Je tiens \`a souligner la contribution importante de plusieurs
personnes \`a ce projet. Merci tout particulier \`a Fran\c cois
Lemay et Bumsoo Kyung qui m'ont fourni une partie des r\'esultats
num\'eriques pr\'esent\'es dans cette th\`ese. Merci \'egalement
\`a David Poulin et Hugo Touchette qui ont contribu\'e \`a l'\'elaboration
du programme de Monte Carlo et de prolongement analytique.
Pour leur soutien, aide et apport au niveau informatique, je
tiens \`a souligner l'apport d'Alain Veilleux et de Michel Barrette.

Je remercie l'institut canadien de recherches avanc\'ees (CIAR) pour
m'avoir permis de participer \`a
un nombre incalculable de conf\'erences et d'\'ecoles d'\'et\'e sur
le th\`eme de la supraconductivit\'e. Je remercie \'egalement le
CERPEMA, la Facult\'e des sciences de l'Universit\'e de Sherbrooke
et le Centre de recherches math\'ematiques qui m'ont \'egalement permis
de participer \`a des
conf\'erences. La participation \`a ces conf\'erences m'a permis
de rencontrer
et d'\'echanger avec des gens qui ont su me conseiller et
m'expliquer un bon nombre de choses. Je remercie Yuri Vilk,
Robert Gooding, Massimiliano Capezzali, Raymond Fr\'esard, Laurent
Raymond, Anne-Marie Dar\'e, Alexandre Zagoskin et Nicolas Dupuis.

Je tiens \`a remercier le CRSNG, le Fonds FCAR et le CERPEMA pour
leur pr\'ecieuses aides financi\`eres tout au long de ces \'etudes.
Pour l'acc\`es \`a des ordinateurs de grande puissance, je remercie le
CRSNG, le CACPUS et la Facult\'e des Sciences pour m'avoir
donn\'e acc\`es \`a un IBM-SP et \`a une
grappe de type Beowulf.

\newpage

\iftdm\immediate\write\tdmL{\par\noexpand\noindent%
Table des mati\`eres\string\pageNO{\folio}}\else\relax\fi
\pageno=-6
\advance\pageno by -1

\centerline{\headingfont LISTE DES ABR\'EVIATIONS}
\vglue.5in\parindent=1cm
\TdMitem{Liste des abr\'eviations}
\+ \kern 2.5cm & \cr
\+ 2D     &Bidimensionnel\cr
\vskip 3truemm
\+ AFM    &Antiferromagn\'etisme\cr
\vskip 3truemm
\+ BCS    &Th\'eorie d\'evelopp\'ee par Bardeen, Cooper et Shrieffer pour\cr
\+        &expliquer la supraconductivit\'e.\cr
\vskip 3truemm
\+ BKT    &Transition sans ordre \`a longue port\'ee dont l'existence a \'et\'e\cr
\+        &d\'emontr\'ee par Berezinsk\u\i i et par Kosterlitz et Thouless.\cr
\vskip 3truemm
\+ FLEX   &Nom donn\'ee \`a une approximation d'apr\`es son appellation en\cr
\+        &anglais ``Fluctuation exchange approximation''.\cr
\vskip 3truemm
\+ MCQ    &Monte Carlo quantique\cr
\vskip 3truemm
\+ MME    &M\'ethode de maximisation d'entropie\cr
\vskip 3truemm
\+ PG     &Pseudogap\cr
\vskip 3truemm
\+ RPA    &Nom d'une approximation dont l'abbr\'eviation provient de son\cr
\+        &nom anglais: ``Random Phase approximation''.\cr
\vskip 3truemm
\+ SC     &Supraconductivit\'e\cr
\vskip 3truemm
\+ SC hTc &Supraconducteur \`a haute temp\'erature critique\cr
\vskip 3truemm
\+ TPSC   &Nom d'une approximation dont l'abbr\'eviation vient de son\cr
\+        &appellation anglaise: ``Two-particle self-consistent''.\cr

\newpage


\iftdmT\centerline{\headingfont LISTE DES TABLEAUX}
\vglue.5in\parindent=1cm\rightskip=1cm
\TdMitem{Liste des tableaux}
\iftdm\immediate\closeout\tdmT\fi\input tables.aux\newpage\fi


\parindent=\indentation
\parskip=4pt
\baselineskip=\single

\iftdmA\centerline{\headingfont LISTE DES ANNEXES}
\vglue.5in\parindent=1cm\rightskip=1cm
\parskip=0pt\chapno=0
\TdMitem{Liste des annexes}
\baselineskip=\single
\iftdm\immediate\closeout\tdmA\fi\input annexes.aux\newpage\fi

\pageno=-5
\global\tdmAfalse
\centerline{\headingfont TABLE DES MATI\`ERES}
\vglue\stdskip
\iftdm\immediate\closeout\tdmP\immediate\closeout\tdmL\fi
\input tdmL.aux\par
\input tdm.aux
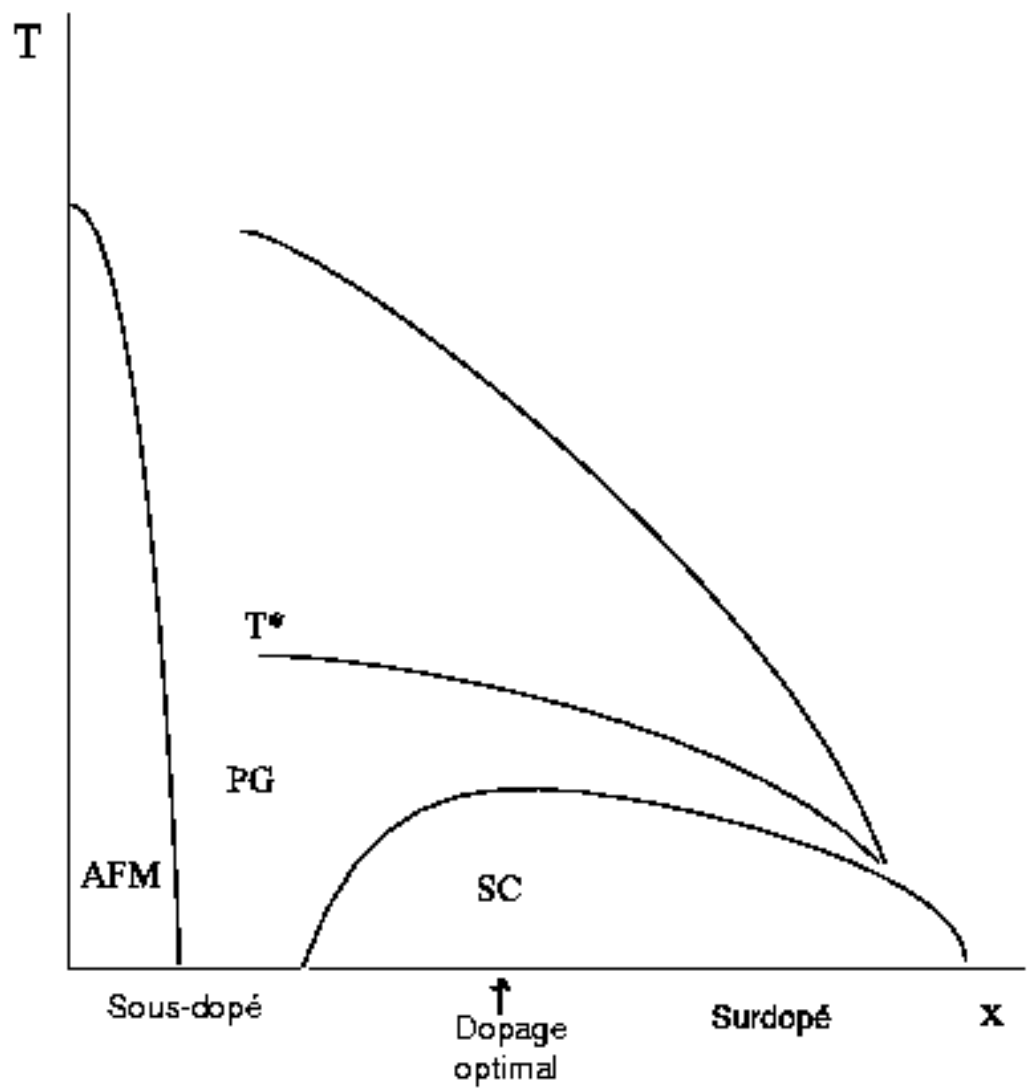
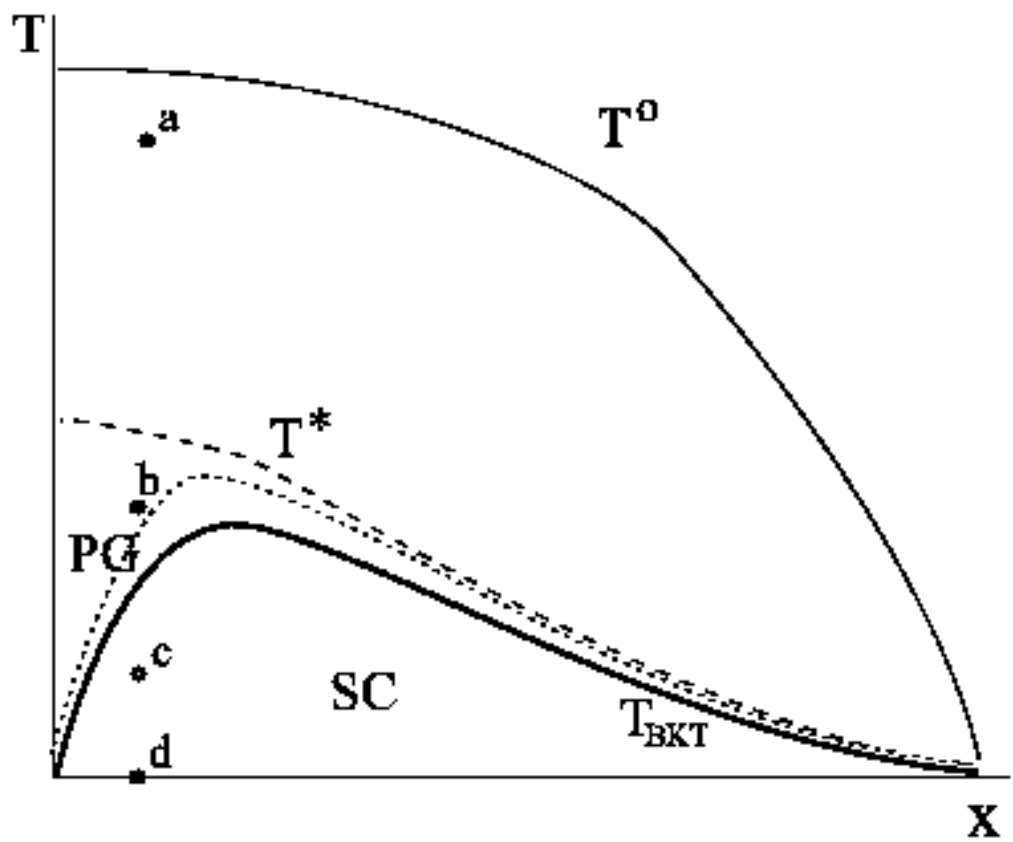
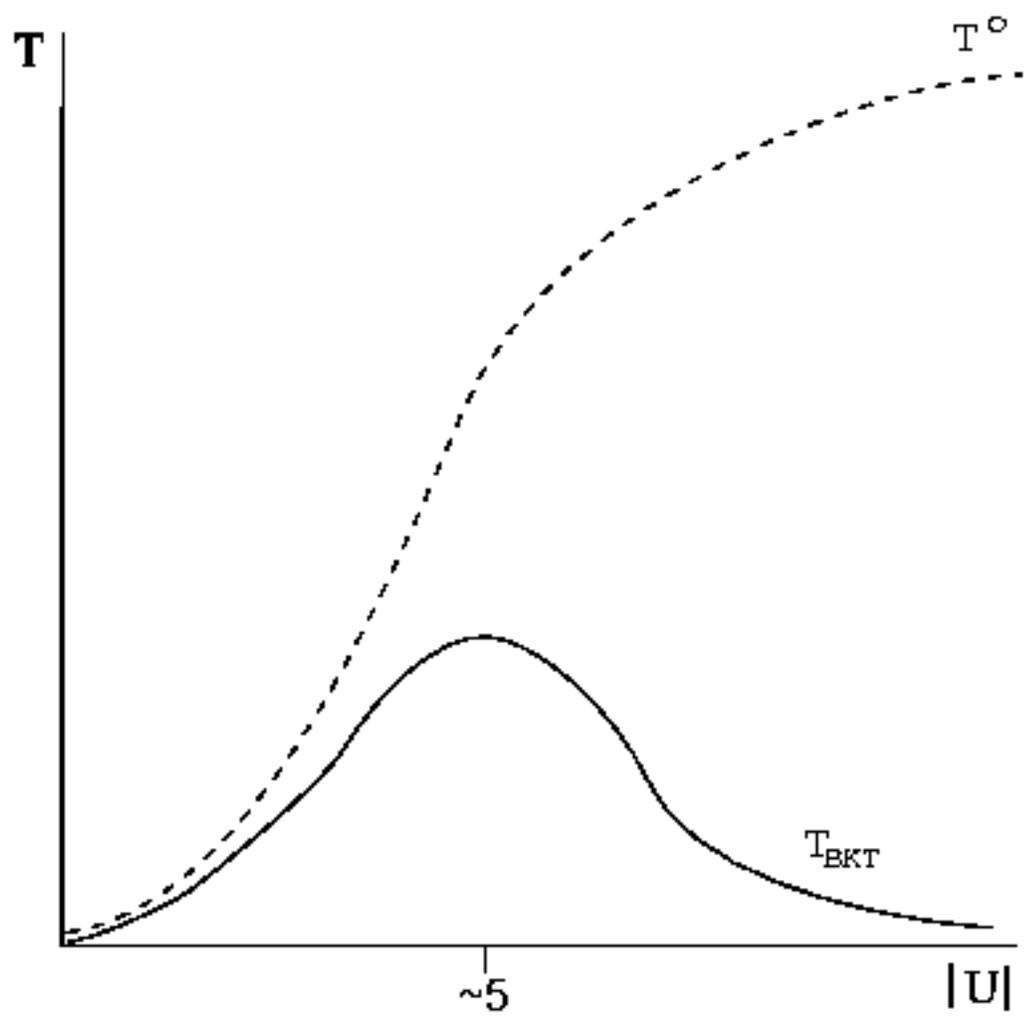
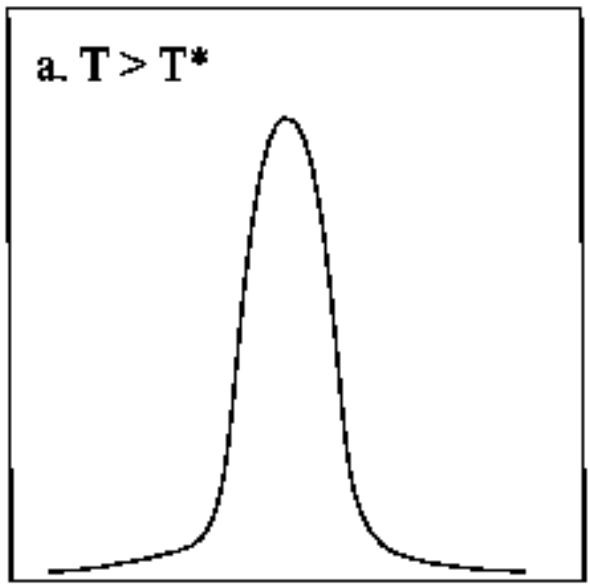
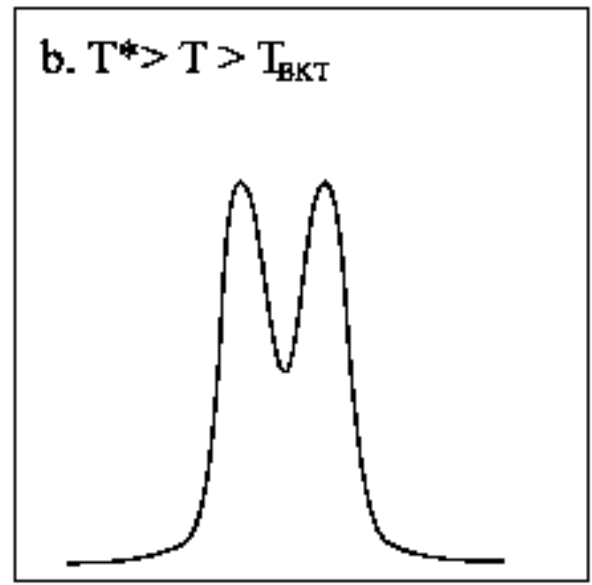
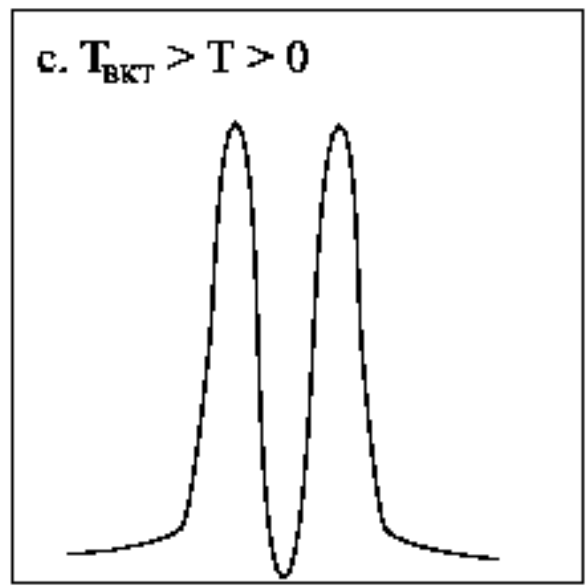
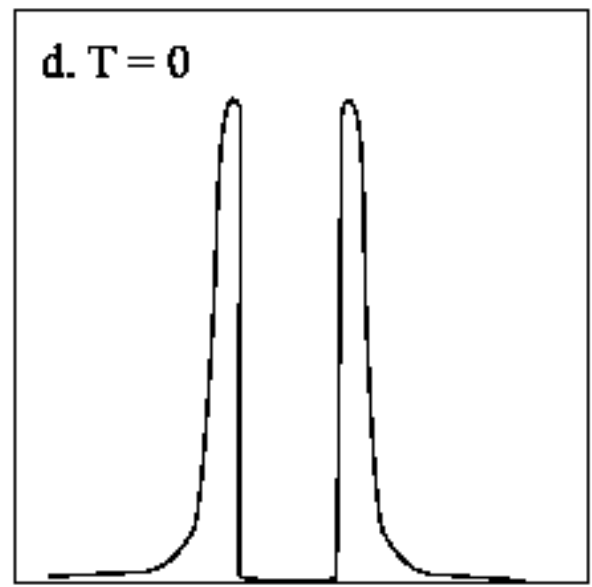
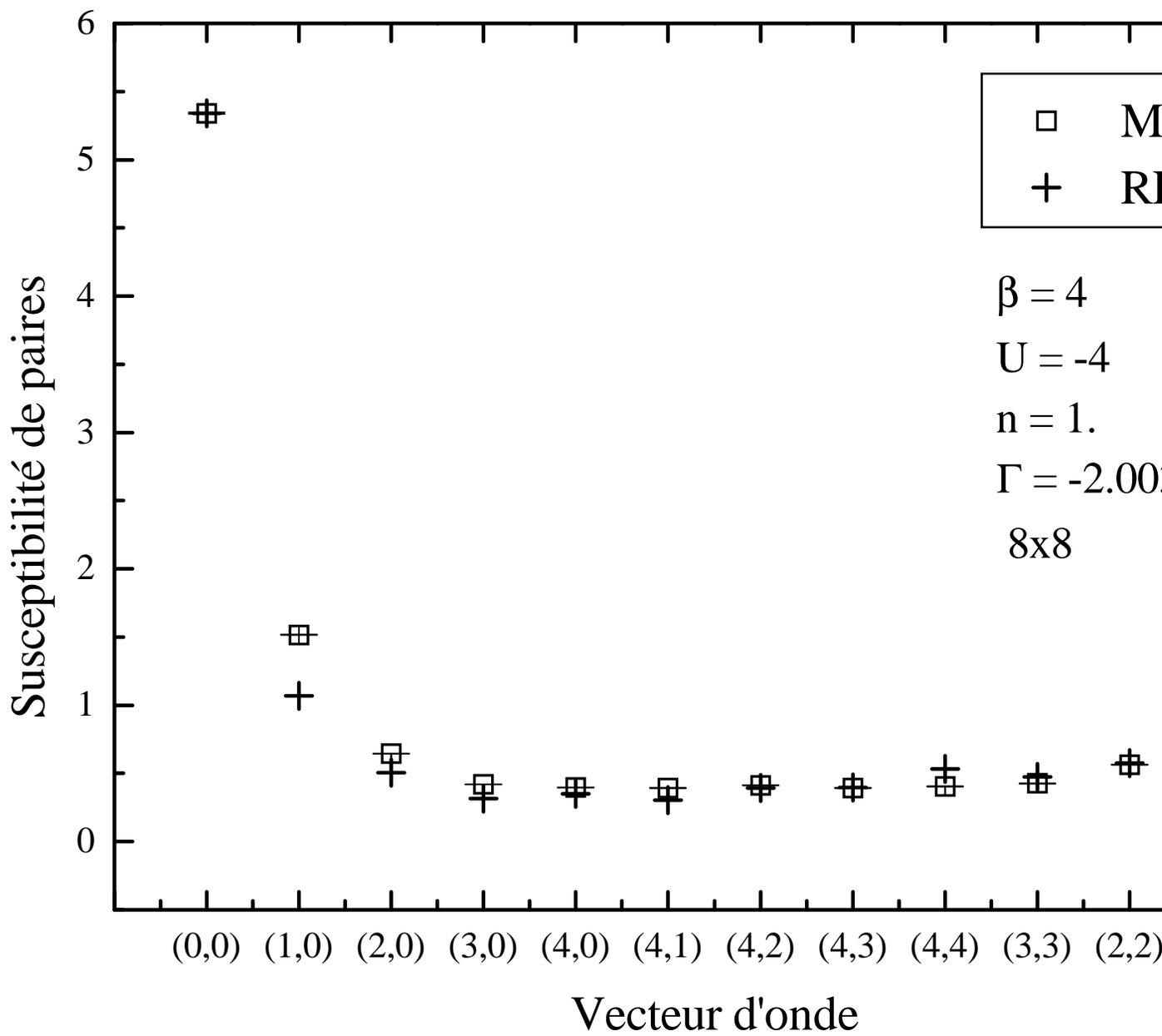
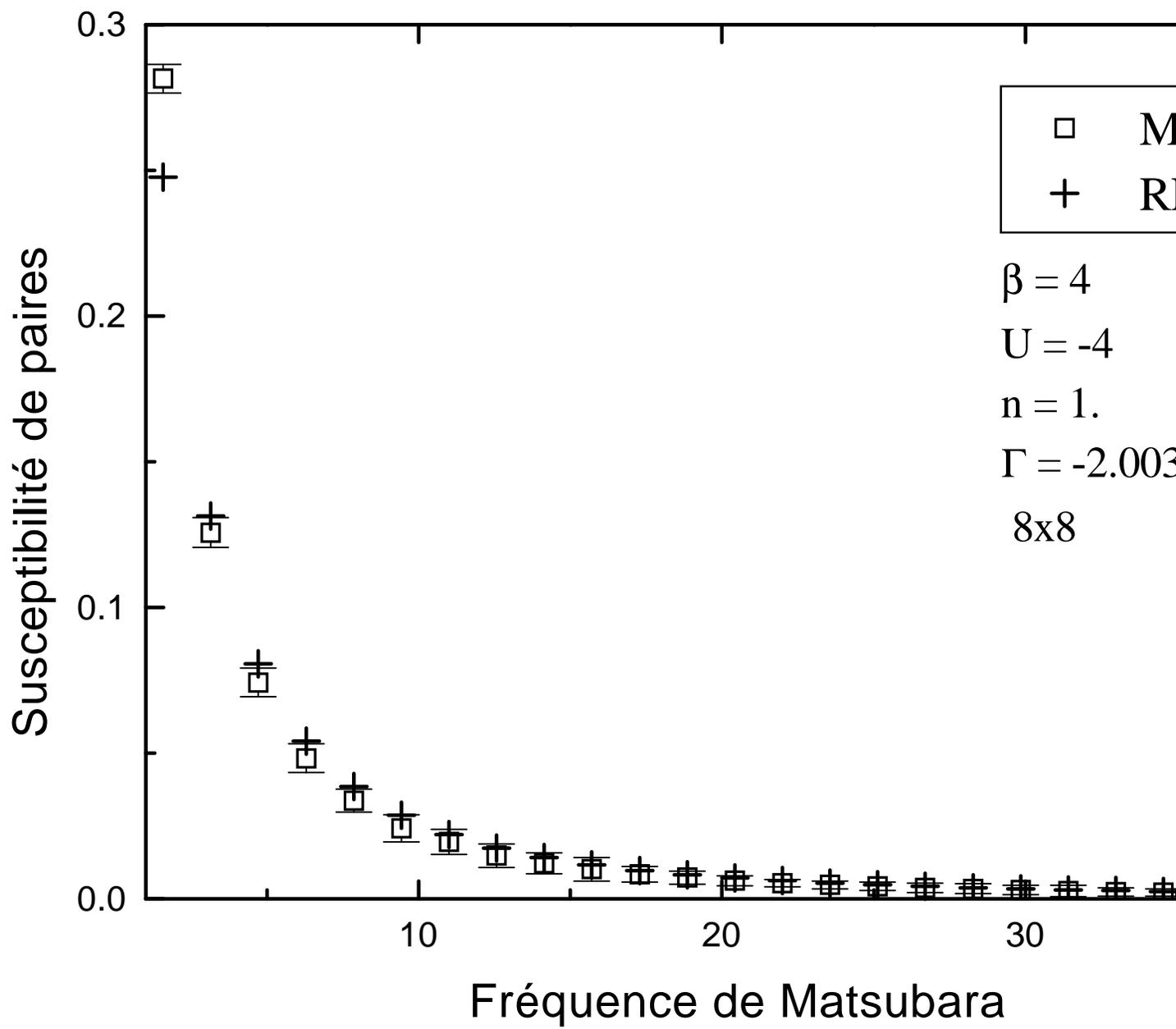
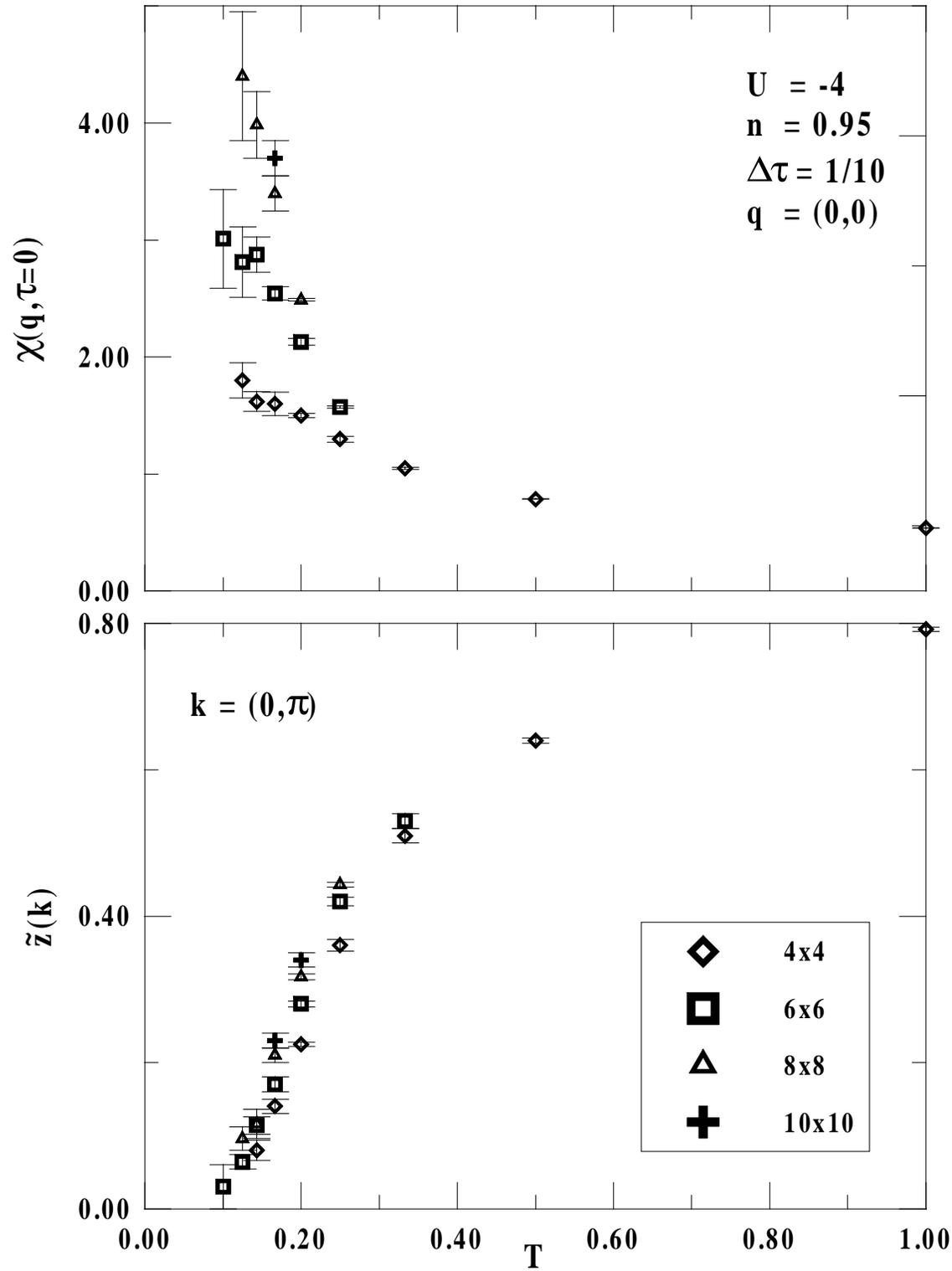
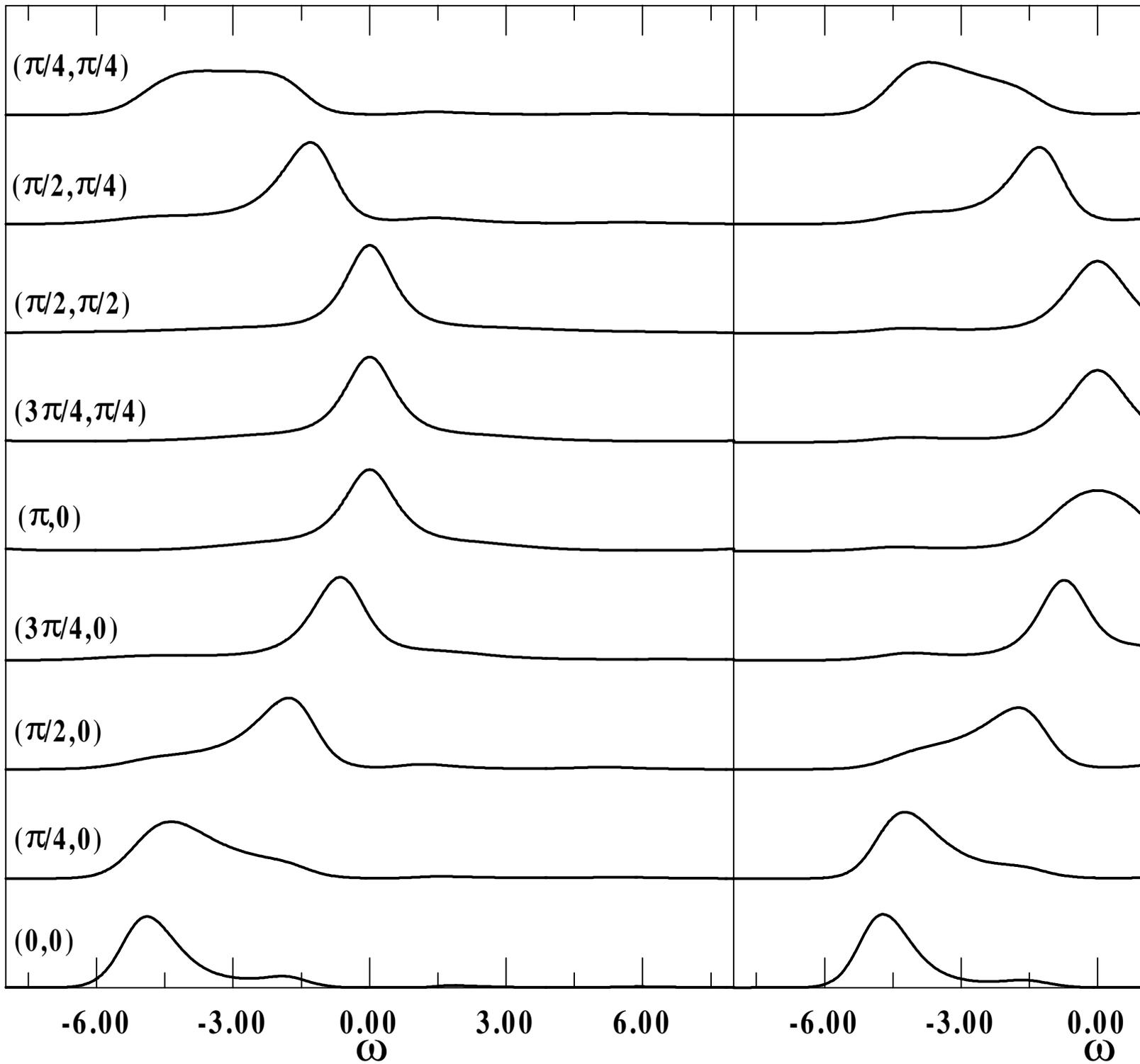
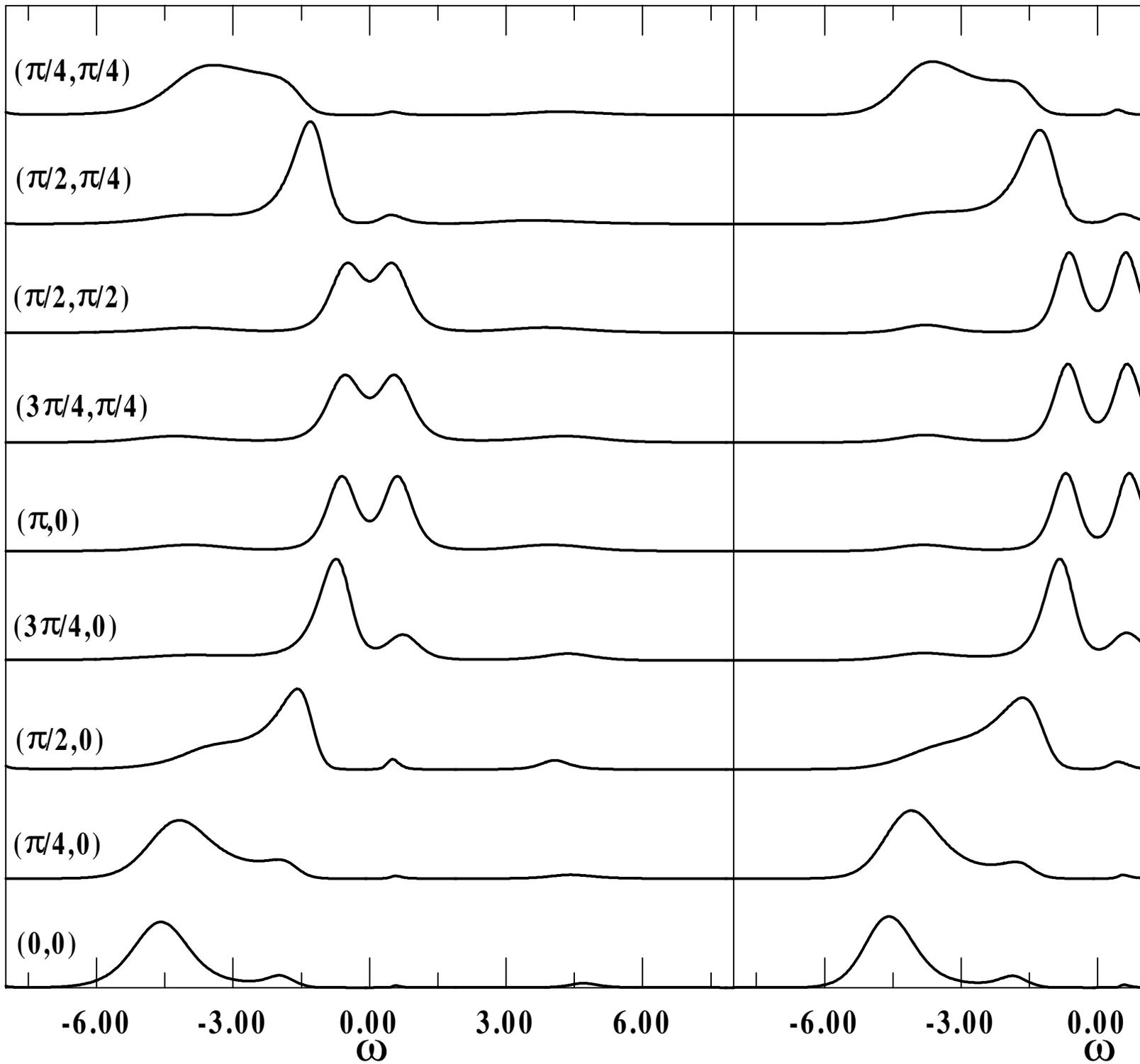
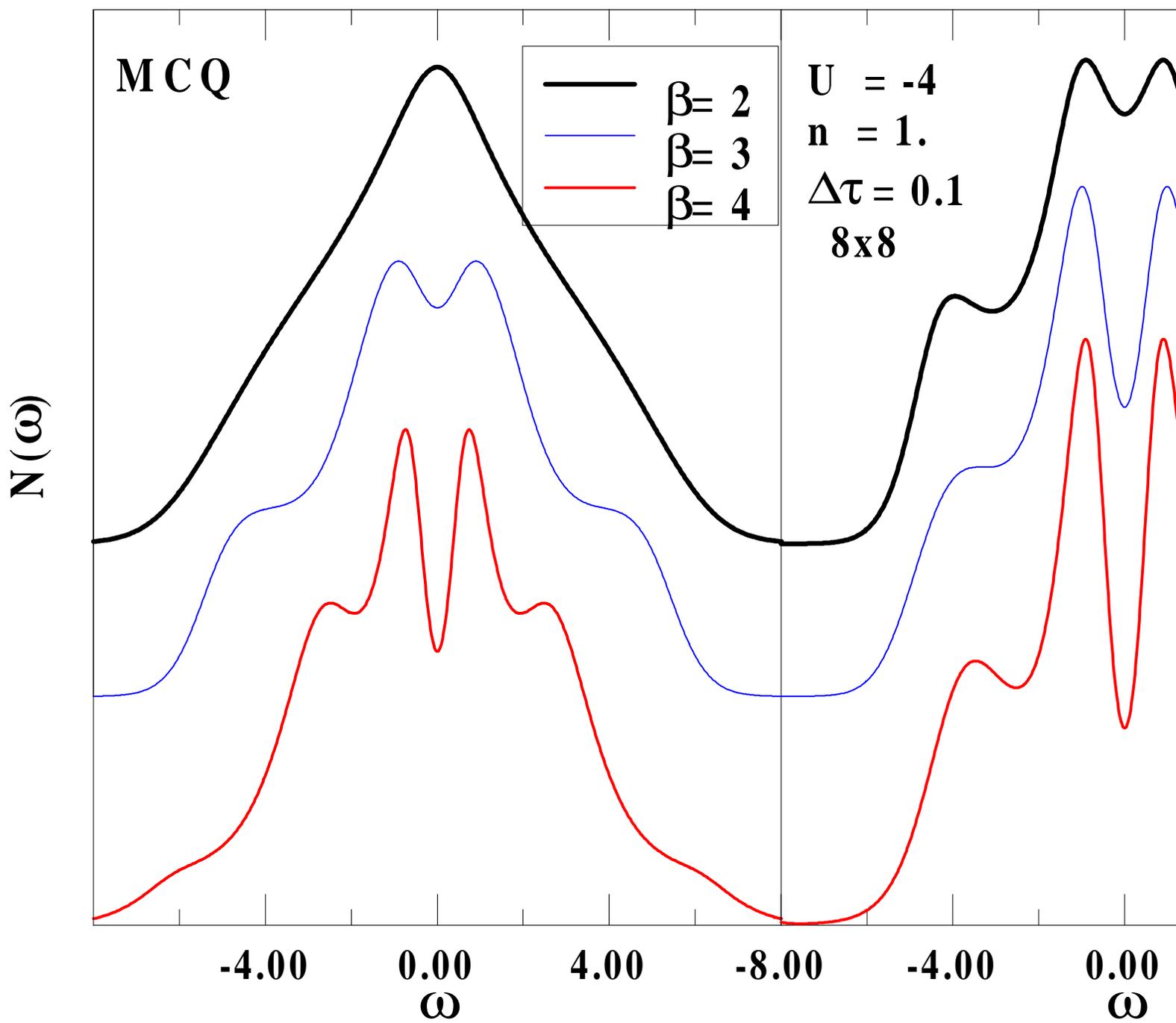
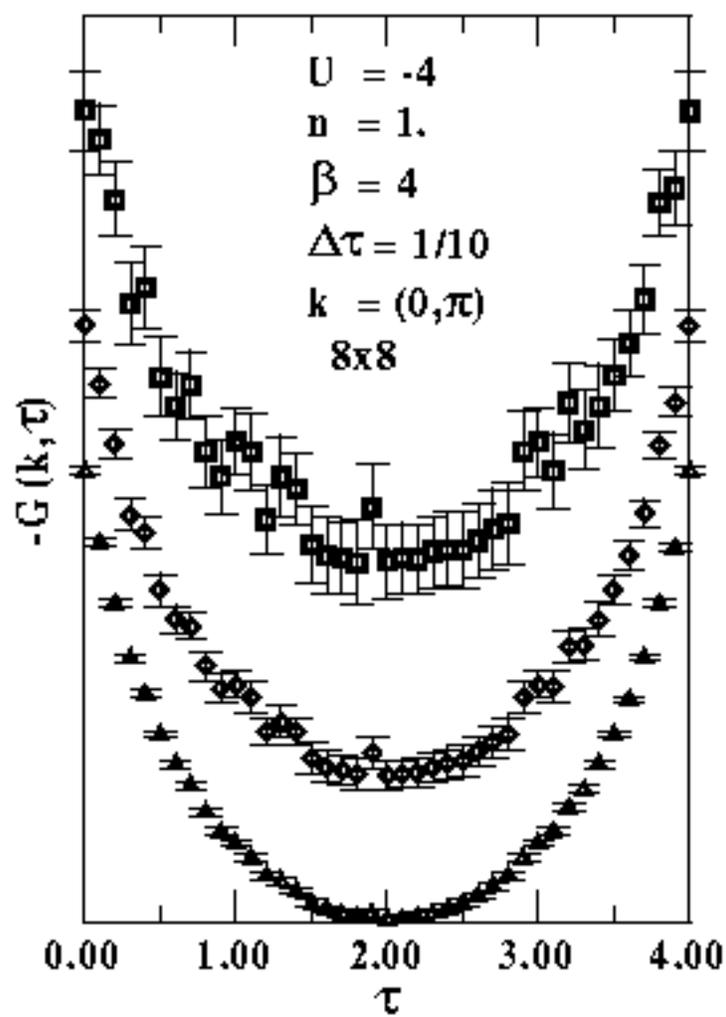
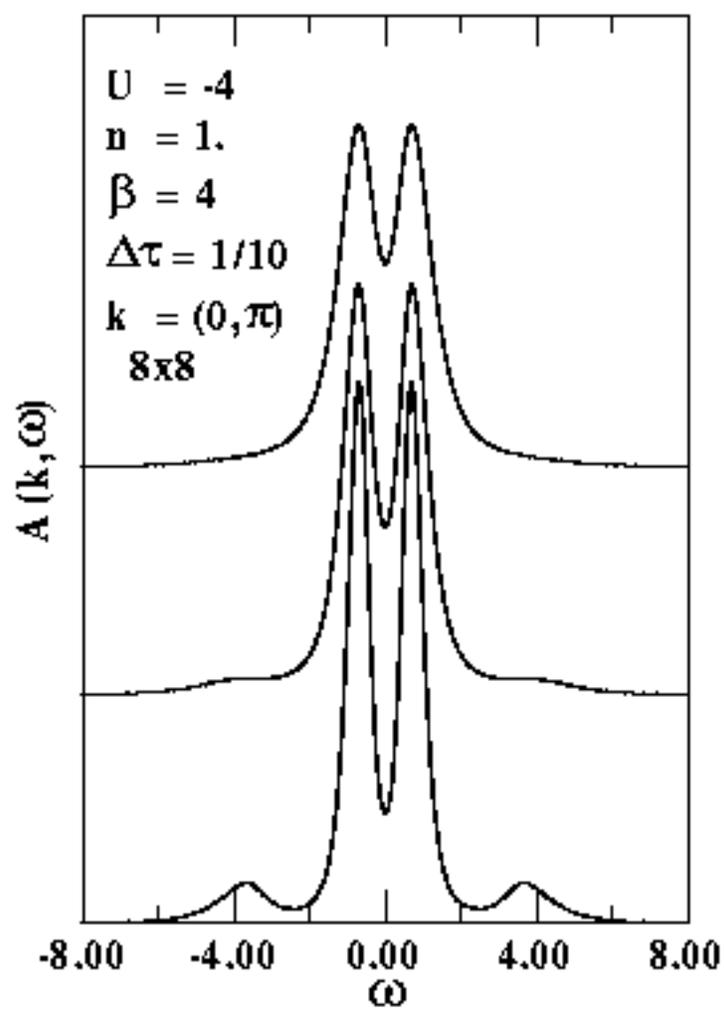
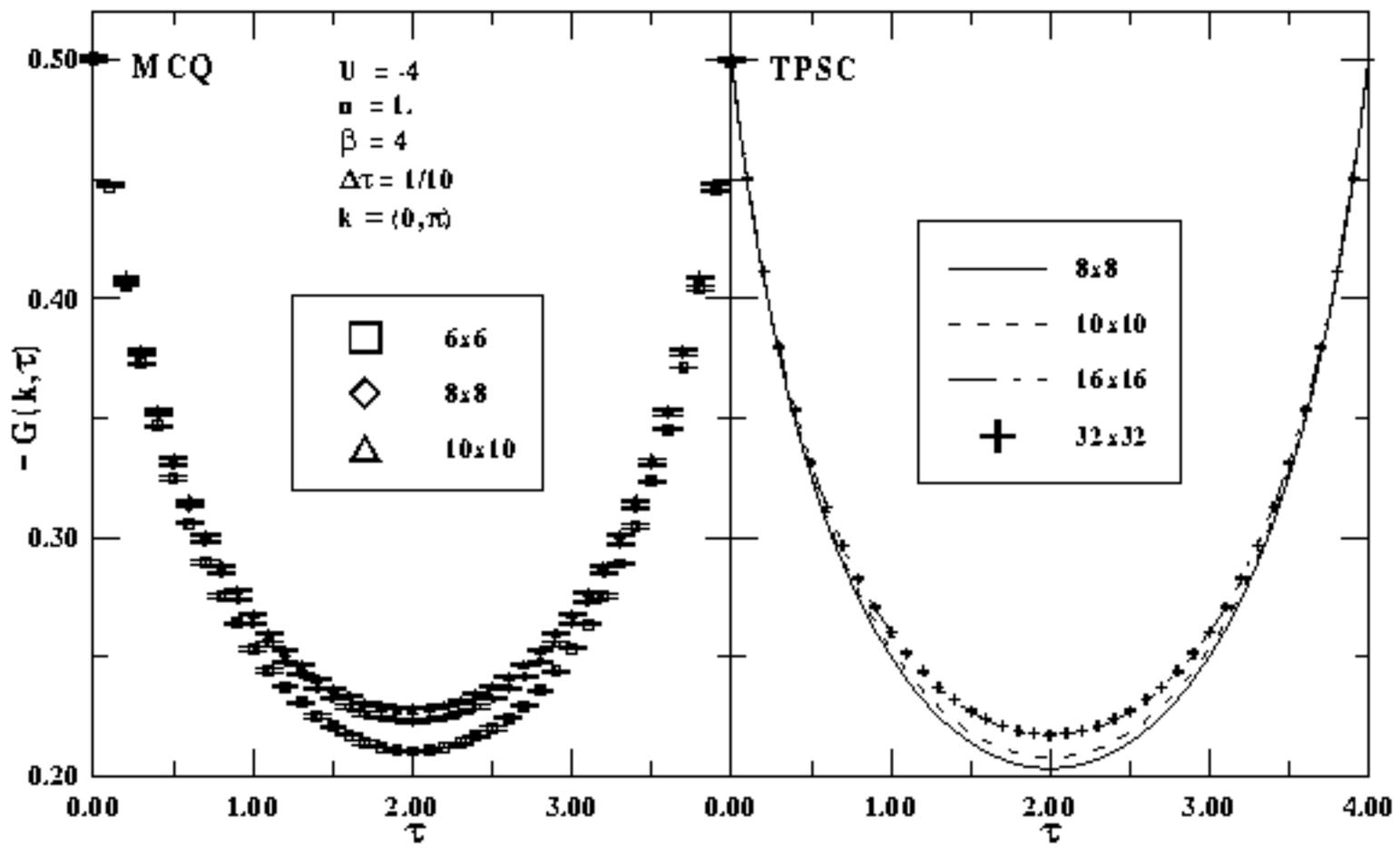
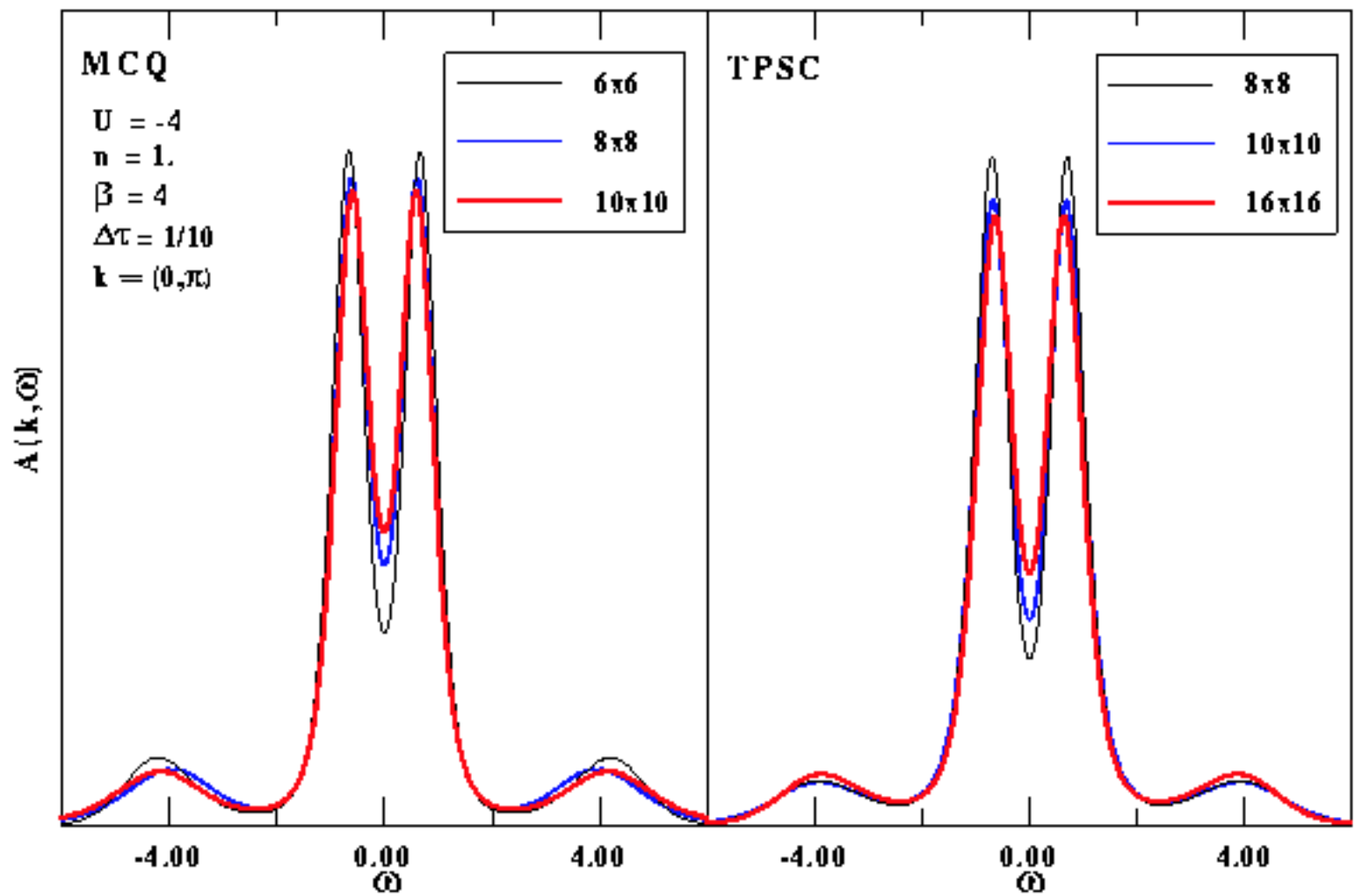
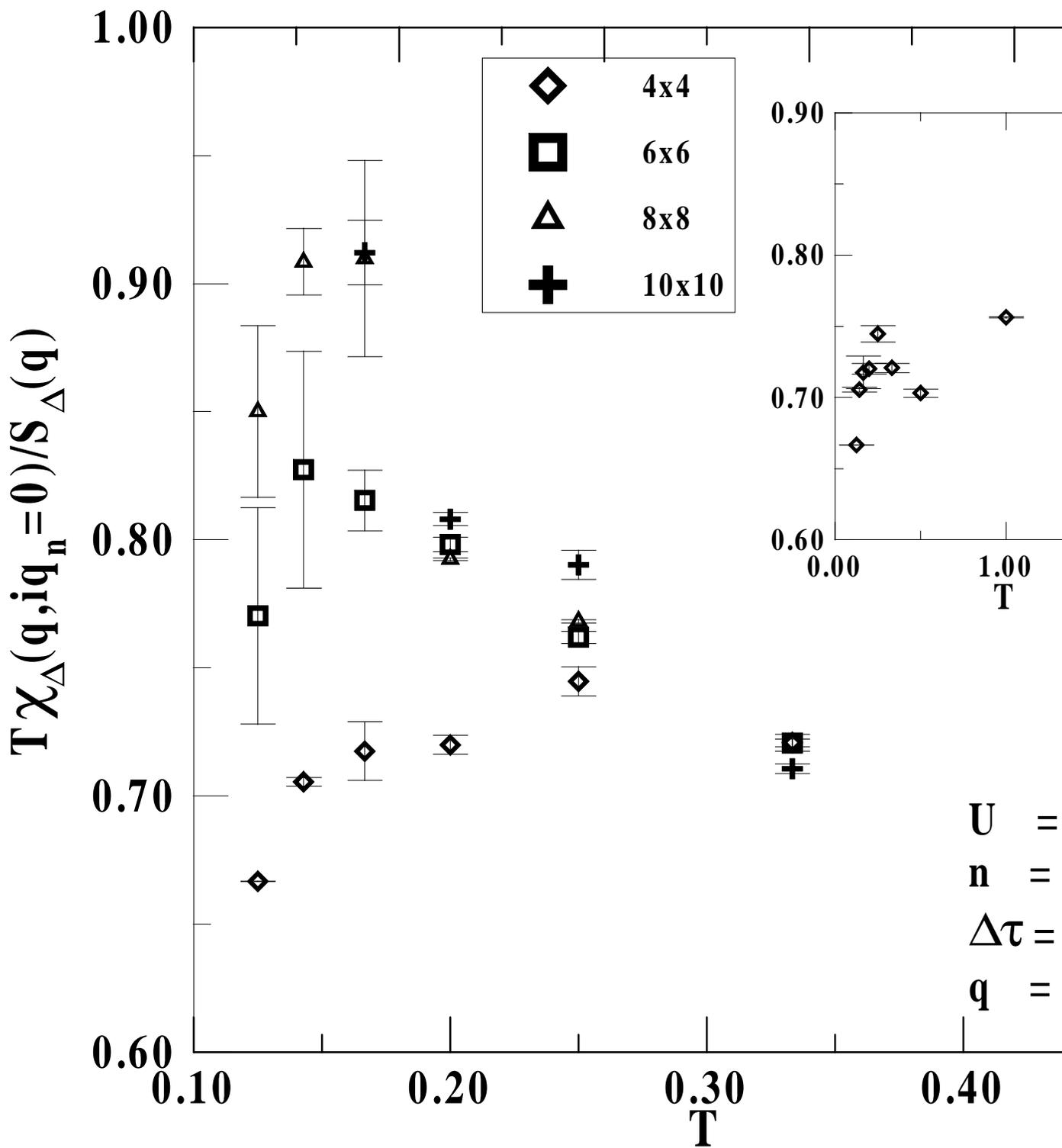
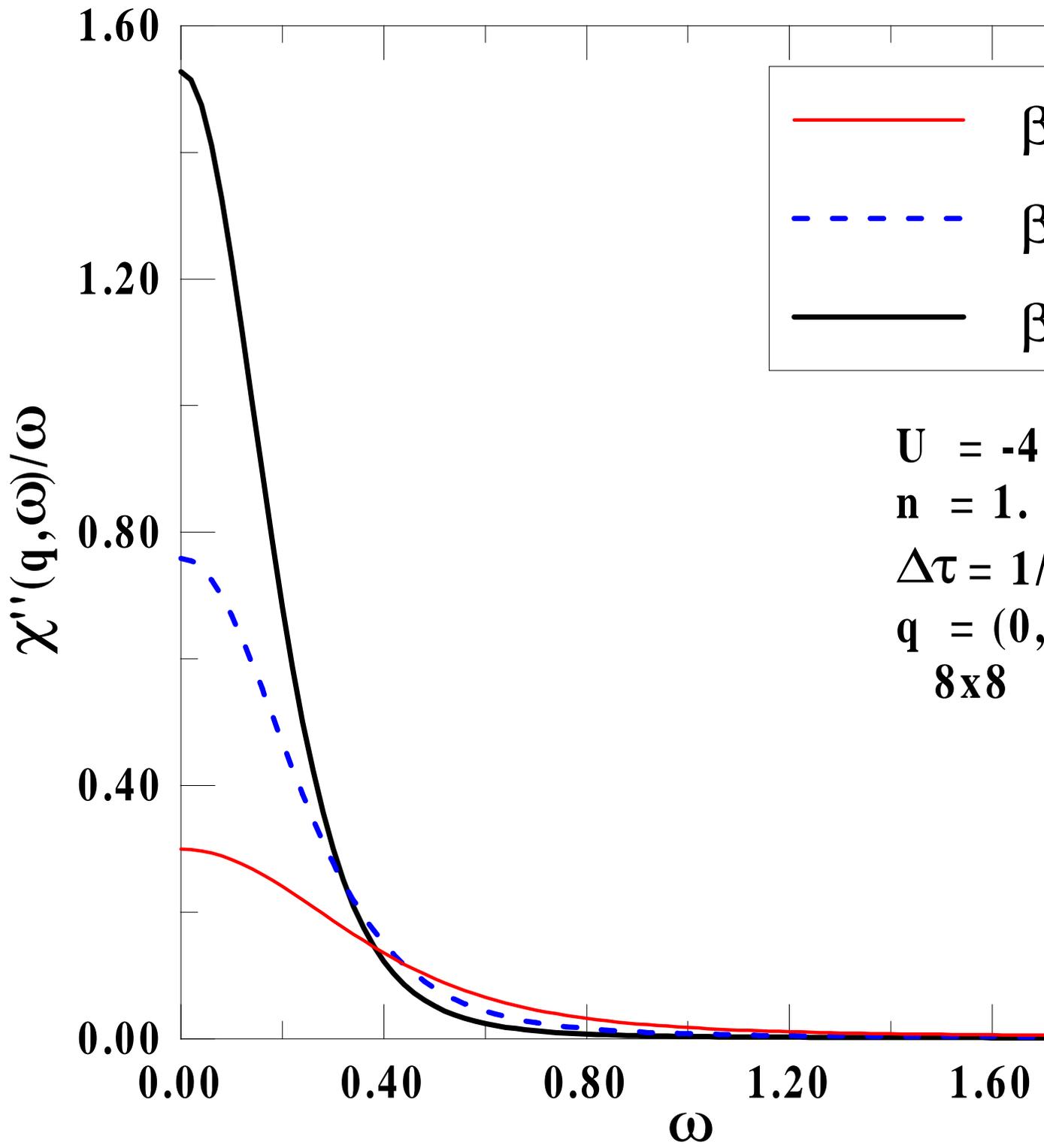
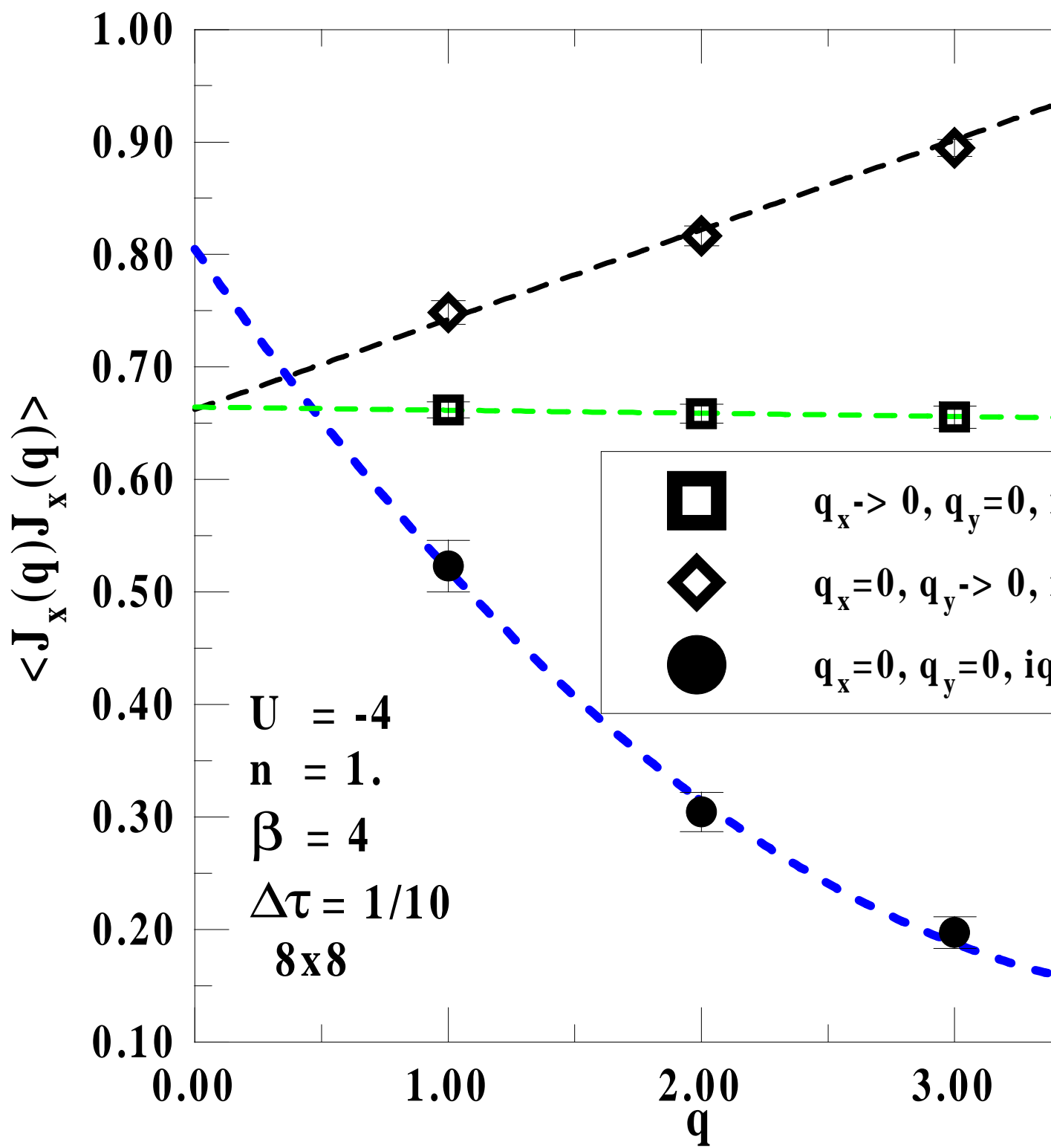
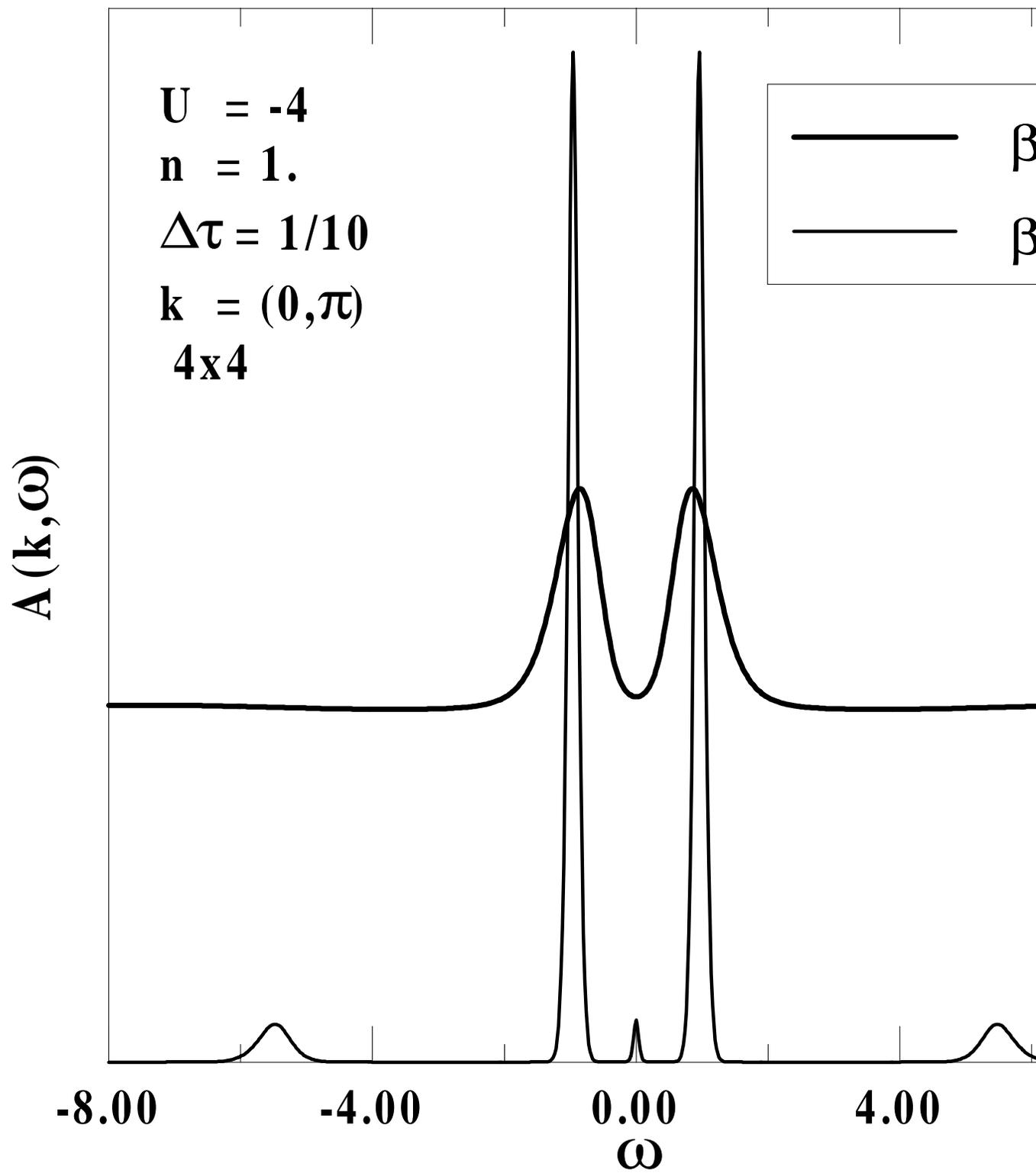
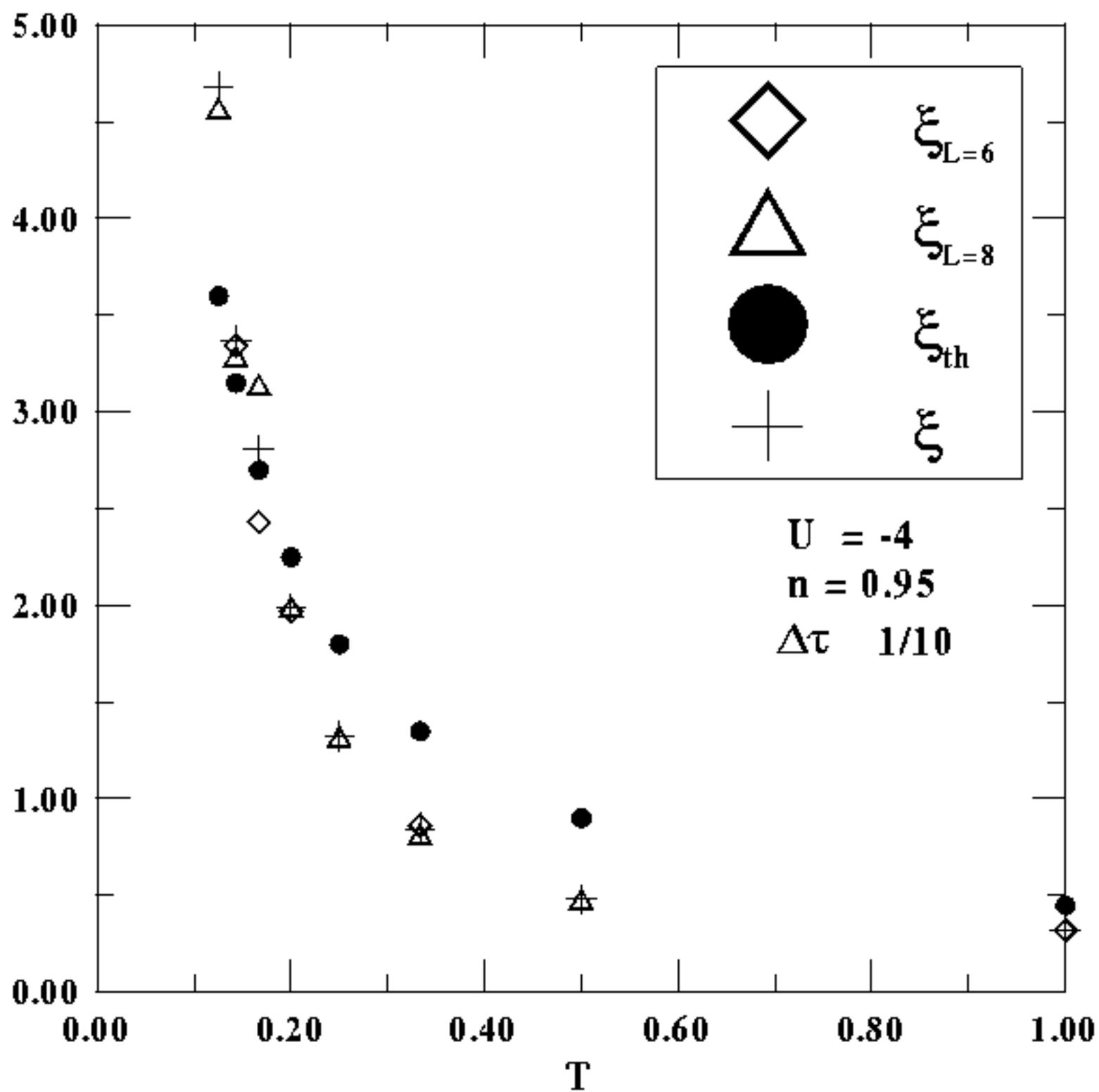
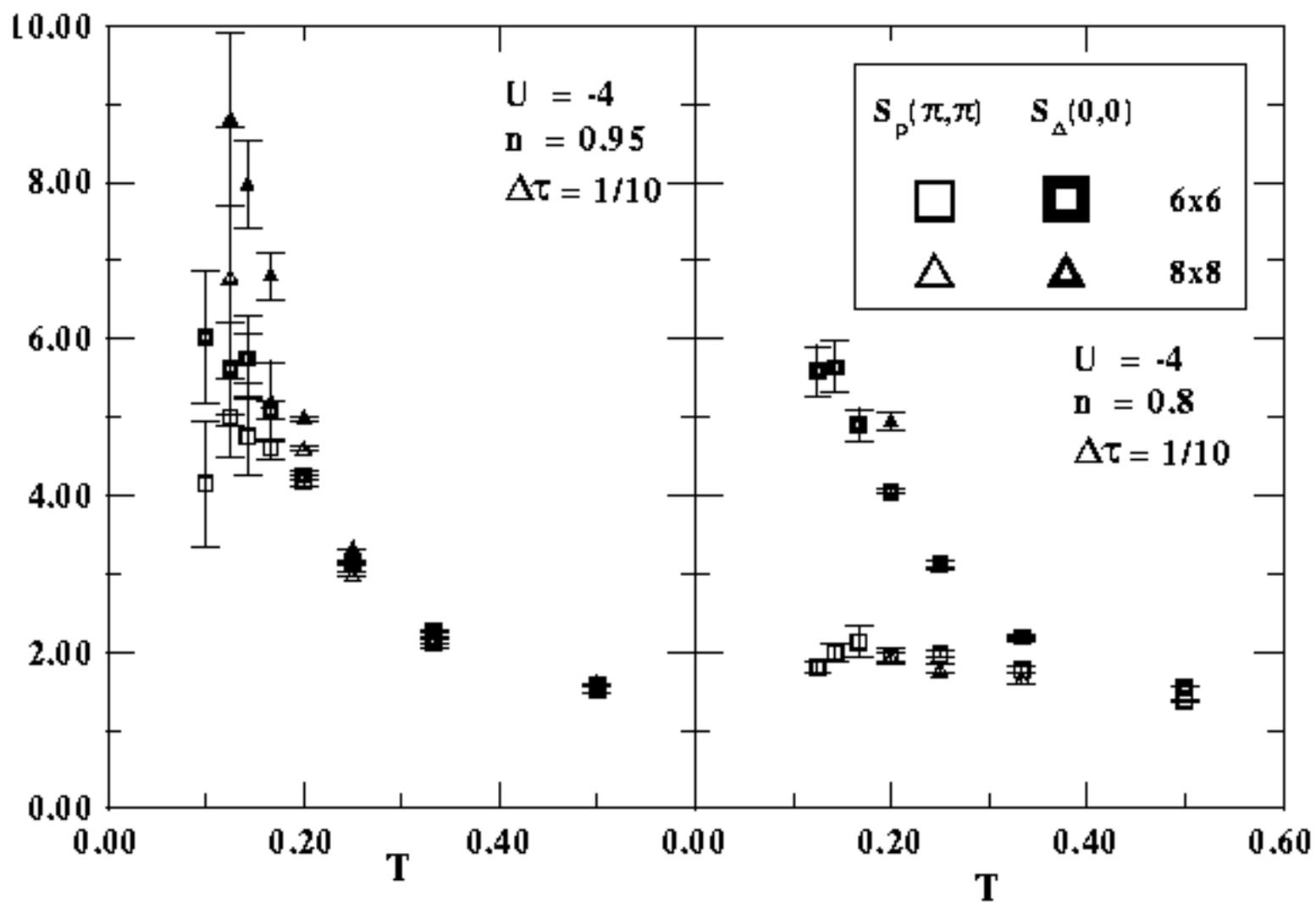
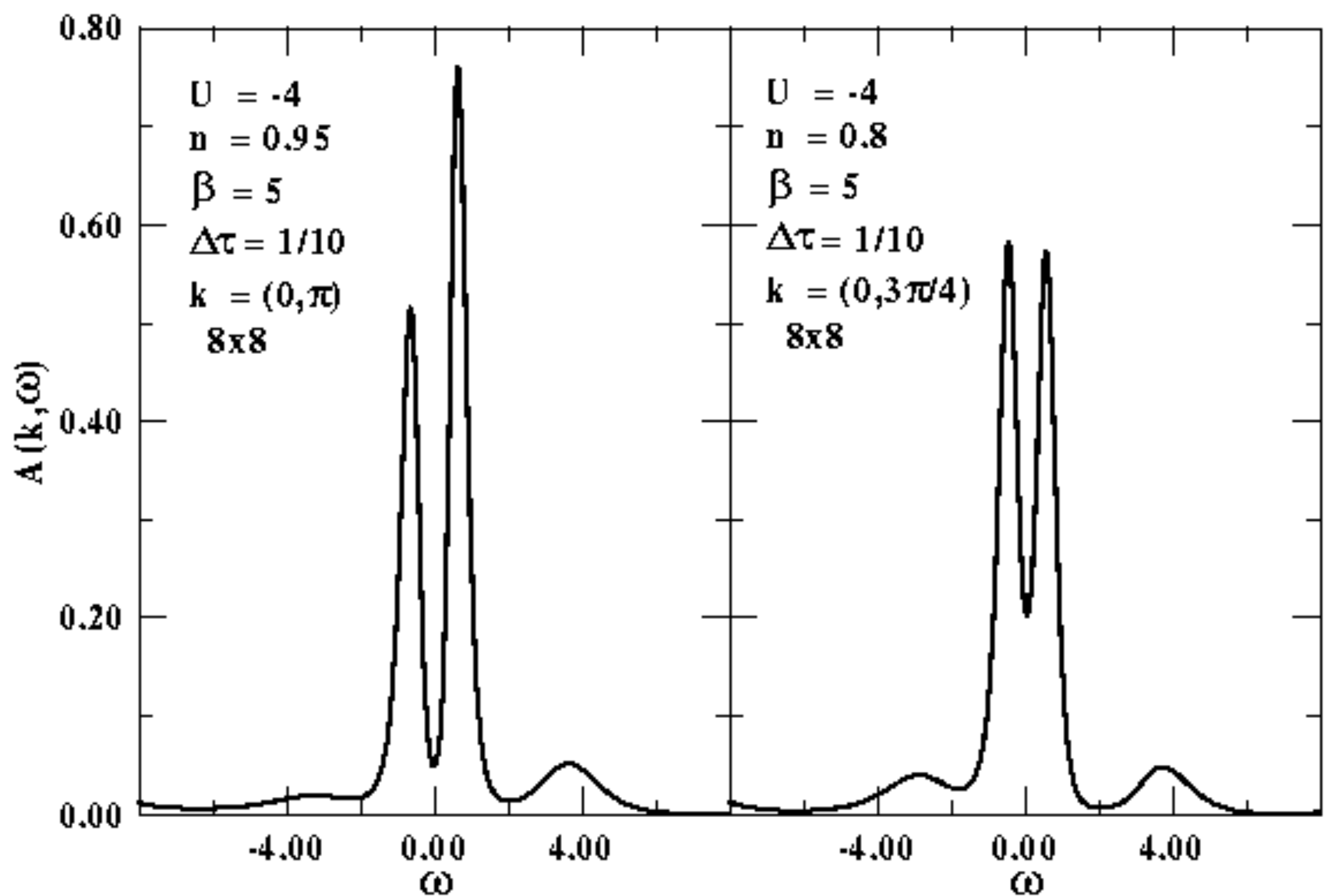
\bye